\documentclass[letter,11pt]{article}
\usepackage{heppub}
\pdfoutput=1
\usepackage{placeins}
\usepackage{braket}
\usepackage{slashed}
\usepackage{xcolor}
\usepackage[normalem]{ulem}
\usepackage{multirow}
\usepackage{xspace}
\usepackage{amsmath}
\usepackage{mathtools}

\usepackage{tikz}
\usepackage[mathscr]{eucal}  
\usepackage[compat=1.0.0]{tikz-feynman}

\newcommand{\df}{\mathrm{d}}
\newcommand{\img}{\mathrm{i}}
\newcommand{\eps}{\epsilon}

\newcommand{\w}{\omega}

\newcommand{\nn}{\nonumber}
\newcommand{\bn}{{\bar n}}
\newcommand{\bq}{{\bar q}}

\newcommand{\bnP}{\overline {\mathcal P}}

\newcommand{\bt}{{\vec b}_T}
\newcommand{\qt}{{\vec q}_T}
\newcommand{\pt}{{\vec p}_T}
\newcommand{\kt}{{\vec k}_T}

\newcommand{\cM}{\mathcal{M}}
\newcommand{\cO}{\mathcal{O}}
\newcommand{\cP}{\mathcal{P}}

\newcommand{\cU}{\mathcal{U}}

\newcommand{\as}{\alpha_s}

\newcommand{\MSbar}{$\overline{\text{MS}}$\xspace}

\newcommand{\sdot}{\!\cdot\!}
\def\Ord{{\cal O}}

\newcommand{\Tr}{\mathrm{Tr}}
\newcommand{\tr}{\operatorname{tr}}

\newcommand{\lp}{ {\tilde{p}} }         

\newcommand{\ldel}{\tilde \delta} 

\newcommand{\na}{n}
\newcommand{\nb}{\bar{n}}

\newcommand{\pa}{p_\ell}         
\newcommand{\pb}{p_{\ell'}}      
\newcommand{\eb}{E_{\ell'}}      
\newcommand{\Pa}{P_N}            
\newcommand{\Pb}{P_h}            
\newcommand{\Eb}{E_h}            
\newcommand{\PTb}{\vec P_{hT}}   
\newcommand{\PX}{P_X}            

\newcommand{\Ma}{M_N}
\newcommand{\Mb}{M_h}

\newcommand{\hard}{\mathrm{hard}}
\newcommand{\phiH}{\phi_h}       
\newcommand{\phiS}{\phi_S}        
\newcommand{\mae}[3]{\langle#1\rvert#2\rvert#3\rangle}
\newcommand{\Mae}[3]{\bigl\langle#1\bigr\rvert#2\bigr\rvert#3\bigr\rangle}
\newcommand{\MAe}[3]{\Bigl\langle#1\Bigr\rvert#2\Bigr\rvert#3\Bigr\rangle}

\DeclareMathOperator*{\SumInt}{%
\mathchoice%
  {\ooalign{$\displaystyle\sum$\cr\hidewidth$\displaystyle\int$\hidewidth\cr}}
  {\ooalign{\raisebox{.14\height}{\scalebox{.7}{$\textstyle\sum$}}\cr\hidewidth$\textstyle\int$\hidewidth\cr}}
  {\ooalign{\raisebox{.2\height}{\scalebox{.6}{$\scriptstyle\sum$}}\cr$\scriptstyle\int$\cr}}
  {\ooalign{\raisebox{.2\height}{\scalebox{.6}{$\scriptstyle\sum$}}\cr$\scriptstyle\int$\cr}}
}

\makeatletter
\def\munderbar#1{\underline{\sbox\tw@{$#1$}\dp\thr@@\z@\box\tw@}}
\makeatother

\newcommand{\SCETi}{\mbox{${\rm SCET}_{\rm I}$}\xspace}
\newcommand{\SCETii}{\mbox{${\rm SCET}_{\rm II}$}\xspace}

\def\cB{\mathcal{B}}
\def\cC{\mathcal{C}}

\def\cF{\mathcal{F}}
\def\cG{\mathcal{G}}
\def\cH{\mathcal{H}}

\def\cL{\mathcal{L}}
\def\cM{\mathcal{M}}

\def\cO{\mathcal{O}}

\def\cP{\mathcal{P}}

\def\cW{\mathcal{W}}

\newcommand{\Sl}[1]{\slashed{#1}}

\newcommand{\sqbra}[1]{[#1|}
\newcommand{\sqket}[1]{|#1]}
\newcommand{\sqbraket}[1]{[#1]}

\makeatletter
\def\munderbarf#1{\underline{\sbox\tw@{}\dp\tw@2pt \wd\tw@4pt\box\tw@}\mkern-7mu #1}
\makeatother

\makeatletter
\def\munderbarh#1{\underline{\sbox\tw@{}\dp\tw@0pt \wd\tw@4pt\box\tw@}\mkern-8mu #1}
\makeatother

\makeatletter
\def\munderbarD#1{\underline{\sbox\tw@{~}\dp\tw@0pt \wd\tw@6pt\box\tw@}\mkern-11mu #1}
\makeatother

\def\l{\langle}
\def\r{\rangle}

\newcommand\newhalfstaple[2][1]
{
 \begin{tikzpicture}[scale=#1]
  \draw [line width=0.25mm] (0,0) -- (0.2,0) -- (0.2,#2*0.2);
 \end{tikzpicture}
}

\newcommand\halfstaple[1][1] { \newhalfstaple[#1]{-1} }     
\newcommand\halfstapleu[1][1]{ \newhalfstaple[#1]{+1} }     

\newcommand\newsoftstaple[2][1]
{
 \begin{tikzpicture}[scale=#1]
  \coordinate (dx) at (#2*0.15, 0);
  \coordinate (dy) at (0   , 0.1);
  \coordinate (bT) at (0.  , 0.05);
  \draw [line width=0.15mm] (0,0) -- ($(dx) + (dy)$) -- ($2*(dy)$) -- ($2*(dy)+(bT)$) -- ($(dx) + (dy) + (bT)$) -- ($(bT)$) -- (0,0);
  \draw [line width=0.12mm] (0,0) -- (0, -0.0055);
 \end{tikzpicture}
}

\newcommand\leftsoftstaple[1][1] { \newsoftstaple[#1]{-1} }     

\usepackage{marginnote}

\newcommand{\colkin}[1]{{\color{blue} #1}}
\newcommand{\colbn}[1]{{\color{purple} #1}}
\newcommand{\colp}[1]{{\color{teal} #1}}

\allowdisplaybreaks[3]

\title{\boldmath Factorization for Azimuthal Asymmetries in SIDIS at Next-to-Leading Power}

\author[a,b]{Markus A.~Ebert,}
\emailAdd{ebert@mpp.mpg.de}

\author[b]{Anjie Gao,}
\emailAdd{anjiegao@mit.edu}

\author[b]{and Iain W.~Stewart}
\emailAdd{iains@mit.edu}

\affiliation[a]{Max-Planck-Institut f\"ur Physik, F\"ohringer Ring 6, 80805 M\"unchen, Germany}
\affiliation[b]{Center for Theoretical Physics, Massachusetts Institute of Technology, Cambridge, MA 02139, USA}

\abstract{%
Differential measurements of the semi-inclusive deep inelastic scattering (SIDIS) process with polarized beams provide important information on the three-dimensional structure of hadrons. Among the various observables are azimuthal asymmetries that start at subleading power, and which give access to novel transverse momentum dependent distributions (TMDs). Theoretical predictions for these distributions are currently based on the parton model rather than a rigorous factorization based analysis. Working under the assumption that leading power Glauber interactions do not spoil factorization at this order, we use the Soft Collinear Effective Theory to derive a complete factorization formula for power suppressed hard scattering effects in SIDIS. This yields generalized definitions of the TMDs that depend on two longitudinal momentum fractions (one of them only relevant beyond tree level), and a complete proof that only the same leading power soft function appears and can be absorbed into the TMD distributions at this order. We also show that perturbative corrections can be accounted for with only one new hard coefficient. Factorization formulae are given for all spin dependent structure functions which start at next-to-leading power. Prospects for improved subleading power predictions that include resummation are discussed.
}

\preprint{\vbox{%
\hbox{MPP--2021--203}
\hbox{MIT--CTP--5373}
}
}

\begin{document}

\maketitle

\section{Introduction}

Deep-inelastic scattering (DIS), where one scatters a lepton off a nucleon,
is a key process for measuring the internal structure of hadrons at collider processes,
namely the parton distribution functions (PDFs) which encode the longitudinal momentum distribution
of quarks and gluons inside hadrons.
In semi-inclusive DIS (SIDIS), one detects a hadron in addition to the scattered lepton,
which similarly gives access to the longitudinal momentum dependence of the fragmentation process
through a fragmentation function (FF).
In the kinematic regime where the transverse momentum $\PTb$ of the detected hadron
is much smaller than the momentum transfer $Q$, one also gains access
to the transverse-momentum distributions inside the colliding and fragmenting hadron 
through transverse-momentum dependent (TMD) PDFs and FFs, respectively.
This makes SIDIS of prime interest to investigate the 3D structure of nuclei,
and thus has been studied extensively at experiments such as COMPASS~\cite{Gautheron:2010wva},
RHIC~\cite{Aschenauer:2015eha} and JLab~\cite{Dudek:2012vr},
see e.g.~\refscite{Avakian:2016rst,Gao:2018grv,Eyser:2019qmp} for experimental reviews.
It is also a key scientific goal of the upcoming EIC~\cite{Accardi:2012qut, AbdulKhalek:2021gbh},
and direct calculations of these functions from lattice have recently attracted much attention, see e.g.~\refcite{Constantinou:2020hdm}.
 
As first pointed out by Cahn, the intrinsic transverse motion of partons inside hadrons gives rise to a nontrivial dependence
on the azimuthal angle $\phiH$ of the scattered hadron~\cite{Cahn:1978se, Cahn:1989yf}.
In polarized SIDIS, additional correlations arise due to the polarization of the incoming lepton and hadron,
and the complete set of independent angular structure functions in SIDIS
were derived a long time ago~\cite{Gourdin:1973qx, Kotzinian:1994dv, Diehl:2005pc}.
For example, in the single-photon exchange approximation,
the SIDIS cross section can be decomposed at small $P_{hT} \ll Q$ as
\begin{align} \label{eq:xs_SIDIS_schematic}
 \df\sigma &\propto
 \big( W_{-1}^U \!+\! \eps W_0^U \bigr)
 + \sqrt{2\eps(1\!+\!\eps)} \cos\phiH W_1^U
 + \lambda_\ell \sqrt{2\eps(1\!-\!\eps)} \sin\phiH W_2^U
 + \eps \cos(2\phiH) W_3^U
 \nn\\
 &\quad + \text{(hadron spin polarization terms)}
\,,\end{align}
where $\eps$ is the ratio of longitudinal to transverse photon flux,
and $\lambda_\ell$ is the polarization of the incoming lepton. 
The individual structure functions $W_i$ are sensitive to different correlations between the spin
of the incoming hadron and the struck quark, and likewise for the outgoing hadron. Thus, their precise determination is of key interest.
For simplicity we only list explicitly the five $W_i^U$ that appear for an unpolarized hadron in \eq{xs_SIDIS_schematic},
leaving a discussion with the full details, including the spin-polarized expression with thirteen more $W_i$s, to the main text.  We will carry out our analysis for the full set of $W_i$s in this paper.

Fully leveraging existing and upcoming measurements of these structures functions
requires a precise theoretical understanding of their relation to TMD correlation functions.
This is achieved via factorization theorems which separate the structure functions
into a process-dependent but calculable hard part as well as the universal but nonperturbative TMD PDFs and FFs.
For example, $W_{-1}^U$ provides access to distributions of unpolarized quarks inside unpolarized hadrons only,
while $W_3^U$ probes the transverse polarization of quarks inside unpolarized hadrons
through the famous Boer-Mulders and Collins functions~\cite{Collins:1992kk, Boer:1997nt}.
In polarized SIDIS, one becomes sensitive to many more TMD correlations
related to the quark and hadron spin~\cite{Mulders:1995dh, Boer:1997nt, Boer:1999uu, Bacchetta:2004zf, Goeke:2005hb},
see \refcite{Bacchetta:2006tn} or \sec{SIDIS} below for an overview.

Establishing these factorization theorems is critical to extracting TMDs from measurements of the structure functions.
For the closely related processes $e^+ e^- \to \gamma^* \to h_1 h_2$ and $p p \to \gamma^* \to \ell^+ \ell^-$ (Drell-Yan),
factorization of the unpolarized structure functions has been derived a long time ago
by Collins, Soper and Sterman (CSS)~\cite{Collins:1981uk, Collins:1981va, Collins:1984kg}.
For SIDIS, this was first achieved in \refscite{Ji:2004wu, Ji:2004xq}.
Modern formalisms for TMD factorization were put forward by Collins~\cite{Collins:2011zzd}
and independently by various groups~\cite{Becher:2010tm, GarciaEchevarria:2011rb, Chiu:2012ir, Li:2016axz}
using soft-collinear effective theory (SCET)~\cite{Bauer:2000ew, Bauer:2000yr, Bauer:2001ct, Bauer:2001yt, Bauer:2002nz}.
Based on these works, TMD factorization of unpolarized structure functions has reached
three-loop accuracy in perturbative QCD~\cite{Li:2016ctv, Luo:2019szz, Luo:2020epw, Ebert:2020qef, Ebert:2020yqt}.

These factorization theorems have only been derived rigorously for the simplest structure functions,
namely those that contribute at \emph{leading power} (LP),%
\footnote{In the literature, one often refers to these as leading-twist structure functions instead of as leading power.
Since twist is also used to classify PDFs and FFs in the region where their dependence on transverse momentum can be treated perturbatively, to avoid any confusion we reserve the notion of twist for this latter case only.
For example, the $W_3^U$ structure function in \eq{xs_SIDIS_schematic} contributes at leading power,
but is given in terms of the Boer-Mulders and Collins functions which themselves are determined by subleading-twist correlation functions for perturbative transverse momentum.}
i.e.~those that scale as $1/P_{hT}^2$ with the hadron transverse momentum $P_{hT}$.
In \eq{xs_SIDIS_schematic}, only $W_{-1}^U$ and $W_3^U$ contribute at leading power, though there are other spin-dependent leading power terms that we discuss later on in the body of the paper.
The structure functions $W_1^U$ and $W_2^U$ contribute at \emph{next-to-leading power} (NLP),
i.e.~they are suppressed as $P_{hT}/Q$ with respect to LP, or equivalently they scale as $1/(P_{hT} Q)$, while the \emph{next-to-next-to-leading power} (NNLP) structure function $W_0^U$ scales as $1/Q^2$.
(Again there are other spin-dependent structure functions that start at NLP that we discuss in the body of the paper.)
Factorization theorems for TMD observables at subleading power have not yet been established.
Broadly speaking, factorization at subleading power is significantly more involved than at LP
since many different mechanisms contribute to the power suppression of subleading-power structure functions,
and only few direct calculations at this order have been carried out so
far~\cite{Balitsky:2017gis, Ebert:2018gsn, Moult:2019vou, Ebert:2020dfc, Moos:2020wvd}.
Nevertheless, these functions have already been studied in great detail in the literature,
building in particular on the tree-level analysis in \refcite{Mulders:1995dh}.
In addition, much insight has been gained concerning the structure of Wilson lines and quark-(gluon-)quark
correlators appearing in the (conjectured) factorized expressions,
see e.g.~\refscite{Collins:2002kn, Bomhof:2004aw, Collins:2004nx, Bacchetta:2005rm}
and \refscite{Boer:2003cm, Bacchetta:2004zf,Goeke:2005hb}, respectively.
These developments were summarized in \refcite{Bacchetta:2006tn}, which still serves
as a useful reference for SIDIS structure functions.
Based on these results, \refcite{Bacchetta:2008xw} proposed a factorization for the $\cos\phi_h$
asymmetry in SIDIS, i.e.~$W_1^U$ in \eq{xs_SIDIS_schematic}, but found that it was incompatible
with collinear factorization that holds at large $P_{hT} \sim Q$.
More recently, \refcite{Bacchetta:2019qkv} proposed to resolve the observed discrepancy
by including the same soft function that appears at leading power in TMD factorization
and captures the effect of soft radiation. However, no proof was given, and the proposal
was validated only at first order in perturbation theory in the limit where collinear factorization can be applied.

In this paper, we initiate a systematic study of TMD factorization at subleading power,
with the aim of deriving factorization theorems for all structure functions in SIDIS and Drell-Yan at NLP.
We employ the Soft Collinear Effective Theory (SCET)~\cite{Bauer:2000ew,Bauer:2000yr,Bauer:2001ct,Bauer:2001yt,Bauer:2002nz}, an effective-field theory obtained by expanding QCD about the soft and collinear limit at the Lagrangian level,
to organize our calculation. 
SCET has already been used to study subleading-power factorization in other processes,
allowing us to leverage various results in the literature~\cite{Manohar:2002fd,Beneke:2002ph,Pirjol:2002km,Beneke:2002ni,Bauer:2003mga,Hill:2004if,Lee:2004ja,Paz:2009ut,
Benzke:2010js,Freedman:2013vya,Freedman:2014uta,Larkoski:2014bxa,Moult:2016fqy,Moult:2017jsg,Goerke:2017lei,Balitsky:2017gis,Beneke:2017vpq,Beneke:2017ztn,Feige:2017zci,Moult:2017rpl,
Chang:2017atu,Beneke:2018gvs,Beneke:2018rbh,Moult:2018jjd,Ebert:2018lzn,Ebert:2018gsn,Bhattacharya:2018vph,Beneke:2019kgv,Moult:2019mog,Beneke:2019slt},  
see also~\cite{Bonocore:2014wua,Bonocore:2015esa,Bonocore:2016awd,Boughezal:2016zws,DelDuca:2017twk,Balitsky:2017flc,Boughezal:2018mvf,vanBeekveld:2019prq,vanBeekveld:2019cks,Bahjat-Abbas:2019fqa,Boughezal:2019ggi}.
Recently, there have been first attempts to study TMD factorization
at subleading power, including analysis at small-$x$~\cite{Balitsky:2020jzt,Balitsky:2021fer}
and investigations of the factorization  structure with SCET and similar techniques~\cite{Inglis-Whalen:2021bea, Vladimirov:2021hdn}.

In particular, within SCET one can immediately identify the minimal set of building blocks required at subleading power,
and it naturally classifies power corrections into different categories.
The SCET Lagrangian can be decomposed as 
\begin{align} \label{eq:L_SCET_schematic}
 \cL_{\rm SCET}
 = \cL_{\rm hard} + \cL_{\rm dyn}
 = \sum_{i\ge0} \cL^{(i)}_{\rm hard}
   + \Big( \sum_{i\ge0} \cL_{\rm dyn}^{(i)}
   + \cL_{\rm G}^{(0)} \Big) 
\,.\end{align} 
Here, $\cL_{\rm hard}$ contains hard scattering operators that mediate the underlying hard interaction,
whereas the dynamics of collinear and soft fields are encoded in $\cL_{\rm dyn}$.
Both ingredients can be expanded in a power counting parameter $\lambda \sim P_{hT} / Q$,
with ${(i)}$ denoting terms contributing at N$^i$LP, i.e.~$\cO(\lambda^i)$.
At leading power, soft and collinear fields are manifestly decoupled in $\cL^{(0)}$,
and can only interact via the LP $\cL_G^{(0)}$ Glauber Lagrangian~\cite{Rothstein:2016bsq}.
Thus, as long as Glauber contributions cancel, factorization at LP is already made manifest at the Lagrangian level.
In the case of TMDs, the cancellation of contributions from the Glauber region was shown in \refscite{Collins:1984kg,Collins:1988ig}, thus completing the proof of TMD factorization at leading power. See \refcite{Collins:2011zzd} for a review of leading power factorization in TMD physics.  

At subleading power in the \SCETii theory that is relevant for TMD physics, one has to consider power corrections from several sources:
\begin{enumerate}
 \item Kinematic power corrections, for example from the observable itself.
 \item Subleading-power hard scattering operators, ${\cal L}_{\rm hard}^{(i)}$ which are generated from the hard region of momentum space with $i\ge 1$ and the hard-collinear region of momentum space with $i\ge 1/2$. 
 \item Subleading-power Lagrangian contributions, ${\cal L}_{\rm dyn}^{(i)}$ with $i\ge 1/2$. 
\end{enumerate}
Further details about SCET at subleading power can be found in \sec{SCET_review} where this list is repeated with additional details.
In this work, we will carry out a subleading power factorization for SIDIS for all of these contributions.
Another key element in the proof of factorization is the behavior of the leading power Glauber Lagrangian, ${\cal L}_G^{(0)}$, and its interaction with various subleading power corrections. In the context of SCET it is clear that the {\it only source} of factorization violating contributions are those induced by ${\cal L}_G^{(0)}$, at both leading and subleading power~\cite{Rothstein:2016bsq}. In particular, while there are power corrections from the Glauber region of momentum space that contribute to ${\cal L}^{(i\ge 1/2)}$, in the absence of effects from ${\cal L}_G^{(0)}$ these terms can be factorized just like other ${\cal L}^{(i\ge 1/2)}$ interactions. This occurs because the power suppression of these Lagrangians ensures that they only enter a fixed number of times, and hence fields that are collinear and soft can always be factorized from each other in independent matrix elements. In general the factorization with subleading power Lagrangians leads to subleading power collinear and soft functions involving time-ordered products of operators, see eg.~\refscite{Mantry:2003uz,Lee:2004ja,Moult:2019mog}.
 
At LP it is known from the proof of factorization by CSS that the effects of ${\cal L}_G^{(0)}$ either cancel out or (in the language of SCET~\cite{Rothstein:2016bsq}) can be absorbed by fixing a proper direction for the Wilson lines that appear at leading power.  
For the purpose of the analysis in this paper, 
we will work under the assumption that contributions from ${\cal L}_G^{(0)}$ cancel similar to LP, and then derive the form that the all orders factorization theorem must take when accounting for contributions 1, 2 and 3 listed above. This provides the target form that any complete proof of factorization would need to find, and allows us to handle hard-collinear and soft-collinear factorization effects at NLP, but it does not suffice to fully prove factorization at subleading power.  A complete analysis of the Glauber region to demonstrate the lack of non-trivial effects from ${\cal L}_G^{(0)}$ will thus be left as the missing ingredient needed to prove factorization for SIDIS at next-to-leading power. 

This paper is structured as follows.
In \sec{SIDIS}, we briefly review the kinematics and tensor decomposition for both unpolarized and polarized SIDIS.
In \sec{SCET_review}, we review the necessary ingredients of the SCET formalism at subleading power.
The factorization of the hadronic tensor for SIDIS through next-to-leading power is derived in \sec{factorization}.
Factorization formula are then given for all the individual structure functions  in \sec{results}, including a comparison to results from prior literature.
We conclude in \sec{conclusion}. In the appendices we provide additional details for various aspects of our analysis.

\section{Semi-Inclusive Deep Inelastic Scattering}
\label{sec:SIDIS}

\subsection{Kinematics}
\label{sec:SIDIS_kinematics}

In this section, we briefly review the kinematics relevant in SIDIS and set up our notation.
We consider the process
\begin{align} \label{eq:SIDIS}
 \ell(\pa,\lambda_\ell) + N(\Pa,S) \to \ell'(\pb,\lambda_{\ell'}) + h(\Pb) + X(\PX)
\,,\end{align}
where $\ell$ is the incoming lepton scattering off the nucleus $N$, $\ell'$ is the detected final-state lepton,
$h$ is the tagged final-state hadron, and $X$ denotes additional radiation that we are inclusive in.
The momentum of particle $i$ is denoted as $p_i$, $\lambda_\ell$ and $\lambda_{\ell'}$ are the lepton helicities,
and $S$ is the spin vector of the target hadron. The outgoing hadrons are always considered to be unpolarized.
For simplicity, we will always assume vanishing lepton masses, $\pa^2 = \pb^2 = 0$,
but for now consider a nonvanishing target mass, $M_N^2 = \Pa^2 > 0$.

We work in the one-photon exchange approximation at leading order in the electroweak interactions.
The process in \eq{SIDIS} is then mediated by a virtual photon of momentum
\begin{align} \label{eq:q}
 q = \pa - \pb \,,\quad Q^2 = -q^2 > 0
\,.\end{align}
In this approximation, $\ell$ and $\ell'$ are of the same flavor, and the matrix element can be factorized as
\begin{align} \label{eq:M_SIDIS}
 \cM(\ell N \to \ell' h X) &
 = \braket{\ell' | J^\mu | \ell} \frac{e^2}{q^2} \braket{h,X | J_{\mu} | N}
\,.\end{align}
Here $1/q^2$ is the photon propagator and the electromagnetic current is 
\begin{align} \label{eq:current}
 J^\mu = \sum_f J_{\bar ff}^\mu
\,,\qquad
 J_{\bar ff}^\mu = Q_f \, \bar q_f \gamma^\mu q_f
\,, \end{align}
where $Q_f$ is the electromagnetic coupling normalized to the unit charge $|e|$,
and the sum runs over all quark and lepton flavors $f$.
One can easily extend the analysis to also include $Z$ or $W$ exchange
by extending \eq{M_SIDIS} to sum over all relevant electroweak currents.

By squaring \eq{M_SIDIS} and integrating over the final-state phase space,
one obtains the fully-differential cross section as
\begin{align} \label{eq:sigma_diff}
 \df\sigma &
 = \frac{\alpha^2}{(\pa\sdot\Pa) Q^4} \frac{\df^3 \pb}{2 \eb} \frac{\df^3 \Pb}{2 \Eb}
   L_{\mu\nu}(\pa, \pb) W^{\mu\nu}(q, \Pa, \Pb)
\,,\end{align}
where the hadronic and leptonic tensors are defined as
\begin{align} \label{eq:tensors}
 W^{\mu\nu}(q, \Pa, \Pb) &
 = \sum_X  \frac{\df^3 p_X}{(2\pi)^3 2 E_X} \delta^4(q + \Pa - \Pb - \PX)
   \braket{N | J^{\dagger \, \mu} | h,X} \braket{h,X | J^\nu | N}
\,,\nn\\
 L^{\mu\nu}(\pa,\pb) &
 = \braket{\ell | J^{\dagger \mu} | \ell'} \braket{\ell' | J^\nu | \ell}
 \nn\\&
 =  2 \delta_{\lambda_\ell \lambda_{\ell'}}
   \bigl[ \bigl( \pa^\mu \pb^\nu + \pa^\nu \pb^\mu  - \pa\sdot\pb \, g^{\mu\nu}\bigr)
         + \img \lambda_\ell \eps^{\mu\nu \rho\sigma} p_{\ell\,\rho} p_{\ell'\,\sigma} \bigr]
\,.\end{align}
They are normalized such that they agree with standard expressions in the literature,
where the antisymmetric tensor is chosen such that $\eps^{0123} = +1$.  Here $\lambda_\ell=\pm 1$ is twice the helicity of the incoming lepton.
In the following, we will suppress the leptonic helicity-conserving Kronecker $\delta_{\lambda_\ell \lambda_{\ell'}}$.

\begin{figure*}
 \centering
 \includegraphics[width=0.8\textwidth]{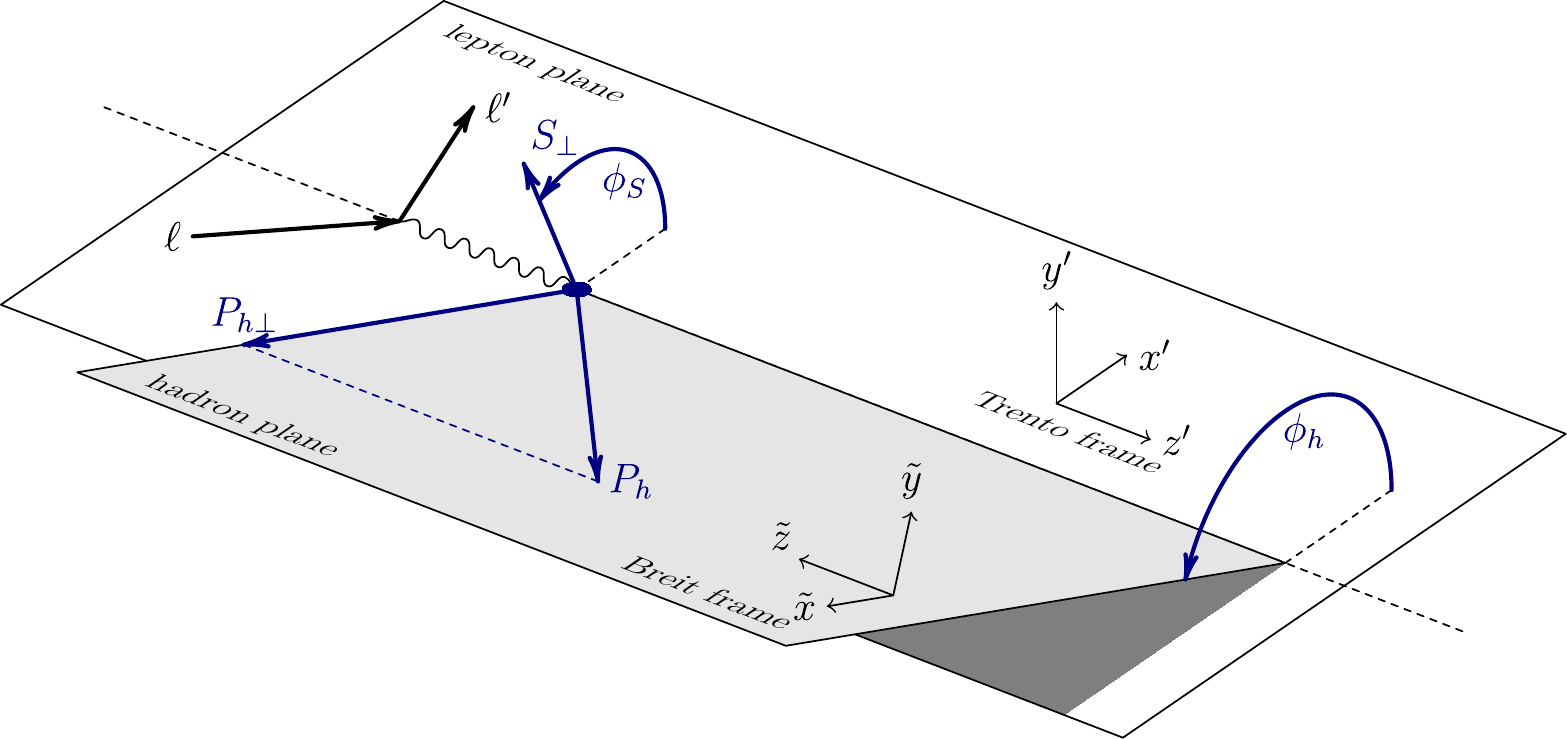}
 \caption{Illustration of the SIDIS kinematics in the target rest frame according to the Trento convention~\cite{Bacchetta:2004jz},
 specified by the unit vectors $x',y',z'$. The momentum transfer $\vec q$ is aligned along the $z$ axis,
 and the hadron transverse momentum $\PTb$ is defined relative to the lepton plane.
 We also show the hadronic Breit frame $\tilde B$ employed in our tensor decomposition, specified by the unit vectors $\tilde x, \tilde y, \tilde z$.
 Its $\tilde z$-axis is opposite to that of the Trento conventions, and the $\tilde x-\tilde z$ plane coincides with the hadron plane rather than the lepton plane.
 }
 \label{fig:trento}
\end{figure*}

As is evident from \eq{sigma_diff}, SIDIS is described by six kinematic variables,
which are typically expressed through the Lorentz invariants
\begin{align} \label{eq:observables}
 x = \frac{Q^2}{2 \Pa \cdot q}
 \,, \quad
 y = \frac{\Pa \cdot q}{\Pa \cdot \pa}
 \,,\quad
 z = \frac{\Pa \cdot \Pb}{\Pa \cdot q}
\,,\end{align}
the transverse momentum $\PTb$ of the final-state hadron $h$ in a suitable reference frame (for which we follow the Trento conventions~\cite{Bacchetta:2004jz} illustrated in \fig{trento}),
and an overall azimuthal angle $\psi$, which we take to be the azimuthal angle between the
outgoing and incoming lepton. Alternatively, for polarized processes one can choose it to
be the angle between the lepton plane and the hadron spin, see \refcite{Diehl:2005pc} for more details.
One obtains%
\footnote{When inverting $\df^3\Pb$ for $\df^2 \PTb \df z$, one obtains a second solution that does not contribute at small $\PTb$,
but rather to the target fragmentation region where $h$ is collinear to $N$, see e.g.~\refcite{Boglione:2016bph}.}
\begin{align} \label{eq:xs_SIDIS}
 \frac{\df\sigma}{\df x \, \df y \, \df z \, \df\psi \, \df^2\PTb} &
 = \kappa_\gamma \frac{\alpha^2}{4 Q^4} \frac{y}{z} L_{\mu\nu}(\pa, \pb) W^{\mu\nu}(q, \Pa, \Pb)
\,.\end{align}
The factor $\kappa_\gamma$ encodes target-masses corrections, and is given by
\begin{align} \label{eq:kappa_gamma}
\kappa_\gamma = \biggl[1 - \Bigl( \frac{m_{Th} \gamma}{z Q}\Bigr)^2 \biggr]^{-1/2}
\,,\quad m_{Th}^2 = M_h^2 + P_{hT}^2
\,,\quad \gamma = \frac{2 x M_N}{Q}
\,.\end{align}
It is often neglected in the literature, but kept for example in \refcite{Kotzinian:1994dv}.
We will mostly consider the massless limit, where $\gamma = 0$ and $\kappa_\gamma =  1$.

\subsection{Tensor Decomposition}
\label{sec:tensor_decomposition}

In this section, we decompose the hadronic tensor into independent structure functions.
We first discuss the general setup for the tensor decomposition in the single vector boson exchange approximation with massless leptons in \sec{tensor_decomposition_setup}, and then
consider in detail the one-photon exchange approximation for the case of fully unpolarized and fully polarized SIDIS in the remaining sections.

\subsubsection{Setup}
\label{sec:tensor_decomposition_setup}

It follows from current conservation that%
\begin{align} \label{eq:W_conservation}
 q_\mu W^{\mu\nu}(q, \Pa, \Pb) = q_\nu W^{\mu\nu}(q, \Pa, \Pb) = 0
\,,\end{align}
and hence $W^{\mu\nu}$ can be decomposed into nine independent structure functions.
To identify these, we follow the strategy of \refscite{Gourdin:1973qx, Kotzinian:1994dv}
and project the hadronic tensor onto a helicity density matrix
constructed from the polarization vectors of the exchanged vector boson.

For this purpose, it is convenient to work in the Breit frame where the momentum transfer is purely along the $z$ axis,
$q^\mu \propto \tilde n_z^\mu$,  and where $\PTb$ is aligned along the $x$ axis.
More precisely, we construct this frame from $q^\mu$ and the hadron momenta $\Pa^\mu$ and $\Pb^\mu$,
and hence refer to it as the \emph{hadronic Breit frame}.
The advantage of this frame is that the ensuing helicity decomposition of the hadronic tensor will only depend on hadronic momenta.

To uniquely construct the hadronic Breit frame, we demand that $q^\mu = (0,0,0,-Q)$ and take $\PTb = (P_{hT},0)$ in the $\tilde n_x$ direction,
and then construct $\tilde n_t$ and $\tilde n_y$ by completing a right-handed coordinate system.
Compared to the Trento conventions, this corresponds to reversing the direction of the $z$ axis
and rotating about the $z$ axis by $\phiH$, the azimuthal angle of the hadron in the Trento conventions.
To illustrate this, we show both frames in \fig{trento}, and provide explicit parameterizations of all particle momenta
in both frames in \app{frames}.
Our main motivation to reverse the orientation of the Breit frame relative to the Trento conventions
is to coincide with the orientation of the rest frame $C''$ of \refcite{Diehl:2005pc},
which will allow us to make immediate contact with their results.

Following the conventions of \refcite{Diehl:2005pc}, we define the basis vectors as
\begin{align} \label{eq:pol_vectors}
 \eps_0^\mu = \tilde n_t^\mu
\,, \qquad
 \eps_\pm^\mu = \frac{1}{\sqrt 2} (\mp\tilde n_x^\mu + \img \tilde n_y^\mu)
\,,\end{align}
where $\eps_0^\mu$ ($\eps_\pm^\mu$) encode longitudinal (transverse) polarizations of the exchanged vector boson,
and $\tilde n_{t,x,y}$ are the unit vectors of the Breit frame.
Due to \eq{W_conservation}, we do not need to consider the fourth independent vector $\tilde n_z^\mu \propto q^\mu$.
We then define projections of the hadronic tensor onto the helicity density matrix as~\cite{Diehl:2005pc}
\begin{align} \label{eq:W_proj}
 W_{\lambda \lambda'}(q, \Pa, \Pb) &
 \equiv \eps^{*\mu}_\lambda \eps^{\nu}_{\lambda'} \, W_{\mu\nu}(q, \Pa, \Pb)
\,,\qquad \lambda \in \{+,-,0\}
\,.\end{align}
Note that this choice of projector coincides with that obtained by contracting $W_{\mu\nu}$ with the tensor $L^{\mu\nu}=\eps^{*\,\mu} \eps^\nu$ relevant for $\gamma^* N$ scattering.

While \eq{W_proj} already yields nine manifestly independent structure functions,
it will be convenient to construct linear combinations of these such
that they will be in one-to-one correspondence to independent angular coefficients.
We thus define the structure functions
\begin{align} \label{eq:def_W_i}
 W_i(q, \Pa, \Pb) = P_i^{\mu\nu}(q, \Pa, \Pb) W_{\mu\nu}(q, \Pa, \Pb)
\,,\end{align}
in terms of the projectors
\begin{alignat}{2} \label{eq:def_P_i_1}
 &P_{-1}^{\mu\nu}= \frac{1}{2} \bigl(\eps_-^{*\mu} \eps_-^\nu + \eps_+^{*\mu} \eps_+^\nu \bigr)
\,,\quad
 && P_0^{\mu\nu} = \eps_0^{*\mu} \eps_0^\nu
\,,\nn\\
 &  P_1^{\mu\nu} = \frac{1}{\sqrt8} \bigl( \eps_0^{*\mu}\eps_-^\nu - \eps_0^{*\mu} \eps_+^\nu + \eps_-^{*\mu} \eps_0^\nu - \eps_+^{*\mu} \eps_0^\nu \bigr)
\,,\quad
 && P_2^{\mu\nu} = \frac{\img}{\sqrt8} \bigl(\eps_0^{*\mu}\eps_-^\nu -\eps_0^{*\mu} \eps_+^\nu - \eps_-^{*\mu}\eps_0^\nu + \eps_+^{*\mu}\eps_0^\nu \bigr)
\,,\nn\\
 &  P_3^{\mu\nu} = -\frac{1}{2} \bigl(\eps_+^{*\mu}\eps_-^\nu + \eps_-^{*\mu}\eps_+^\nu \bigr)
\,,\quad
 && P_4^{\mu\nu} = \frac{1}{2} \bigl( \eps_+^{*\mu}\eps_+^\nu - \eps_-^{*\mu}\eps_-^\nu\bigr)
\,,\nn\\
 &P_5^{\mu\nu}   = \frac{\img}{\sqrt8} \bigl( \eps_-^{*\mu}\eps_0^\nu + \eps_+^{*\mu} \eps_0^\nu - \eps_0^{*\mu}\eps_-^\nu - \eps_0^{*\mu}\eps_+^\nu \bigr)
\,,\quad
 && P_6^{\mu\nu} = \frac{-1}{\sqrt8} \bigl( \eps_0^{*\mu}\eps_-^\nu + \eps_0^{*\mu}\eps_+^\nu + \eps_-^{*\mu}\eps_0^\nu + \eps_+^{*\mu}\eps_0^\nu \bigr)
\,,\nn\\
 & P_7^{\mu\nu}  = \frac{\img}{2} \bigl (\eps_-^\nu \eps_+^{*\mu}-\eps_+^\nu \eps_-^{*\mu}\bigr)
\,.\end{alignat}
Equivalently, we can express these projectors more compactly
in terms of the unit vectors $\tilde n_i^\mu$ of the hadronic Breit frame,
\begin{alignat}{3} \label{eq:def_P_i}
 &P_{-1}^{\mu\nu}= \frac12 (\tilde n_x^\mu \tilde n_x^\nu + \tilde n_y^\mu \tilde n_y^\nu )
\,,\quad~
 && P_0^{\mu\nu} = \tilde n_t^\mu \tilde n_t^\nu
\,,\quad~
 && P_1^{\mu\nu} = \frac12 (\tilde n_t^\mu \tilde n_x^\nu + \tilde n_x^\mu \tilde n_t^\nu)
\,,\nn\\
 &  P_2^{\mu\nu} = \frac\img2 \bigl( \tilde n_t^\mu \tilde n_x^\nu - \tilde n_x^\mu \tilde n_t^\nu\bigr)
\,,\quad~
 && P_3^{\mu\nu} = \frac12 (\tilde n_x^\mu \tilde n_x^\nu - \tilde n_y^\mu \tilde n_y^\nu)
\,,\quad~
 && P_4^{\mu\nu} = \frac{\img}{2}  (\tilde n_y^\mu \tilde n_x^\nu - \tilde n_x^\mu \tilde n_y^\nu)
\,,\nn\\
 &  P_5^{\mu\nu} = \frac12 (\tilde n_t^\mu \tilde n_y^\nu + \tilde n_y^\mu \tilde n_t^\nu)
\,,\quad~
 && P_6^{\mu\nu} = \frac{\img}{2} (\tilde n_y^\mu \tilde n_t^\nu - \tilde n_t^\mu \tilde n_y^\nu)
\,,\quad~
 && P_7^{\mu\nu} = \frac12 (\tilde n_x^\mu \tilde n_y^\nu + \tilde n_y^\mu \tilde n_x^\nu)
\,.\end{alignat}
Compared to \eq{def_P_i_1}, this form will be more convenient for calculating contractions
with the hadronic tensor.

The projections in \eq{def_W_i} can be inverted as
\begin{align} \label{eq:W_inv_proj}
 W^{\mu\nu}(q, \Pa, \Pb) &= \sum_{i=-1}^7 (P_i^{-1})^{\mu\nu}(q, \Pa, \Pb) W_i(q, \Pa, \Pb)
\,,\end{align}
up to additional terms proportional to $q^\mu$ and $q^\nu$,
which however vanish when contracted with the conserved leptonic current.
The inverse projectors $P_i^{-1}$ are identical to the projectors $P_i$
up to a trivial change in normalization, $(P_i^{-1})^{\mu\nu} = P_i^{\mu\nu} / (P_i^{\alpha\beta} P_{i\,\alpha\beta})$.
We define the \emph{leptonic structure functions} analogous to \eq{def_W_i} as
\begin{align} \label{eq:def_L_i}
 L_i(\pa,\pb,\Pa,\Pb) = (P_i^{-1})^{\mu\nu}(q, \Pa, \Pb) \, L_{\mu\nu}(\pa,\pb)
\,.\end{align}
Note that since the inverse projectors depend on the hadron momenta $\Pa$ and $\Pb$,
so do the $L_i$. Using \eq{tensors} for $L^{\mu\nu}$ and \eq{momenta_Breit_had}
for the lepton momenta, \eq{L_i_results} can be evaluated as
\begin{align} \label{eq:L_i_results}
 L_i(\pa,\pb,\Pa,\Pb) = \frac{2 Q^2}{1-\eps}\, g_i(\eps, \lambda_\ell, \phiH)
\,,\end{align}
where we pulled out the common kinematic prefactor $2 Q^2 / (1-\eps)$ with $\eps$ defined as
the usual ratio of the longitudinal to the transverse photon flux
\begin{align} \label{eq:eps}
 \eps = \frac{1 - y - \frac14 y^2 \gamma^2}{1 - y + \frac14 y^2 \gamma^2 + \frac12 y^2 }
\,.\end{align}
This makes the remaining angular functions $g_i(\eps, \lambda_\ell, \phi_h)$
only depend on $\eps$, the lepton helicity $\lambda_\ell = \lambda_{\ell'}$,
and the azimuthal angle $\phi_h$.
These coefficient functions are given by
\begin{alignat}{3} \label{eq:def_g_i}
 g_{-1}(\eps, \lambda_\ell,\phiH) &= 1
\,, \qquad &
 g_0(\eps, \lambda_\ell,\phiH) &= \eps
\,, \nn \\
 g_1(\eps, \lambda_\ell,\phiH) &= \sqrt{2\eps(1+\eps)} \cos\phiH
\,, \qquad &
 g_2(\eps, \lambda_\ell,\phiH) &= \lambda_\ell \sqrt{2\eps(1-\eps)} \sin\phiH
\,, \nn \\
 g_3(\eps, \lambda_\ell,\phiH) &= \eps \cos(2\phiH)
\,, \qquad &
 g_4(\eps, \lambda_\ell,\phiH) &= \lambda_\ell \sqrt{1-\eps^2}
\,, \nn \\
 g_5(\eps, \lambda_\ell,\phiH) &= \sqrt{2\eps(1+\eps)} \sin\phiH
\,, \qquad &
 g_6(\eps, \lambda_\ell,\phiH) &= \lambda_\ell \sqrt{2\eps(1-\eps)} \cos\phiH
\,, \nn \\
 g_7(\eps, \lambda_\ell,\phiH) &= \eps \sin(2\phiH)
\,,\end{alignat}
where only $g_{2,4,6}$ actually depend on the lepton helicity.

The benefit of the above construction is that the SIDIS cross section in \eq{xs_SIDIS}
can now be written as
\begin{align} \label{eq:xs_SIDIS_2}
 \frac{\df\sigma}{\df x \, \df y \, \df z \, \df\psi \, \df^2\PTb} &
 = \kappa_\gamma \frac{\alpha^2}{4 Q^4} \frac{y}{z} L_{\mu\nu}(\pa, \pb) W^{\mu\nu}(q, \Pa, \Pb)
 \nn\\&
 = \kappa_\gamma \frac{\alpha^2}{4 Q^4} \frac{y}{z} \sum_{i=-1}^7  L_i(\pa,\pb,\Pa,\Pb) W_i(q, \Pa, \Pb)
 \nn\\&
 = \frac{\alpha^2}{2 Q^2} \frac{y}{z} \frac{\kappa_\gamma}{1-\eps} \sum_{i=-1}^7  g_i(\eps, \lambda_\ell, \phiH) W_i(q, \Pa, \Pb)
\,.\end{align}
Since the $g_i$ map onto independent angular coefficients as given in \eq{def_g_i},
this illustrates that we have achieved a decomposition of the SIDIS cross section
into hadronic structure functions which are in one-to-one correspondence with all possible independent angular coefficients.
In the following
we specialize to unpolarized SIDIS in \sec{tensor_decomposition_unpolarized} and polarized SIDIS in \sec{tensor_decomposition_spin}, in both cases with the one-photon exchange approximation.

\subsubsection{Unpolarized SIDIS}
\label{sec:tensor_decomposition_unpolarized}

We now restrict ourselves to working in the single-photon exchange approximation with an unpolarized target hadron.
Since the hadronic tensor is hermitian and obeys parity invariance, we can impose
\begin{align} \label{eq:hermiticity_parity}
W_{\lambda\lambda'}^* =  W_{\lambda'\lambda}~\text{(hermiticity)}
\quad\text{and}\quad
 W_{\lambda\lambda'} = (-1)^{\lambda+\lambda'}W_{-\lambda,-\lambda'}~\text{(parity)}
\,.\end{align}
For unpolarized SIDIS this eliminates the structure functions $W_{4, \dots, 7}$ leaving only the following
manifestly real structure functions:
\begin{alignat}{5} \label{eq:def_W_i_U}
 W_{UU,T}
   &\equiv W_{-1}^U
   &&= P_{-1}^{\mu\nu}\, W_{\mu\nu}
   &&= W_{++}
\,, \nn \\
 W_{UU,L}
   &\equiv W_0^U
   &&= P_0^{\mu\nu} \, W_{\mu\nu}
   &&= W_{00}
\,, \nn \\
 W_{UU}^{\cos\phiH}
   &\equiv W_1^U
   &&= P_1^{\mu\nu} \, W_{\mu\nu}
   && =  -\sqrt{2} \Re(W_{+0})
\,, \nn \\
 W_{LU}^{\sin\phiH}
   &\equiv W_2^U
   &&= P_2^{\mu\nu} \, W_{\mu\nu}
   && =  -\sqrt{2} \Im(W_{+0})
\,, \nn \\
 W_{UU}^{\cos2\phiH}
   &\equiv W_3^U
   &&= P_3^{\mu\nu} \, W_{\mu\nu}
   &&  = -\Re\bigl(W_{+-}\bigr)
\,,\end{alignat}
where $\Re$ and $\Im$ give the real and imaginary parts, respectively.
Here, we use two different notations for the independent structure functions.
In the first column, we follow the notation of \refcite{Bacchetta:2006tn} to label the structure functions
by a superscript denoting which angular coefficient the structure function maps onto.
The first (second) subscript denotes the beam (target) polarization, where $U$ and $L$ denote unpolarized and longitudinally polarized respectively.
The third subscript on $W_{UU,T}$ and $W_{UU,L}$ denotes the transverse or longitudinal polarization of the virtual photon.
In the second column of \eq{def_W_i_U}, we simply enumerate the structure functions as $W_i^U$ with $i=-1,\dots,3$,
which will allow for compact expressions in the following, with $U$ denoting the unpolarized structure function.
The third column in \eq{def_W_i} defines the $W_i^U$ through projections on the $P_i^{\mu\nu}$ defined in \eq{def_P_i}.
In the last column, we provide compact results in terms of the helicity-projections $W_{\lambda\lambda'}$
defined in \eq{W_proj} after applying \eq{hermiticity_parity}.

Inserting the above results into \eq{xs_SIDIS_2}, we obtain 
\begin{align} \label{eq:xs_SIDIS_unpolarized}
 \frac{\df\sigma}{\df x \, \df y \, \df z \, \df\psi \, \df^2\PTb} &
 = \frac{\alpha^2}{2 Q^2} \frac{y}{z} \kappa_\gamma \frac{\delta_{\lambda\lambda'}}{1-\eps} \Bigl[
      \big( W_{-1}^U + \eps W_0^U \bigr)
      +  \eps \cos(2\phiH) W_3^U
\\\nn&\hspace{2.8cm}
      + \sqrt{2\eps(1+\eps)} \cos\phiH W_1^U
      + \lambda_\ell \sqrt{2\eps(1-\eps)} \sin\phiH W_2^U
      \Bigr]
\,.\end{align}
All structure functions are even in $\phiH$, except for $W_2^U$ which is odd in $\phiH$.
It is proportional to $\lambda_\ell$, and thus only arises for polarized lepton beams.

Comparing \eq{xs_SIDIS_unpolarized} with the corresponding result in \refcite{Bacchetta:2006tn},
we can relate our structure functions $W_i^U$ with their structure functions $F_i$ through
\begin{align} \label{eq:relation_W_F}
 W_i^U  &
 = \frac{2}{\kappa_\gamma} \Bigl(1 + \frac{\gamma^2}{2x}\Bigr) \frac{z}{x} F_i
 \stackrel{M_N \to 0}{=} \frac{2 z}{x} F_i
\,.\end{align}

To explore TMD distribution and fragmentation functions, we consider the limit $P_{hT} \ll Q$ with $x$, $y$, $z$, and $\psi$ treated as fixed ${\cal O}(1)$ variables.  In this limit the structure functions $W_{-1}^U$ and $W_3^U$ are leading power, scaling as $\sim 1/P_{hT}^2$, the structure functions $W_2^U$ and $W_3^U$ are next-to-leading power, scaling as $\sim 1/P_{hT}$, and $W_0^U$ is next-to-next-to-leading power, scaling as ${\cal O}(P_{hT}^0)$, see for example Ref.~\cite{Bacchetta:2006tn}. 
In the helicity notation $W_{\lambda\lambda'}$ this corresponds with a power suppression by $P_{hT}/Q$ for each $\lambda=0$ or $\lambda'=0$. 

\subsubsection{Polarized SIDIS}
\label{sec:tensor_decomposition_spin}

Here, we extend our previous setup to also allow for polarized targets.
We decompose the hadron spin vector as~\cite{Bacchetta:2006tn}
\begin{align} \label{eq:S}
 S^\mu &
 = S_L \frac{2x \Pa^\mu - \gamma^2 q^\mu}{\gamma Q \sqrt{1+\gamma^2}} + S_\perp^\mu
 = \Bigl( 0 \,,\, S_T \cos\phiS \,,\, S_T \sin\phiS \,,\, - S_L\Bigr)_T
\,,\end{align}
where the parameterization is given in the hadron rest frame corresponding to the Trento convention.
Following the strategy of \refcite{Diehl:2005pc}, we then define the spin-density matrix as
\begin{align} \label{eq:rho}
 \rho_{mn} &
 = \frac12 \bigl( 1 + \vec S \cdot \vec\sigma \bigr)_{mn}
 = \frac12 \begin{pmatrix}
            1 + S_L & S_T e^{-\img (\phiH-\phiS)} \\
            S_T e^{\img (\phiH - \phiS)} & 1 - S_L
           \end{pmatrix}_{\!H}
\,,\end{align}
where $\vec\sigma$ are the Pauli matrices. This form of the spin-density matrix
only holds in a hadron rest frame, with the three-dimensional spin vector $\vec S$ defined accordingly.
In the last equality in \eq{rho}, we have used the expression
$\vec S = ( S_T \cos(\phiH-\phiS), S_T \sin(\phiH-\phiS), S_L )$
for $\vec S$ in the hadronic rest frame that is obtained by a boost of the hadronic Breit frame along the $z$-axis, and denoted by a supscript $H$. 
\Eq{rho} applies in a basis of polarization states specified by the two-component spinors
\begin{align}
 \chi_{+\frac12} = \begin{pmatrix} 1 \\ 0 \end{pmatrix}
\,,\qquad
 \chi_{-\frac12} = \begin{pmatrix} 0 \\ 1 \end{pmatrix}
\,,\end{align}
which describe positive and negative helicity along the $\tilde z$ axis.
Since only the two values $\pm \frac12$ can arise, we will simply label $\chi_m$ with $m\in\{+,-\}$,
with the understanding that $m=\pm$ corresponds to a spin label of $\pm1/2$.

Using \eq{rho}, the hadronic tensor can  be written as~\cite{Diehl:2005pc}
\begin{align} \label{eq:W_pol}
 W_{\mu\nu}(q, \Pa, \Pb) &
 = \sum_{m,n} \rho_{nm} W^{mn}_{\mu\nu}(q, \Pa, \Pb)
\,,\\\nn
 W^{mn}_{\mu\nu}(q, \Pa, \Pb) &
 = \sum_X  \frac{\df^3 p_X}{(2\pi)^3 2 E_X} \delta^4(q + \Pa - \Pb - \PX)
 \braket{N(m) | J^{\dagger}_\mu | h,X} \braket{h,X | J_\nu | N(n)}
\,,\end{align}
where the sum over the polarizations of the final state $hX$ is kept implicit.

The hadronic structure functions defined in \eq{def_W_i} can thus be evaluated as
\begin{align} \label{eq:def_W_i_pol}
 W_i(q, \Pa, \Pb) &
 = P_i^{\mu\nu}(q, \Pa, \Pb) \sum_{m,n} \rho_{nm} \, W_{\mu\nu}^{mn}(q, \Pa, \Pb)
 \nn\\&
 = W_i^U + S_L W_i^L
 + S_T \cos(\phiH - \phiS) W_i^{Tx}
 + S_T \sin(\phiH - \phiS) W_i^{Ty}
\,,\end{align}
where in the second line we suppressed the arguments, and defined the abbreviations
\begin{align} \label{eq:W_proj_3}
 W_i^{U\phantom{-}} &= \frac12 \bigl(W^{++}_i + W^{--}_i \bigr)
\,,\qquad
 W_i^{L\phantom{-}} = \frac12 \bigl(W^{++}_i - W^{--}_i \bigr)
\,,\nn\\
 W_i^{Tx} &= \frac12 \bigl(W^{+-}_i + W^{-+}_i \bigr)
\,,\qquad
 W_i^{Ty} = \frac{\img}{2} \bigl( W^{+-}_i - W^{-+}_i \bigr)
\,.\end{align}
Here, $W_i^U$ and $W_i^{L}$ describe an unpolarized or longitudinally polarized target,
while $W_i^{Tx}$ and $W_i^{Ty}$ correspond to a hadron polarized
transversely in the $x$ and $y$ direction in the hadronic Breit frame, respectively.
They are independent of the hadron spin, as the dependence of $W_i$ on $\vec S$
has been made explicit in \eq{def_W_i_pol}. The $W_i^{mn}$ in \eq{W_proj_3}
are defined analogous to \eq{def_W_i} as
\begin{align}
 W_i^{mn}(q, \Pa, \Pb) = P_i^{\mu\nu}(q, \Pa, \Pb) W_{\mu\nu}^{mn}(q, \Pa, \Pb)
\,,\end{align}
which as before are linear combinations of the helicity projections
\begin{align} \label{eq:W_proj_pol}
 W^{mn}_{\lambda \lambda'}(q, \Pa, \Pb) &
 \equiv \eps^{*\mu}_\lambda \eps^{\nu}_{\lambda'} \, W^{mn}_{\mu\nu}(q, \Pa, \Pb)
\,,\qquad \lambda \in \{+,-,0\}
\,.\end{align}

By inserting \eq{def_W_i_pol} into \eq{xs_SIDIS_2}, one obtains the tensor
decomposition for the polarized SIDIS process. We write it as
\begin{align} \label{eq:xs_SIDIS_pol}
 \frac{\df\sigma}{\df x \, \df y \, \df z \, \df\psi \, \df^2\PTb}
 = \frac{\alpha^2}{2 Q^2} \frac{y}{z}  \frac{\kappa_\gamma}{1-\eps}
   \Bigl[ &
       (W{\cdot}L)_{UU} + S_L (W{\cdot}L)_{UL}
 + \lambda_\ell (W{\cdot}L)_{LU} 
   \\\nn&
    + \lambda_\ell S_L (W{\cdot}L)_{LL}
     + S_T (W{\cdot}L)_{UT} + \lambda_\ell S_T (W{\cdot}L)_{LT}
   \Bigr]
\,,\end{align}
where the $(W{\cdot}L)_{XY}$ are the components of $W \cdot L$
corresponding to beam and target polarization $X$ and $Y$, respectively,
and normalized to the common prefactor $2 Q^2 / (1-\eps)$ appearing \eq{L_i_results}
as well as the lepton helicity $\lambda_\ell$ and target spin $S_{T,L}$.

To reduce the number of independent structure functions, we impose~\cite{Diehl:2005pc}
\begin{alignat}{2} \label{eq:hermiticity_parity_pol}
 &\text{hermiticity:} \qquad && W^{mn}_{\lambda\lambda'} = (W^{nm}_{\lambda'\lambda})^*
\,,\nn\\
&\text{parity:}  \qquad && W^{mn}_{\lambda\lambda'} = (-1)^{\lambda+\lambda' + (n-m)/2} \, W^{-m,-n}_{-\lambda,-\lambda'}
\,.\end{alignat}
The explicit factor of $(-1)^{(n-m)/2}$ accounts for the spin vector being a pseudovector,
with the factor of $1/2$ compensating for our notation where $m,n\in\{\pm\}$ rather than $m,n\in\{\pm1/2\}$.
Reducing all appearing angular dependencies to a minimal basis,
this leaves 18 independent angular structures. They are given by
\begin{subequations}
\begin{align} \label{eq:coeffs_full}
 (W{\cdot}L)_{UU} &
 = W_{UU,T} + \eps W_{UU,L}
 + \sqrt{2\eps(1+\eps)} \cos(\phiH) W_{UU}^{\cos(\phiH)}
 \nn\\&\quad
 + \eps \cos(2\phiH) W_{UU}^{\cos(2\phiH)}
\,,\\
 (W{\cdot}L)_{UL} &
 = \sqrt{2\eps(1+\eps)} \sin(\phiH) W_{{UL}}^{\sin (\phiH)}
 + \eps \sin(2\phiH) W_{{UL}}^{\sin (2\phiH)}
\,,\\
 (W{\cdot}L)_{LU} &
 = \sqrt{2\eps(1-\eps)} \sin(\phiH) W_{LU}^{\sin(\phiH)}
\,,\\
(W{\cdot}L)_{LL} &
 =  \sqrt{1-\eps^2} W_{LL} + \sqrt{2\eps(1-\eps)} \cos(\phiH) W_{LL}^{\cos(\phiH)}
\,,\\
 (W{\cdot}L)_{UT} &
 =  \sin(\phiH -\phiS) \left[ W_{UT,T}^{\sin(\phiH -\phiS)} + \eps W_{UT,L}^{\sin(\phiH -\phiS)} \right]
 \nn\\&\quad
 + \eps \left[ \sin(\phiH +\phiS) W_{UT}^{\sin(\phiH +\phiS)} + \sin(3 \phiH -\phiS) W_{UT}^{\sin(3 \phiH -\phiS)} \right]
 \nn\\&\quad
 + \sqrt{2\eps(1+\eps)} \left[ \sin(\phiS) W_{UT}^{\sin(\phiS)} + \sin(2 \phiH -\phiS) W_{UT}^{\sin(2 \phiH -\phiS)} \right]
\,,\\
(W{\cdot}L)_{LT} &
 = \sqrt{1-\eps^2} \cos(\phiH -\phiS) W_{LT}^{\cos(\phiH -\phiS)}
 \nn\\&\quad
   + \sqrt{2\eps(1-\eps)} \left[ \cos(\phiS) W_{LT}^{\cos(\phiS)} + \cos(2\phiH-\phiS) W_{LT}^{\cos(2\phiH -\phiS)} \right]
\,.\end{align}
\end{subequations}
The fundamental hadronic structure functions are defined as
\begin{subequations} \label{eq:def_W_polarized}
\begin{alignat}{4}
&W_{{UU,T}} &&= W_{-1}^{{U}} &&= \frac{1}{2} \left(W_{{++}}^{{++}}+W_{{++}}^{{--}}\right)
\,, \nn \\
&W_{{UU,L}} &&= W_0^{{U}} &&= W_{00}^{{++}}
\,, \nn \\
&W_{{UU}}^{\cos (\phiH)} &&= W_1^{{U}} &&= -\frac{1}{\sqrt2} \Re\left(W_{+0}^{{++}}+W_{+0}^{{--}}\right)
\,, \nn \\
&W_{{UU}}^{\cos (2 \phiH)} &&= W_3^{{U}} &&= -\Re\left(W_{{+-}}^{{++}}\right)
\,,\\
&W_{{UL}}^{\sin (\phiH)} &&= W_5^{{L}} &&= \frac{-1}{\sqrt2} \Im\left(W_{+0}^{{++}} - W_{+0}^{{--}}\right)
\,, \nn \\
&W_{{UL}}^{\sin (2 \phiH)} &&= W_7^{{L}} &&= -\Im\left(W_{{+-}}^{{++}}\right)
\,,\\
&W_{{LU}}^{\sin (\phiH)} &&= W_2^{{U}} &&= -\frac{1}{\sqrt2} \Im\left(W_{+0}^{{++}}+W_{+0}^{{--}}\right)
\,,\\
&W_{{LL}} &&= W_4^{{L}} &&= \frac{1}{2} \left(W_{{++}}^{{++}}-W_{{++}}^{{--}}\right)
\,, \nn \\
&W_{{LL}}^{\cos (\phiH)} &&= W_6^{{L}} &&= \frac{-1}{\sqrt2} \Re\left(W_{+0}^{{++}} - W_{+0}^{{--}}\right)
\,,\\
&W_{{UT,T}}^{\sin (\phiH-{\phiS})} &&= W_{-1}^{{Ty}} &&= -\Im\left(W_{{++}}^{{+-}}\right)
\,, \nn \\
&W_{{UT,L}}^{\sin (\phiH-{\phiS})} &&= W_0^{{Ty}} &&= -\Im\left(W_{00}^{{+-}}\right)
\,, \nn \\
&W_{{UT}}^{\sin (2 \phiH-{\phiS})} &&= \frac12 (W_1^{{Ty}} + W_5^{{Tx}}) &&= -\frac{1}{\sqrt2}  \Im\left(W_{+0}^{{-+}}\right)
\,, \nn \\
&W_{{UT}}^{\sin (3 \phiH-{\phiS})} &&= \frac12 (W_3^{{Ty}} + W_7^{{Tx}}) &&= -\frac{1}{2} \Im\left(W_{{+-}}^{{-+}}\right)
\,, \nn \\
&W_{{UT}}^{\sin ({\phiS})} &&= \frac12 (W_5^{{Tx}} - W_1^{{Ty}}) &&= -\frac{1}{\sqrt2} \Im\left(W_{+0}^{{+-}}\right)
\,, \nn \\
&W_{{UT}}^{\sin (\phiH+{\phiS})} &&= \frac12 (W_7^{{Tx}} - W_3^{{Ty}}) &&= -\frac{1}{2} \Im\left(W_{{+-}}^{{+-}}\right)
\,,\\
&W_{{LT}}^{\cos (\phiH-{\phiS})} &&= W_4^{{Tx}} &&= \Re\left(W_{{++}}^{{+-}}\right)
\,, \nn \\
&W_{{LT}}^{\cos (2 \phiH-{\phiS})} &&= \frac12 (W_6^{{Tx}} - W_2^{{Ty}}) &&= -\frac{1}{\sqrt2} \Re\left(W_{+0}^{{-+}}\right)
\,, \nn \\
&W_ {{LT}}^{\cos ({\phiS})} &&= \frac12 (W_ 2^{{Ty}} + W_ 6^{{Tx}}) &&= -\frac{1}{\sqrt2} \Re\left(W_ {+0}^{{+-}}\right)
\,.\end{alignat}
\end{subequations}
Each $W_{XY}^\Omega$ is labeled by the beam and target polarizations $X$ and $Y$,
and the angular distribution $\Omega$ it multiplies.
In the case of $W_{XY,T}^\Omega$ and $W_{XY,L}^\Omega$, $T$ and $L$ refer to the
transverse or longitudinal polarization of the virtual photon.
The second column shows its definition in terms of the helicity projections defined in \eq{W_proj_3},
while the last column shows their expression in terms of the helicity projections
in \eq{W_proj_pol} after applying the hermiticity and parity constraints from \eq{hermiticity_parity_pol}.
These results precisely agree with those in appendix A of \refcite{Bacchetta:2006tn},
as we have used the same conventions for the fundamental helicity projectors.

In the limit $P_{hT} \ll Q$ the structure functions again enter at different orders in this power expansion. Just like in the unpolarized case, in the helicity decomposition $W_{\lambda\lambda'}^{mn}$ this corresponds to having a power suppression by $P_{hT}/Q$ for each $\lambda=0$ or $\lambda'=0$, see for example Ref.~\cite{Bacchetta:2006tn}.  Again the leading power structure functions scale as $\sim 1/P_{hT}^2$, the next-to-leading power structure functions scale as $\sim 1/P_{hT}$, etc.

\subsection{Lightcone Coordinates and Factorization Frame}
\label{sec:factorization_frame}

It was natural to work in the hadronic Breit frame for the decomposition of the
hadronic tensor into independent structure functions.
In contrast, the factorization of the hadronic tensor is most naturally addressed
using lightcone coordinates.

\paragraph{Conventions.}
Our conventions for lightcone coordinates follow the SCET literature,
as our treatment of subleading-power factorization relies on various results that make use of SCET.
We define two lightlike reference vectors $n^\mu$ and $\bn^\mu$ normalized such that
\begin{align}
 n^2 = \bn^2 = 0 \,,\qquad n \cdot \bn = 2
\,.\end{align}
Any four vector $p^\mu$ can then be decomposed as
\begin{align} \label{eq:lc_coord}
 p^\mu = p^-  \frac{n^\mu}{2} + p^+ \frac{\bn^\mu}{2} + p_\perp^\mu
 \equiv (p^+, p^-, p_\perp)
\,,\quad\text{where}\quad
 p^- = \bn \cdot p \,, \quad p^+ = n \cdot p
\,.\end{align}
It is also useful to define the transverse metric and antisymmetric tensor,
\begin{align} \label{eq:gperp_F}
 g_\perp^{\mu\nu} &
 = g^{\mu\nu} - \frac12 \bigl( n^\mu \bn^\nu + n^\mu \bn^\nu\bigr)
\,,\qquad
 \eps_\perp^{\mu\nu} = \frac12 {\eps^{\mu\nu}}_{\rho\sigma} n^\rho \bn^\sigma
\,.\end{align}
Transverse vectors can then be defined as
\begin{align}
 p_\perp^\mu \equiv g_\perp^{\mu\nu} p_\nu
\,.\end{align}
This definition implicitly depends on the choice of $n^\mu$ and $\bn^\mu$.
Unless stated otherwise, we will use the $\perp$ notation exclusively
for transverse vectors as specified by the following choice of $n^\mu$ and $\bn^\mu$.
It is also convenient to identify the Minkowski transverse vector $p_\perp^\mu$
in terms of its Euclidean components $\pt$ as $p_\perp^\mu = (0, \pt, 0)$.

\paragraph{Factorization frame.}
A convenient choice for the reference vectors is to align $n^\mu$ and $\bn^\mu$
with the target hadron $N$ and the detected final-state hadron $h$.
From now on, we will always neglect hadron masses, $M_N^2 = M_h^2 = 0$.
We choose $n^\mu$ and $\bn^\mu$ such that%
\footnote{The measurement only fixes the product $2 \Pa \cdot \Pb = \Pa^- \Pb^+ = z Q^2 / x$,
such that one can freely choose the ratio $\Pa^- / \Pb^+$.
}
\begin{align} \label{eq:Pa_Pb}
 \Pa^\mu &= \Pa^- \frac{n^\mu}{2} = \frac{Q}{x}  \frac{n^\mu}{2}
\,,\qquad
 \Pb^\mu = \Pb^+ \frac{\bn^\mu}{2} = z Q \frac{\bn^\mu}{2}
\,.\end{align}
We define the \emph{factorization frame} such that these unit vectors take the standard form
\begin{align}
 n^\mu = (1, 0, 0, 1)_F
\,,\qquad
 \bn^\mu = (1, 0, 0, -1)_F
\,.\end{align}
To fix a reference transverse direction, we first construct the transverse metric
\begin{align}
 g_\perp^{\mu\nu}
 = g^{\mu\nu} -  \frac{2x}{z Q^2} (\Pa^\mu \Pb^\nu + \Pa^\nu \Pb^\mu)
\,.\end{align}
Since $\Pa^\mu$ and $\Pb^\mu$ have no transverse component in this frame,
it is natural to define the reference direction in terms of
\begin{align}
 q_\perp^\mu &
 = q^\mu + \Bigl(1 - \frac{q_T^2}{Q^2} \Bigr) x \Pa^\mu  - \frac{\Pb^\mu}{z}
\,,\qquad
 q_T^2 \equiv -q_\perp^2 = \frac{P_{hT}^2}{z^2}
\,.\end{align}
To complete the construction of the unit vectors $n_i^\mu$ in the factorization frame,
we choose $n_x^\mu$ such that $q_\perp^\mu = -q_T (0, 1, 0, 0)_F$, which yields
\begin{align} \label{eq:lc_vectors}
 n_t^\mu &= \frac12\bigl(n^\mu + \bn^\mu)
\,,\quad
 n_x^\mu =
 - \frac{q_\perp^\mu}{q_T}
\,,\quad
 n_z^\mu = \frac12\bigl(n^\mu - \bn^\mu)
\,,\quad
 n_y^\mu = \eps^{\mu\nu\rho\sigma} n_{t\,\nu} n_{x\,\rho} n_{z\,\sigma}
\,.\end{align}
Explicit expressions in terms of the hadron momenta are given by
\begin{align} \label{eq:lc_vectors_explicit}
 n_t^\mu = \frac{x \Pa^\mu}{Q} + \frac{\Pb^\mu}{z Q}
\,,\quad
 n_x^\mu = \frac{1}{q_T} \biggl[\Bigl(\frac{q_T^2}{Q^2} - 1\Bigr) x \Pa^\mu  + \frac{\Pb^\mu}{z} - q^\mu\biggr]
\,,\quad
 n_z^\mu = \frac{x \Pa^\mu}{Q} - \frac{\Pb^\mu}{z Q}
\,.\end{align}

\paragraph{Relation to other frames.}
In the lightcone coordinates of the factorization frame, the momentum transfer $q^\mu = (q^+, q^-, q_\perp)$ reads
\begin{align} \label{eq:q_ff}
 q^+ = Q
\,,\qquad
 q^- = Q \Bigl( \frac{q_T^2}{Q^2} - 1 \Bigr)
\,,\qquad
 q_\perp^2 = -\frac{P_{hT}^2}{z^2}
\,.\end{align}
It is also instructive to relate $q_\perp^\mu$ to the Trento rest frame and the hadronic Breit frame,
which can easily be obtained by inserting \eqs{momenta_Trento}{momenta_Breit_had}
into $n_x^\mu$ in \eq{lc_vectors_explicit},%
\footnote{Note that as a hadron rest frame, the Trento frame is defined with $M_N \ne 0$. The results in \eq{qperp_relations} follow by taking the $M_N\to 0$ limit of the corresponding $M_N\ne 0$ relations.
}
\begin{align} \label{eq:qperp_relations}
 q_\perp^\mu &
 = - q_T \bigl(0, 1, 0, 0\bigr)_F
 = \Bigl(0 \,,\, -\frac{\PTb}{z} \,,0 \Bigr)_T
 = - q_T \Bigl(  \frac{q_T}{Q} \,,\, 1  \,,\, 0 \,,\,  \frac{q_T}{Q} \Bigr)_{B}
\,.\end{align}
Due to the simple relation between the components in the factorization and Trento frame,
in the literature it is often stated that $\qt = -\PTb / z$.%
\footnote{In our case, the relation is given by $\qt = +\PTb / z$ due to defining
$n_x^\mu = -q_\perp^\mu/q_T$ in \eq{lc_vectors}. This choice is made such that
\eq{rel_lc_unit_vectors} does not receive a relative minus sign between $n_x^\mu$ and $\tilde n_x^\mu$.}
However, this is a bit misleading, as it relates $\qt$ as defined in the factorization frame
to $\PTb$ as defined in the Trento frame. (In the Trento frame, one has $\qt = 0$.)
Instead, this should be understood as a relation between components in two different frames
as in \eq{qperp_relations}.

It will also be useful to relate the unit vectors $n_i^\mu$ of the factorization frame
to the unit vectors $\tilde n_i^\mu$ in the hadronic Breit frame,
\begin{align} \label{eq:rel_lc_unit_vectors}
 \tilde n_t^\mu &= n_t^\mu - \frac{q_T}{Q} n_x^\mu + \frac12 \frac{q_T^2}{Q^2}  n^\mu
\,,\qquad
 \tilde n_x^\mu = n_x^\mu + \frac{q_T}{Q} n^\mu
\,,\nn\\
 \tilde n_z^\mu &= n_z^\mu  + \frac{q_T}{Q} n_x^\mu - \frac12 \frac{q_T^2}{Q^2}  n^\mu
\,,\qquad
 \tilde n_y^\mu =n_y^\mu = \frac12 \eps^{\mu\nu\rho\sigma} n_\nu \bn_\mu n_{x\,\rho} 
\,.\end{align}

\paragraph{Spin vector in the factorization frame.}

Earlier we separated the target spin vector $S^\mu$ into the longitudinal and transverse components in the Trento frame as in \eq{S} or equivalently to the rest frame obtained from the hadronic Breit frame as in \eq{rho}, and carried out the tensor decomposition using this separation. However, when deriving factorization for spin dependent structure functions later in \sec{results}, we will decompose the quark-quark and quark-gluon-quark correlators into different Dirac structures in the factorization frame. Therefore, we need to address the issue of conversion of the spin vector in the factorization frame to that in the Breit frame as above.

To this end, we look at the corresponding target-rest frames for the factorization frame and the Breit frame. The difference for these two target-rest frames is the choice of the longitudinal direction: one is determined by $P_h^\mu$, while the other is determined by $q^\mu$. The conversion of these two coordinate systems is characterized by a rotation of small angle $P_{hT}/P_{hL}\sim P_{hT}/E_{h}$. Here $P_{hL}$ is the longitudinal momentum of the outgoing hadron in the target rest frame, and $E_h$ is its energy. Notice that $E_h$ is
\begin{equation}
  E_h=\frac{\Pb\cdot\Pa}{\Ma}\sim \Ord(Q^2/\Ma)\,.
\end{equation}
As a consequence, the angle $P_{hT}/P_{hL}$, as well as the difference of longitudinal/transverse separation of $S^\mu$ in the two different frames,  is of order $\Ord(P_{hT} M_N/Q^2)$ suppressed, which is beyond the level considered in this paper.  The change from the Breit and factorization target rest frames involves a rotation around the $y$-axis, and hence also modifies the meaning of $\phi_S$, however again this modification is power suppressed by $\Ord(P_{hT} M_N/Q^2)$.
Therefore, from now on, we ignore these differences and use $S_L$, $S_T$, and $\phi_S$ for both frames.

\section{SCET Ingredients at Subleading Power}
\label{sec:SCET_review}

To describe the dynamics of the collinear and soft particles in the presence of a hard interaction, we make use of SCET~\cite{Bauer:2000ew, Bauer:2000yr, Bauer:2001ct, Bauer:2001yt, Bauer:2002nz}, a top-down effective field theory that is derived from QCD. For our analysis of SIDIS at subleading power, the relevant theory is known as \SCETii~\cite{Bauer:2002aj}, which involves collinear and soft particles whose transverse momenta are of the same parametric size.  In SCET the importance of operators is classified by a dimensionless power counting parameter $\lambda\ll 1$. For SIDIS this encodes the expansion in small transverse momentum $q_T \ll Q$ with $\lambda\sim q_T/Q$ (where there may or may not be an additional hierarchy between $q_T$ and $\Lambda_{\rm QCD}$). \SCETii includes interactions of $n_i$-collinear particles with momentum $p$ close to the $\vec n_i$ direction, where $n_i=(1,\vec n_i)$.  The momenta for $n_i$-collinear particles scale as $(n_i\cdot p, \bar n_i\cdot p,p_{n_i\perp})\sim Q (\lambda^2,1,\lambda)$, where $Q\sim \lambda^0$ is a generic hard momentum scale. Here $\bn_i$ is an auxiliary light-cone vector satisfying $n_i\cdot \bar n_i=2$, and is often for simplicity chose to be $\bn_i=(1,-\vec n_i)$. When the underlying choice of the collinear direction $n_i$ is clear from the context, we refer to $n_i\cdot p=p^+$ as the small momentum component, and $\bn_i\cdot p=p^-$ as the large momentum component. \SCETii also includes soft particles with momenta scaling as $k^\mu\sim Q\lambda$.  Modes in SCET are infrared in origin, extending from their scaling dimension down to zero momentum, and the double counting of infrared regions is avoided by the presence of zero-bin subtractions~\cite{Manohar:2006nz} (which are referred to as soft subtractions  in CSS~\cite{Collins:1982wa,Collins:2011zzd}).  For the TMD distributions that appear in SIDIS these subtractions lead to division by additional vacuum matrix elements of Wilson lines, and their precise form depends on the invariant mass and rapidity regulators that are used to define and renormalize functions, see~\refcite{Ebert:2019okf} for a review of various common constructions.

For SIDIS within \SCETii there will be two relevant collinear directions, for the incoming and outgoing hadrons. When we wish to have a generic notation for these two directions we will denote them by $n_1$ and $n_2$. For specific calculations it is often useful to go to the back-to-back frame where the directions can be taken as $n$ and $\bn$ (corresponding to the specialization to $n_1=\bar n_2=n$ and $n_2=\bar n_1=\bar n$).  For simplicity we will use the more common notation of the back-to-back frame for our presentation of SCET ingredients in this section. The generalization to arbitrary $n_1$ and $n_2$ is quite straightforward and will be used in \sec{factorization} below.
The Lagrangian for \SCETii can be decomposed as
\begin{align} \label{eq:SCETLagExpand}
\cL_{\text{SCET}_{\rm II}}=\cL_\hard+\cL_{\rm dyn}= \Big(\sum_{i\geq0} \cL_\hard^{(i)} \Big) + \Big( \sum_{i\geq0} \cL_{\rm dyn}^{(i)} +\cL_G^{(0)} \Big) \,,
\end{align}
with each term having a definite power counting ${\cal O}(\lambda^i)$ as indicated by the superscripts $(i)$. As written, the \SCETii Lagrangian is divided into three different contributions. The $\cL_\hard^{(i)}$ term contains operators that mediate the hard scattering process, and can be derived from QCD by matching calculations. For cases like the SIDIS process treated here, the $\cL_\hard^{(i)}$ Lagrangians also always involve the external current that couples to the leptons.  The $\cL_{\rm dyn}^{(i)}$ describe the long wavelength dynamics of soft and collinear modes in the effective theory. At leading power for SIDIS we have
\begin{align} \label{eq:Ldyn0}
  \cL^{(0)}_{\rm dyn} = \cL_{n}^{(0)} + \cL_{\bn}^{(0)}
     + \cL_{s}^{(0)} 
 = \sum_{i=q,g} \bigl[ \cL_{ni}^{(0)} + \cL_{\bn i}^{(0)}
     + \cL_{s i}^{(0)}  \bigr] \,.
\end{align}
This Lagrangian already carries a factorized structure since $\cL_{n}^{(0)}$ only has interactions between $n$-collinear quarks and gluons, $\cL_{\bn}^{(0)}$ has interactions between $\bn$-collinear quarks and gluons, and $\cL_s^{(0)}$ has interactions between soft quarks and gluons. 
Finally, the Glauber Lagrangian~\cite{Rothstein:2016bsq} $\cL_G^{(0)}$ describes interactions between soft and collinear fields that are induced by instantaneous off-shell Glauber potentials $\sim 1/k_\perp^2$.  These contributions are leading order in the power counting expansion, and spoil factorization unless they can be shown to cancel out for a given process. For leading power SIDIS, it is known that contributions from the Glauber region of momentum space either cancel out~\cite{Collins:2011zzd}, or (in the SCET Language~\cite{Rothstein:2016bsq}) can be absorbed by a proper choice of the direction of the collinear and soft Wilson lines appearing in TMD distribution functions~\cite{Collins:2002kn,Collins:2004nx}.

An important property of SCET is that the Lagrangian $\cL_G^{(0)}$ provides the only mechanism by which factorization can be violated, even at subleading power~\cite{Rothstein:2016bsq}.  The subleading power ${\cal L}_{\rm dyn}^{(i>0)}$ and $\cL_{\rm hard}^{(i>0)}$ Lagrangians do involve interactions between $n$-collinear, $\bn$-collinear, and soft fields, but on their own these can always be factorized into independent time-ordered products in the $n$, $\bn$, and soft sectors since these power suppressed Lagrangians are only inserted a finite number of times at a given order in the power expansion. Only $\cL_G^{(0)}$ can be inserted any number of times without changing the order in the power counting, and hence only $\cL_G^{(0)}$ will violate factorization. Nevertheless, the proof of cancellation of $\cL_G^{(0)}$ interactions is challenging, and will not be taken up here. Since the proof that Glauber effects cancel out in leading power SIDIS is simpler than in Drell-Yan~\cite{Collins:2011zzd}, and since this arises from the incoming and outgoing hadron kinematics, it is likely that this will remain true at NLP. 
In this paper we will simply  {\it make the assumption} that the cancellation of Glauber effects occurs at NLP in SIDIS, and then derive formula for the NLP factorization theorem for SIDIS in this context. This amounts to ignoring $\cL_G^{(0)}$ for our analysis.

Collinear operators are constructed out of products of fields and Wilson lines
that are invariant under collinear gauge
transformations~\cite{Bauer:2000yr,Bauer:2001ct}.  The smallest building blocks
are collinear gauge-invariant quark and gluon fields, which we define here as
\begin{align} \label{eq:chiB}
	\chi_{n,\w}(x) &= \Bigl[\delta(\w - \bnP_n)\, W_n^\dagger(x)\, \xi_n(x) \Bigr]
	\,,\\
	\cB_{n\perp,\w}^\mu(x)
	&= \frac{1}{g}\Bigl[\delta(\w + \bnP_n)\, W_n^\dagger(x)\,\img D_{n\perp}^\mu W_n(x)\Bigr]
     = \frac{\img}{g} \biggl[ \delta(\w + \bnP_n)\,\frac{1}{\bnP_n} \bn_\nu G_n^{B\nu\mu_\perp} W_{n\,{\rm adj}}^{BA} \biggl]
	\,.\nn
\end{align}
Here $\xi_n(x)$ is the $n$-collinear quark field, 
which obeys $\frac14 \slashed{n}\slashed{\bn} \xi_n = \xi_n$
and $\frac14 \slashed{\bn}\slashed{n} \xi_n = 0$
and thus constitutes the ``good components'' of the quark field.
The transverse collinear covariant derivative is defined as
\begin{equation}
	\img D_{n\perp}^\mu = \cP^\mu_{n\perp} + g A^\mu_{n\perp}\,,
\end{equation}
where $\cP_{n\perp}^\mu \equiv i\partial_{n\perp}^\mu$ is the collinear transverse momentum operator, and $A_n^\mu$ are the $n$-collinear gluon fields which also form a field strength through $\img g G_n^{B\nu\mu} T^B =[\img D_n^\nu,\img D_n^\mu]$.
The collinear quark and gluon fields in \eq{chiB} carry the large momentum $\w n^\mu/2$,
where $\w$ is fixed by the $\delta$-functions involving the label momentum operator $\overline\cP_n\equiv i\bn\cdot\partial_n \sim \lambda^0$.
With this definition of $\chi_{n,\w}$, we have $\w > 0$ for an incoming quark and $\w < 0$ for an outgoing antiquark. Note that we will also use the notation $\bar \chi_{n,\w}= (\chi_{n,-\w})^\dagger \gamma^0$, so that  $\bar \chi_{n,\w}$ has $\w<0$ for an outgoing quark, and $\w>0$ for an incoming antiquark. (Note that this definition of $\bar\chi_{n,\omega}$ differs from the most common convention in SCET~\cite{SCETreview}, but agrees with the convention used in \refcite{Moult:2015aoa,Feige:2017zci}.)
For $\cB_{n,\w\perp}$, $\w > 0$ ($\w < 0$) corresponds to an outgoing (incoming) gluon.  
In \eq{chiB}
\begin{equation} \label{eq:Wn}
	W_n(x) = \biggl[\sum_\text{perms} \exp\Bigl(-\frac{g}{\overline {\cal P}_n}\,\bn\cdot A_n(x)\Bigr)\biggr]
\end{equation}
is a Wilson line of $n$-collinear gluons in label momentum space. 
We use $W_n$ for the Wilson line in the fundamental color representation, 
and $W_{n\,{\rm adj}}$ for the same Wilson line in the adjoint color representation.
Note that the subscript $n$ indicates the type of fields this collinear Wilson line is built out of, rather than the direction of its path. 
The label operator $\overline\cP_n$ picks out the large momentum components of $A_n(x)$ in \eq{Wn},
while the position $x = (x^+, x^-, x_\perp)$ corresponds to the residual momentum (via a Fourier transform).
When carrying zero residual momentum, via the restriction to $x=(0,x^-,x_\perp)$,
this $W_n(x)$ is simply the Fourier transform (between $b^+\leftrightarrow p^-$) 
of a standard position space Wilson line starting at $b=(b^+,x^-,x_\perp)$,
\begin{align} \label{eq:Wn_FT}
  W_n(b;0,\infty) = \bar P \, \exp\Big( -\img g \int_0^\infty \!\! \df s \: \bn\cdot A_n(b+\bn s) \Big) \,.
\end{align}
Here the anti-path ordering $\bar P$ orders the colored matrices so that higher values of $s$ stand to the right.

In general the construction of the hard-scattering operators in $\cL_{\rm hard}^{(i)}$ results from integrating out offshell fluctuations with momenta that are further offshell than the soft and collinear modes in \SCETii, namely $p^2 \gg Q^2 \lambda^2$.  These offshell modes include the hard region of momentum space where $p^2\sim Q^2$, which is induced whenever collinear particles from two different sectors interact, such as $(p_n+p_\bn)^2\sim Q^2$. They also include the hard-collinear region of momentum space where $p^2\sim Q^2\lambda$, which is induced by interactions of collinear and soft particles, for example $(p_n + p_s)^2\sim Q^2\lambda$.  At leading power, integrating out these momentum regions leads to the presence of the collinear Wilson lines in \eq{Wn} as well as soft Wilson lines $S_n$ and $S_\bn$~\cite{Bauer:2001yt,Bauer:2002nz}, where
\begin{align} \label{eq:softW}
  S_n(b;0,\infty) &= \bar P \exp\Big( -ig \int_0^\infty \!\! \df s \: n\cdot A_s(b+n s) \Big) \,, \\
  S_n^\dagger(b;0,\infty) &= P \exp\Big( ig \int_0^\infty \!\! \df s \: n\cdot A_s(b+n s) \Big) \,. \nn
\end{align}
Here $P$ and $\bar P$ are path and anti-path ordering for the color matrices (with higher values of $s$ standing to the left and right respectively), the subscript $n$ indicates the path direction and we use $S_n$ and ${\cal S}_n$ for the fundamental and adjoint representations, respectively.  Our focus will be on SIDIS for electron scattering induced by a virtual photon, $e^- p \to e^- h X$. For this process there is a single operator obtained by integrating out offshell modes at leading power in SCET and to all orders in $\alpha_s$, which is\footnote{Throughout the body of the paper we work with equations that are gauge invariant under covariant gauge transformations, where the gauge fields vanish at infinity. To restore complete gauge invariance, such as in light-cone gauges, requires including transverse Wilson lines~\cite{Ji:2002aa,Garcia-Echevarria:2011ewi,Idilbi:2010im}.  We discuss these transverse Wilson lines at both LP and NLP in \app{Twilsonlines}.} 
\begin{align} \label{eq:Lhard_0}
 \cL_{\rm hard}^{(0)}(0) = \frac{\img e^2}{Q^2} J_{\bar e e}^\mu\,
  \sum_{f}     \int\!\! \df\omega_1\df\omega_2\, 
   C_f^{(0)}(\omega_1\omega_2) \: 
   (\bar{\chi}^f_{\bn,-\w_2})\,(S_{\bn}^\dagger S_{n})
   \gamma^\perp_\mu (\chi^f_{n,-\w_1}) 
 \,.
\end{align}
Here we sum over quark flavors $f$ and $J^\mu_{\bar e e} =  (-1) \bar e \gamma^\mu e$ is the electron vector current.
The tree level process $e^- q^f \to e^- q^f$ gives $C_f^{(0)}=Q_f +{\cal O}(\alpha_s)$, where the quarks $q^f$ have charge $Q_f|e|$.
The dimensionless Wilson coefficient $C_f^{(0)}(\omega_1\omega_2)$ encodes virtual corrections that arise from the hard scale $\omega_1\omega_2\sim Q^2$,
and is related to the space-like massless quark form factor.
$C_f^{(0)}$ is manifestly real for the SIDIS space-like kinematics where $\omega_1\omega_2<0$.
For example, after renormalization in the \MSbar scheme it is a function $C_f^{(0)}(\omega_1\omega_2,\mu)$ involving a perturbative series in $\alpha_s(\mu)$ and real logarithms $\ln(-\omega_1\omega_2/\mu^2)$.

The formalism for constructing SCET operators at any order in the power expansion is well developed. 
In particular the complete set of field-products that serve as operator building blocks has been worked out. For collinear fields this set is simply $\{\chi_{n,\omega}, {\cal B}_{n\perp,\omega}^\mu, \cP_\perp^\mu\}$, where all three of these are ${\cal O}(\lambda)$ in the power counting, see~\refcite{Marcantonini:2008qn}. For soft fields the \SCETii building blocks are quark and gluon fields dressed by soft Wilson lines
\begin{align} \label{eq:softPsiB}
   \psi_{s(n)} = S_{n}^\dagger q_s \sim \lambda^{3/2} 
   \,,\qquad
   {\cal B}_{s(n)}^{A\mu} = \biggl[ \frac{\img}{g} \frac{1}{in\cdot\partial_s} n_\nu G_s^{B\nu\mu} S_{n\,{\rm adj}}^{BA} \biggl] \sim \lambda \,,
\end{align}
together with soft derivatives $\img\partial_s^\mu\sim \lambda$. Note that there is only one type of transverse derivative in \SCETii, $i\partial_{s\perp}^\mu = \cP_\perp^\mu$. Also note that $n_\mu {\cal B}_{s(n)}^{A\mu} =0$, and that operators can also involve $\psi_{s(\bn)}\sim \lambda^{3/2}$ and $ {\cal B}_{s(\bn)}^{A\mu}\sim \lambda$.
To determine what operators are needed for a desired order in the power counting, one can make use of the SCET power counting theorem which only depends on the power counting order of operators and some topological properties of graphs. First results for this formula were obtained in Ref.~\cite{Bauer:2002uv}, and then extended to a complete theorem that fully accounts for Glauber induced operators in Ref.~\cite{Rothstein:2016bsq}. For \SCETii the theorem states that a graph will scale as $\lambda^\delta$ where
\begin{align} \label{eq:pc}
  \delta &= 6 - N^n -N^\bn -N^{nS}-N^{\bn S} 
  \\
  &\ + \sum_k \bigl[ (k-4) (V_k^n + V_k^\bn + V_k^S) + (k-3)(V_k^{nS}+V_k^{\bn S}) + (k-2) V_k^{n\bar n}  \bigr]
  \,. \nn
\end{align} 
Here $V_k^X$ counts the number of operators in the graph containing fields of the types in $X$ and having scaling $\lambda^k$. (The one exception is $V_k^{n\bar n}$ which can contain soft fields in addition to its $n$ and $\bn$ collinear fields.) The topological factors $N^X$ in \eq{pc} count the number of disconnected components obtained from a graph if all lines of types other than those in $X$ are erased.  For SCET we can have half-integer values for $k$ when there are an odd number of soft fermion fields, balanced in fermion number by a collinear fermion.   The leading power Lagrangians in \eq{Ldyn0} contribute to $V_4^n$, $V_4^\bn$, and $V_4^S$ respectively. The leading power Glauber Lagrangian $L_G^{(0)}$ has operators that contribute to $V_3^{nS}$, $V_3^{\bn S}$, and $V_2^{n\bn}$. 
When we say that we will study NLP corrections we mean those suppressed by a full integer power of $\lambda$, namely those that are ${\cal O}(\lambda)$ suppressed relative to LP. This is because the half-integer powers must always come together with another half-integer power term in order to not violate fermion number conservation in the factorized soft matrix elements.
A powerful method for counting the number of independent operators with a given field content and order in $\lambda$ is to make use of scalar operators of definite helicity, see \refscite{Moult:2015aoa,Moult:2016fqy}. For two collinear directions this enumeration of operators has been carried out up to ${\cal O}(\lambda^2)$ for scalar~\cite{Moult:2017rpl,Chang:2017atu} and vector and axial-vector currents~\cite{Feige:2017zci} in \SCETi. The \SCETi theory shares many of the features described above, but has ultrasoft modes with momentum $p_{\rm us}^\mu\sim Q\lambda^2$ instead of the soft modes. As explained in \refcite{Feige:2017zci}, many of the results described there carry over directly to the case of \SCETii, and we will make use of this in our analysis.

In particular, a nice way of obtaining results in \SCETii is to make use of \SCETi as an intermediate theory~\cite{Bauer:2002aj,Bauer:2003mga}, and thus carry out the matching in two stages, QCD $\to$ \SCETi $\to$ \SCETii. In this way the soft Wilson lines $S_n$ and $S_\bn$ of \SCETii are derived by exploiting the simple BPS field redefinition~\cite{Bauer:2001yt}. This field redefinition decouples ultrasoft gluons from the leading power collinear \SCETi Lagrangians and induces ultrasoft Wilson lines $Y_n$ in other \SCETi operators and Lagrangians. The definition of $Y_n$ is the same as in \eq{softW} but with $n\cdot A_{us}$ fields. For the collinear building blocks the field redefinition gives
\begin{align} \label{eq:BPS}
  \chi_{n,\omega} \to Y_n \chi_{n,\omega}\,,\qquad
  \cB_{n\perp,\omega}^\mu \to Y_n \cB_{n\perp,\omega}^\mu Y_n^\dagger \,,
\end{align}
and $\cP_\perp^\mu$ commutes with $Y_n$ since the fields in $Y_n$ do not carry ${\cal O}(\lambda)$ perpendicular momenta.
For example, the leading power quark current in \SCETi is 
\begin{align} \label{eq:SCET1O0}
  {\cal O}_{{\rm I}\,\mu}^{(0)} 
 &= \sum_f \int\!\! \df\omega_1\df\omega_2\, 
   C_f^{(0)}(\omega_1\omega_2) \: 
   (\bar{\chi}^f_{\bn,-\w_2})\,
   \gamma^\perp_\mu (\chi^f_{n,-\w_1}) 
   \\
 &= \sum_f \int\!\! \df\omega_1\df\omega_2\, 
   C_f^{(0)}(\omega_1\omega_2) \: 
   (\bar{\chi}^f_{\bn,-\w_2})\,(Y_{\bn}^\dagger Y_{n})
   \gamma^\perp_\mu (\chi^f_{n,-\w_1}) 
   \,. \nn
\end{align}
To obtain the second line we made the BPS field redefinition, which induces the ultrasoft Wilson lines $Y_{\bn}^\dagger Y_{n}$. Matching this operator to \SCETii we modify the scaling of momentum for the collinear fields and relabel the ultrasoft Wilson lines as soft, giving \eq{Lhard_0}.\footnote{More specifically, the statement is that all time-ordered products of the leading power \SCETi and \SCETii hard scattering operators computed with leading power Lagrangians and the same \SCETii states, are identical.} 

Within this setup at subleading power, the \SCETii Lagrangians that involve offshell hard-collinear propagators can be derived from time-ordered products of the simpler hard scattering and dynamical Lagrangians in \SCETi, as discussed in~\cite{Bauer:2003mga}. A nice feature is that the power counting of a \SCETi contribution immediately constrains the resulting order in \SCETii as follows:
\begin{align}  \label{eq:SCETitoiipc}
  \text{\SCETi at }\: {\cal O}(\lambda^k)
  \quad \Longrightarrow \quad
  \text{\SCETii at }\: {\cal O}(\lambda^{k/2+E}) \ \text{with}\ E\ge 0 
  \,.
\end{align}
This enables us to enumerate a finite number of terms that must be considered to obtain results at a desired order in the power expansion.  For our ${\cal O}(\lambda)$ analysis of SIDIS, we can infer from this construction that we will need to consider \SCETi operators up to ${\cal O}(\lambda^2)$ and the resulting power suppressed operators in \SCETii will either be ${\cal O}(\lambda^{1/2})$ or ${\cal O}(\lambda)$.  Finally, within this setup we will be able to easily exploit SCET reparameterization invariance~\cite{Manohar:2002fd}, which gives relations for Wilson Coefficients, and which have been extensively consider in \SCETi.

The \SCETii dynamical Lagrangian describes interactions between soft and collinear particles, and has a power expansion in $\lambda\ll 1$ of the form
\begin{align} \label{eq:Lscetii}
  {\cal L}_{\rm dyn} = {\cal L}_{\rm dyn}^{(0)} 
    +  {\cal L}_{\rm dyn}^{(1/2)} + {\cal L}_{\rm dyn}^{(1)} + \ldots \,,
\end{align}
where the ellipses denote terms that only contribute beyond NLP. Here the ${\cal L}_{\rm dyn}^{(1/2)}$ Lagrangian involves a single soft fermion field. For example, Ref.~\cite{Moult:2017xpp} constructed the contribution to this Lagrangian from Glauber quark exchange mediating an interaction between two collinear and two soft fields, 
\begin{align} \label{eq:L2half}
  {\cal L}_{Gns}^{(1/2)} = -4\pi\alpha_s e^{-ix\cdot\cP} \sum_n
  \Big( \bar \psi_s^{(n)} {\slashed \cB}_{s\perp}^{(n)} \frac{1}{\slashed\cP_\perp} {\slashed\cB}_{n\perp} \chi_n
  + \bar\chi_n {\slashed\cB}_{n\perp}\frac{1}{\slashed\cP_\perp} 
   {\slashed \cB}_{s\perp}^{(n)}\psi_s^{(n)}  \Big)
  \,.
\end{align}
A general feature of all subleading power \SCETii dynamic Lagrangians is that they must involve at least two $n$-collinear building block fields and two soft building block fields, or two $n$-collinear and two $\bn$-collinear building block fields. This is required in order to conserve momentum.  An example of an ${\cal L}_{\rm dyn}^{(1)}$ Lagrangian with a non-trivial hard-collinear coefficient function, obtained by matching \SCETi$\to$ \SCETii, was given in Ref.~\cite{Mantry:2003uz}, which constructed
\begin{align}\label{eq:L2one}
  {\cal L}_{\xi\xi q q}^{(1)} &= \int\!\! \df\omega \df k^+  
   \frac{\bar J_{\xi\xi q q}^{(0)}(\omega k^+)}{\omega k^+}
   \Big[\bar \chi_{n,-\omega} \slashed\bn P_L \chi_{n,\omega} \Big]
   \Big[ \bar \psi_{s,-k^+}^{(n)} \slashed n P_L \psi_{s,k^+}^{(n)} \Big] 
   +\ldots \,,
\end{align}
where $P_L=(1-\gamma_5)/2$ and we have made the tree level dependence on the hard-collinear scale $\omega k^+\sim Q^2\lambda$ explicit.  For brevity we have suppressed flavor indices, and left other spin and color combinations discussed in~\cite{Mantry:2003uz} in the ellipses.  A general lesson of the examples in \eqs{L2half}{L2one} is that the subleading power \SCETii dynamic Lagrangians are generated both by Glauber potentials as well as offshell hard-collinear propagators.
In Ref.~\cite{Chang:2022unpub} a rather extensive determination of ${\cal L}_{\rm dyn}^{(1/2)}$ and ${\cal L}_{\rm dyn}^{(1)}$ terms in \SCETii is given, however for reasons that will become clear below in our analysis in \secs{fact_vanishing_Lag}{fact_vanishing_soft}, we will not need the full results of that work here.

To organize the subleading power hard scattering operators in \SCETii we find it convenient to divide them into two categories
\begin{align} \label{eq:Lhard}
  \cL_{\rm hard}^{(i>0)} = \cL_{\rm h}^{(i)} + \cL_{\rm hc}^{(i)} \,.
\end{align}
Here the Wilson coefficients of the operators in $\cL_{\rm h}^{(i)}$ only contain physics from the hard scale $p^2\sim Q^2$, while the Wilson coefficients of the operators in $\cL_{\rm hc}^{(i)}$ can also contain contributions from the hard-collinear scale and thus can dependent on soft momenta. Since there are no $\cL_{\rm h}^{(1/2)}$ terms, the results we need to consider for our next-to-leading power analysis of SIDIS to ${\cal O}(\lambda)$ in the cross section, are $\cL_{\rm h}^{(1)}$, $\cL_{\rm hc}^{(1/2)}$, and $\cL_{\rm hc}^{(1)}$. The desired operators in $\cL_{\rm h}^{(1)}$ can be directly obtained from those in \SCETi, which have been enumerated to the required ${\cal O}(\lambda^2)$ order in \refcite{Feige:2017zci}.  Furthermore, the terms in $\cL_{\rm hc}^{(i)}$ are always obtained from a time-ordered product in \SCETi that involves at least one subleading power \SCETi Lagrangian.\footnote{This follows because the offshell hard-collinear propagators that generate the desired Wilson coefficients come from propagating collinear degrees of freedom in the \SCETi, and time-ordered products with leading power Lagrangians only involve fields in a single sector, and hence when considering the matching will not leave behind contributions that are pinned at the hard-collinear scale.} Unlike the dynamic \SCETii Lagrangians, the power suppressed ${\cal L}_{\rm hc}^{(i)}$ terms can involve only one soft building block field, since the soft momentum can flow into the leptonic current. The \SCETii terms $\cL_{\rm hc}^{(1/2)}$ and $\cL_{\rm hc}^{(1)}$ that we need to consider for our analysis are obtained from the following time-ordered products in \SCETi
\begin{align}  \label{eq:SCET1Tproducts}
  T\big[ {\cal O}_{\rm I}^{(0)} {\cal L}_{\rm I}^{(1)} \big] 
  \,,\qquad
  T\big[ {\cal O}_{\rm I}^{(0)} {\cal L}_{\rm I}^{(2)} \big] 
  \,,\qquad
  T\big[ {\cal O}_{\rm I}^{(0)} {\cal L}_{\rm I}^{(1)} {\cal L}_{\rm I}^{(1)} 
  \big] 
  \,,\qquad
  T[ {\cal O}_{\rm I}^{(1)} {\cal L}_{\rm I}^{(1)} ]
  \,,
\end{align}
where the subscript ${\rm I}$ indicates \SCETi operators and Lagrangians, with ${\cal O}_{\rm I}^{(0)}$ given in \eq{SCET1O0}. The required subleading power \SCETi Lagrangians $\cL_{\rm I}^{(1)}$ and $\cL_{\rm I}^{(2)}$ are given in \refcite{Bauer:2003mga}, and the ones we need will be given in later sections.  The relevant ${\cal O}_{\rm I}^{(1)}$ operators can be found in \refcite{Feige:2017zci}, and will also be given in \secs{RPI}{operators_NLP_B}.

With this formalism in hand, we can revisit the summary of the different sources for NLP contributions to SIDIS that we must consider for our analysis. They are:
\begin{itemize}
 \item Kinematic power corrections, for example from expanding the projectors in \eq{def_P_i} in the factorization frame
  \item Hard scattering power corrections from the hard region through ${\cal L}_{\rm h}^{(1)}$
  \item Hard scattering power corrections from the hard-collinear region through ${\cal L}_{\rm hc}^{(1)}$ and $T\big[ {\cal L}_{\rm hc}^{(1/2)} {\cal L}_{\rm dyn}^{(1/2)}\big]$
  \item Lagrangian insertions involving the leading power hard scattering operator, through the time-ordered products $T\big[ {\cal L}_\hard^{(0)} {
\cal L}_{\rm dyn}^{(1/2)} {\cal L}_{\rm dyn}^{(1/2)}\big]$ and $T\big[ {\cal L}_\hard^{(0)} {\cal L}_{\rm dyn}^{(1)}\big]$
\end{itemize}
Except for the kinematic corrections, all sources of power corrections are related to the \SCETii Lagrangians at subleading power.
All four of these sources for power corrections will be analyzed in \sec{factorization}, and final results for the NLP $W_i$s will be given in \sec{results}.

\section{SIDIS Factorization to Next-to-Leading Power (NLP)}
\label{sec:factorization}

In this section, we derive the factorization formula for SIDIS in the limit of small transverse momentum.
More precisely, we study the hadronic structure functions $W_i$
that enter the angular decomposition of the SIDIS cross section in \eqs{xs_SIDIS_unpolarized}{xs_SIDIS_pol}.
The $W_i = P_i^{\mu\nu} W_{\mu\nu}$ are projections of the hadronic tensor $W^{\mu\nu}$
onto the basis of projectors defined in \eq{def_P_i}, and thus studying their factorization
is equivalent to factorizing $W^{\mu\nu}$ itself. The advantage of considering the $W_i$
is that it takes into account power corrections from the projectors themselves,
thereby yielding the power expansion of the SIDIS cross section.

Our derivation is based on SCET, reviewed in \sec{SCET_review}, and follows the procedure of \refcite{Stewart:2009yx} for the treatment of label and residual momentum in the multipole expansion.
We first define a symbolic power counting parameter
\begin{align}
 \lambda \sim P_{hT}/Q
\,,\end{align}
in terms of which we can expand the structure functions as
\begin{align}
 W_i &
 = W_i^{(0)} + W_i^{(1)} + \cdots
\,.\end{align}
Leading-power (LP) contributions $W_i^{(0)}$ scale as $\cO(\lambda^{-2})$,
while next-to-leading power (NLP) contributions $W_i^{(1)}$ have a relative suppression by one power of $\lambda$
and thus scale as $\cO(\lambda^{-1})$, and so on.
We will only consider structure functions up to NLP,
as starting at NNLP most structure functions receive corrections from subleading Lagrangian contributions
that are beyond the scope of this paper.

While the power expansion of the projectors $P_i^{\mu\nu}$ is straightforward,
the factorization of the hadronic tensor is highly nontrivial.
Recall its definition in \eq{tensors},
\begin{align}\label{eq:W}
 W^{\mu\nu}(q, \Pa, \Pb) &
 = \SumInt_X \, \delta^4(q + \Pa  - \Pb - \PX) \,
   \braket{N | J^{\dagger\,\mu}(0) | h,X} \braket{h,X | J^\nu(0) | N}
\nn\\&
 = \SumInt_X \int\!\frac{\df^4 b}{(2\pi)^4} \, e^{\img b\cdot q}
   \braket{N | J^{\dagger\,\mu}(b) | h,X} \braket{h,X | J^\nu(0) | N}
\,,\end{align}
where for the moment we neglect polarizations of the target nucleon $N$.
In \eq{W}, we abbreviate the sum over all states $X$ and the corresponding phase space integral by $\SumInt_X$,
and in the second line used momentum conservation to shift the position of the first current.

While \eq{W} is manifestly Lorentz covariant, explicit expressions do depend on the choice of frame
in which the factorization it is discussed. From now on, we will always neglect hadron masses
and work in the factorization frame characterized by \eq{Pa_Pb},
\begin{align} \label{eq:Pa_Pb_rep}
 \Pa^\mu &= \frac{Q}{x} \frac{n^\mu}{2}
\,,\qquad
 \Pb^\mu = z Q \frac{\bn^\mu}{2}
\,,\end{align}
where $n^\mu$ and $\bn^\mu$ are lightlike reference vectors. This $n$ and $\bn$ define the reference vectors for our two collinear SCET sectors in this back-to-back frame.  In addition we have a soft sector as discussed in \sec{SCET_review}.
For most of our results, the precise form of $P_N^-$ and $P_h^+$ in \eq{Pa_Pb_rep}
will not matter, but occasionally we will explicitly make use of this particular choice to simplify results.

As already mentioned in the introduction, in this paper we make the assumption that  Glauber interactions from the SCET Lagrangian ${\cal L}_G^{(0)}$ do not spoil factorization at NLP.  We then work out the full form that the factorization formula must take for the set of $W_i$s which themselves start off at NLP.

The organization of our analysis below is as follows.
We start in \sec{operators} by discuss the hard scattering operators needed to NLP in \SCETii,
with the general setup in \sec{operators_setup}, operators induced by integrating out hard interactions
in sections~\ref{sec:operators_LP}--\ref{sec:operators_NLP_Bs},
and from integrating out hard-collinear interactions using \SCETi as an intermediate theory in \sec{operators_NLP_Bs}.
In \sec{factorization_LP} we review LP factorization using SCET, keeping track of the NLP contributions that come from implementing momentum conservation.
In \sec{factorization_kinematic_corr} we discuss kinematic power corrections that enter from subleading contributions to the projectors $P_i^{\mu\nu}$ (associated to differences between the Trento and factorization frames).
In \sec{fact_vanishing_Lag} we show that contributions from \SCETii Lagrangian insertions vanish for SIDIS structure functions that start at NLP.
In \sec{fact_vanishing_soft} we demonstrate that soft contributions involving derivatives $n\cdot i\partial_s$, $\bn\cdot i\partial_s$ or the soft gluon building blocks $\bn\cdot {\cal B}_{s}^{(n)}$, $n\cdot {\cal B}_{s}^{(\bn)}$, and $\cB_{s\perp}^{(n_i)\mu}$ vanish for all $W_i$ at NLP,
and also discuss the vanishing at NLP of operators induced by the hard-collinear region of momentum space.
Finally, we discuss the nonvanishing contributions from operators including $\cP_\perp$ and $\cB_{n_i\perp}$ insertions
in \sec{factorization_P} and \sec{factorization_B}, respectively.

\subsection{Hard Operators in SCET}
\label{sec:operators}

\subsubsection{General Setup}
\label{sec:operators_setup}

The currents $J^\mu$ in \eq{W} are the currents in full QCD. In the case considered here,
only the vector current $J_{\bq q}^\mu  = \bar q \gamma^\mu q$ contributes, see \eq{current}.
This current has to be matched onto the corresponding current in \SCETii,
which contains operators built from soft and collinear quark and gluon fields and the corresponding Wilson coefficients.
Schematically, this reads
\begin{align} \label{eq:current_matching_schematic}
 J_{\bq q}^\mu(x) = \sum_\cO \sum_{j=0}^\infty J_{\cO}^{(j)\,\mu}
\,,\end{align}
where we sum over all relevant SCET operators $\cO$, and where $J_\cO^{(j)}$
denotes the corresponding current at $\cO(\lambda^j)$ in the power expansion.

The current in \eq{current_matching_schematic} couples to the corresponding
leptonic vector current, see \eq{M_SIDIS}. Since we work at tree level
in the electroweak theory, the leptonic current does not receive any corrections
and thus has trivial matching in the effective field theory (EFT).
This also implies that the photon field is not dynamic, and formally can be integrated out of the theory.
Thus, rather than performing the matching onto SCET at the level of the QCD current
as illustrated in \eq{current_matching_schematic}, we can equivalently consider the hard scattering Lagrangian
\begin{align}
 \cL_{\rm hard} = \frac{\img e^2}{Q^2}\, J_{\bar \ell \ell\,\mu} \sum_q J_{\bq q}^\mu
\,,\end{align}
whose power expansion follows from \eq{current_matching_schematic} as
\begin{align} \label{eq:Lhard_withJ}
 \cL_\hard = \sum_{j=0}^\infty \cL_\hard^{(j)}
\,,\qquad
 \cL_\hard^{(j)} = \frac{\img e^2}{Q^2}\, \sum_\cO J_{\bar \ell \ell\,\mu} J_{\cO}^{(j)\,\mu}
\,.\end{align}
Note that matrix elements like $\langle \img {\cal L}_{\rm hard} \rangle$ give rise to the amplitude, and the $e^2/Q^2$ prefactors are from the virtual photon exchange as in \eq{M_SIDIS}. The conjugate amplitudes are instead obtained from the matrix elements $\langle -\img {\cal L}_{\rm hard}^\dagger \rangle$. We do not impose the restriction of hermiticity on ${\cal L}_{\rm hard}$ in this setup.\footnote{The fact that ${\cal L}_{\rm hard}^\dagger\ne {\cal L}_{\rm hard}$ is clear from the example of leading power TMDs in DY where $C^{(0)}(\omega_1\omega_2)$ in \eq{Lhard_0} is complex even though the operator is related by hermitian conjugation together with flipping $n_1\leftrightarrow n_2$ and $\omega_1\to\omega_2$. These flips do not modify the Wilson coefficient.}
This approach has several advantages.
First, the contraction with the leptonic current eliminates some contributions to $J_{\bq q}^\mu$,
and thus facilitates the construction of a minimal basis for the hard matching.
Second, higher-order QCD corrections to the required amplitudes are typically
evaluated using the spinor-helicity formalism which is more naturally used with $\cL_\hard$.
For this reason, this approach was previously used both at LP and NLP~\cite{Moult:2015aoa,Feige:2017zci}.

The power-suppressed hard Lagrangians receive contributions which from \eq{Lhard} can be divided into two categories: $\cL_{\rm h}^{(i)}$ whose Wilson coefficients contain contributions from the hard scale $\sim Q^2$, and $\cL_{\rm hc}^{(j)}$ whose Wilson coefficients also get contributions from the  hard-collinear scale.  The decomposition in \eq{Lhard_withJ} applies equally well to both of these contributions, so they both can be discussed in the language of hadronic currents.
In the sections below we will first focus on enumerating all $\cL_{\rm h}^{(j)}$ contributions, before turning in \sec{operators_NLP_Bs} to $\cL_{\rm hc}^{(j)}$ contributions. It is important to note that at leading power there is only a $\cL_{\rm h}^{(0)}$ contribution, whereas the hard-collinear contributions start with $\cL_{\rm hc}^{(j\ge 1/2)}$.
For ${\cal L}_{\rm h}^{(j)}$ we can make \eq{Lhard_withJ} more explicit by factoring out the hard $p^2\sim Q^2$ flucutuations into Wilson coefficients $C^{(j)}$ with convolutions involving ${\cal O}(\lambda^0)$ momenta~\cite{Bauer:2001yt}, by writing
\begin{align} \label{eq:Lsub}
 \cL_{\rm h}^{(j)}(0) &
 = \sum_{\{n_i\}} \sum_{\{\lambda_i\}} \int\!\df\w_1 \cdots \df\w_n \,
   \nn\\&\qquad\times
   \cO^{A_1 \cdots a_n\,(j)}_{\{\lambda_i\}} \bigl( \{n_i\}; \{\w_i\} \bigr)
   \, C_{\{\lambda_i\}}^{A_1 \cdots a_n\,(j)} \bigl( \{n_i\}; \{\w_i\} \bigr)
  \,.
\end{align}
Here the Lagrangian is taken at position $b=0$ since this is what we will need when making use of the hard scattering currents (and we can easily translate to include the full position space, respecting the SCET multipole expansion).  The operators
are built from fields with fixed label momenta
\begin{align}
 \tilde p_i^\mu = \w_i \frac{n_i^\mu}{2}
\,.\end{align}
In \eq{Lsub}, the first sum runs over all possible collinear reference directions $\{n_i\}$,
which will be fixed by the hard process in consideration.
The second sum runs over all operators contributing at this order,
which are labeled by the set of parameters $\{\lambda_i\}$.
For example, if one organizes the operators by particle helicities,
the labels would correspond to the individual helicities.
For an operator consisting of $n$ fields (counting the lepton fields as well),
the $\w_1 \cdots \w_n$ are the label momenta of the fields, and are integrated over.
The operators $\cO^{A_1 \cdots a_n\,(j)}_{\{\lambda_i\}}$
and Wilson coefficients $C_{\{\lambda_i\}}^{A_1 \cdots a_n\,(j)}$
are correspondingly labeled by the $\{\lambda_i\}$, and depend on both
the reference directions $\{n_i\}$ and the label momenta $\{\w_i\}$.
The superscripts denote the color index of the $n$ particles,
with $A_i, a_i$ and $\bar a_i$ denoting adjoint, fundamental
and antifundamental colors, and summation over colors is implied.
With respect to color, \eq{Lsub} can be interpreted as having distinct
Wilson coefficients for each color configuration.  

For more complicated operators involving multiple colored particles,
the color structure in \eq{Lsub} becomes quite involved, and it is convenient
to express operators and Wilson coefficients as vectors in color space,
see \refscite{Moult:2015aoa,Feige:2017zci} for a discussion in the context of SCET matching.
In our case, the color space will be trivial.
At LP, the hard scattering operator is built from $q^a$ and $\bar q^{\bar b}$ fields,
and thus the unique color structure is
\begin{align}
 \hat T^{a \bar b} = \bigl(\delta_{a\bar b}\bigr)
\,.\end{align}
At NLP, we will at most have operators built from $g^A$, $q^a$ and $\bar q^{\bar b}$ fields,
which is still described by a unique color structure,
\begin{align}
 \hat T^{A\,a\bar b} = \bigl(T^A_{a\bar b}\bigr)
\,.\end{align}
Due to this simplicity, we refrain from using more advanced treatments of the appearing color structures,
but note that starting at NNLP operators with four colored fields arise,
in which case a nontrivial color algebra becomes necessary.

\subsubsection{Leading Power}
\label{sec:operators_LP}

At leading power, the only hard operator is~\cite{Stewart:2009yx}
\begin{equation}\label{eq:op_lp}
  \cO_f^{(0)\alpha\beta}(\lp_1,\lp_2)=\bar{\chi}_{n_2,-\w_2}^{f \alpha \bar a}\, 
   \left[ S_{n_2}^\dagger S_{n_1}\right]^{a \bar b}\chi_{{n_1},-\w_1}^{f \beta b}
\,,\end{equation}
where we sum over the spin indices $\alpha$ and $\beta$ and the color indices $\bar a$, $b$.
They depend on the light-cone directions $n_1$ and $n_2$, which so far are still arbitrary,
and the corresponding light-cone momenta
\begin{equation}
 \lp_1^\mu = \w_1 \frac{n_1^\mu}{2}
\,,\qquad
 \lp_2^\mu = \w_2\frac{{n}_2^\mu}{2}
\,.\end{equation}

Using the notation in \eq{Lhard_withJ}, the leading-power hard current corresponding to \eq{op_lp} is given by
\begin{equation} \label{eq:crrnt_lp}
 J^{(0)\,\mu}(0)
 = \sum_{n_1,n_2} \int\!\df\w_1\,\df\w_2\, 
   \sum_f C_{f}^{(0)\,\mu\,\alpha\beta}(\lp_1,\lp_2) \, \cO_f^{(0)\alpha\beta}(\lp_1,\lp_2)
\,.\end{equation}
Here, the sum runs over all possible light-cone directions, which will be fixed once acting with this operator on specific states,
as well as the flavor $f$ of the quark initiating the hard scattering.
The Wilson coefficient $C_f^{(0)\mu\,\alpha\beta}$ in \eq{crrnt_lp} is written as
\begin{align}\label{eq:hard_lp}
 C_{f}^{(0)\,\mu\,\alpha\beta}(\lp_1,\lp_2) &
 = (\gamma_{\perp}^\mu)^{\alpha\beta}\, C_f^{(0)}(\tilde q^2)
 \nn\\&
 = (\gamma_{\perp}^\mu)^{\alpha\beta}\, \Bigl[ Q_f C_{q}^{(0)}(\tilde q^2) 
    + \sum_{f'} Q_{f'} C_{vf'}^{(0)}(\tilde q^2) \Bigr]
\,.\end{align}
Here, we made explicit that by Lorentz invariance the right-hand side can only depend on $\tilde  q^2 = (\lp_1+\lp_2)^2= 2 \lp_1 \cdot \lp_2 = \w_1 \w_2\,  (n_1\cdot n_2)/2 $.
At LP we have the Dirac structure $\gamma_{\perp}^\mu$, which is orthogonal to both $n_1$ and $n_2$, so $\gamma_\perp^\mu = \gamma^\mu - (n_1^\mu \slashed{n}_2+n_2^\mu\slashed{n}_1)/(n_1\cdot n_2)$.
In \eq{hard_lp}, $C_{vf'}^{(0)}$ encodes contributions to the quark form factor from closed quark loops,
such that the quark coupling to the virtual photon is not identical to the quark extracted from the proton.
It first contributes at three loops, $C_{vf'}^{(0)} = \cO(\as^3)$.

From now on, we will mostly leave implicit the quark flavor $f$ superscript on the collinear quark fields using $\chi_n$ in place of $\chi_n^f$, and do the same for Wilson coefficients and operators, eg.~writing $C^{(0)}$ instead of $C_f^{(0)}$.  We will also suppress color indices when the contractions are clear. 

We also define the conjugate operator and Wilson coefficient such that the usual factor of $\gamma^0$ are included in the latter,
\begin{align} \label{eq:Odagger}
 \cO^{(0)\,\dagger\beta\alpha}(\lp_1,\lp_2) &
 = \bar{\chi}_{{n}_1,\w_1}^{\beta}\, S_{n_1}^\dagger S_{n_2}  \,\chi_{n_2,\w_2}^{\alpha}
 =\cO^{(0)\beta\alpha}(-\lp_2,-\lp_1)
\,,\nn\\
 \bar C_{\mu}^{(0)\,\mu\,\beta\alpha}(\lp_1,\lp_2) &
 = \Bigl[ \gamma^0 C^{(0)\,\mu\,\dagger}(\lp_1,\lp_2) \gamma^0 \Bigr]^{\beta\alpha}= (\gamma_{\perp}^\mu)^{\beta\alpha}\, C^{(0)}(\tilde q^2)
\,.\end{align}
The Wilson coefficient $C^{(0)}$ in the \MSbar scheme can be calculated from the quark form factor in pure dimensional regularization. For the space-like case $\tilde q^2<0$ that is relevant here, both the form factor and $C^{(0)}(\tilde q^2)$ are real, explaining why there is no need for a complex conjugation in \eq{Odagger}.

\subsubsection{NLP Operators involving ${\cal P}_\perp$, $\partial_s$, ${\cal B}_s^{(n)}$, ${\cal B}_s^{(\bn)}$ and RPI Constraints}
\label{sec:RPI}

We now extend our discussion of hard scattering operators to subleading power for ${\cal L}_h^{(1)}$,
i.e.~operators suppressed by $\cO(\lambda)$ that appear after integrating out hard scale fluctuations. Our analysis is based on \refcite{Feige:2017zci},
which derived the complete set of subleading operators using the spinor-helicity formalism.  The desired NLP \SCETii operators that appear at ${\cal O}(\lambda)$ involve the operator structures:
\begin{align}
  \bar\chi_{n_1} \cP_\perp^\mu \chi_{n_2} \,, \quad
  \bar\chi_{n_1} {\cal B}_{n_i\perp}^\mu \chi_{n_2} \,, \quad   
  \bar\chi_{n_1} \img n_1\cdot \partial_s \chi_{n_2} \,, \quad
  \bar\chi_{n_1} \img  n_2\cdot\partial_s \chi_{n_2} \,, \quad
  \bar\chi_{n_1} {\cal B}_s^{(n_i)\mu} \chi_{n_2} \,. 
\end{align}
In this section we will obtain the operators that involve ${\cal P}_\perp^\mu$, $n_1\cdot\partial_s$, $n_2\cdot\partial_s$, $\bn_2\cdot {\cal B}_s^{(n_1)}$, and $\bn_1\cdot {\cal B}_s^{(n_2)}$, for which it turns out that the Wilson coefficients are related to the leading power Wilson coefficient $C^{(0)}(\tilde q^2)$ in \eq{hard_lp}. 
There is also a partial constraint on the operator ${\cal B}_{s\perp}^{(n_i)\mu}$ discussed here, and this operator will be taken up in full in \sec{operators_NLP_Bs}.  Finally, there is no constraint on operators involving 
${\cal B}_{n_1\perp}^\mu$, which will be taken up independently in \sec{operators_NLP_B} below.

We start by considering the analog hard scattering operators in \SCETi, and constraints on the Wilson coefficients of these operators from the reparameterization invariance (RPI) symmetry of SCET~\cite{Manohar:2002fd}. The results for RPI relation for subleading power \SCETi dijet operators are available in the literature, see eg.~\refscite{Marcantonini:2008qn,Larkoski:2014bxa,Feige:2017zci}, and we will review results we need here. 
Performing the matching of \SCETi $\to$ \SCETii we then obtain the analogous constraints on operators for \SCETii, since the hard Wilson coefficients for operators in ${\cal L}_{\rm h}^{(1)}$ are not changed by this matching. 

Reparameterization invariance (RPI) symmetry in SCET arises because of two freedoms in the construction: i) the freedom of how to divide up hierarchically large and small momentum components, and ii) the freedom to modify the choice of the reference vectors $n$ and $\bn$ for each collinear sector~\cite{Manohar:2002fd}. This RPI symmetry relates the properties of operators that appear at different orders in the power expansion. In \SCETi, freedom i) connects operators with collinear and ultrasoft derivatives. When combined with gauge symmetry it can be implemented by the replacements
\begin{align}
 W_n^\dagger \img D_{n\perp}^\mu W_n & \to 
  W_n^\dagger \img {\cal D}_{n\perp}^\mu W_n \equiv
  W_n^\dagger \img D_{n\perp}^\mu W_n + \img D_{{\rm us}\perp}^\mu\,,
 \\
  \bar\cP_n & \to 
   \img \bn\cdot \partial_{n}^{\rm RPI}\equiv 
    \bar\cP_n + i\bn\cdot D_{\rm us}
 \,,\nn 
\end{align}
in collinear operators, with a subsequent expansion in $\lambda$ giving  operators with the ultrasoft derivatives that are constrained to have the same hard Wilson coefficients (unless there is more than one source for the specific operator with ultrasoft fields).  To implement constraints from ii) there are two possible equivalent methods, working out the RPI transformations on a basis of operators and then constructing linear combinations of operators that are invariant as in \refcite{Manohar:2002fd}, or constructing RPI-invariant operators and then expanding them in $\lambda$ as in \refcite{Marcantonini:2008qn}. We will use this second approach here.

The RPI and gauge invariant operator whose power expansion yields the field structure of the leading power \SCETi hard scattering operator in \eq{SCET1O0} is given by
\begin{align} \label{eq:Orpi}
  &C^{(0)}(\tilde q^2) \bigl(\bar \psi_{n_2}\cW_{n_2}\bigr)
 \delta\bigl(\omega_2-\img\, \bar n_1\! \cdot\! \overleftarrow \partial_{n_2}^{\rm RPI} \bigr) 
  \gamma_\mu \,
  \delta(\omega_1+\img\, \bar n_1 \!\cdot \partial_{n_1}^{\rm RPI} \bigr) 
  \bigl(\cW_{n_1}^\dagger \psi_{n_1} \bigr)
  \nn\\
 &\quad 
 = C^{(0)}(\tilde q^2)\, (\bar\chi_{n_2,-\omega_2}) \gamma^\perp_\mu (\chi_{n_1,-\omega_1}) + \ldots \,,
\end{align}
where $\tilde q^2=\omega_1\,\omega_2\, n_1\cdot n_2/2$,  $\psi_n$ is an RPI-invariant fermion field, ${\cal W}_n$ is an RPI invariant Wilson line, and the $\delta$-functions have been made invariant under the RPI of type i). We can also consider additional terms to make the $\delta$-functions invariant under type ii) RPI, as done for example in Ref.~\cite{Larkoski:2014bxa}, but the additional terms do not play a role for the relations between \SCETii operators that we derive here.  The ellipses in \eq{Orpi} denote power suppressed terms that may be connected by RPI to this leading power operator, and thus be constrained to depend on the same Wilson coefficient $C^{(0)}(\tilde q^2)$.

To obtain the terms needed for the translation to constraints in \SCETii we should expand the operator in \eq{Orpi} to ${\cal O}(\lambda)$ for subleading collinear terms, and to ${\cal O}(\lambda^2)$ for subleading ultrasoft terms. 
The field $\psi_{n_i}$ has the expansion~\cite{Marcantonini:2008qn}
\begin{align}  \label{eq:psin}
  \psi_{n_i}&=\left(1 +\frac{1}{\img \bn_i\cdot D}\img\slashed {\cal D}_{n_i\perp} \frac{\slashed\bn_i}{2}\right) \xi_{n_i}+\cO(\lambda^3)\nn\\
&=\xi_{n_i}
 +W_{n_i}\frac{1}{\overline\cP_{n_i} }
  \left(\slashed\cP_{n_i\perp}+g\slashed\cB_{n_i\perp} 
   + \img \slashed D_{{\rm us}\perp} \right) \frac{\slashed\bn_i}{2} W_{n_i}^\dagger \xi_{n_i}+\cO(\lambda^3)\,.
\end{align}
The RPI version collinear Wilson line is given by
\begin{align} \label{eq:cW}
  \cW_{n_i}=W_{n_i}e^{-\img R_{n_i}} 
    = W_{n_i} \big( 1 - \img R_{n_i}^{(1)} +\ldots \big)
    \,,
\end{align}
where $R_{n_i}^{(1)}\sim\Ord(\lambda)$ and full expressions for the expansion are given in \refcite{Marcantonini:2008qn}. We note that in terms of field structures, $R_{n_i}^{(1)}$ contains a $(1/\bnP_{n_i} )\cB_{n_i\perp}$.
Finally, we need to expand the $\delta$-functions in \eq{Orpi}, which gives
\begin{align}
 \delta(\omega_i+\img\, \bar n_i \!\cdot \partial_{n_i}^{\rm RPI} \bigr) 
 &=\Bigl(1+\img\bn_i\cdot D_{us}\frac{\partial}{\partial\omega_i}\Bigr)
 \delta(\omega_i+\img\bn_i\cdot\partial_{n_i}) 
  +{\cal O}(\lambda^4) 
\,.
\end{align}
Here the $\partial/\partial\omega_i$ derivative can be integrated by parts onto the Wilson coefficient.
Combining these expansions together to determine the ellipses in \eq{Orpi}, and making the BPS field redefinition, we find that the terms to the order we need are
\begin{align}  \label{eq:Orpiexpn}
 &C^{(0)}(\tilde q^2) \bigl(\bar \psi_{n_2}\cW_{n_2}\bigr)
 \delta\bigl(\omega_2-\img\, \bar n_1\! \cdot\! \overleftarrow \partial_{n_2}^{\rm RPI} \bigr) 
  \gamma_\mu \,
  \delta(\omega_1+\img\, \bar n_1 \!\cdot \partial_{n_1}^{\rm RPI} \bigr) 
  \bigl(\cW_{n_1}^\dagger \psi_{n_1} \bigr)
  \\
 &\quad 
 = C^{(0)}(\tilde q^2)\, 
  (\bar\chi_{n_2,-\omega_2})
  \big(1-\img R_{n_2}^{(1)} \big)
  (Y_{n_2}^\dagger Y_{n_1}) \gamma^\perp_\mu 
  \big(1+\img R_{n_1}^{(1)\dagger} \big)  
  (\chi_{n_1,-\omega_1})  
  \nn\\
 &\quad +C^{(0)}(\tilde q^2)\, \bigg\{
  (\bar\chi_{n_2,-\omega_2})
  (Y_{n_2}^\dagger Y_{n_1}) 
   \bigg[ \gamma^\mu \frac{1}{\bnP_{n_1}}\slashed\cP_{n_1\perp} 
    \frac{\slashed\bn_1}{2} + \frac{\slashed\bn_2}{2} \slashed\cP_{n_2\perp}^\dagger 
    \frac{1}{\bnP_{n_2}^\dagger} \gamma^\mu 
   \bigg] (\chi_{n_1,-\omega_1})  
  \nn\\
  &\qquad\qquad
  +(\bar\chi_{n_2,-\omega_2})
  (Y_{n_2}^\dagger Y_{n_1}) 
   \bigg[ \gamma^\mu \frac{1}{\bnP_{n_1}} g\slashed\cB_{n_1\perp}
    \frac{\slashed\bn_1}{2} + \frac{\slashed\bn_2}{2} g\slashed\cB_{n_2\perp}
    \frac{1}{\bnP_{n_2}^\dagger} \gamma^\mu 
   \bigg] (\chi_{n_1,-\omega_1}) 
 \nn\\
  &\qquad\qquad
  +(\bar\chi_{n_2,-\omega_2})
   \bigg[ \gamma^\mu \frac{1}{\bnP_{n_1}}  
    \Big(Y_{n_2}^\dagger \img \slashed D_{{\rm us}\perp} Y_{n_1} \Big)  
    \frac{\slashed\bn_1}{2} + \frac{\slashed\bn_2}{2} 
    \Big(Y_{n_2}^\dagger (-\img)  \overleftarrow {\slashed D}_{{\rm us}\perp} Y_{n_1} \Big)  
    \frac{1}{\bnP_{n_2}^\dagger} \gamma^\mu 
   \bigg] (\chi_{n_1,-\omega_1}) 
  \bigg\}
 \nn\\
 &\quad - \frac{\partial C^{(0)}(\tilde q^2)}{\partial\omega_1} \, 
  (\bar\chi_{n_2,-\omega_2})
  \big(Y_{n_2}^\dagger \img\bar n_1\cdot D_{us} Y_{n_1}\big) 
  \gamma^\perp_\mu 
  (\chi_{n_1,-\omega_1})  
 \nn\\
 &\quad - \frac{\partial C^{(0)}(\tilde  q^2)}{\partial\omega_2} \, 
  (\bar\chi_{n_2,-\omega_2})
  \big(Y_{n_2}^\dagger (-\img)\bar n_2\cdot \overleftarrow D_{us} Y_{n_1}\big) 
  \gamma^\perp_\mu 
  (\chi_{n_1,-\omega_1})   \,.
  \nn
\end{align}

To determine whether \eq{Orpiexpn} constrains the Wilson coefficients of subleading power operators we need to know whether there are other RPI invariant operators or mechanisms by which these operators can be generated. From \refcite{Marcantonini:2008qn} we know there are other RPI operators whose leading expansion gives terms involving $\cB_{n_i\perp}$, so these structures are not constrained. This means we can simply ignore the $R_{n_2}^{(1)}$, $R_{n_1}^{(1)\dagger}$, $\slashed \cB_{n_1\perp}$ and $\slashed \cB_{n_2\perp}$ terms in \eq{Orpiexpn}. We will study the $\cB_{n_i\perp}$ operators independently using the helicity basis in \sec{operators_NLP_B}.  A more tricky set of operators are the two that involve $\img\slashed D_{{\rm us}\perp}$.  In \SCETi these operators are constrained by RPI with the Wilson coefficient given in \eq{Orpiexpn}.  However, in \SCETii there is another source that contributes to the matching onto these operators from the hard-collinear scale through ${\cal L}_{\rm hc}^{(1)}$, and hence their Wilson coefficients are not simply constrained by \eq{Orpiexpn} alone.  We will leave these operators for now, and take them up again in detail in \sec{operators_NLP_Bs} below.

The remaining operators in \eq{Orpiexpn} are constrained by RPI when matched onto \SCETii. In this procedure ultrasoft fields become soft, so the Wilson lines $Y_{n_i}\to S_{n_i}$, and  $\bn_i\cdot D_{\rm us} \to \bn_i\cdot D_s\sim \lambda$, so the last two terms in \eq{Orpiexpn} also contribute at NLP. For the \SCETii operators with $\cP_\perp$, writing the contributions to ${\cal L}_h^{(1)}$ with currents following \eq{Lhard_withJ} we find
 \begin{align}\label{eq:crrnt_P1}
  J^{(1)\mu}_{\cP_\perp}(0) 
   &= \sum_{n_1,n_2}\int\!\df\w_1\,\df\w_2\,
  \sum_f C^{(0)}(\tilde q^2)\, 
  \cO_{\cP_\perp }^{(1)\mu}(\tilde p_1,\tilde p_2)
   \,,\\
  J^{(1)\mu}_{\cP^\dagger_\perp}(0)
   &= \sum_{n_1,n_2}\int\!\df\w_1\,\df\w_2\,
  \sum_f C^{(0)} (\tilde q^2)\,
  \cO_{\cP^\dagger_\perp }^{(1)\mu}(\tilde p_1,\tilde p_2)
  \,, \nn
\end{align}
where
\begin{align}\label{eq:op_P1}
  \cO_{\cP_\perp }^{(1)\mu}
   &=-\frac{1}{2\w_1} \big[\bar{\chi}_{n_2,-\w_2} (S_{n_2}^\dagger S_{n_1}) \,\gamma^\mu\, \slashed{\cP}_{n_1\perp}\,\slashed\bn_1  \chi_{n_1,-\w_1} \big]
  \,,\\
  \cO_{\cP_\perp^\dagger }^{(1)\mu}
   &=\frac{1}{2\w_2}\big[\bar{\chi}_{n_2,-\w_2} \,\slashed\bn_2\slashed{\cP}^\dagger_{n_2\perp}\,\gamma^\mu\, (S_{n_2}^\dagger S_{n_1}) \chi_{n_1,-\w_1} \big]
  \,.\nn
\end{align}
Once we specialize to the back-to-back frame $n_2=\bar n_1$, which is needed for our factorization analysis, we have $\cP_{n_1\perp}=\cP_{n_2\perp}=\cP_\perp$  and these two operators give the complete basis of operators involving $\cP_\perp$ at this order for a vector current. This follows from \refcite{Feige:2017zci} (appendix C), which counts four independent helicity operators, and the constraint from parity for our case reduces this down to two operators.  The choice of the two operators in \eq{op_P1} thus suffices.%
\footnote{Note that we do not need to include operators with $\cP_\perp$ acting on solely on the soft Wilson lines, as in $[[\cP_\perp^\mu S_{n_2}^\dagger S_{n_1}]]$, since such terms can be written in terms of $\cB_{s\perp}^{(n_i)\mu}$s, and hence it is enough to include a complete basis of operators involving $\cB_{s\perp}^{(n_i)\mu}$s, which we will do.} 
An example of this for a different operator is given in greater detail in \sec{operators_NLP_B}.
We also note that 
$\big(J^{(1)\mu}_{\cP_\perp}\big)^\dagger=J^{(1)\mu}_{\cP_\perp^\dagger}$.
The extra ${\cal P}_\perp$s do not change the renormalization structure, so just like at leading power we will not need to consider evanescent operator extensions of these operators at higher orders~\cite{Moult:2015aoa}.
The contributions of these operators to the NLP factorization formula will be derived in \sec{factorization_P}. 

For the $\img \bn_i\cdot D_{s}$ terms in \eq{Orpiexpn} it is useful to use 
\begin{equation} \label{eq:nbDrelation}
  S_{n_i}^\dagger\img \bn_i\cdot D_{s}
  S_{n_i} =    \bn_i\cdot \img\partial_s + g \bn_i\cdot \cB_{s}^{(n_i)}\,,
\end{equation}
where $\cB_{s}^{(n_i)}$ is given in \eq{softPsiB}. This allows us to write
\begin{align} \label{eq:softmanip}
  S_{n_2}^\dagger \img\bar n_1\cdot D_{us} S_{n_1}
 &=  S_{n_2}^\dagger S_{n_1} S_{n_1}^\dagger \img\bar n_1\cdot D_{us} S_{n_1}
 = S_{n_2}^\dagger S_{n_1} \big( \bn_1\cdot \img\partial_s 
    + g \bn_1\cdot \cB_{s}^{(n_1)} \big) \,,
  \\
 -\img S_{n_2}^\dagger \bar n_2\cdot \overleftarrow D_{us} S_{n_1}
 &= -\img S_{n_2}^\dagger \bar n_2\cdot \overleftarrow D_{us} S_{n_2} S_{n_2}^\dagger S_{n_1}
  = \big(- \bn_2\cdot \img\overleftarrow\partial_s 
      + g\bn_2\cdot \cB_{s}^{(n_2)}\big) S_{n_2}^\dagger S_{n_1} 
\,. \nn
\end{align}
Again following the notation in \eq{Lhard_withJ}, this gives the subleading power currents
\begin{align}  \label{eq:J1nnb}
  J^{(1)\mu}_{\bn_i\cdot {\cal B}_{s} }(0) &= -\sum_{n_1,n_2}\int\!\df\w_1\,\df\w_2\,
   \sum_f \bigg[ \frac{\partial C_f^{(0)}(\tilde q^2)}{\partial \omega_1}\, \cO_{\bn_1\cdot {\cal B}_s}^{(1)\mu}(\tilde p_1,\tilde p_2)
  + 
  \frac{\partial C_{f}^{(0)}(\tilde q^2)}{\partial \omega_2}\, \cO_{\bn_2\cdot {\cal B}_s}^{(1)\mu}(\tilde p_1,\tilde p_2)
   \bigg]
   , \nn \\
 J^{(1)\mu}_{\bn_i\cdot \img \partial_s }(0) &= -\sum_{n_1,n_2}\int\!\df\w_1\,\df\w_2\,
  \sum_f \bigg[ \frac{\partial C_{f}^{(0)}(\tilde q^2)}{\partial \omega_1}\, \cO_{\bn_1\cdot \img \partial_s}^{(1)\mu}(\tilde p_1,\tilde p_2)
  + 
  \frac{\partial C_{f}^{(0)}(\tilde q^2)}{\partial \omega_2}\, \cO_{\bn_2\cdot \img \partial_s}^{(1)\mu}(\tilde p_1,\tilde p_2)
   \bigg]
   \,,
\end{align}
with the operators
\begin{align} \label{eq:n1n2dotops}
\cO_{\bn_1\cdot {\cal B}_s}^{(1)\mu}
 &=  \bar\chi_{n_2,\w_2}
  \big( S_{n_2}^\dagger S_{n_1} \bn_1\cdot\cB_s^{(n_1)}\big)
  \gamma_\perp^\mu \chi_{n_1,\w_1}
  ,
  \quad
 \cO_{\bn_2\cdot {\cal B}_s}^{(1)\mu}
 = \bar\chi_{n_2,\w_2}
  \big(\bn_2\cdot\cB_s^{(n_2)}S_{n_2}^\dagger S_{n_1}\big) 
  \gamma_\perp^\mu \chi_{n_1,\w_1}
  , \nn\\
\cO_{\bn_1\cdot \img \partial_s}^{(1)\mu}
  &= \bar\chi_{n_2,\w_2} \big( S_{n_2}^\dagger S_{n_1}
  \img\bn_1\cdot \partial_s \big)
  \gamma_\perp^\mu \chi_{n_1,\w_1} 
  \,, \quad
\cO_{\bn_2\cdot \img \partial_s}^{(1)\mu}
  = \bar\chi_{n_2,\w_2}
  \big(-\img\bn_2\cdot\overleftarrow\partial_s
  S_{n_2}^\dagger S_{n_1} \big) 
  \gamma_\perp^\mu \chi_{n_1,\w_1} 
 \,.
\end{align}
Here the $\bn_i\cdot\partial_s$ operators pick out residual soft momentum of ${\cal O}(\lambda)$ that are carried by the collinear fields, beneath their large $n_i$-collinear momenta  $\bn_i \cdot p_{n_i}\sim \lambda^0$.  They parameterize the NLP impact of subleading soft momenta that flow through the hard loops in QCD which were integrated out to obtain $C^{(0)}$.
Again for the case of the back-to-back frame $n_2=\bn_1$ and $\bn_2=n_1$, these operators give a complete set for the vector current. This again follows from the enumeration of all allowed helicity operators in \refcite{Feige:2017zci} and imposing parity. Since in this frame $\omega_1=\omega_2=Q$ and $\tilde q^2=\omega_1\omega_2$, the derivatives of the Wilson coefficients will also (eventually, after taking the matrix elements) become identical, and hence can be pulled out as a common factor, giving
\begin{align} \label{eq:bbJ1nnb}
 J^{(1)\mu}_{\stackrel{\mbox{\tiny $(-)$}}{n} \cdot {\cal B}_{s} }(0) &= -\sum_{n_1}\int\!\df\w_1\,\df\w_2\,
   \sum_f \frac{\partial C_{f}^{(0)}(\tilde q^2)}{\partial \omega_1}\,
    \cO_{\stackrel{\mbox{\tiny $\!\!\!(-)$}}{n_1}\cdot {\cal B}_s}^{(1)\mu}(\tilde p_1,\tilde p_2)
   \,,  \\
 J^{(1)\mu}_{\stackrel{\mbox{\tiny $(-)$}}{n} \cdot \img \partial_s }(0) &= -\sum_{n_1}\int\!\df\w_1\,\df\w_2\,
  \sum_f \frac{\partial C_{f}^{(0)}(\tilde q^2)}{\partial \omega_1}\, 
   \cO_{\stackrel{\mbox{\tiny $\!\!\!(-)$}}{n_1} \cdot \img \partial_s}^{(1)\mu}(\tilde p_1,\tilde p_2)
   \,, \nn
\end{align}
with 
\begin{align} \label{eq:bbJ1nnbops}
\cO_{\stackrel{\mbox{\tiny $\!\!\!(-)$}}{n_1} \cdot {\cal B}_s}^{(1)\mu}
 &=  \bar\chi_{\bn_1,\w_2}
  \Big( S_{\bn_1}^\dagger S_{n_1} \bn_1\cdot\cB_s^{(n_1)}
  + n_1\cdot\cB_s^{(\bn_1)}S_{\bn_1}^\dagger S_{n_1}
  \Big)
  \gamma_\perp^\mu \chi_{n_1,\w_1}
  , \\
\cO_{\stackrel{\mbox{\tiny $\!\!\!(-)$}}{n_1}\cdot \img \partial_s}^{(1)\mu}
  &= \bar\chi_{\bn_1,\w_2} 
  \Big( S_{\bn_1}^\dagger S_{n_1}
  \img\bn_1\cdot \partial_s
  -\img n_1\cdot\overleftarrow\partial_s
  S_{\bn_1}^\dagger S_{n_1}
  \Big)
  \gamma_\perp^\mu \chi_{n_1,\w_1} 
 \,. \nn
\end{align}
The contributions of the currents in \eq{bbJ1nnb} to the NLP factorization formula will be considered in \sec{fact_vanishing_soft}.

\subsubsection{NLP Operators with a Collinear $\cB_{n_i\perp}$}
\label{sec:operators_NLP_B}

In this section we construct a complete basis of \SCETii operators involving the field structures  $\bar\chi_{n_1} {\cal B}_{n_i\perp}^\mu \chi_{n_2}$. Since the allowed form of these operators is not constrained by RPI, we will rely more heavily on the helicity operator basis, and then convert these to a useful form for our analysis.

We begin by briefly reviewing the spinor helicity formalism and corresponding conventions
employed in \refscite{Moult:2015aoa,Feige:2017zci} that are needed for our analysis here.
We use the standard spinor helicity notation
\begin{align} \label{eq:braket_def}
 \ket{p} \equiv \ket{p+} &= \frac{1 + \gamma_5}{2}\, u(p)
  \,,
 & \sqket{p} & \equiv \ket{p-} = \frac{1 - \gamma_5}{2}\, u(p)
  \,, \\
 \bra{p} \equiv \bra{p-} &= \mathrm{sgn}(p^0)\, \bar{u}(p)\,\frac{1 + \gamma_5}{2}
  \,, 
 & \sqbra{p} & \equiv \bra{p+} = \mathrm{sgn}(p^0)\, \bar{u}(p)\,\frac{1 - \gamma_5}{2}
  \,, \nn 
\end{align}
with $p$ lightlike. Spinor products of two lightlike vectors $p$ and $q$ are denoted as
\begin{align}
 \braket{pq} = \braket{p{-}|q{+}}
\,,\qquad
 \sqbraket{pq} = \braket{p{+}|q{-}}
\,.\end{align}
The polarization vector of an outgoing gluon with momentum $p$ can be written as
\begin{equation}
  \varepsilon_+^\mu(p,r) = \frac{\mae{p+}{\gamma^\mu}{r+}}{\sqrt{2} \langle r p \rangle}
 \,,\qquad
  \varepsilon_-^\mu(p,r) = - \frac{\mae{p-}{\gamma^\mu}{r-}}{\sqrt{2} [r p]}
 \,,\end{equation}
where $r\neq p$ is an arbitrary lightlike reference vector.

Next, we define collinear quark and gluon fields of definite helicity as
\begin{align} \label{eq:field_pm_def}
 \cB^A_{n_i\pm} &
 = -\varepsilon_{\mp\mu}(n_i, \bn_i)\,\cB^{A\mu}_{n_i\perp,\w_i}
\,,\quad
 \chi_{n_i \pm}^a
 = \frac{1\,\pm\, \gamma_5}{2} \chi_{n_i,  -\w_i}^a
\,,\quad
 \bar{\chi}_{n_i \pm}^{\bar a}
 = \bar{\chi}_{n_i,  -\w_i}^{\bar a} \frac{1\,\mp\, \gamma_5}{2}
\,,\end{align}
where $i=1,\,2$ and $A,\,a,\,\bar a$ are color indices.
Note that in SCET we use $n_i$ for each collinear direction in spinors, rather than the momentum for each particle.
Since fermions always come in pairs, it is also useful to define currents of definite helicity,
\begin{align} \label{eq:hel_current_chi}
 J_{n_2 n_1\pm}^{\bar a b} &
 = \mp \sqrt{\frac{2}{\w_2 \w_1}} \, \frac{\varepsilon_\mp^\mu(n_2, n_1) }{\braket{n_1{\mp}|n_2{\pm}}}
   \, \bar{\chi}^{\bar a}_{n_2\pm} \gamma_\mu \chi^b_{n_1\pm}
\,.\end{align}
The leptonic currents are defined analogously as
\begin{equation} \label{eq:hel_current_el}
 J_{e\, n_5n_4\pm}
 = \mp \sqrt{\frac{2}{\w_5 \w_4}} \, \frac{\varepsilon_\mp^\mu(n_5, n_4) }{\braket{n_4{\mp}|n_5{\pm}}}
   \, \bar{e}_{n_5\pm} \gamma_\mu e_{n_4\pm}
\,.\end{equation}
Since the directions of the leptons are fixed by the process,
we often suppress the explicit dependence on the lightcone vectors
and abbreviate $J_{e\,\pm} \equiv J_{e\, n_5n_4 \pm}$, with the understanding
that the reference vectors $n_{4,5}$ in \eq{hel_current_el} are identified
with those of the incoming and outgoing lepton, respectively.

In total, there are eight operators involving a single collinear gluon field $\cB_\pm$~\cite{Feige:2017zci} which matter for our analysis at NLP,%
\footnote{There are additional NLP operators in~\cite{Feige:2017zci} which involve two quark fields in the same collinear direction or three $\cB$ fields. For our NLP analysis these operators do not contribute since when combined with the leading power operator the factorized matrix elements do not conserve fermion number.}
\begin{alignat}{2} \label{eq:Z1_basis}
  &\cO_{n_1\, n_2 n_1 -(-;\pm)}^{(1)A \bar a b}
  = g\,\cB_{n_1-}^A\, J_{n_2 n_1\,-}^{\bar a b} \,J_{e\pm }
  \,,\qquad &
  &\cO_{n_1\, n_2 n_1 +(+;\pm)}^{(1)A \bar a b}
  = g\,\cB_{n_1 +}^A \, J_{n_2 n_1\,+}^{\bar a b}\,J_{e\pm }
\,,\nn\\
  &\cO_{n_2\, n_2 n_1 +(-;\pm)}^{(1)A \bar a b}
  = g\,\cB_{n_2+}^A\, J_{n_2 n_1\,-}^{\bar a b} \,J_{e\pm }
  \,,\qquad &
  &\cO_{n_2\, n_2 n_1 -(+;\pm)}^{(1)A \bar a b}
  = g\,\cB_{n_2-}^A \, J_{n_2 n_1\,+}^{\bar a b}\,J_{e\pm }
\,.\end{alignat}
The operators $\cO_{n_i\, n_2 n_1 \lambda_3 (\lambda_{21}; \pm)}$ are labelled
by the direction $n_i$ with $i=1,2$ of the gluon field,
the directions $n_2 n_1$ of the quark fields,
the helicity $\lambda_3$ of the gluon and helicity $\lambda_{21}$ of the quark current,
and the helicity $\pm$ of the lepton current.
For brevity, we have suppressed the explicit arguments of the operators,
which encompass the directions $n_{1,2}$, the large ${\cal O}(\lambda^0)$ label momenta $\{\w_{1,\cdots,5}\}$
and the residual position $b^\mu$.

The unique color structure of all operators in \eq{Z1_basis} is
$\bar T^{A\,a\bar b} = \bigl(T^A_{a\bar b}\bigr)$.
In this one-dimensional color space, the Wilson coefficients
$C^{(1)}_{\lambda_3(\lambda_{21}:\lambda_{45})}(n_1,n_2;\{\w_i\})$
are simple functions rather than vectors in color space,
such that we can drop the color indices on the $C^{(1)}$.
A priori, the eight operators in \eq{Z1_basis} are independent of each other.
However, they are related to each other using discrete symmetries, yielding only one independent coefficient for the vector current relevant for photon exchange.
Parity and charge conjugation invariant of QCD imply the following relations respectively~\cite{Feige:2017zci}, where here we are being careful about the overall phase induced from the definition of the spinors,
\begin{subequations} \label{eq:paritycharge}
\begin{align} \label{eq:parity}
 C^{(1)}_{\lambda_3(\lambda_{21}:\lambda_{45})}(\omega_3;\omega_1,\omega_2;\omega_4,\omega_5) &
= -C^{(1)}_{-\lambda_3(-\lambda_{21}:-\lambda_{45})}(\omega_3;\omega_1,\omega_2;\omega_4,\omega_5)\Big|_{\Braket{\cdot\cdot} \leftrightarrow [\cdot\cdot]}
\,,\\ \label{eq:charge}
 C^{(1)}_{\lambda_3(\lambda_{21}:\lambda_{45})}(\omega_3;\omega_1,\omega_2;\omega_4,\omega_5) &
 = -C^{(1)}_{\lambda_3(-\lambda_{21}:-\lambda_{45})}(\omega_3;\omega_2,\omega_1;\omega_5,\omega_4)
\,.\end{align}
\end{subequations}
Since we only consider the leading order in the electromagnetic coupling, the leptons couple only through the vector currents which satisfy
\begin{align}\label{eq:lo_em}
  \mae{n_5\,\pm}{\gamma^\mu}{n_4\, \pm} = \mae{n_4\,\mp}{\gamma^\mu}{n_5\,\mp}
\,.\end{align}
Thus, the Wilson coefficients obey
\begin{align}
  C^{(1)}_{\lambda_3(\lambda_{21}:\lambda_{45})}(\omega_3;\omega_1,\omega_2;\omega_4,\omega_5) &
  = C^{(1)}_{\lambda_3(\lambda_{21}:-\lambda_{45})}(\omega_3;\omega_1,\omega_2;\omega_5,\omega_4)\,.
\end{align}
Since the leptons couple only through the vector current in \eq{lo_em},
it can be factored out from the Wilson coefficients, allowing us to write
\begin{equation}\label{eq:lep_reduce}
  C^{(1)}_{\lambda_3(\lambda_{21}:\lambda_{45})}(\omega_3;\omega_1,\omega_2;\omega_4,\omega_5)
  = \mae{n_5\,\lambda_{45}}{\gamma_\mu}{n_4\,\lambda_{45}}\, \braket{n_1{\lambda_3}|n_2{-\lambda_3}}
  \,  C^{(1)\mu}_{\lambda_3 \lambda_{21}}(\lp_1,\lp_2,\lp_3)\,.
\end{equation}
For reasons discussed later, we also factored out $\braket{n_1{\lambda_3}|n_2{-\lambda_3}}$.
$C^{(1)\mu}_{\lambda_3 \lambda_{21}}$ is now a vector which does not have dependence on the leptonic parts.

To further constrain $C^{(1)}_{\lambda_3(\lambda_{21}:\lambda_{45})}$, we also make use of the little group property. Instead of discussing the little group for each individual particle as usually done in the helicity amplitude literature, in SCET we consider little group scaling for each collinear sector.
Under the little group scaling for the $n_i$ sector,
\begin{align}
  |n_i\rangle\to t_i |n_i\rangle\,, \qquad |n_i]\to t_1^{-1}|n_i]\,,
\end{align}
the one particle state with momentum $p$ in the collinear-$n_i$ sector with helicity $h$ scales like
\begin{equation}
  \left|p, h\right>\to t_i^{2h} \left|p, h\right>\,.
\end{equation}
The Feynman rules for $\chi_{n_i\pm}$ and $\cB_{n_i\pm}$ (with outgoing momenta and positive outgoing collinear momentum labels) were given in Ref.~\cite{Moult:2015aoa}
\begin{align} \label{eq:quarkbaseFR}
  \Mae{0}{\chi^b_{n_i\pm}}{q_\pm^{\bar a}(-p)}
  &= \delta^{b{\bar a}}\,\ldel(\lp_i - p)\, \ket{(-p_i)\pm}_{n_i} \,,\\
  \Mae{q_\pm^a(p)}{\bar\chi^{\bar b}_{n_i\pm}}{0}
  &= \delta^{a\bar b}\,\ldel(\lp_i - p)\, {}_{n_i\!}\bra{p_i\pm}
  \,, \nn \\
  \Mae{0}{\bar\chi^{\bar b}_{n_i\pm}}{\bar{q}_\mp^a(-p)}
  &= \delta^{a\bar b}\,\ldel(\lp_i - p)\,{}_{n_i\!}\bra{(-p_i)\pm}
  \,, \nn \\
  \Mae{\bar{q}_\mp^{\bar a}(p)}{\chi^b_{n_i\pm}}{0}
  &= \delta^{b{\bar a}}\,\ldel(\lp_i - p)\, \ket{p_i\pm}_{n_i}
 \,,\nn\\
  \l g_\pm^A(p) | \cB_{n_i\pm}^B |0 \r & = \delta^{AB} \ldel (\lp_i -p)\,,  \\
  \l 0 | \cB_{n_i\pm}^B |g_\mp^A(-p) \r & = \delta^{AB} \ldel (\lp_i -p) .\nn
\end{align}
From these we can conclude that the quark fields $\chi_{n_i\pm}$ does not scale since quark state scales in the same way as spinor, while the gluon field should scale in the same way as the gluon state, $\cB_{n_i\pm} \to t_i^{\pm2} \cB_{n_i\pm}$.

The subleading power currents must be invariant under the little group scaling for each collinear sector,
so the scaling of the Wilson coefficients must be opposite to that of their corresponding operators.
Specifically, under independent actions of the little group,
\begin{align}
  |n_1\rangle\to t_1 |n_1\rangle\,, \qquad |n_1]\to t_1^{-1}|n_1]\,,\qquad
  |n_2\rangle\to t_2 |n_2\rangle\,, \qquad |n_2]\to t_2^{-1}|n_2]\,,
\end{align}
the polarization vectors, gluon and quark fields, and quark currents transform as
\begin{alignat}{2} \label{eq:scaling_epsilon}
 \varepsilon_\pm^\mu(n_i,r) &\to t_i^{\mp2} \varepsilon_\pm^\mu(n_i,r)
\,,\qquad
 \cB_{n_i \pm}^a &&\to t_i^{\pm 2}\cB_{n_i \pm}^a
\,,\nn\\
  \chi_{n_i\pm}^{a} &\to \chi_{n_i\pm}^{a}
\,,\hspace{1.7cm}
  J_{n_2 n_1\pm}^{\bar a b} &&\to t_2^\pm t_1^{\mp}  J_{n_2 n_1\pm}^{\bar a b}
\,.\end{alignat}
The eight helicity operators \eq{Z1_basis} then transform as
\begin{align}\label{eq:scaling_op}
  &\cO_{n_1\, n_2 n_1 -(-;\pm)}^{(1)A \bar a b}
 \to \frac{1}{t_2 t_1} \, \cO_{n_1\, n_2 n_1 -(-;\pm)}^{(1)A \bar a b}
  \,,\quad &
  &  \cO_{n_1\, n_2 n_1 +(+;\pm)}^{(1)A \bar a b}
  \to  t_2 t_1 \, \cO_{n_1\, n_2 n_1 +(+;\pm)}^{(1)A \bar a b}
\,,\nn\\
  &\cO_{n_2\, n_2 n_1 +(-;\pm)}^{(1)A \bar a b}
  \to t_2 t_1\, \cO_{n_2\, n_2 n_1 +(-;\pm)}^{(1)A \bar a b}
  \,,\quad &
  &\cO_{n_2\, n_2 n_1 -(+;\pm)}^{(1)A \bar a b}
  \to \frac{1}{t_2 t_1} \, \cO_{n_2\, n_2 n_1 -(+;\pm)}^{(1)A \bar a b}\,.
\end{align}
Notice that the scaling of these helicity operators is exactly canceled by the scaling of $\braket{n_1{\lambda_3}|n_2{-\lambda_3}}$ in \eq{lep_reduce}. Therefore, $C^{(1)\mu}_{\lambda_3 \lambda_{21}}$ does not scale under the little group scaling.

The possible structure for the $\mu$ index in $C^{(1)\mu}_{\lambda_3 \lambda_{21}}$ includes $\varepsilon_\pm^\mu(n_1,n_2)$, $\varepsilon_\pm^\mu(n_2,n_1)$, $n_1^\mu$ and $n_2^\mu$, together with coefficients that are functions of spinor brackets $\left<n_1 n_2\right>$ and $\left[n_1n_2\right]$, and label momenta $\w_1$, $\w_2$, $\w_3$. 
However, the scaling of $\varepsilon_\pm^\mu$ in \eq{scaling_epsilon} can never be canceled by coefficients constructed using spinor brackets $\left<n_1 n_2\right>$ and $\left[n_1n_2\right]$ which scale like $(t_2 t_1)^\pm$. Therefore, $C^{(1)\mu}_{\lambda_3 \lambda_{21}}$ can only contain terms proportional to $n_1^\mu$ and $n_2^\mu$, which do not scale under the little group.

Further, we notice that $\braket{n_1 n_2} \sqbraket{n_1 n_2}=-2(n_1\cdot n_2)$. That $C^{(1)\mu}_{\lambda_3 \lambda_{21}}$ does not scale then also imply that it is invariant when exchanging $\ket{\cdot\cdot}$ and $\left|\cdot\cdot\right]$.
Applying \eq{parity} to \eq{lep_reduce}, we obtain
\begin{align}
  C^{(1)}_{\lambda_3(\lambda_{21}:\lambda_{45})}&= -\left[ \mae{n_5\,-\!\!\lambda_{45}}{\gamma_\mu}{n_4\,-\!\!\lambda_{45}}
  \braket{n_1\,{-\lambda_3}|n_2\,{\lambda_3}}
   C^{(1)\mu}_{-\lambda_3 -\lambda_{21}}(\lp_1,\lp_2,\lp_3) \right]
  _{\ket{\cdot\cdot} \leftrightarrow \left|\cdot\cdot\right]}\nn\\
  &= -\mae{n_5\,\lambda_{45}}{\gamma_\mu}{n_4\,\lambda_{45}}
  \braket{n_1\,{\lambda_3}|n_2\,{-\lambda_3}}
  C^{(1)\mu}_{-\lambda_3 -\lambda_{21}}(\lp_1,\lp_2,\lp_3)\nn\\
  &= -\frac{\braket{n_1{\lambda_3}|n_2{-\lambda_3}}}{\braket{n_1{-\lambda_3}|n_2{\lambda_3}}}
  C^{(1)}_{-\lambda_3(-\lambda_{21}:\lambda_{45})}
\,,\end{align}
where we suppressed the arguments of $C^{(1)}$.
Combining all Wilson coefficients with the corresponding operator from \eq{Z1_basis},
we obtain the combinations
\begin{align} \label{eq:linear_combination_1}
 & C^{(1)}_{-({-}:\pm)} g \cB_{n_1-} J_{n_2n_1 -} J_{e\pm} - C^{(1)}_{+(+:\pm)} g \cB_{n_1+} J_{n_2n_1 +} J_{e\pm}
 \nn\\&
 = \frac{C^{(1)}_{-({-}:\pm)}}{\braket{n_1 n_2}} \Bigl( \braket{n_1 n_2} g \cB_{n_1-} J_{n_2n_1-} - \sqbraket{n_1 n_2}\, g \cB_{n_1+} J_{n_2n_1+}\Bigr)J_{e\pm}
\,,\\
 &  C^{(1)}_{-(+:\pm)} g \cB_{n_2-} J_{n_2n_1 +} J_{e\pm}-C^{(1)}_{+({-}:\pm)} g \cB_{n_2+} J_{n_2n_1 -} J_{e\pm} 
 \nn\\&
 = \frac{C^{(1)}_{-(+:\pm)}}{\braket{n_1 n_2}} \Bigl(\braket{n_1 n_2} g \cB_{n_2-} J_{n_2n_1 +} - \sqbraket{n_1 n_2} g \cB_{n_2+} J_{n_2n_1 -} \Bigr)J_{e\pm}
\end{align}
As a consequence, the $g \cB_{n_1 \lambda_3} J_{n_2n_1 \lambda_{12}}$ currents only appear in the linear combinations
\begin{alignat}{2} \label{eq:linear_combination_2}
 \cO^{(1)}_{\cB 1}
 &\equiv
\sqrt{\frac{\omega_1 \omega_2}{2}} \bigl(\braket{n_1 n_2} g \cB_{n_1-} J_{n_2n_1-} - \sqbraket{n_1 n_2} \, g \cB_{n_1+} J_{n_2n_1+}\bigr)
 &&= \bar \chi_{\bar n_1,-\omega_2}\, g \slashed{\cB}_{n_1\perp,\omega_3} \chi_{n_1,-\omega_1}
\,,\nn\\
 \cO^{(1)}_{\cB 2}
 &\equiv
\sqrt{\frac{\omega_1 \omega_2}{2}} \bigl( \braket{n_1 n_2} g \cB_{n_2 -} J_{n_2n_1+}-\sqbraket{n_1 n_2}\, g \cB_{n_2+} J_{n_2n_1-}\bigr)
 &&=  \bar \chi_{\bar n_1,-\omega_2} \, g \slashed{\cB}_{\bar n_1,\omega_3} \chi_{n_1,-\omega_1}
\,,\end{alignat}
with Wilson coefficients specified by \eq{linear_combination_1}.
As usual, the quark flavors are kept implicit both in the operator and the quark fields.
For the final equalities in \eq{linear_combination_2} we have taken $n_1$ and $n_2$ to be back-to-back, so that $n_2=\bar n_1$, which enables us to use the completeness relation for polarization vectors
\begin{align}\label{eq:completeness}
\sum\limits_{\lambda=\pm} \epsilon^\lambda_\mu(n_i,\bar n_i) \left (\epsilon^\lambda_\nu(n_i,\bar n_i) \right )^*=-g_{\mu \nu}+\frac{n_{i\mu} \bar n_{i\nu}+n_{i\nu} \bar n_{i\mu}}{n_i\cdot \bar n_i} = - g^\perp_{\mu \nu} ( n_i, \bn_i)\,.
\end{align}

To summarize the discussion so far, starting from the helicity operator basis with a priori eight independent operators as given in \eq{Z1_basis},
using C/P invariance, the choice of vector current, and little group scaling we have shown that there are only two independent operators given in \eq{linear_combination_1}.
Falling back to the standard SCET notation, we define the corresponding hard currents as
\begin{subequations}  \label{eq:JB1B2}
\begin{align} \label{eq:matched_1gluon_barn1}
  J_{\cB 1}^{(1)\mu}(0) &
  =\! \sum_{n_1,n_2} \int\!\!\df \w_1\df\w_2\df\w_3 \, 
    \,\sum_f C^{(1)\mu}_{1}(\lp_1,\lp_2,\lp_3)\,\cO^{(1)}_{\cB 1}\left(\lp_1,\lp_2,\lp_3\right)
 ,\\ \label{eq:matched_1gluon_barn2}
  J_{\cB 2}^{(1)\mu}(0) &
  =\! \sum_{n_1,n_2} \int\!\!\df \w_1\df\w_2\df\w_3 \, 
    \,\sum_f C^{(1)\mu}_{2}(\lp_1,\lp_2,\lp_3)\, \cO^{(1)}_{\cB 2}\left(\lp_1,\lp_2,\lp_3\right)
,\end{align}
\end{subequations}
where the sum runs over all quark flavors $f$.
The label momenta are given by $\lp_1^\mu=\w_1\frac{n_1^\mu}{2}$ and $\lp_2^\mu=\w_2\frac{n_2^\mu}{2}$,
with $\lp_3^\mu=\w_3\frac{ n_1^\mu}{2}$ in \eq{matched_1gluon_barn1} and $\lp_3^\mu=\w_3\frac{n_2^\mu}{2}$ in \eq{matched_1gluon_barn2}.

As argued above, $C^{(1)\mu}_{i}$ can only depend on $n_1^\mu$ and $n_2^\mu$.
Another constraint arises from current conservation $q_\mu J_i^{(1)\mu}=0$,
where at this order $q^\mu= \tilde p_1^\mu+\tilde p_2^\mu+\tilde p_3^\mu+\Ord(\lambda)$.
This implies that $C^{(1)\mu}_{1}$ is proportional to $\tilde p_1^\mu-\tilde p_2^\mu+\tilde p_3^\mu$,
while $C^{(1)\mu}_{2}$ is proportional to $\tilde p_1^\mu-\tilde p_2^\mu-\tilde p_3^\mu$.
Factoring these coefficients out, we have
\begin{align} \label{eq:C11C12}
 C^{(1)\mu}_{1}(\lp_1,\lp_2,\lp_3) &
 = \frac{\w_2 n_2^\mu - (\w_1+\w_3) n_1^\mu}{(\w_1+\w_3)\w_2}  C^{(1)}_{1}(\tilde q^2,\xi_1)
 = \frac{2\left[\lp_2^\mu - (\lp_1^\mu + \lp_3^\mu)\right]}{\tilde q^2}  C^{(1)}_{1}(\tilde q^2,\xi_1)
\,,\\
 C^{(1)\mu}_{2}(\lp_1,\lp_2,\lp_3) &
 = \frac{(\w_2+\w_3) n_2^\mu - \w_1 n_1^\mu}{\w_1(\w_2+\w_3)} C^{(1)}_{2}(\tilde q^2,\xi_2)
 = \frac{2\left[(\lp_2^\mu + \lp_3^\mu) - \lp_1^\mu\right]}{\tilde q^2}C^{(1)}_{2}(\tilde q^2,\xi_2)  \,. \nn
\end{align}
The normalization is chosen such that the scalar coefficients $C^{(1)}_{i}(\tilde q^2, \xi_i)$ are dimensionless.
Momentum conservation and RPI imply that they only depend on $\tilde q^2$ and $\xi_i$, where
\begin{align} \label{eq:def_xi_i}
 C^{(1)}_{1} : &\qquad \tilde q^2 = (\w_1+\w_3)\w_2 \frac{n_1\cdot n_2}{2}
                   \,,\qquad \xi_1 = \frac{\w_3}{\w_1+\w_3}
\,,\nn\\
 C^{(1)}_{2} : &\qquad \tilde q^2 = (\w_2+\w_3)\w_1 \frac{n_1\cdot n_2}{2}
                   \,,\qquad \xi_2 = \frac{\w_3}{\w_2+\w_3}
\,.\end{align}

Now using charge invariance, \eq{charge}, we obtain
\begin{align}
  C^{(1)\mu}_{2}(\lp_1,\lp_2,\lp_3)=-C^{(1)\mu}_{1}(\lp_1,\lp_2,\lp_3)\Big|_{ \w_1 \leftrightarrow \w_2,\, n_1\leftrightarrow n_2  }\,.
\end{align}
This implies that the two Wilson coefficients are in fact equal!  We 
will denote the single independent coefficient as
\begin{equation}\label{eq:C1}
 C^{(1)}(\tilde q^2,\,\xi) \equiv C^{(1)}_{1}(\tilde q^2,\,\xi)
  = C^{(1)}_{2}(\tilde q^2,\,\xi)
\,.\end{equation}
So far we have been suppressing the dependence of the coefficient on the quark flavor.
Restoring the flavor index $f$, we can extract the
quark charge similar to the leading-power Wilson coefficient in \eq{hard_lp},
\begin{align}\label{eq:hard_nlp}
  C_{f}^{(1)} (\tilde q^2,\xi) = Q_{f}\,C_{q}^{(1)}(\tilde q^2,\xi)+\sum_{f'} Q_{f'} C_{vf'}^{(1)}(\tilde q^2,\xi)
\,.\end{align}
Since the flavor index is easy to restore, we will continue keep it implicit below.

The results for the form of these Wilson coefficients must also be consistent with those obtained purely in \SCETi.
The form of \eq{C11C12} is consistent with the tree-level matching in eqs.~(5.16) and (5.21) of \refcite{Feige:2017zci}. At tree-level we simply have $C_q^{(1)}(\tilde q^2,\xi)=1$ and $C_{vf'}^{(1)}(\tilde q^2,\xi)=0$. 
Results for the anomalous dimension of these coefficients have also been obtained at NLL in Ref.~\cite{Beneke:2017ztn,Beneke:2018rbh}, which determines the form of scale dependent logarithmic terms in the matching coefficients at ${\cal O}(\alpha_s)$.  

We have not yet considered whether the Wilson coefficients $C_q^{(1)}$ are real. For SIDIS it is known that the leading power coefficient $C^{(0)}$ is real due to the space-like kinematics $\tilde q^2<0$. The coefficient is dimensionless and only involves logarithms of $\ln[(-\tilde q^2-i0)/\mu^2]$ which are real. This is also clear from the relation of $C^{(0)}$ to the space-like form factor in pure dimensional regularization. (In contrast, for Drell-Yan and $e^+e^-$ annihilation the coefficient $C^{(0)}$ is related to the time-like form factor and is complex.) 
For the dimensionless coefficient $C_q^{(1)}$ the anomalous dimension is real at LL~\cite{Beneke:2017ztn,Beneke:2018rbh}, but it is known that complex contributions begin to appear at NLO~\cite{Vladimirov:2021hdn}.  We will assume it is complex below.

So far, we have manipulated the operators and Wilson coefficients as in a \SCETi situation.
As described in \sec{SCET_review}, to obtain the corresponding \SCETii operators,
we match this intermediate \SCETi theory  onto \SCETii, which induces soft Wilson lines
analogous to the BPS field redefinition in \eq{BPS}, but with soft Wilson lines. For subleading power \SCETi operators like these that contribute directly to ${\cal L}_h^{(1)}$ without involving any time-ordered products, the (pre-BPS) \SCETi $\to$ \SCETii matching amounts to making the replacements
$\chi_{n,-\omega} \to S_n \chi_{n,-\omega}$ and
$\cB_{n\perp,\omega_3}^{\mu}
 \to S_n \cB_{n\perp,\omega_3}^{\mu} S_n^{\dagger}$.
Furthermore the structure of Wilson coefficients and operators for ${\cal L}_h^{(1)}$ is exactly as in \eq{JB1B2}, where we express the results with hadronic currents following the notation in \eq{Lhard_withJ}. 
The corresponding Wilson coefficients are given by combining \eq{C11C12},
with \eq{C1}.
In the back-to-back frame with $n_2=\bar n_1$ we have
\begin{align}  \label{eq:JB1B2scetii}
  J_{\cB 1}^{(1)\mu}(0) &
  =\! \sum_{n_1} \int\!\!\df \w_1\df\w_2\df\w_3 \,
    \,\sum_f 
\frac{2\left(\lp_2^\mu - \lp_1^\mu - \lp_3^\mu\right)}{\tilde q^2}  C^{(1)}(\tilde q^2,\xi_1) \,
  \cO^{(1)}_{\cB 1}\left(\lp_1,\lp_2,\lp_3\right)
 , \\
  J_{\cB 2}^{(1)\mu}(0) &
  =\! \sum_{n_1} \int\!\!\df \w_1\df\w_2\df\w_3 \,
    \,\sum_f 
  \frac{2\left(\lp_2^\mu + \lp_3^\mu - \lp_1^\mu\right)}{\tilde q^2}C^{(1)}(\tilde q^2,\xi_2) \,
  \cO^{(1)}_{\cB 2}\left(\lp_1,\lp_2,\lp_3\right)
  \,.\nn
\end{align}
with $\tilde q^2, \xi_1$ and $\xi_2$ defined in \eq{def_xi_i}, and 
where \eq{linear_combination_2} yields the \SCETii operators
\begin{subequations} \label{eq:Bf}
\begin{align} 
 \label{eq:O11}
 \cO^{(1)}_{\cB 1}(\lp_1,\lp_2,\lp_3) 
 &= \big[ \bar \chi_{\bar n_1,-\omega_2}
   S^{\dagger}_{\bar n_1} S_{n_1} \,g
   \slashed{\cB}_{n_1\perp,\omega_3}
   S_{n_1}^\dagger S_{n_1}  \chi_{n_1,-\omega_1} \big]
 \nn\\
 &= \big[ \bar \chi_{\bar n_1,-\w_2}\,(S_{\bar n_1}^\dagger S_{n_1})
 \,g \slashed{\cB}_{n_1\perp,\w_3}\, \chi_{ n_1,-\w_1} \big]
  \,,\\
  \label{eq:O12}
\cO_{\cB 2}^{(1)}\left(\lp_1,\lp_2,\lp_3\right) 
  &= \big[ \bar \chi_{\bar n_1,-\w_2}\,S_{\bar n_1}^\dagger S_{\bar n_1}\,g \slashed{\cB}_{\perp \bar n_1,\w_3}
   \, S_{\bar n_1}^\dagger S_{n_1}\, \chi_{ n_1,-\w_1} \big]
 \nn\\
  &= \big[ \bar \chi_{\bar n_1,-\w_2}\,g\slashed{\cB}_{\perp \bar n_1,\w_3}
   \, (S_{\bar n_1}^\dagger S_{n_1})\, \chi_{ n_1,-\w_1} \big]
  \,.
\end{align}
\end{subequations}
Notice that the soft Wilson lines between the quark and gluon fields in the same collinear direction 
cancel out as $S_{n}^\dagger S_{n} = 1$, so that overall the soft gluons couple to only the global color charge of the combined set of collinear fields in that direction. This 
leaves only the combination $(S_{\bar n_1}^\dagger S_{n_1})$ that already appear at LP, which is a key step in the proof that these NLP operators will lead to a factorization formula having the same soft factor as at LP, rather than a genuinely new soft factor.

It is natural to ask whether loop corrections to the operators in \eq{Bf} will require the introduction of evanescent operators that vanish in 4-dimensions. The results of Ref.~\cite{Beneke:2017ztn} imply that such operators are not needed for the one-loop anomalous dimension of our ${\cal O}(\lambda)$ operators ${\cal O}^{(1)}_{{\cal B}i}$, which are multiplicatively renormalized. We anticipate that this will continue to the be the case at higher orders because when additional Dirac matrices from collinear loops are inserted in the product of operators $\slashed{\cB}_{n\perp}\chi_{n}$, they can be reduced back to this form with $d$-dimensional Dirac algebra.

Note that unlike $C^{(0)}$, the new hard coefficient $C^{(1)}$ in \eq{JB1B2scetii} depends on extra momentum fractions $\xi_i$ that determine the split of the large momentum between quark and gluon fields in the same collinear direction. Examples of these convolutions are known from subleading power factorization for light-cone distribution functions in $B$-physics, see for eg.~\cite{Bauer:2002aj,Beneke:2003pa,Hill:2004if,Lee:2004ja,Beneke:2004in}. Once the SIDIS matrix elements have been factorized at NLP, we can anticipate from the field structure that the corresponding subleading power TMD will also depends on $\xi$. Thus we can already infer the presence of an additional convolution integral over $\xi$, between the hard coefficient and subleading power TMD. 

\subsubsection{NLP Operators with a $\cB_{s\perp}$, Hard and Hard-Collinear Contributions}
\label{sec:operators_NLP_Bs}

Next we turn to deriving the hard scattering Lagrangians for the NLP \SCETii operators with the field structures $\bar\chi_{n_1} {\cal B}_{s\perp}^{(n_1)\mu} \chi_{n_2}$ and $\bar\chi_{n_1} {\cal B}_{s\perp}^{(n_2)\mu} \chi_{n_2}$.  It turns out that there are two sources for these operators, from integrating out fluctuations at the hard scale, ${\cal L}_h^{(1)}$, and at the hard-collinear scale, ${\cal L}_{\rm hc}^{(1)}$. 

\paragraph{Hard scale contribution}
For the hard scale contribution the structure of the operators is constrained by RPI, and follows from the \SCETi operators given in the fifth line of \eq{Orpiexpn}. 
To discuss these terms it will be useful to first observe a few relations between operators in \SCETii.
In the back-to-back frame where $n_2=\bn_1$ and $\bn_2=n_1$ the following relation between forms of the soft operator will be useful
\begin{align}\label{eq:Os_relation}
 &  \bar \chi_{\bn_1,-\w_2}\,\big(S_{\bn_1}^\dagger \img\slashed{D}_{s\perp} S_{n_1} \big)\,  \chi_{ n_1,-\w_1}
   \\
  &\ \ 
  =\bar \chi_{\bn_1,-\w_2}\,\big(S_{\bn_1}^\dagger S_{n_1}\, \slashed{\cB}_{s\perp}^{(n_1)} \big)\,  \chi_{ n_1,-\w_1}
  + \bar \chi_{\bn_1,-\w_2}\,\big(S_{\bn_1}^\dagger S_{n_1}\, \big)\,  
   \slashed{\cP}_{\perp} \chi_{ n_1,-\w_1}
  \nn \\
  &\ \
  = \bar \chi_{\bn_1,-\w_2}\,\big(S_{\bn_1}^\dagger \img \slashed{D}_{s\perp} S_{\bn_1} S_{\bn_1}^\dagger S_{n_1} \big)\,  \chi_{ n_1,-\w_1}
  = \bar \chi_{\bn_1,-\w_2}\, \big(g\slashed{\cB}_{s\perp}^{(\bn_1)}+\slashed\cP_\perp\big) S_{\bn_1}^\dagger S_{n_1} \,  \chi_{ n_1,-\w_1}
  \nn\\
  &\ \
  = \bar \chi_{\bn_1,-\w_2}\,\big( g\slashed{\cB}_{s\perp}^{(\bn_1)} S_{\bn_1}^\dagger S_{n_1} \big)\,  \chi_{ n_1,-\w_1}
  + \bar \chi_{\bn_1,-\w_2}\, \big[\!\big[ \slashed{\cP}_{\perp}
    S_{\bn_1}^\dagger S_{n_1} \big]\!\big]\,  \chi_{ n_1,-\w_1}
   \nn\\
  &\qquad\quad
  + \bar \chi_{\bn_1,-\w_2}\,\big( S_{\bn_1}^\dagger S_{n_1} \big)\,
     \slashed{\cP}_{\perp}  \chi_{ n_1,-\w_1}
 \nn\\
  &\ \
  = \frac12 \bar \chi_{\bn_1,-\w_2}\,\big( g\slashed{\cB}_{s\perp}^{(\bn_1)}
      S_{\bn_1}^\dagger S_{n_1} + S_{\bn_1}^\dagger S_{n_1} 
     g\slashed{\cB}_{s\perp}^{(n_1)} \big)\,  \chi_{ n_1,-\w_1}
  \nn\\
  &\qquad\quad
  + \frac12 \bar \chi_{\bn_1,-\w_2}\,\big[\!\big[ \slashed{\cP}_{\perp}
    S_{\bn_1}^\dagger S_{n_1} \big]\!\big]\,  \chi_{ n_1,-\w_1}
  + \bar \chi_{\bn_1,-\w_2}\,\big( S_{\bn_1}^\dagger S_{n_1} \big)\,
     \slashed{\cP}_{\perp}  \chi_{ n_1,-\w_1}
  \,. \nn
\end{align}
To obtain the last equality we have symmetrized over the results in the first and third equalities.
Similarly we have
\begin{align} \label{eq:Os_relation2}
 &   \bar \chi_{\bn_1,-\w_2}\,\big(S_{\bn_1}^\dagger(-\img)
   \overleftarrow {\slashed{D}}_{s\perp} S_{n_1} \big)\,  \chi_{ n_1,-\w_1}
   \\
  &\ \
  = \bar \chi_{\bn_1,-\w_2}\,\big( g\slashed{\cB}_{s\perp}^{(\bn_1)} S_{\bn_1}^\dagger S_{n_1} \big)\,  \chi_{ n_1,-\w_1}
  + \bar \chi_{\bn_1,-\w_2} \slashed{\cP}_{\perp}^\dagger \,\big( 
     S_{\bn_1}^\dagger S_{n_1} \big)\,  \chi_{ n_1,-\w_1}
  \nn\\
  &\ \
  = \bar \chi_{\bn_1,-\w_2}\,\big( S_{\bn_1}^\dagger S_{n_1} 
     g\slashed{\cB}_{s\perp}^{(n_1)} \big)\, \chi_{ n_1,-\w_1} 
  - \bar \chi_{\bn_1,-\w_2}\,\big[\!\big[   \slashed{\cP}_{\perp} 
   S_{\bn_1}^\dagger S_{n_1} \big]\!\big]\,  \chi_{ n_1,-\w_1}
   \nn\\
  &\qquad\quad
  + \bar \chi_{\bn_1,-\w_2}     \slashed{\cP}_{\perp}^\dagger
   \,\big( S_{\bn_1}^\dagger S_{n_1} \big)\,  \chi_{ n_1,-\w_1}
  \nn\\
  &\ \
  = \frac12 \bar \chi_{\bn_1,-\w_2}\,\big( g\slashed{\cB}_{s\perp}^{(\bn_1)}
      S_{\bn_1}^\dagger S_{n_1} + S_{\bn_1}^\dagger S_{n_1} 
     g\slashed{\cB}_{s\perp}^{(n_1)} \big)\, \chi_{ n_1,-\w_1} 
   \nn\\
  &\qquad\quad
  - \frac12 \bar \chi_{\bn_1,-\w_2}\,\big[\!\big[   \slashed{\cP}_{\perp} 
   S_{\bn_1}^\dagger S_{n_1} \big]\!\big]\,  \chi_{ n_1,-\w_1}
  + \bar \chi_{\bn_1,-\w_2}     \slashed{\cP}_{\perp}^\dagger
   \,\big( S_{\bn_1}^\dagger S_{n_1} \big)\,  \chi_{ n_1,-\w_1}
  \,. \nn
\end{align}
In \eqs{Os_relation}{Os_relation2} the double brackets in $\big[\!\big[ \slashed \cP_\perp S_{\bn_1}^\dagger S_{n_1} \big]\!\big]$ indicate that the $\cP_\perp^\alpha$ derivative only acts inside, and for the latter equation we have used $\big[\!\big[  S_{\bn_1}^\dagger S_{n_1} \slashed \cP_\perp^\dagger \big]\!\big]= - \big[\!\big[ \slashed \cP_\perp S_{\bn_1}^\dagger S_{n_1} \big]\!\big]$.

For the contribution from the hard region, the Wilson coefficient of the \SCETii operator is fixed by the reparameterization invariance arguments in \sec{RPI} (which fixed the coefficients of the $\img \slashed D_{{\rm us}\perp}$ terms) to be the LP coefficient $C^{(0)}$.
These constraints are inherited from the constraint in \SCETi because the physics from integrating out the hard scale fluctuations is the same in the two theories.
To obtain this relation we consider the fifth line of \eq{Orpiexpn} and make use of \eq{Os_relation} and \eq{Os_relation2}.
When matching onto \SCETii the terms with $\slashed\cP_\perp \chi_{n_1,-\omega_1}$ in \eq{Os_relation} and $\bar\chi_{\bn_1,-\omega_2}\slashed\cP_\perp^\dagger$ in \eq{Os_relation2} must be dropped since they are already included explicitly on the third line of \eq{Orpiexpn}. 
When we add \eq{Os_relation} and \eq{Os_relation2} the terms involving $\big[\!\big[ \slashed \cP_\perp S_{\bn_1}^\dagger S_{n_1} \big]\!\big]$ cancel out.\footnote{Note that we do not have to consider these terms as a potential additional independent operator that is not constrained by RPI, due to the integration by parts argument explained in \sec{RPI}.}  Thus the sum of operators on the fifth line of \eq{Orpiexpn} purely involves terms with a $\cB_{s\perp}^{(n_1)}$ or $\cB_{s\perp}^{(n_2)}$.
The Dirac structure appearing for these terms in \eq{Orpiexpn} can also be simplified further in the back-to-back frame, where after the above considerations we now have
\begin{align}
& \frac12 C^{(0)}(\tilde q^{\,2})\: \bar\chi_{\bn_1,-\omega_2}
   \bigg[ \gamma^\mu \frac{1}{\bnP}  
    \gamma_\perp^\alpha  
    \frac{\slashed\bn_1}{2} 
    + \frac{\slashed n_1}{2} \gamma_\perp^\alpha 
    \frac{1}{\bnP^\dagger} \gamma^\mu 
   \bigg] \Big(g \cB_{s\perp\alpha}^{(\bn_1)} S_{\bn_1}^\dagger S_{n_1} + 
   S_{\bn_1}^\dagger S_{n_1} g \cB_{s\perp\alpha}^{(n_1)}\Big) 
   \chi_{n_1,-\omega_1}
 \nn\\
 &\ \
 = \Big( \frac{\bn_1^\mu}{2\omega_1} - \frac{n_1^\mu}{2\omega_2}\Big) 
  C^{(0)}(\tilde q^{\,2})\:
\bar\chi_{\bn_1,-\omega_2}
    \Big( g \slashed \cB_{s\perp}^{(\bn_1)} S_{\bn_1}^\dagger S_{n_1} 
    + S_{\bn_1}^\dagger S_{n_1} g \slashed \cB_{s\perp}^{(n_1)} \Big)
    \chi_{n_1,-\omega_1} \,.
\end{align}
To obtain the last line we used the projection relations on the SCET fermion fields given below \eq{chiB}.  Taken all together, we obtain the following hadronic current
\begin{align}\label{eq:current_Bs}
  J_{h\,{\cB}_s^\perp}^{(1)\mu}(0)
  &=\sum_{n_1}\int\!\df \w_1\df\w_2\,\,\sum_f 
  \frac{2(\tilde p_2^\mu-\tilde p_1^\mu)}{\tilde q^2}\, C^{(0)}(\tilde q^2)\,\cO^{(1)}_{\cB_s^\perp}\left(\lp_1,\lp_2\right)
  \,,
\end{align}
where the subscript $h$ indicates it arises as an ${\cal L}_h^{(1)}$ contribution. Here $\tilde q^2 = \omega_1 \omega_2$ and the operator 
\begin{align} \label{eq:OBs}
  \cO^{(1)}_{\cB_s^\perp}\left(\lp_1,\lp_2\right)\equiv 
  \frac12 
\bar \chi_{\bn_1,-\w_2}\,\Big( g \slashed \cB_{s\perp}^{(\bn_1)} S_{\bn_1}^\dagger S_{n_1} 
    + S_{\bn_1}^\dagger S_{n_1} g \slashed \cB_{s\perp}^{(n_1)} \Big)\,  \chi_{ n_1,-\w_1}
  \,.
\end{align}
Not that this operator is hermitian on its own if we take the dagger and use the freedom to exchange $n_1\leftrightarrow \bn_1$ and $\omega_1\leftrightarrow \omega_2$ due to the sum and integrals. Indeed this is the same combination that we saw for the final ${\slashed \cB}_{n_i\perp}$ operators in \eq{JB1B2scetii}, except that now the two terms have a common Wilson coefficient $C^{(0)}$ that does not depend on the soft gluon's momentum. By the same counting of helicity operators and use of symmetries carried out in \sec{operators_NLP_B}, we obtain that \eq{OBs} is the unique operator of this type at this order.

\paragraph{Hard-collinear scale contribution}
Next we consider contributions to the $\bar\chi_{n_1} {\cal B}_{s\perp}^{(n_1)\mu} \chi_{n_2}$ and $\bar\chi_{n_1} {\cal B}_{s\perp}^{(n_2)\mu} \chi_{n_2}$ operators from integrating out fluctuations at the hard-collinear scale $p^2\sim Q^2\lambda$, denoted as ${\cal L}_{\rm hc}^{(1)}$ contributions. 
Here the analysis for the number and Dirac structure for the operators exactly parallels that with ${\cal B}_{n_i\perp}^\mu$ from \sec{operators_NLP_B}. All the symmetry arguments that have been applied to determine the structure of $\cB_{n_i\perp}$-type operators in \sec{operators_NLP_B} can also be applied to ${\cal B}_{s\perp}^{(n_i)}$. The key difference for the operators is that the soft Wilson lines appear in the definition of ${\cal B}_{s\perp}^{(n_i)}$ itself, as well as multiplying it, whereas for the collinear ${\cal B}_{s\perp}^{(n_i)}$ the definition contains collinear Wilson lines, and it is multiplied by soft Wilson lines. This does not change the constraints on the Dirac structures, so the operators obtained from ${\cal L}_{\rm hc}^{(1)}$ must yield contributions of the same form as \eq{OBs}.

To derive results in the form of a factorization theorem valid to all orders in $\alpha_s$ for this ${\cal L}_{\rm hc}^{(1)}$ term, we again utilize the framework of matching from \SCETi onto \SCETii. Using \SCETi all the hard-collinear contributions can be obtained by factorizing the time-ordered products in \eq{SCET1Tproducts} when they are considered with external \SCETii kinematics.  In order to contribute to an offshell hard-collinear contribution, these time-ordered products must (after the BPS field redefinition) involve \SCETi Lagrangians ${\cal L}_{\rm I}^{(1,2)}$ with an explicit ultrasoft quark or gluon field that can inject an ultrasoft momentum. These terms can be obtained from Ref.~\cite{Bauer:2003mga}. At the order we are working there are no contributions with ultrasoft quark fields since $\psi_{us}\sim \lambda^3$, and the power counting would limit us to having at most one such ultrasoft quark field, which then leads to vanishing contributions in the soft matrix element by fermion number conservation. Thus only Lagrangians with ultrasoft gluon fields ${\cal B}_{us\perp}^{(n_i)\mu}\sim \lambda^2$ will contribute. The decomposition of the subleading power \SCETi Lagrangians using these ultrasoft gluon fields can be found in Ref.~\cite{Moult:2019mog}.  

Although it appears that this analysis could be fairly complicated, it turns out that for the NLP terms in SIDIS there is a significant simplification. As explained in detail in \sec{fact_vanishing_soft} below, any term involving a 
${\cal O}_{\rm I}^{(0)}$ does not contribute to the $W_i$ structure functions that are of interest here, namely those which obtain their first non-zero contributions at NLP. Thus the only non-trivial time-ordered product that we will need from the list in \eq{SCET1Tproducts} is, in the back-to-back frame where $n_2=\bar n_1$, 
\begin{align}   \label{eq:Tprod11}
T[ {\cal O}_{\rm I}^{(1)} {\cal L}_{\rm I}^{(1)} ]
  &=  \sum_{n_1}  \int\!\! \df^4x \: 
   T\, {\cal O}_{\rm I}^{(1)}(0) \Big( {\cal L}_{n_1\,\rm I}^{(1)}(x)
   + {\cal L}_{\bn_1\,\rm I}^{(1)}(x) \Big)
  \,.
\end{align}
Here the ${\cal O}_{\rm I}^{(1)}$ operators are just the $\bar\chi_{\bn_1} {\cal P}_{\perp}^{\mu} \chi_{n_1}$ and $\bar\chi_{\bn_1} {\cal B}_{n_i\perp}^{\mu} \chi_{n_1}$ terms given in \eqs{crrnt_P1}{JB1B2} with \SCETi collinear fields.  The relevant terms in the subleading power Lagrangian are
\begin{align}
  {\cal L}_{n\,{\rm I}}^{(1)} &=  {\cal L}_{\chi_n \cB_{us}^\perp}^{(1)}  
      + {\cal L}_{A_n \cB_{us}^\perp}^{(1)}  
  = g {\cB}_{us\perp \alpha}^{(n) a} \Big( K_{\chi_n}^{\alpha a (1)} 
   +  K_{A_n}^{\alpha a(1)} \Big)
  \,,
\end{align}
where we can read off the required formula from Ref.~\cite{Moult:2019mog},
\begin{align}
  K_{\chi_n}^{\alpha a(1)}
  &= \bar \chi_n T^a \cP_{\perp}^\alpha \frac{\slashed \bn}{\bar \cP} \chi_n  
   +\bar \chi_n \Bigl(T^a\gamma^\alpha_\perp\frac{1}{\bar \cP} g\Sl{\cB}_{n\perp} 
   +g\Sl{\cB}_{n\perp}\frac{1}{\bar \cP} \gamma_\perp^\alpha T^a \Bigr) \frac{\Sl{\bar n}}{2} \chi_n 
  \,,\nn \\
  K_{A_n}^{\alpha a(1)} 
 &= -if^{abc} \left\{ 
  \cB^{\mu b}_{n\perp}[\cP^\mu_\perp \cB^{\alpha c}_{n\perp}]
  -\cB^{\mu b}_{n\perp}[\cP^\alpha_\perp \cB^{\mu c}_{n\perp}]
  + [\bar \cP \cB^{\alpha c}_{n\perp}] n\cdot \cB^b_n  \right\}
  +g f^{cbe} f^{ade} \cB^{\mu c}_{n\perp}   \cB^{\alpha b}_{n\perp} 
 \cB^{\mu d}_{n\perp} 
  \nn \\
  &~~~\,  
  +2 \text{Tr}  \left(    \bar c_n [ T^a,[i D^\perp_{n\alpha}, c_n]]        \right)  
  +2 \text{Tr}  \left(    \bar c_n  [ \cP_{\perp \alpha}  , [W_n   T^a   W^\dagger_n, c_n] ]   \right)  
 \nn \\
&~~~\,
  + 2\zeta \text{Tr}\left( [T^a,A_{n\perp \alpha}][i \partial_n^\nu,A_{n\nu}]     \right) 
  \,. 
\end{align}
Here the contributions to the pure gluon action are written in a general covariant gauge with gauge parameter $\zeta$, and $c_n$ are $n$-collinear ghost fields. Note that only transverse ultrasoft gluon fields, ${\cB}_{us\perp}^{(n)}$, appear in the \SCETi Lagrangian at this order, explaining why we only have the possibility of matching onto the \SCETii operators of the form $\bar\chi_{\bn_1} \cB_{s\perp}^{(n_i)}\chi_{n_1}$.

Consider first the \SCETi $\to$ \SCETii matching for the $\bar\chi_{\bn_1} {\cal P}_{\perp}^{\mu} \chi_{n_1}$ terms in ${\cal O}_{\rm I}^{(1)}$.  We find that all such contributions enter beyond ${\cal O}(\eta)$, ie. with $E\ge 1/2$ in \eq{SCETitoiipc}. This occurs because even in the presence of $n$-collinear propagators and loops in \SCETi, the $\cP_\perp^\mu$ always gives an external momentum, and hence an extra suppression factor when matching to \SCETii.  For example, consider the following tree level and one-loop contractions
\begin{align}  \label{eq:Pexamples}
 \int\!\!\df^4 x \ 
 & [\bar\chi_{\bn_1} {\slashed \cP}_\perp\! 
  \color{orange} \underbrace{ \color{black}
  \,\,\,\overbracket{ \!\!\! 
  \chi_{n_1}] (0) \Big[ \bar\chi_{n_1} 
  \!\!\!\! }\,\,\,\,
  } \color{black}
 g {\slashed B}_{us\perp} \frac{1}{\bnP} (\slashed\cP_\perp + g {\slashed B}_{n_1\perp} )
  \frac{\slashed\bn}{2} \chi_n \Big](x)
  \,,\\[-5pt]
 & \color{orange}{ 
  \  \lambda \hspace{0.4cm}  \lambda \hspace{1cm}  \lambda^{-1} 
  \hspace{0.8cm}  \lambda \hspace{1.7cm}  \lambda \hspace{1.8cm} \lambda \hspace{1cm}  = {\cal O}(\lambda^4) }
  \nn\\[5pt]
 \int\!\!\df^4 x\, \df^4 y \ 
 & [\bar\chi_{\bn_1} {\slashed \cP}_\perp\!
  \color{orange} \underbrace{ \color{black}  
 \,\,\,\overbracket{ \!\!\! \chi_{n_1}] (0) {\cal L}_n^{(0)}(y) 
  \hspace{-1cm} } \hspace{1cm}
  \hspace{-0.7cm} \overbracket{ \hspace{0.7cm} 
  \hspace{-0.6cm}  \overbracket{ \hspace{0.6cm} \Big[ \bar\chi_{n_1}
  \hspace{-0.4cm}  } \hspace{0.4cm}
   (\slashed\cP_\perp + g {\slashed B}_{n_1\perp} ) 
  \hspace{-0.7cm} } \hspace{0.7cm}
    } \color{black}
  \frac{1}{\bnP} g{\slashed B}_{us\perp}
  \frac{\slashed\bn}{2} \chi_n \Big](x)
  \,,\nn\\
 & \color{orange}{ 
  \  \lambda \hspace{0.4cm}  \lambda \hspace{2.7cm}  \lambda^{-1/2} 
  \hspace{2.7cm}  \lambda \hspace{1.1cm} \lambda \hspace{0.5cm}  = {\cal O}(\lambda^{7/2}) }
  \nn
\end{align} 
In orange we show the power counting obtained when matching to \SCETii, which is ${\cal O}(\lambda^2)$ and ${\cal O}(\lambda^{3/2})$ suppressed relative to the leading power hard scattering operator. The power counting for the contractions is given together and includes factors of $\lambda$ from the hard-collinear space-time integrations. A similar suppression is obtained for any other contractions involving the $\bar\chi_{\bn_1} {\cal P}_{\perp}^{\mu} \chi_{n_1}$ current, so these terms only start to contribute beyond the ${\cal O}(\lambda)$ NLP corrections that we are considering. 

Next we consider the \SCETi $\to$ \SCETii matching from the $\bar\chi_{n_1} {\slashed \cB}_{n_i\perp} \chi_{\bn_1}$ terms in ${\cal O}_{\rm I}^{(1)}$.
Just like the previous case, at tree level the time-ordered products here again only yield contributions beyond leading power, with $E\ge 1$. This occurs because only one pair of fields are contracted from each of ${\cal O}_{\rm I}^{(1)}$ and ${\cal L}_{\rm I}^{(1)}$ in \eq{Tprod11}, so we end up with the same type of suppression observed in the first example of \eq{Pexamples}. On the other hand, once loops are considered a contribution at NLP can now be obtained. This occurs because we can carry out contractions of all fields in loops at the hard-collinear scale except for the fields $\bar\chi_{\bn_1} \cB_{us\perp}^\mu \chi_{n_1}$, and hence end up with a ${\cal L}_{\rm hc}^{(1)}$ contribution that matches onto the \SCETii operators $\bar\chi_{n_1} {\slashed \cB}_{s\perp}^{(n_i)}\chi_{\bn_1}$.  As an example of a one-loop contraction that contributes at NLP we have
\begin{align}  \label{eq:Bexample}
\int\!\!\df^4 x \ 
 & [\bar\chi_{\bn_1} 
  \color{orange} \underbrace{ \color{black}  
 \hspace{0.3cm} \overbracket{ \hspace{-0.3cm} {\slashed \cB}_{n_1\perp} 
 \,\,\,\overbracket{ \!\!\! \chi_{n_1}] (0) \Big[ \bar\chi_{n_1}
  \hspace{-0.4cm}  } \hspace{0.4cm}
   (\slashed\cP_\perp + g {\slashed B}_{n_1\perp} ) 
  \hspace{-0.7cm} }     \hspace{0.7cm}
   } \color{black} 
  \frac{1}{\bnP} g{\slashed B}_{us\perp}
  \frac{\slashed\bn}{2} \chi_n \Big](x)
  \,,\\*[-3pt]
 & \color{orange}{ 
  \  \lambda \hspace{2.85cm}   \lambda^{0} 
   \hspace{2.95cm}  \lambda\hspace{0.8cm} \lambda \hspace{0.5cm}  = {\cal O}(\lambda^3) }
  \nn
\end{align} 
where again we have shown the power counting in $\lambda$ after the result is matched to \SCETii.
This example has contractions that yield an $n_1$-hard-collinear loop.  There are mirror contributions from $\bn_1$-hard-collinear loops as well involving the operator $\bar\chi_{\bn_1} {\slashed \cB}_{\bn_1\perp} \chi_{n_1}$. 
If we consider the complete basis of \SCETii operators that can be obtained from this hard-collinear matching, then we can use the fact that the result must involve a ${\cB}_{s\perp}^{(n_i)\mu}$, and is constrained by the same symmetries that we already considered when carrying out the matching for ${\cal L}_h^{(1)}$.  This implies that these ${\cal L}_{\rm hc}^{(1)}$ contributions will match onto the same two operators that appears above in \eq{OBs}.
We have confirmed that the sum of hard-collinear one-loop diagrams in \SCETi is non-zero from contractions like the one in \eq{Bexample}, and that it yields a non-zero contribution to the NLP matching onto \SCETii that gives either one of the two operators in \eq{OBs}.  

To formulate the result from the hard-collinear loop corrections in a form that is valid to all orders in $\alpha_s$, we first note that the ${\cal O}_{\rm I}^{(1)}$ operators $\bar\chi_{\bn_1,\omega_1} {\slashed \cB}_{n_i\perp} \chi_{n_1,\omega_2}$ have hard coefficients $C^{(1)}(\tilde q^2,\xi)$ whose form was determined in \sec{operators_NLP_B}, see \eq{JB1B2} with \eqs{C11C12}{C1}. 
For an $n_1$-hard-collinear loop the matching result depends on $\xi_1$, and on the $p_{n_1}^-=\omega_1\sim \lambda^0$ and $p_s^+\sim \lambda$ momentum of the $n$-collinear and soft fields in \eq{OBs}. For an $\bn_1$-hard-collinear loop the result depends on $\xi_2$, and the $p_{\bn_1}^+=\omega_2\sim \lambda^0$ of the $\bn_1$-collinear field and $p_s^-\sim \lambda$ momentum of the soft field.
These hard-collinear loops give rise to a non-trivial scalar matching coefficient function $J_{\cB_s}$.  Thus the all orders form for the matching is
\begin{align}  \label{eq:J1hcBs}
  & J^{(1)\mu}_{{\rm hc}\cB_s^\perp}(0) 
   = \sum_{n_1} \int\!\! \df\omega_1\df\omega_2\:
  \sum_f \frac{(\tilde p_2^\mu-\tilde p_1^\mu)}{\tilde q^2}\,
   \\
  &\times \bigg\{
    \int\!\! \df\xi_1\  C^{(1)}(\tilde q^2,\xi_1) 
    \int\!\! \df p_s^+ \: J_{\cB_s^\perp}(\omega_1 p_s^+,\xi_1)\,
    \bar\chi_{\bn_1,-\omega_2} \big[ S_{\bn_1}^\dagger S_{n_1} g {\slashed\cB}_{s\perp}^{(n_1)} \big]_{p_s^+}\, \chi_{n_1,-\omega_1}
   \nn\\
 &\ \ \ + 
  \int\!\! \df\xi_2\ C^{(1)}(\tilde q^2,\xi_2) 
  \int\!\! \df p_s^- \,  J_{\cB_s^\perp}(\omega_2 p_s^-,\xi_2) \,
   \bar\chi_{\bn_1,-\omega_2} \big[ g {\slashed\cB}_{s\perp}^{(\bn_1)}  S_{\bn_1}^\dagger S_{n_1} \big]_{p_s^-}\, \chi_{n_1,-\omega_1}
   \bigg\}
  \nn\\
 &= \sum_{n_1} \int\!\! \df\omega_1\df\omega_2 \: 
 \sum_f  \frac{(\tilde p_2^\mu-\tilde p_1^\mu)}{\tilde q^2}\,
  \nn \\
  &\times \bigg\{ 
    \int\!\! \df\xi_1\  C^{(1)}(\tilde q^2,\xi_1)
    \int\!\! \frac{\df b_s^-}{\omega_1} \, 
    \tilde J_{\cB_s^\perp}(b_s^-/\omega_1,\xi_1)\,
    \bar\chi_{\bn_1,-\omega_2} \big[ S_{\bn_1}^\dagger S_{n_1} g {\slashed\cB}_{s\perp}^{(n_1)} \big](b_s^-)\, \chi_{n_1,-\omega_1}
   \nn\\
 &\ \ \ + 
  \int\!\! \df\xi_2\ C^{(1)}(\tilde q^2,\xi_2) 
   \int\!\! \frac{\df b_s^+}{\omega_2} \, 
   \tilde J_{\cB_s^\perp}(b_s^+/\omega_2,\xi_2) \,
  \bar\chi_{\bn_1,-\omega_2}\big[g {\slashed\cB}_{s\perp}^{(\bn_1)}  S_{\bn_1}^\dagger S_{n_1}  \big](b_s^+)\, \chi_{n_1,-\omega_1}
   \bigg\} .
 \nn 
\end{align}
Here the subscripts $p_s^\pm$ fix the fields in the $[\cdots]_{p_s^\pm}$ to have these momenta, for example 
\phantom{stuff}
$\big[ S_{\bn_1}^\dagger S_{n_1} g {\slashed\cB}_{s\perp}^{(n_1)} \big]_{p_s^+} = \big[\delta(p_s^+-  n_1\cdot\img \partial_s) S_{\bn_1}^\dagger S_{n_1} g {\slashed\cB}_{s\perp}^{(n_1)} \big]$.
Only a single dimensionless function $J_{\cB_s}(s,\xi)$ appears in \eq{J1hcBs} due to the symmetry between $n_1$ and $\bn_1$ hard-collinear loops. When renormalized in the ${\overline {\rm MS}}$ scheme $J_{\cB_s}(s,\xi,\mu)$ will involve logarithms like $\ln(s/\mu^2)$, and we note that the scale of its first argument, $s\sim Q^2\lambda$, is hard-collinear.
The dimension-$(-2)$ matching coefficients $\tilde J_{\cB_s}$ are the Fourier transform of the momentum space coefficients $J_{\cB_s}$ with respect to the soft momenta,
\begin{align}
 \frac12 \int\!\! \df p_s^+\: e^{-i b_s^- p_s^+/2}\: J_{\cB_s^\perp}(p_s^+\omega_1,\xi_1)
  &= \frac{1}{\omega_1} \tilde J_{\cB_s^\perp}(b_s^-/\omega_1,\xi_1) 
\,.
\end{align} 

\paragraph{Total Contribution} To conclude this section we add the hard and hard-collinear contributions from \eqs{current_Bs}{J1hcBs} to obtain the final \SCETii current with ${\cal B}_{s\perp}^{(n_i)}$ insertions in the following form:
\begin{align}\label{eq:current_Bs_full}
  J_{\cB_s^\perp}^{(1)\mu}(0) 
  &=J_{h\,\cB_s^\perp}^{(1)\mu}(0) + J_{{\rm hc}\,\cB_s^\perp}^{(1)\mu}(0)
   \\
  &=\sum_{n_1}\int\!\df \w_1\df\w_2\,\sum_f 
  \frac{(\tilde p_2^\mu-\tilde p_1^\mu)}{\tilde q^2}\,
  \nn\\
 &\quad\times 
\bigg\{ 
   \int\!\! \frac{\df b_s^-}{\omega_1} \,  
  C_{\cB_s^\perp}^{(1)}\big(\tilde q^2, b_s^-/\omega_1\big) \,
  \bar\chi_{\bn_1,-\omega_2} \big[ S_{\bn_1}^\dagger S_{n_1} g {\slashed\cB}_{s\perp}^{(n_1)} \big](b_s^-)\, \chi_{n_1,-\omega_1}
  \nn\\
 &\quad \ \ \ +
  \int\!\! \frac{\df b_s^+}{\omega_2} \,  
  C_{\cB_s^\perp}^{(1)}\big(\tilde q^2, b_s^+/\omega_2\big) \,
  \bar\chi_{\bn_1,-\omega_2} \big[g {\slashed\cB}_{s\perp}^{(\bn_1)}  S_{\bn_1}^\dagger S_{n_1}  \big](b_s^+)\, \chi_{n_1,-\omega_1} 
    \bigg\} 
  \nn\\
 &= \sum_{n_1}\int\!\df \w_1\df\w_2\,\sum_f 
  \frac{(\tilde p_2^\mu-\tilde p_1^\mu)}{\tilde q^2}\,
  \int\!\!\df \hat b_s \,   
  C_{\cB_s^\perp}^{(1)}\big(\tilde q^2, \hat b_s \big) \,
  \nn\\
 &\quad\times 
  \bar\chi_{\bn_1,-\omega_2}  \Big\{  
  \big[ S_{\bn_1}^\dagger S_{n_1} g {\slashed\cB}_{s\perp}^{(n_1)} \big](\omega_1\hat b_s)
  +
  \big[g {\slashed\cB}_{s\perp}^{(\bn_1)}  S_{\bn_1}^\dagger S_{n_1}  \big](\omega_2\hat b_s)
    \Big\} \, \chi_{n_1,-\omega_1}
  \,, \nn
\end{align}
where it was convenient to introduce the common integration variable $\hat b_s$, which has dimension-$-2$. The combined matching coefficients containing both hard and hard-collinear contributions are
\begin{align}
  C_{\cB_s^\perp}^{(1)}\big(\tilde q^2, \hat b_s \big)
  &= 
C^{(0)}(\tilde q^2)\, \delta( \hat b_s) 
   \ +\ \int\!\! \df\xi\:  C^{(1)}(\tilde q^2,\xi)\,
    \tilde J_{\cB_s^\perp}(\hat b_s,\xi)
  \,.
\end{align}

\subsection{Factorization with Leading Power Currents}
\label{sec:factorization_LP}

In this section, we calculate the contributions from the LP current $J^{(0)}$ to the factorization theorem.
This will both reproduce the well-known LP factorization theorem and induce kinematic corrections at NLP.
We start from the definition of the hadronic tensor in \eq{W}, evaluated with the LP current $J^{(0)}$,
\begin{align} \label{eq:W_rep}
 W_{J^{(0)}}^{\mu\nu} &
 = \SumInt_X \, \delta^4(q + \Pa  - \Pb - \PX)
   \braket{N | J^{(0)\,\dagger\,\mu}(0) | h,X} \braket{h,X | J^{(0)\,\nu}(0) | N}
\,.\end{align}
Note that here, we have not yet absorbed the $\delta$ function by shifting the current
as in \eq{W}, as it will be illustrative to manipulate the $\delta$ function directly,
rather than having to insert the hard current to gain access to its $b$ dependence.
Of course, in the end both approaches give the same result.

The state $X$ and its phase space integral $\SumInt_X$ are factorized into the two collinear and one soft sectors,
\begin{equation} \label{eq:factorize_sectors}
  \ket{X} = \ket{X_n} \ket{X_\bn} \ket{X_s}
  \,,\qquad\qquad
  \SumInt_{X} = \SumInt_{X_n} \SumInt_{X_\bn} \SumInt_{X_s}\,,
\end{equation}
so that the collinear and soft fields in the SCET operators overlap only with
the corresponding counterpart of the state $X$.
We also decompose $P_X$ as
\begin{equation}
  P_X = P_{X_n}+P_{X_\bn}+P_{X_s}
\,,\end{equation}
and define the momenta of the initial and final state partons as
\begin{equation} \label{eq:pa_pb}
  p_a=P_N-P_{X_n}\,,\qquad p_b=P_h+P_{X_\bn}
\,.\end{equation}
This definition is motivated by the observation that the $n$-collinear radiation $X_n$
dominantly arises from radiation off the incoming hadron $N$, and thus by momentum conservation
the struck parton carries all momentum of $P_N$ that is not radiated away by $X_n$,
and similarly for the final-state momentum $p_b$.
The $\delta$ function in \eq{W_rep} is not homogeneous in the power counting,
as is easy to see from the scalings $q \sim Q(1, 1, \lambda)$, $p_a \sim Q(\lambda^2, 1, \lambda)$ and $p_b \sim Q( 1, \lambda^2,\lambda)$.
Expanding it to NLP, we obtain\footnote{There are two alternative ways to treat the expansion in \eq{delta_expanded} which we discuss in \sec{nbpartials}.}
\begin{align} \label{eq:delta_expanded}
 \delta^4(q + \Pa  - \Pb - \PX) &
 = \delta^4(q +p_a - p_b - P_{X_s})
\\\nn&
 = \bigl[ 1+\bn\cdot P_{X_s} \partial_{\w_a}-n\cdot P_{X_s} \partial_{\w_b} + \cO(\lambda^2) \bigr]
   \\\nn&\quad  \times
   2 \delta(\w_a- \bn\cdot p_a) \, \delta(\w_b-n\cdot p_b)\, \delta^{(2)}(\vec q_T+\vec p_{a,T}-\vec p_{b,T}- \vec P_{X_s,T})
\,.\end{align}
Here, we parameterized $q^\mu$ as
\begin{equation}
 q^\mu = (q^+, q^-, q_\perp) = (\w_b,  -\w_a, q_\perp)
\end{equation}
with $\w_{a,b} > 0$ and expanded the $\delta$ functions around these $\cO(1)$ components $\w_{a,b}$.

From \sec{operators_LP}, the LP current and its conjugate are given by
\begin{align} \label{eq:crrnt_lp_rep}
 J^{\,\nu}(0) &
 = \sum_{n_1,n_2} \sum_f (\gamma_{\perp}^\nu)^{\alpha\beta}
   \int\!\df\w_1\,\df\w_2\, C^{(0)}(2 \lp_1 \cdot \lp_2)
   \bar{\chi}_{n_2,-\w_2}^{\alpha \bar a} \bigl(S_{n_2}^\dagger S_{n_1}\bigr)^{a \bar b}\chi_{{n_1},-\w_1}^{\beta b}
\,,\\\nn
 J^{\dagger \,\mu}(0) &
 = \sum_{n'_1,n'_2} \sum_{f'}  (\gamma_{\perp}^\mu)^{\beta'\alpha'}
   \int\!\df\w'_1\,\df\w'_2\, C^{(0)}(2 \lp'_1 \cdot \lp'_2)
   \bar\chi_{{n'_1},\w'_1}^{\beta' \bar b'} \bigl( S_{n'_1}^\dagger S_{n'_2} \bigr)^{b' \bar a'} \chi_{n'_2,\w'_2}^{\alpha' a'}
\,,\end{align}
where for brevity we continue to suppress the flavor and explicit position of the fields at the origin.
In the following, fields and Wilson coefficients with prime momenta depend on $f'$,
while the ones without primes depend on $f$, and ultimately will be fixed to equal, $f = f'$. 
Inserting \eq{crrnt_lp_rep} together with \eq{delta_expanded} into \eq{W_rep}, we obtain
\begin{align} \label{eq:W_J0_1}
 W_{J^{(0)}}^{\mu\nu} &
 = \sum_{n'_1,n'_2,n_1,n_2} \sum_{f,f'}
   \int\!\df\w_1\,\df\w_2\, \int\!\df\w'_1\,\df\w'_2\,
   C^{(0)}(2 \lp_1 \cdot \lp_2) C^{(0)}(2 \lp'_1 \cdot \lp'_2)
   (\gamma_{\perp}^\nu)^{\alpha\beta} (\gamma_{\perp}^\mu)^{\beta'\alpha'}
   \nn\\&\quad\times
   \SumInt_{X_n} \SumInt_{X_\bn} \SumInt_{X_s} \, \delta^{(2)}(\vec q_T+\vec p_{a,T}-\vec p_{b,T}- \vec P_{X_s,T})
   \nn\\&\quad\times
   \bigl\{ \bigl[ 1+\bn\cdot P_{X_s} \partial_{\w_a}-n\cdot P_{X_s} \partial_{\w_b} + \cO(\lambda^2) \bigr]
           \, 2 \, \delta(\w_a- \bn\cdot p_a) \, \delta(\w_b-n\cdot p_b)  \bigr\}
   \nn\\&\quad\times
   \Braket{N | \bar\chi_{{n'_1},\w'_1}^{\beta' \bar b'} \bigl(S_{n'_1}^\dagger S_{n'_2} \bigr)^{b' \bar a'} \chi_{n'_2,\w'_2}^{\alpha' a'} | h,X_n,X_\bn,X_s}
 \nn\\&\quad\times
   \Braket{h,X_n,X_\bn,X_s |\bar{\chi}_{n_2,-\w_2}^{\alpha \bar a} \bigl(S_{n_2}^\dagger S_{n_1} \bigr)^{a \bar b}\chi_{{n_1},-\w_1}^{\beta b} | N}
\,.\end{align}
Here, the directions $n_i$ and momenta $\w_i$ are still arbitrary and thus summed and integrated over, respectively.
By employing that the hadron states are collinear and that $X$ factorizes as in \eq{factorize_sectors},
we can factorize the matrix elements into their $n$-collinear, $\bn$-collinear and soft sectors,
which then fixes these directions and momenta.
For example, one of the matrix elements we encounter is
\begin{align}
 \braket{N | \bar\chi_{{n'_1},\w'_1}^{\beta' \bar b'} | X_n} \braket{X_n | \chi_{{n_1},-\w_1}^{\beta b} | N}
 &= \delta_{ff'} \delta_{n'_1, n} \delta_{n_1,n} \theta(-\w_1) \delta(\w'_1 - \w_1)
 \nn\\&\quad\times
   \braket{N | \bar\chi_{n}^{\beta' \bar b'} | X_n}
   \braket{X_n | \chi_{n,-\w_1}^{\beta b} | N}
\,.\end{align}
Here, we used that by flavor conservation the (suppressed) flavor indices must be equal,
and that the $n$-collinear nucleon state $N$ fixes $n_1 = n'_1 = n$.
Momentum conservation implies that both quark fields have equal momenta $\w_1 = \w'_1$,
such that we let the second quark field be unconstrained, and the $\theta$ function
fixes the correct sign of the quark momentum following our conventions of $\chi_{n,-\w}$.
Note that by fermion number conservation, we must have a $\bar\chi_n$
and a $\chi_n$ field in the $n$-collinear matrix element, and combinations
such as $\chi_n$ and $\chi_n$ are excluded.
Carrying out this straightforward algebra, we arrive at
\begin{align} \label{eq:W_J0_2}
 W_{J^{(0)}}^{\mu\nu} &
 = \sum_f \int\!\df\w_1\,\df\w_2 \, |C^{(0)}(\w_1 \w_2)|^2
   (\gamma_{\perp}^\nu)^{\alpha\beta} (\gamma_{\perp}^\mu)^{\beta'\alpha'}
   \nn\\&\quad\times
   \SumInt_{X_n} \SumInt_{X_\bn} \SumInt_{X_s} \delta^{(2)}(\vec q_T+\vec p_{a,T}-\vec p_{b,T}- \vec P_{X_s,T})
   \nn\\&\quad\times
   \bigl\{ \bigl[ 1+\bn\cdot P_{X_s} \partial_{\w_a}-n\cdot P_{X_s} \partial_{\w_b} + \cO(\lambda^2) \bigr]
   2 \, \delta(\w_a+\w_1) \, \delta(\w_b-\w_2) \bigr\}
  \nn\\&\quad\times \Bigl\{
   \theta(-\w_1)\braket{N | \bar\chi_{n}^{\beta' \bar b'} | X_n} \braket{X_n | \chi_{n,-\w_1}^{\beta b} | N}
   \times
   \theta(\w_2) \braket{0 | \chi_{\bn}^{\alpha' a'} | h, X_\bn} \braket{h, X_\bn |\bar{\chi}_{\bn,-\w_2}^{\alpha \bar a} | 0}
   \nn\\&\qquad\qquad\times
   \braket{0 | \bigl( S_n^\dagger S_{\bn} \bigr)^{b' \bar a'} | X_s} \braket{X_s | \bigl(S_{\bn}^\dagger S_{n}\bigr)^{a \bar b} | 0}
   \nn\\&\qquad+
   \theta(-\w_1) \braket{N | \chi_{n}^{\alpha' a'} | X_n} \braket{X_n |\bar{\chi}_{n,-\w_1}^{\alpha \bar a}  | N}
   \times
   \theta(\w_2) \braket{0 | \bar\chi_{\bn}^{\beta' \bar b'}  | h, X_\bn} \braket{h, X_\bn | \chi_{\bn,-\w_2}^{\beta b} | 0}
   \nn\\&\qquad\qquad\times
   \braket{0 | \bigl( S_{\bn}^\dagger S_n \bigr)^{b' \bar a'} | X_s} \braket{X_s | \bigl(S_{n}^\dagger S_\bn\bigr)^{a \bar b} | 0}
   \Bigr\}
\,.\end{align}
(Note that we have swapped $\w_1\leftrightarrow\w_2$ labels for the antiquark contribution.)
Here, the two terms in the curly brackets encode the contributions from quarks $f$ and antiquarks $\bar f$, respectively.
We have used that by momentum conservation $-\w_1 = \bn \cdot (P_N - P_{X_n}) = \bn \cdot p_a$
and $\w_2 = n \cdot (P_h + P_{X_\bn}) = n \cdot p_b$.
The derivatives only act on the $\delta$ functions, and the sum over complete states is still pulled out
as the $\delta$ functions act on the momenta of the $X_i$. To resolve this, we first rewrite
\begin{align}
 \delta^{(2)}(\qt +\vec p_{a,T}-\vec p_{b,T}- \vec P_{X_s,T}) &
 = \int\!\frac{\df^2\bt}{(2\pi)^2}\, e^{-\img \bt \cdot (\qt - \vec P_{X_n,T} - \vec P_{X_\bn,T} - \vec P_{X_s,T})}
\,,\end{align}
where we used \eq{pa_pb} to set $\vec p_{a,T} = - \vec P_{X_n,T}$ and $\vec p_{b,T} = \vec P_{X_\bn,T}$.
We can then absorb the phases using the standard trick
\begin{align}
 e^{\img \bt \cdot \vec P_{X_n,T}} \braket{N | \bar\chi_{n}^{\beta' \bar b'}(0) | X_n} 
 = \braket{N | \bar\chi_{n}^{\beta' \bar b'}(b_\perp) | X_n} 
\,,\end{align}
and likewise for the other matrix elements. For the soft matrix element, we also have to take care of
the $P_{X_s}^\pm$ terms in the square brackets, which arise at NLP.
This can be achieved using the additional transformation
\begin{align} \label{eq:subl_soft}
 P_{X_s}^+
 \braket{0 | (S_n^\dagger S_{\bn})^{b' \bar a'}(0,0,b_\perp)| X_s} &
 = 2\img \frac{\partial}{\partial b_s^-} e^{-\frac{\img}{2} b_s^- P_{X_s}^+} \braket{0 | (S_n^\dagger S_{\bn})^{b' \bar a'}(0,0,b_\perp)| X_s} \Bigr|_{b_s^- = 0}
 \nn\\&
 = 2\img \frac{\partial}{\partial b_s^-} \braket{0 | (S_n^\dagger S_{\bn})^{b' \bar a'}(0,b_s^-,b_\perp)| X_s} \Bigr|_{b_s^- = 0}
\,.\end{align}
Here, the first line is a trivial equality, while in the second line we again used momentum conservation
to shift the soft current to the residual position $b_s^-$.
In this fashion, all dependence on the momenta $P_{X_i}$ are expressed through either the labels
or the position of the quark fields and soft Wilson lines, and \eq{W_J0_2} becomes
\begin{align} \label{eq:W_J0_3}
 W_{J^{(0)}}^{\mu\nu} &
 = (\gamma_{\perp}^\nu)^{\alpha\beta} (\gamma_{\perp}^\mu)^{\beta'\alpha'}
   \int\!\frac{\df^2\bt}{(2\pi)^2}\, e^{-\img \bt \cdot \qt}
   \biggl[ 1 + 2\img \frac{\partial}{\partial b_s^+} \partial_{\w_a} - 2\img \frac{\partial}{\partial b_s^-} \partial_{\w_b} + \cO(\lambda^2) \biggr]
  \nn\\&\quad\times
   \sum_f 2 \cH_f^{(0)}(-\w_a \w_b) \Bigl[
   \hat B_{f/N}^{\beta\beta'\,\bar b'b}(\w_a, \bt)
   \cG_{h/f}^{\alpha'\alpha\,a'\bar a}(\w_b, \bt)
   \hat S_f^{b' \bar a' a \bar b}(b_T, b_s^+ b_s^-)
   \nn\\&\hspace{3.2cm}+
   \hat B_{\bar f/N}^{\alpha'\alpha\,a'\bar a}(\w_a, \bt)
   \hat \cG_{h/\bar f}^{\beta\beta'\,\bar b'b}(\w_b, \bt)
   S_{\bar f}^{b' \bar a' a \bar b}(b_T, b_s^+ b_s^-)
   \Bigr]_{b_s^\pm = 0}
\,.\end{align}
Here, we defined the LP hard function as
\begin{align}
 \cH_f^{(0)}(\tilde q^2) = |C_f^{(0)}(\tilde q^2)|^2
\,,\end{align}
where here $\tilde q^2 = -\w_a \w_b = q^+ q^- < 0$.
The correlators for the quark contribution are given by
\begin{align} \label{eq:quark_correlators_LP}
 \hat B_{f/N}^{\beta\beta'\,\bar b'b}(\w_a, \bt) &
 = \theta(\w_a) \braket{N | \bar\chi_{n}^{\beta' \bar b'}(b_\perp) \chi_{n,\w_a}^{\beta b}(0) | N}
\,,\nn\\
 \hat \cG_{h/f}^{\alpha'\alpha\,a'\bar a}(\w_b, \bt) &
 = \theta(\w_b) \SumInt_{X_\bn} \braket{0 | \chi_{\bn}^{\alpha' a'}(b_\perp) | h, X_\bn} \braket{h, X_\bn |\bar{\chi}_{\bn,-\w_b}^{\alpha \bar a} | 0}
\,,\nn\\
  S_f^{b' \bar a' a \bar b}(b_T, b_s^+ b_s^-) &
 = \braket{0 | \bigl( S_n^\dagger S_{\bn} \bigr)^{b' \bar a'}(b_s^+, b_s^-, b_\perp) \bigl(S_{\bn}^\dagger S_{n}\bigr)^{a \bar b}(0) | 0}
\,,\end{align}
and those for the antiquark contribution read
\begin{align} \label{eq:antiquark_correlators_LP}
 \hat B_{\bar f/N}^{\alpha'\alpha\,a'\bar a}(\w_a, \bt) &
 = \theta(\w_a) \braket{N | \chi_{n}^{\alpha' a'}(b_\perp) \bar{\chi}_{n,\w_a}^{\alpha \bar a}(0)  | N}
\,,\nn\\
 \hat \cG_{h/\bar f}^{\beta\beta'\,\bar b'b}(\w_b, \bt) &
 = \theta(\w_b) \SumInt_{X_\bn} \braket{0 | \bar\chi_{\bn}^{\beta' \bar b'}(b_\perp)  | h, X_\bn} \braket{h, X_\bn | \chi_{\bn,-\w_b}^{\beta b}(0) | 0}
\,,\nn\\
 S_{\bar f}^{b' \bar a' a \bar b}(b_T, b_s^+  b_s^-) &
 = \braket{0 | \bigl( S_{\bn}^\dagger S_n \bigr)^{b' \bar a'}(b_s^+, b_s^-, b_\perp) \bigl(S_{n}^\dagger S_\bn\bigr)^{a \bar b}(0) | 0}
 \,.
\end{align}
In $\hat B_{i/N}$ and $ S_f$, we eliminated the sum over $X_n$ and $X_s$ using the completeness relation,
which is not possible for $\hat \cG$,
unless we introduce creation and annihilation operators for the hadron $h$.
The Dirac indices on $\hat B_{i/N}$ and $\hat \cG_{h/i}$ are chosen such that the index of the $\chi$
field always comes before that of the $\bar\chi$ field, which ensures that the Dirac indices in \eq{W_J0_3} form a trace.
While we express the functions in terms of the Euclidean vector $\bt$,
the fields are evaluated at position $b_\perp^\mu = (0, \bt, 0)$.
In the case of the soft function, azimuthal symmetry implies that it only depends
on the magnitude $b_T \equiv |\bt|$. 
We also note that there is no time ordering in the soft function since the fields are spacelike separated and commute.

Next, we employ that the hadronic matrix elements are diagonal in color to define the color-traced objects
\begin{align} \label{eq:def_B_hat}
 \hat B_{f/N}^{\beta\beta'}(x = \w_a/P_N^-, \bt) &
 = \theta(\w_a) \, \braket{N | \bar\chi_{n}^{\beta'}(b_\perp) \chi_{n,\w_a}^{\beta}(0) | N}
\,,\nn\\
 \hat \cG_{h/f}^{\alpha'\alpha}(z = P_h^+/\w_b, \bt) &
 = \frac{\theta(\w_b)}{2 z N_c} \SumInt_{X_\bn} \tr \braket{0 | \chi_{\bn}^{\alpha'}(b_\perp) | h, X_\bn} \braket{h, X_\bn |\bar{\chi}_{\bn,-\w_b}^{\alpha} | 0}
\,,\nn\\
 S(b_T, b_s^+ b_s^-) &
 = \frac{1}{N_c} \tr \braket{0 | \bigl( S_n^\dagger S_{\bn} \bigr)(b_s^+, b_s^-, b_\perp) \bigl(S_{\bn}^\dagger S_{n}\bigr)(0) | 0}
\,,\end{align}
where the normalization factors are chosen as in the literature to ensure proper normalization at tree level.
We also defined the momentum fractions $x$ and $z$ for the collinear matrix elements.
Similar definitions arise for the antiquark distributions. Note that the soft function agrees
for quarks and antiquarks by symmetry under $n \leftrightarrow \bn$, so we identify $S \equiv  S_f = S_{\bar f}$.  Although we are working in a general covariant gauge here, without explicit transverse Wilson lines at light-cone infinity, it is straightforward to generalize the steps above to include them, and we so in \app{Twilsonlines}, and give the full gauge invariant version of \eq{def_B_hat} in \eq{def_B_hat_trans}.

In \eq{W_J0_3}, the soft functions involve Wilson lines at position $b = (b_s^+, b_s^-, b_\perp)$,
but by RPI-III invariance (longitudinal boosts) only depend on the combination $b_s^+ b_s^-$.
The residual positions $b_s^\pm$ are set to zero at the end, which at LP can be done immediately.
At NLP, it will allow us to evaluate the power corrections from the $\delta$ function using derivatives w.r.t.~$b_s^\pm$.
In the rest of this section, we focus on LP only, and will return to study the NLP terms in \sec{nbpartials}.

Plugging this into \eq{W_J0_3} and dropping the NLP terms,
we finally arrive at the LP hadronic structure function
\begin{align} \label{eq:W_J0}
 W^{(0)\,\mu\nu} &
 = 4 z \int\!\frac{\df^2\bt}{(2\pi)^2}\, e^{-\img \bt \cdot \qt}
   \sum_f \cH_f^{(0)}(q^+ q^-)\,
   \Tr\Bigl[ \hat B_{f/N}(x, \bt) \gamma_{\perp}^\mu \hat \cG_{h/f}(z, \bt) \gamma_{\perp}^\nu \Bigr] S(b_T)
 \nn\\&\quad+ (f \to \bar f \,,\, \mu \leftrightarrow \nu)
\,,\end{align}
where $\Tr$ traces over Dirac indices, and the $2z$ compensates for the factor in \eq{def_B_hat}.
Here $S(b_T)=S(b_T,b_s^+b_s^-=0)$.
In the literature, it is common to absorb the soft function into the collinear matrix elements, which can be achieved as%
\footnote{For many rapidity regulators, one has that $B = \hat B \sqrt{S} = \hat B^{\rm (u)} / \sqrt{S}$, where $\hat B^{\rm (u)}$ is calculated as an unsubtracted matrix element (ie.~as a naive collinear matrix element without zero-bin subtractions~\cite{Manohar:2006nz}). 
The precise equation here depends on the form of the zero-bin subtraction,
which often is equivalent to dividing by the full soft function~\cite{Idilbi:2007ff}, $S^{\rm{z{-}b}} = S$,
such that $(\hat B^{\rm (u)} / S^{\rm{0bin}}) \sqrt{S} = \hat B^{\rm (u)} / \sqrt{S}$.
Technically, for renormalized beam functions as in \eq{tildeB_absorbed},
one has to apply the zero-bin subtraction prior to renormalization,
explaining why $S^{\rm{0bin}}$ is already subtracted in the definition of the  renormalized $\hat B$.}
\begin{align} \label{eq:BG_soft}
 B_{f/N} (x,\bt) &\equiv \hat B_{f/N}(x,\bt)\sqrt{S(b_T)}
\,,\\
 \cG_{h/f} (z,\bt) &\equiv \hat\cG_{h/f}(z,\bt)\sqrt{S(b_T)}
\,,\end{align}
and similarly for $B_{\bar f/N}$ and $\cG_{h/\bar f}$.
\Eq{W_J0} then becomes
\begin{align} \label{eq:hadronicts_LP}
 W^{(0)\,\mu\nu} &
 = 4 z \int\!\frac{\df^2\bt}{(2\pi)^2}\, e^{-\img \bt \cdot \qt}
   \sum_f \cH_f^{(0)}(q^+ q^-)\,
   \Tr\Bigl[ B_{f/N}(x, \bt) \gamma_{\perp}^\mu \cG_{h/f}(z, \bt) \gamma_{\perp}^\nu \Bigr]
 \nn\\&\quad+ (f \to \bar f \,,\, \mu \leftrightarrow \nu)
\,.\end{align}
Here, the sum runs over quark flavors $f$, and the antiquark $\bar f$ contribution
is added explicitly with flipped indices $\mu,\nu \to \nu,\mu$.
This only affects hadronic structure functions with antisymmetric dependence on $\mu\nu$,
which is however compensated by explicit sign change in the decomposition
of the corresponding antiquark correlators relative to the quark correlator,
such that the final structure functions will always be symmetric in $f \leftrightarrow \bar f$.

Note that \eq{hadronicts_LP} holds with both bare and renormalized quantities, but is most useful in its renormalized form. Here we have suppressed dependence on the renormalization scale $\mu$, which will appear in renormalized expressions for ${\cal H}_f^{(0)}$, $B_{f/N}$, and ${\cG}_{h/f}$, as well as dependence on the Collins-Soper scales $\zeta_a=(x P_N^- e^{y_n})^2$ in $B_{f/N}$ and $\zeta_b=(P_h^+ e^{-y_n}/z)^2$ in $\cG_{h/f}$. 

\subsection{Kinematic Corrections at NLP}
\label{sec:factorization_kinematic_corr}

The structure functions $W_i = P_i^{\mu\nu} W_{\mu\nu}$ are defined by projecting
the hadronic tensor $W^{\mu\nu}$ onto suitable projectors $P_i^{\mu\nu}$.
While the factorization of $W^{\mu\nu}$ is derived in the factorization frame,
the $P_i^{\mu\nu}$ are constructed in \eq{def_P_i} in terms of the unit vectors in the hadronic Breit frame.
As discussed in \sec{factorization_frame}, the unit vectors between the two frames
differ by $\Ord(\lambda)$ corrections, which thus also affect the $W_i$ themselves.
In this section, we discuss this source of \emph{kinematic} power corrections.
To do so, it is easiest to expand the projectors the factorization frame up to next-to-leading power,
\begin{equation}
  P_i^{\mu\nu}=P_i^{(0)\mu\nu}+P_i^{(1)\mu\nu}+\dots\,.
\end{equation}
This can be achieved by plugging \eq{rel_lc_unit_vectors} into \eq{def_P_i}.
The leading power projectors $P_i^{(0)\mu\nu}$ are simply $P_i^{\mu\nu}$ with $\tilde n_i$ replaced by $n_i$,
which gives rise to the LP structure functions $W_i^{(0)} = P_i^{(0)\mu\nu} W^{(0)}_{\mu\nu}$.
At NLP, one has to take power corrections from the hadronic tensor and the projector into account,
such that 
\begin{align}
 W_i^{(1)} = P_i^{(1)\mu\nu} W^{(0)}_{\mu\nu}+P_i^{(0)\mu\nu} W^{(1)}_{\mu\nu}
 \,.
\end{align}
Since the first term here arises from the difference between the factorization and Breit frame,
we refer to this contributions as the kinematic power corrections. The second term $P_i^{(0)\mu\nu} W^{(1)}_{\mu\nu}$
comprises the genuine NLP correction to the hadronic tensor and will be discussed in the sections below.

From \eq{hadronicts_LP}, we see that the Lorentz indices of $W^{(0) \,\mu\nu}$ are purely transverse
in the factorization frame, i.e.~transverse with respect to $n$ and $\bn$.
When contracting a projector with \eq{hadronicts_LP}, only the transverse part of the projector yields a nonzero contribution.
\Eq{rel_lc_unit_vectors} implies that the $\tilde n_{x,y}$ are purely transverse at leading power
and do not receive an additional transverse contribution at $\cO(\lambda)$,
while $\tilde n_{t,z}$ gain a transverse component at subleading power.
Since $\tilde n_z$ does not contribute to any projector due to current conservation,
we conclude that only projectors containing one $\tilde n_t$ induce kinematic power corrections at NLP, namely $P_{1,2,5,6}^{\mu\nu}$. We find
\begin{align} \label{eq:P_pcs}
 P_1^{(0)\mu\nu} &
 = \frac12 (n_t^\mu n_x^\nu+n_x^\mu n_t^\nu) \,,
 \qquad\quad 
 P_1^{(1)\mu\nu} = -\frac{q_T}{Q}n_x^\mu n_x^\nu+\dots
\,,\nn\\
 P_2^{(0)\mu\nu} & =\frac12 (n_t^\mu n_x^\nu-n_x^\mu n_t^\nu)\,,
 \qquad\quad 
 P_2^{(1)\mu\nu} = \dots 
\,,\nn\\
 P_5^{(0)\mu\nu} &
 = \frac12 (n_t^\mu n_y^\nu + n_y^\mu n_t^\nu)\,,
 \qquad\quad
 P_5^{(1)\mu\nu} = -\frac{q_T}{2Q} (n_x^\mu n_y^\nu + n_y^\mu n_x^\nu)+\dots
\,,\nn\\
 P_6^{(0)\mu\nu} &
 = \frac{\img}{2} ( n_y^\mu  n_t^\nu -  n_t^\mu  n_y^\nu)\,,
 \qquad\quad
 P_6^{(1)\mu\nu} = -\frac{\img q_T}{2Q} ( n_y^\mu  n_x^\nu -  n_x^\mu  n_y^\nu)+\dots
\,.\end{align}
Here $\dots$ are terms proportional to $n^\mu$ or $n^\nu$, which do not contribute when contracting with the leading power hadronic tensor.
Thus the structure functions $W_1^X$, $W_2^X$ and $W_6^X$ will receive kinematic power corrections.  These kinematic power corrections are one of the nonzero contributions to the final NLP results given in \sec{results} below.

\subsection{$J^{(0)}$ with \SCETii Lagrangian Insertions at NLP}
\label{sec:fact_vanishing_Lag}

As our first potential non-trivial source of NLP corrections, lets continue to consider two leading power currents $J^{(0)\mu}$ but now dress them with power suppressed insertions from the dynamic \SCETii Lagrangian, discussed in \eq{Lscetii}. The relevant terms at NLP involve two insertions of ${\cal L}_{\rm dyn}^{(1/2)}$ or one insertion of ${\cal L}_{\rm dyn}^{(1)}$.  The resulting NLP corrections to the hadronic tensor are
\begin{align}  \label{eq:W1cL}
W^{(1)\mu\nu}_{J^{(0)}\cL}
  =& \SumInt_X \, \delta^4(q \!+\! \Pa \!-\! \Pb \!-\! \PX) 
 \bigg\{\!
 \int\!\!\df^4x\, \Big[ \braket{N | J^{(0)\dagger\,\mu}(0) | h,X} \braket{h,X |  T\, J^{(0)\nu}(0)\cL_{\rm dyn}^{(1)}(x) | N}
 \nn\\
  &\quad
   + \braket{N |T\,  J^{(0)\dagger\,\mu}(0)\cL_{\rm dyn}^{(1)}(x) | h,X} \braket{h,X |  J^{(0)\nu}(0) | N}   \Big]  
  \nn\\
  &+ \int\!\df^4x\, \df^4y\,  \Big[
  \braket{N | J^{(0)\dagger\,\mu}(0) | h,X} 
  \braket{h,X | T\, J^{(0)\nu}(0)\cL_{\rm dyn}^{(1/2)}(x)\cL_{\rm dyn}^{(1/2)}(y) | N}
   \nn\\
  &\quad 
  + \braket{N |  T\, J^{(0)\dagger\,\mu}(0)  \cL_{\rm dyn}^{(1/2)}(x) | h,X} 
  \braket{h,X | T\, J^{(0)\nu}(0)\cL_{\rm dyn}^{(1/2)}(y) | N} 
   \nn \\
  &\quad
  +\braket{N | T\,  J^{(0)\dagger\,\mu}(0) \cL_{\rm dyn}^{(1/2)}(x)\cL_{\rm dyn}^{(1/2)}(y) | h,X} 
  \braket{h,X | J^{(0)\nu}(0) | N}
  \Big]
  \bigg\}
  \,.
\end{align}

The terms we are interested in calculating at NLP are the structure functions that first obtain non-zero contributions at this order, which are $W_1^X$, $W_2^X$, $W_5^X$, $W_6^X$ for various polarizations $X$, see \sec{tensor_decomposition}. To compute the NLP contribution to these structure functions we contract the leading power projectors $P_{i\mu\nu}^{(0)}$ from \eq{P_pcs} with this subleading power hadronic tensor $W^{(1)\mu\nu}_{\cL}$ from \eq{W1cL} for $i=1,2,5,6$. However, from \eq{P_pcs} we see that for $i=1,2,5,6$ at least one of the indices $\mu$ or $\nu$ is contracted by a $n_t$ vector.  For the leading power current, the indices $\mu$ and $\nu$ are purely transverse, with $n_t^\mu J_\mu^{(0)\dagger} = 0 = n_t^\nu J_\nu^{(0)}$, and hence all potential contributions explicitly vanish at NLP.  We note that this would no longer be the case at NNLP.

Thus we conclude that there are no contributions from Lagrangian insertions taken with the leading power \SCETii current to the SIDIS structure functions that start at NLP.

\subsection{
Soft Contributions at NLP}
\label{sec:fact_vanishing_soft}

In this section we consider the various sources of power suppressed soft operators that could give contributions at NLP.  This includes 
\begin{itemize}
 \item[i)] contributions involving subleading power hard Lagrangians ${\cal L}_{\rm hc}^{(i)}$ which are obtained through the \SCETi time ordered products
 $T\big[ {\cal O}_{\rm I}^{(0)} {\cal L}_{\rm I}^{(1)} \big]$,
 $T\big[ {\cal O}_{\rm I}^{(0)} {\cal L}_{\rm I}^{(2)} \big]$,
 $T\big[ {\cal O}_{\rm I}^{(0)} {\cal L}_{\rm I}^{(1)} {\cal L}_{\rm I}^{(1)} 
  \big]$,
 \item[ii)] operators from \sec{operators_NLP_Bs} involving a $\cB_{s\perp}^{(n_i)\mu}$, whose final form was given in \eq{current_Bs_full},
 \item[iii)] operators which involve a $\bn_i \cdot \partial_s$ and $\bn_i\cdot {\cal B}_s^{(n_i)}$.  Such terms appeared in the subleading power currents, given by the RPI protected result in \eq{bbJ1nnb}, and from the higher order terms documented in \sec{factorization_LP} that arise from expanding the momentum conserving $\delta$-function in the LP factorization theorem. These two types of contributions are related, for reasons we will explain. 
\end{itemize}
We consider each of these contributions in turn. 

\subsubsection{Hard-Collinear Terms obtained from \SCETi with ${\cal O}_{\rm I}^{(0)\mu}$}
In general the \SCETi time ordered products $T\big[ {\cal O}_{\rm I}^{(0)} {\cal L}_{\rm I}^{(1)} \big]$,
 $T\big[ {\cal O}_{\rm I}^{(0)} {\cal L}_{\rm I}^{(2)} \big]$,
 $T\big[ {\cal O}_{\rm I}^{(0)} {\cal L}_{\rm I}^{(1)} {\cal L}_{\rm I}^{(1)} 
  \big]$
will give rise to \SCETii hard scattering operators in ${\cal L}_{\rm hc}^{(1/2)}$ and ${\cal L}_{\rm hc}^{(1)}$, which then will contribute to observables at NLP. In the squared amplitude this occurs either through the \SCETii 
${\cal L}_{\rm hc}^{(1/2)}$ taken with ${\cal L}_{\rm dyn}^{(1/2)}$ and a ${\cal L}_{\rm hard}^{(0)}$, or through one ${\cal L}_{\rm hc}^{(1)}$ and one ${\cal L}_{\rm hard}^{(0)}$. However for the structure functions that we are interested in, which start at NLP, these contributions vanish for the same reason as the terms considered in \sec{fact_vanishing_Lag}. Here it is the presence of two \SCETi leading power operators ${\cal O}_{\rm I}^{(0)}$ that are common to all these contributions, and which carry transverse indices $\mu$ and $\nu$. The transverse nature of these $\mu$ and $\nu$ indices in \SCETi is retained by the matching onto \SCETii.

Thus the resulting contributions again have vanishing contraction with the leading power projectors $P_i^{(0)\mu\nu}$ for $i=1,2,5,6$, yielding no contribution at NLP.  Non-zero contributions can occur at NNLP where this argument breaks down.

\subsubsection{Operators involving a $\cB_{s\perp}^{(n_i)\mu}$}

In this section, we consider contributions from the \SCETii currents involving $\cB_{s\perp}^{(n_i)\mu}$ given in \eq{current_Bs_full}.  For these contributions it will be necessary to carry out a factorization of the collinear and soft matrix elements in order to assess these potential NLP contributions.

The $\cB_{s\perp}^{(n_i)\mu}$ contributions enter through the current $J_{\cB_s^\perp}^{(1)\mu}$ given in \eq{current_Bs_full}.  Calculating the matrix elements of
\begin{equation}
  J^{(0)\mu\dagger} J_{\cB_s^\perp}^{(1)\nu}
  + J_{\cB_s^\perp}^{(1)\mu\dagger} J^{(0)\nu} \,,
\end{equation}
using the procedure similar to the LP section, we get the factorized formula for the $\cB_{s\perp}$ operator contribution
\begin{align}\label{eq:W_s}
  W_{\cB_s^\perp}^{(1)\mu\nu}& = -\frac{4z}{Q}\sum_f \int\! \frac{\df^2b_T}{(2\pi)^2}\, e^{-\img \vec q_T \cdot \bt}\, 
  \int\!\!\df \hat b_s \,   
  C^{(0)}(\tilde q^2)
  \\
  \times\biggl\{ &  (n^\nu\!+\!\bn^\nu)\, C_{\cB_s^\perp}^{(1)}\big(\tilde q^2, \hat b_s \big)\,\Tr\left[\hat B_{f/N}(x,\bt)\, \gamma_\perp^\mu \, \hat \cG_{h/f}(z,\bt)\, \gamma_{\perp\rho}\right] \hat S_1^\rho(b_\perp,Q\hat b_s,Q\hat b_s)
  \nn\\
 + &  
  (n^\mu\!+\!\bn^\mu)\, C_{\cB_s^\perp}^{(1)*}\big(\tilde q^2, \hat b_s \big) \,\Tr\left[\hat B_{f/N}(x,\bt)\, \gamma_{\perp\rho} \, \hat \cG_{h/f}(z,\bt)\, \gamma_\perp^\nu\right] 
  \hat S_2^\rho(b_\perp,Q\hat b_s,Q\hat b_s) \biggr\} \nn
  \\
  +&
  (f\to\bar f,\mu\leftrightarrow\nu\,,C_{\cB_s^\perp}^{(1)} \leftrightarrow C_{\cB_s^\perp}^{(1)*} )
  \nn
\end{align}
Here, the soft matrix elements $\hat S_1^\rho$ and $\hat S_2^\rho$ are defined as\footnote{From direct calculations, the soft matrix elements for antiquark would be \eq{S1S2def} with $n$ and $\bar n$ switched. A minus sign from the fact that $(\lp_2-\lp_1)^\mu$ in \eq{J1nnb} picks out opposite signs for quark and antiquark, is then canceled by the minus sign due to \eqs{flip}{S_decomp}. Therefore we don't need to define extra soft functions for antiquarks. }
\begin{align}\label{eq:S1S2def}
  \hat S_1^\rho(b_\perp,b_s^+,b_s^-)\equiv\,&\frac{1}{N_c}\tr\, 
  \MAe{0}{
  \bigl[S^\dagger_{\na}(b_\perp)\, S_{\nb}(b_\perp)\bigr]  \bigl[S^\dagger_{\nb}(b_s^-)\,S_{\na}(b_s^-) g\cB_{s\perp}^{(n)\rho}(b_s^-) \bigr] 
   }{0}
  \\*
  &\ +\frac{1}{N_c}\tr\, 
  \MAe{0}{
  \bigl[S^\dagger_{\na}(b_\perp)\, S_{\nb}(b_\perp)\bigr]  \bigl[g\cB_{s\perp}^{(\bn)\rho}(b_s^+) S^\dagger_{\nb}(b_s^+)\,S_{\na}(b_s^+) \bigr] 
   }{0}
 \,,\nn \\
  \hat S_2^\rho(b_\perp,b_s^+,b_s^-)\equiv\,&\frac{1}{N_c}\tr\, \MAe{0}{
  \bigl[g\cB_{s\perp}^{(n)\rho}(b_\perp,b_s^-)\,S^\dagger_{\na}(b_\perp,b_s^-)\, S_{\nb}(b_\perp,b_s^-)\bigr]  \bigl[S^\dagger_{\nb}(0)\,S_{\na}(0)  \bigr]  
  }{0}
   \nn \\*
  &\ +\frac{1}{N_c}\tr\, 
  \MAe{0}{
   \bigl[S^\dagger_{\na}(b_\perp,b_s^+)\, S_{\nb}(b_\perp,b_s^+)\,g\cB_{s\perp}^{(\bn)\rho}(b_\perp,b_s^+)\bigr]  \bigl[ S^\dagger_{\nb}(0)\,S_{\na}(0) \bigr]  
   }{0}
   \,.\nn
\end{align}
Using the fact that the vacuum is parity invariant we find
{\small
\begin{align} \label{eq:S1_P}
 & \hat S_1^\rho(b_\perp,b_s^+,b_s^-) 
 \nn\\
 & = \frac{1}{N_c} \tr \MAe{0}{P^{-1} P  S^{\dagger}_{\na}(b_\perp)
  P^{-1} P   S_{\nb}(b_\perp) P^{-1}  
  P S^{\dagger}_{\nb}(b_s^-) P^{-1} P  S_{\na}(b_s^-) 
  P^{-1} P  g\cB_{s\perp}^{(n)\rho}(b_s^-) P^{-1}   P   }{0}
 \nn\\&
 + \frac{1}{N_c} \tr \MAe{0}{P^{-1} P S^{\dagger}_{\na}(b_\perp) 
  P^{-1} P  S_{\nb}(b_\perp) P^{-1} P g\cB_{s\perp}^{(\bn)\rho}(b_s^+)
  P^{-1} P S^{\dagger}_{\nb}(b_s^+) P^{-1} P S_{\na}(b_s^+) P^{-1} P  }{0}
 \nn\\&
 =
 \frac{-1}{N_c} \tr \MAe{\!0}{  S^{\dagger}_{\nb}(-b_\perp\!)
  S_{\na}(-b_\perp\!)  
  S^{\dagger}_{\na}(b_s^+) S_{\nb}(b_s^+)
   g\cB_{s\perp}^{(\bn)\rho}(b_s^+)  
  \!+\! S^{\dagger}_{\nb}(-b_\perp\!) 
  S_{\na}(-b_\perp\!)  g\cB_{s\perp}^{(n)\rho}(b_s^-)
  S^{\dagger}_{\na}(b_s^-)  S_{\nb}(b_s^-)
  }{0\!}
 \nn\\&
 =
 \frac{-1}{N_c} \tr \MAe{0}{ 
   S^{\dagger}_{\na}(b_\perp,b_s^+) S_{\nb}(b_\perp,b_s^+)
   g\cB_{s\perp}^{(\bn)\rho}(b_\perp,b_s^+)
   S^{\dagger}_{\nb}(0) S_{\na}(0) 
  \nn\\
  & \qquad\qquad\quad
  + g\cB_{s\perp}^{(n)\rho}(b_\perp,b_s^-)
  S^{\dagger}_{\na}(b_\perp,b_s^-)  S_{\nb}(b_\perp,b_s^-)
  S^{\dagger}_{\nb}(0)  S_{\na}(0)  
  }{0}
 \nn\\&
 = -\hat S_2^\rho(b_\perp,b_s^+,b_s^-)
\,.
\end{align}
}%
In the second equality we used that under parity there is an overall sign flip associated to the spatial index $\rho$, and that $n\leftrightarrow\bn$ and $b_s^+\leftrightarrow b_s^-$. In the third equality we did a cyclic reordering of the fields in the trace (since they are spacelike separated and hence commute), and then used translation invariance to shift all fields by $+b_\perp$. 

We also note that from the definitions and our ability to make cyclic reorderings and translations, that we have
\begin{align} \label{eq:flip}
  \hat S_1^\rho(b_\perp,b_s^+,b_s^-) \Big|_{n \leftrightarrow \bn}  
  = \hat  S_2^\rho(-b_\perp,b_s^+,b_s^-) \,,
 \quad
 \hat S_2^\rho(b_\perp,b_s^+,b_s^-) \Big|_{n \leftrightarrow \bn}  
  = \hat  S_1^\rho(-b_\perp,b_s^+,b_s^-) 
 \,.
\end{align}
Next we consider how these soft matrix elements are constrained by charge conjugation. Recall that the gluon field transforms under charge conjugation as
\begin{align} \label{eq:AunderC}
 A_\mu = A_\mu^A T^A \stackrel{C}{\longrightarrow} A_\mu^A (\bar T)^A = -A_\mu^A (T^A)^T = - A_\mu^T
\,,\end{align}
where the superscript $T$ denotes the transpose.
Under this transformation, a soft Wilson line in the fundamental representation transforms as
\begin{align} \label{eq:SunderC}
 S_n^{ab}(b;0,\infty)
 =&~ \bar P \exp\Big[ -i g \int_0^\infty \!\! \df s \: n\cdot A_s(b+n s)^T \Big]^{ab}
 \nn\\
 \stackrel{C}{\longrightarrow}&~
 P \exp\Big[ +i g \int_0^\infty \!\! \df s \: n\cdot A_s(b+n s) \Big]^{ba}
 = S_n^{\dagger ba}(b;0,\infty)
 \,.
\end{align}
Here, $ab$ are the color indices of the Wilson line, and the transpose of the product of color matrices reverses the path ordering.
For the soft gluon building block field
\begin{align}
 g\cB_{s\perp}^{(n)\rho\,ab} 
  = \big[ S_n^\dagger i D_{s\perp}^\rho S_n \big]^{ab}
 \stackrel{C}{\longrightarrow}
  \big[ S_n^\dagger (i \overleftarrow D_{s\perp}^\rho) S_n \big]^{ba}
  = - g\cB_{s\perp}^{(n)\rho\,ba}
\,,\end{align}
exactly like the gluon field in \eq{AunderC}. Using the fact that the vacuum is invariant under charge conjugation $C$, we find that
{\small
\begin{align} \label{eq:S1_C}
 & \hat S_1^\rho(b_\perp,b_s^+,b_s^-) 
 \nn\\
  &= \frac{1}{N_c} \MAe{0}{C^{-1} C  S^{\dagger ab}_{\na}(b_\perp)
  C^{-1} C   S^{bc}_{\nb}(b_\perp) C^{-1}  
  C S^{\dagger cd}_{\nb}(b_s^-) C^{-1} C  S^{de}_{\na}(b_s^-) 
  C^{-1} C  g\cB_{s\perp}^{(n)\rho\,ea}(b_s^-) C^{-1}   C   }{0}
 \nn\\&
 + \frac{1}{N_c}  \MAe{0}{C^{-1} C S^{\dagger ab}_{\na}(b_\perp) 
  C^{-1} C  S^{bc}_{\nb}(b_\perp) C^{-1} C g\cB_{s\perp}^{(\bn)\rho\,cd}(b_s^+)
  C^{-1} C S^{\dagger de}_{\nb}(b_s^+) C^{-1} C S^{ea}_{\na}(b_s^+) C^{-1} C 
  }{0}
 \nn\\&
 =
 \frac{-1}{N_c} \MAe{0}{ S^{ba}_{\na}(b_\perp) 
  S^{\dagger cb}_{\nb}(b_\perp)
  \big[ S^{dc}_{\nb}\,S^{\dagger ed}_{\na}
  g\cB_{s\perp}^{(n)\rho\,ae} \big](b_s^-) 
 +
 S^{ba}_{\na}(b_\perp) 
  S^{\dagger cb}_{\nb}(b_\perp)
  \big[ g\cB_{s\perp}^{(\bn)\rho\,dc}
  S^{ed}_{\nb}\,S^{\dagger ae}_{\na} \big](b_s^+) 
  }{0}
 \nn\\&
 = \frac{-1}{N_c} \tr \MAe{0}{
  g\cB_{s\perp}^{(n)\rho}(b_s^-) S^\dagger_n(b_s^-) S_{\nb}(b_s^-) 
  S^{\dagger}_{\nb}(b_\perp) S_{\na}(b_\perp)  
 + 
  S^{\dagger}_{\na}(b_s^+) S_{\nb}(b_s^+) g\cB_{s\perp}^{(\bn)\rho}(b_s^+)
  S^{\dagger}_{\nb}(b_\perp) S_{\na}(b_\perp)
  }{0}
 \nn\\&
 = \frac{-1}{N_c} \tr \MAe{0}{
  S^{\dagger}_{\nb}(b_\perp) S_{\na}(b_\perp)  
 g\cB_{s\perp}^{(n)\rho}(b_s^-) S^\dagger_n(b_s^-) S_{\nb}(b_s^-) 
 + 
  S^{\dagger}_{\nb}(b_\perp) S_{\na}(b_\perp)
 S^{\dagger}_{\na}(b_s^+) S_{\nb}(b_s^+) g\cB_{s\perp}^{(\bn)\rho}(b_s^+)
  }{0}
 \nn\\&
 = -\hat S_1^\rho(b_\perp,b_s^+,b_s^-)\bigg|_{n \leftrightarrow \bn}
 = -\hat S_2^\rho(-b_\perp,b_s^+,b_s^-)
\,.\end{align}
}%
In the third equality we reordered the fields from the second equality into a trace (since they are spacelike separated and hence commute).  To obtain the fourth equality we used the cyclic property of the trace.

Since the index $\rho$ in $\hat S_i^\rho$ is transverse  we can decompose $\hat S_i^\rho(b_\perp,b_s^+,b_s^-)$ as
\begin{align}\label{eq:S_decomp}
  \hat S_i^\rho(b_\perp,b_s^+,b_s^-)
   =b_\perp^\rho\,  S_i^\parallel(b_T,b_s^+,b_s^-) 
    + \epsilon_\perp^{\rho\sigma}b_{\perp\sigma}\,  S^\top_i\!(b_T,b_s^+,b_s^-)  \,, 
   \qquad i=1,\,2
   \,.
\end{align}
Here $ S_i^\parallel$ and $ S_i^\top$ are scalar functions and hence depend only on the magnitude  $b_T=|\bt|$. The parity relation in \eq{S1_P} then implies
\begin{align}
   S_1^\parallel(b_T,b_s^+,b_s^-) = -  S_2^\parallel(b_T,b_s^+,b_s^-)
  \,,\qquad 
   S_1^\top(b_T,b_s^+,b_s^-) = -  S_2^\top(b_T,b_s^+,b_s^-) 
  \,.
\end{align}
On the other hand, the charge conjugation relation in \eq{S1_C} gives
\begin{align}
   S_1^\parallel(b_T,b_s^+,b_s^-) = +  S_2^\parallel(b_T,b_s^+,b_s^-)
  \,,\qquad 
   S_1^\top(b_T,b_s^+,b_s^-) = +  S_2^\top(b_T,b_s^+,b_s^-) 
  \,.
\end{align}
Taken together these are contradictory, and hence the soft matrix elements defined in \eq{S1S2def} must vanish
\begin{align}
 \hat S_1^\rho(b_\perp,b_s^+,b_s^-)  = 0 \,,\qquad\qquad 
  \hat S_2^\rho(b_\perp,b_s^+,b_s^-)  = 0 \,.
\end{align}
We have carried out a cross check on this symmetry argument by verifying that the perturbative calculation of these soft matrix element vanishes at the integrand level at one-loop.
As a further cross-check we have also considered various two loop contributions that were easy to analyze, including the fully abelian $C_F^2$ terms, and terms involving quark, gluon, or ghost vacuum polarization type graphs. Again these all vanish at the integrand level, giving us full confidence in the all orders in $\alpha_s$ symmetry argument presented here.

Thus at NLP there is no contribution from the $J_{\cB_s^\perp}^{(1)\mu}$ current to any $W_i^X$ in SIDIS. The arguments based on discrete symmetries presented here will no longer carry through at NNLP, where we expect contributions from $J_{\cB_s^\perp}^{(1)\dagger\mu}J_{\cB_s^\perp}^{(1)\nu}$ for example.

The argument given above for the vanishing of the NLP soft functions, based on the application of parity, charge conjugation, and other symmetries on the vacuum matrix elements, actually works just as well to eliminate the $\bn_i\cdot {\cal B}_s^{(n_i)}$ contributions from the current in \eq{bbJ1nnb}. 
The sum of $\bn\cdot {\cal B}_s^{(n)}$ and $n\cdot {\cal B}_s^{(\bn)}$ terms appearing in \eq{bbJ1nnb} is just like the sum of terms appearing in \eq{current_Bs_full}, and it is straightforward to show that the same symmetry relations in \eqs{flip}{S1_C}) apply, and a parity relation like in \eq{S1_P} applies but without the extra multiplicative minus sign on the RHS. This again leads to contradictory equations, and vanishing soft function for this current. 
Rather than spelling out this argument in detail, we give an alternate argument in the next section, whereby these operators are considered together with others, and again give a vanishing soft NLP contribution.

\subsubsection{Operators involving a $n\cdot \partial_s$, $\bn\cdot \partial_s$, $\bn\cdot {\cal B}_s^{(n)}$, or $n\cdot {\cal B}_s^{(\bn)}$} \label{sec:nbpartials}

In this section we consider NLP contributions to $W^{\mu\nu}$ that come from either a single $\bn\cdot \partial_s$,   $n\cdot\partial_s$,  $\bn\cdot {\cal B}_s^{(n)}$, or $n\cdot {\cal B}_s^{(\bn)}$.  There are two places that we have such contributions appeared: 
\begin{itemize}
\item[a)] in the RPI expansion leading to the operators in \eq{bbJ1nnb}, and
\item[b)] from expanding the momentum conserving $\delta$-function in the analysis of the leading power factorization theorem in \sec{factorization_LP}, shown in \eq{W_J0_3}.
\end{itemize}
Both of these results had their hard coefficient functions determined by a derivative of the leading power Wilson coefficient, $\partial C^{(0)}/\partial\omega_i$. In fact, it turns out that the results in a) and b) give identical contributions, and are just two different ways of constructing the same contribution. In both cases we can follow the same steps for the factorization as carried out in \sec{factorization_LP}, but with a difference in one step of the procedure.

In approach a) we setup the SCET field theory with exact momentum conservation for momenta of all sizes, which in particular implies momentum conservation for both ${\cal O}(\lambda^0)$ large $p_n^-$ and $p_\bn^+$ momenta, as well as for ${\cal O}(\lambda)$ soft $p_s^-$ and $p_s^+$ momenta which appear as residual momenta in the $n$-collinear and $\bn$-collinear fields respectively.
In this approach, when we carry out the leading power factorization analysis there is never an expansion of the type carried out in \eq{delta_expanded}, since instead the momentum conserving $\delta$-function is written as a product of distributions that implement momentum conservation at each of the scales. In this approach residual momentum conserving $\delta$-functions shift the coordinates of fields to account for the residual momentum flow. This arises because the formulation of SCET implements an exact power expansion, with terms that are always homogeneous in $\lambda$, without the need for re-expansion to obtain subleading terms. In this approach, only the operators in \eq{bbJ1nnb}  give contributions of the type considered in this section.

In approach b) we instead expand the full momentum conserving $\delta$-function in powers of $\lambda$, as in \eq{delta_expanded}. In this approach the resulting collinear fields are already evaluated with spacetime arguments that force them to carry zero residual momenta. So for example an $n$-collinear field has a $p_n^-$ momenta of ${\cal O}(\lambda^0)$, but no residual minus-momentum of ${\cal O}(\lambda)$. Therefore, in this approach the operators in  \eq{bbJ1nnb} give zero contribution, either explicitly because the $n_1\cdot\partial_s$ and $\bn\cdot\partial_s$ derivatives vanish, or due to momentum conservation, since no object in the factorization can balance the soft momentum carried by the one or more soft gluons emitted by the operator $\cO_{\stackrel{\mbox{\tiny $\!\!\!(-)$}}{n_1} \cdot {\cal B}_s}^{(1)\mu}$ in \eq{bbJ1nnbops}. Thus, in this approach only the NLP terms in \eq{W_J0_3} give contributions.

To see that the two methods give exactly the same contribution, we can start with \eq{W_J0_3}. Combining \eq{W_J0_3} with the rearrangement of color indices that leads to \eq{def_B_hat}, we see that the result for the NLP contribution involves either a $\img\bn\cdot \partial_s =2\img \partial/\partial b_s^+$ or a $\img n\cdot \partial_s = 2\img \partial/\partial b_s^-$ derivative of the leading power soft function $S(b_T,b_s^+b_s^-)$, after which we must send $b_s^\pm \to 0$.  Looking inside this soft function, we see that the derivative acts on the Wilson lines as in \eq{subl_soft}
\begin{align} \label{eq:relateab}
 i n\cdot\partial_s \big( S_n^\dagger S_\bn \big)(b_s^+) 
  &= S_n^\dagger(b_s^+)\Big\{ in\cdot \overleftarrow\partial_s + in\cdot\partial_s \Big\} S_\bn(b_s^+) 
  \nn\\
  &= S_n^\dagger(b_s^+)\Big\{ in\cdot \overleftarrow D_s + in\cdot D_s \Big\} S_\bn(b_s^+) 
  \nn\\
  &= S_n^\dagger(b_s^+)\,  in\cdot D_s \, S_\bn(b_s^+) \,,
\end{align}
where in the last line we used the Wilson line equation of motion, $in\cdot D_s S_n =0$. The final operator in \eq{relateab} is the same as the one which led to the $in\cdot \partial_s$ and $g n\cdot \cB_{s}^{(\bn)}$ operators in \eq{bbJ1nnbops} through the use of \eq{nbDrelation} (taken here with $n\leftrightarrow \bn$).  The same steps apply for the $i\bn\cdot\partial_s(S_n^\dagger S_\bn)$ term in \eq{W_J0_3}, where this time the covariant derivative $i\bn\cdot D_s S_\bn=0$, and this leads to the $i\bn\cdot \partial_s$ and $g \bn\cdot \cB_s^{(n)}$ terms appearing in \eq{bbJ1nnbops}.  Thus the two approaches are equivalent and lead to the same result. 

Adopting the already quite compact result from approach b), 
we denote the NLP term from \eq{W_J0_3} as $W^{(1)\mu\nu}_\delta$, and see that it corresponds to
\begin{align}
    W^{(1)\mu\nu}_\delta &=
    2\left(\img\bn\cdot\partial_s \partial_{\w_a} - \img n\cdot\partial_s\, \partial_{\w_b}\right)  
  \cH^{(0)}_f(\w_a\w_b)\,  S(b_T,b_s^+b_s^-)
    \\
  &\qquad\times
    \biggl\{\Tr\left[\hat B_{f/N}(x,\bt)\, \gamma_\perp^\mu \, \hat\cG_{h/f}(z,\bt)\, \gamma_\perp^\nu\right]
  +(f\to \bar f,\, \mu\leftrightarrow\nu) \left.\biggr\}\right|_{b_s^+=b_s^-=0}
  \,. \nn
\end{align}
It is interesting to note that in this subleading TMD factorization formula that the $\omega_{a,b}$ derivatives will hit both the hard function and the collinear functions (in contrast to \SCETi, where the analogous contributions typically only act on the hard function). Nevertheless, here $W_\delta^{(1)\mu\nu}$ actually evaluates to zero due to the action of the derivatives and limits of the generalized soft function $S(b_T,b_s^+b_s^-)$. Essentially, this occurs because if we consider the possible form of results for 
\begin{align} \label{eq:Sdeltas}
  S_\delta^-(b_T) = 2\img \frac{\partial}{\partial b_s^+} S(b_T,b_s^+b_s^-) 
  \bigg|_{b_s^\pm \to 0} \,,
  \qquad
  S_\delta^+(b_T) = 2\img \frac{\partial}{\partial b_s^-} S(b_T,b_s^+b_s^-) 
  \bigg|_{b_s^\pm \to 0} \,,
\end{align}
then we must obtain a result for $S_\delta^-(b_T)$ that scales linear in $\bn$ (ie.~as an inverse plus variable) under an RPI-III transformation, and for $S_\delta^+(b_T)$ that scales linear in $n$. However, there are no available quantities left to carry these scaling properties after we set $b_s^\pm =0$, thus the functions $S_\delta^\pm(b_T)=0$.

To make the argument technically sound we have to consider the fact that the soft function has rapidity divergences associated to its dependence on light-cone positions in the limit $b_s^\pm\to 0$. 
In particular, analytic perturbative results for the unregulated fully differential soft function $S(b_T,b_s^+b_s^-)$ can be found in Refs.~\cite{Li:2016axz,Li:2016ctv}, and can easily be used to confirm that the functions in \eq{Sdeltas} exhibit power law divergences in the desired limit.
Therefore we must confirm that the above argument still applies in the presence of one or more of the possible rapidity regulators that are used for leading (and subleading) power TMDs in the literature~\cite{Collins:2011zzd,  Becher:2010tm, Becher:2011dz, GarciaEchevarria:2011rb, Chiu:2011qc, Chiu:2012ir, Li:2016axz, Ebert:2018gsn}. 
We anticipate that the vanishing result will still be achieved for any reasonable rapidity regulator.\footnote{To achieve a non-zero result the rapidity regulator would need to cause a hard breaking of rapidity scaling laws like a hard cutoff, as opposed to a soft breaking like those that mimic dimensional regularization for invariant-mass singularities. Even in the hard cutoff case it becomes clear that a non-zero result would be a regulator artifact, rather than a required contribution.} For illustration it is useful to choose a regulator that does not modify the fully-differential soft function $S(b_T,b_s^+b_s^-)$ when it is removed in cases where $b_s^+b_s^-\ne 0$, and one that gives a nonvanishing soft function $S(b_T)=S(b_T,0)$ at LP, so that it is clear that the regulator is not simply shifting the question we are aiming to answer into the collinear sector.
We will use the $\eta$ regulator of \refscite{Chiu:2011qc, Chiu:2012ir}, which is known to correctly regulate Glauber interactions in \SCETii~\cite{Rothstein:2016bsq},
and which has already been studied at subleading power for the $q_T$ distribution in \refcite{Ebert:2018gsn}.%
\footnote{An important detail is that the $\eta$ regulator acts on the \emph{group momentum} $k_g$,
which is defined as the total momentum flowing into a connected web (CWEB), see appendix A in \refcite{Chiu:2012ir}.  This is required in order to not spoil non-abelian exponentiation for Wilson line operators~\cite{Gatheral:1983cz, Frenkel:1984pz}.
It is the sum of these CWEBs that appear in the logarithm of the LP soft function, and which have only single primitive rapidity divergences, $\propto 1/\eta$, regardless of the order considered in perturbation theory. At subleading power the form of non-abelian exponentiation is also known for Wilson line operators~\cite{Laenen:2008gt}.} 
In momentum space the primitive regulated integral which corresponds to $S_\delta^\pm$ are given by an integral over a (connected web) group momentum $k$, where the $\partial/\partial b_s^\mp$ derivatives produce an extra $k^+$ or $k^-$ in the integrand:
\begin{align} \label{eq:soft_relation_sub}
 I_\delta^\pm(k_T) &
 = \int_0^\infty\!\! \df k^+ \df k^- \: k^\pm\: I(k^+ k^-, k_T^2,\epsilon)
   \: |k^+-k^-|^{-\eta} \nu^\eta
  \nn\\
 & = \int\df s \, s \, I(s, k_T^2,\epsilon) \Big|\frac{\nu}{\sqrt{s}}\Big|^{-\eta}
 \int\df y \, e^{\mp y} \, |e^y-e^{-y}|^{-\eta}
\,.
\end{align}
Here $I(k^+k^-,k_T^2,\epsilon)$ is the RPI-III invariant LP integrand for the soft contribution at whatever order in $\alpha_s$ is being considered, and contains within it all integrals other than that over the group momentum. It is defined to also includes dependence on the dimensional regularization parameter $\epsilon=(d-4)/2$ and $\mu^{\epsilon}$ factors. In addition $\nu$ is a renormalization scale associated with rapidity divergences, and results are to be considered as expanded in the limit that the rapidity regulator is removed, as $\eta\to 0$. 
In the second line of \eq{soft_relation_sub} we have made a change of variable to an invariant mass variable $s=k^+k^-$ and rapidity $y=\frac12\ln(k^-/k^+)$, noting that $dk^+ dk^- = ds dy$. The fact that dimensional regularization does not suffice to regulate rapidity divergences is manifest in \eq{soft_relation_sub} due to the fact that the $\epsilon$ dependence does not influence the $y$ integral. In general the rapidity integral
\begin{align} \label{eq:eta_regulator}
 I_\eta^{(\alpha)}(s/\nu^2) &
  = \Bigl(\frac{\nu^2}{s}\Bigr)^{\eta/2} \int\df y \, e^{\alpha y} \,  |e^y - e^{-y}|^{-\eta}
 \nn\\&
 = \frac{1}{\pi} \Bigl(\frac{\nu^2}{s}\Bigr)^{\eta/2} \cos\Bigl(\frac{\alpha \pi}{2}\Bigr) \sin\Bigl(\frac{\eta \pi}{2}\Bigr)
   \Gamma(1-\eta) \Gamma\Bigl(\frac{\eta-\alpha}{2}\Bigr) \Gamma\Bigl(\frac{\eta+\alpha}{2}\Bigr)
\,,\end{align}
which is obtained by the appropriate analytic
continuation in $\eta$~\cite{Ebert:2018gsn}. The remaining integration over $s$ in \eq{soft_relation_sub} does not add any additional $1/\eta$ divergences. This is the case because the $s$ integral is regulated by $\epsilon$, and one must consider $\eta\to 0$ for fixed $\epsilon$ so that rapidity divergences are handled independent from invariant mass divergences~\cite{Chiu:2012ir}.
Expanding \eq{eta_regulator} in $\eta\to0$, we obtain $I_\eta^{(\alpha)}(s/\nu^2) =0$ for all odd $\alpha$, and in particular for the integral that appears in \eq{soft_relation_sub} for the NLP result, $I_\eta^{(\pm 1)}(s/\nu^2)=0$. This vanishing of the NLP terms was also discussed explicitly for Drell-Yan at NLO in \refcite{Ebert:2018gsn}.
In contrast, the integral that appears at leading power is $I_\eta^{(0)}(s/\nu^2)\ne 0$ and this regulator gives a non-trivial leading power soft function. We refer the reader to~\refcite{Ebert:2018gsn} for a more detailed discussion of regulating rapidity divergences at subleading power, including effects at NNLP which can generically be non-zero, but which also must be considered hand in hand with collinear effects at that order.  

Thus at NLP, i.e.~$\alpha = \pm 1$, the soft functions vanish analytically even in the presence of a rapidity regulator, $S_\delta^\pm(k_T)=0$,
and hence $W_\delta^{(1)\mu\nu}=0$, and does not contribute to the SIDIS factorization formula at this order.

\subsection{NLP Contributions from the $\cP_\perp$ Operators}
\label{sec:factorization_P}

We next consider contributions from the operators with a $\cP_\perp$ insertion, given in \eq{crrnt_P1}.
Since these operators have the same field content and Wilson coefficient as the leading-power currents,
calculating their contribution to the hadronic tensor proceeds analogous to the LP case discussed in \sec{factorization_LP}.
Schematically, one inserts the currents
\begin{equation}
 \Bigl(J_\cP^{(1)\mu\dagger}+J_{\cP^\dagger}^{(1)\mu\dagger}\Bigr)J^{(0)\nu}
 \quad{\rm and}\quad
 J^{(0)\mu\dagger} \Bigl(J_\cP^{(1)\nu}+J_{\cP^\dagger}^{(1)\nu}\Bigr)
\end{equation}
into \eq{W}. The explicit derivation parallels that in \sec{factorization_LP},
so we only discuss the final result,
\begin{align} \label{eq:WP_hat}
 \hat W^{(1)\mu\nu}_{\cP} &
 = 4z \int\frac{\df^2\bt}{(2\pi)^2} e^{-\img \bt \cdot \qt}
   \sum_f \,\cH^{(0)}_f(q^+ q^-) \, S( b_T)
  \\\nn&
  \times \Big\{\Tr\left[\hat B_{\cP\, f/N}(x,\bt)\, \gamma^\mu \, \hat\cG_{h/f}(z,\bt)\, \gamma^\nu+ \hat B_{f/N}(x,\bt)\, \gamma^\mu \, \hat\cG_{\cP\, h/f}(z,\bt)\, \gamma^\nu\right]
 \nn\\
 & \quad +(f\to \bar f,\, \mu\leftrightarrow \nu)\Big\}
\,.\nn
\end{align}
Here, $\hat B_{f/N}$ and $\hat \cG_{h/f}$ are the LP correlators defined in \eq{def_B_hat},
and the new correlators involving the $\cP_\perp$ operator are defined as
\begin{subequations} \label{eq:def_BP_f}
\begin{align} \label{eq:def_BP_f_a}
 \hat B_{\cP\, f/N}^{\beta\beta'}(x,\bt) &
 \equiv \frac{\theta(\w_a)}{2Q} \biggl\{
         \MAe{N}{\bar\chi_{\na}^{\beta'}(b_\perp) \left[\slashed\cP_\perp\slashed\bn\, \chi_{\na,\w_a}(0)\right]^{\beta}}{N}
 \nn\\&\hspace{1.8cm}
         + \MAe{N}{\left[\bar\chi_{\na}(b_\perp)\,\slashed\bn\,\slashed\cP_\perp^\dagger\right]^{\beta'} \,
  \chi_{\na,\w_a}^{\beta}(0)}{N}\biggr\}
 \\&  \label{eq:def_BP_f_b}
 = \frac{-\img}{2Q} \frac{\partial}{\partial b_\perp^\rho}
   \left[ \gamma_\perp^\rho\, \slashed\bn\,,\,\hat B_{f/N}(x,\bt) \right]^{\beta\beta'}
\,.\end{align}
\end{subequations}
and
\begin{subequations} \label{eq:def_GP_f}
\begin{align} \label{eq:def_GP_f_a}
 \hat\cG_{\cP\, h/f}^{\alpha'\alpha}(z,\bt) &
 \equiv \frac{\theta(\w_b)}{4z N_c Q} \SumInt_{X_\bn} \biggl\{
   \MAe{0}{\left[\slashed\cP_\perp\slashed n\, \chi_{\nb}(b_\perp)\right]^{\alpha'}}{h,X_\bn}
   \MAe{h,X_\bn}{\bar\chi_{\nb,-\w_b}^{\alpha}(0)}{0}
 \nn\\&\hspace{2cm}
   + \MAe{0}{\chi_{\nb}^{\alpha'}(b_\perp)}{h,X_\bn}
     \MAe{h,X_\bn}{\left[\bar\chi_{\nb,-\w_b}(0)\, \slashed n\,\slashed\cP_\perp^\dagger\right]^{\alpha}}{0}
 \biggr\}
 \\&  \label{eq:def_GP_f_b}
 = \frac{\img}{2Q}\frac{\partial}{\partial b_\perp^\rho}
   \left[ \gamma_\perp^\rho\, \slashed n\,,\,\hat\cG_{h/f}(x,\bt) \right]^{\alpha'\alpha}
\,.\end{align}
\end{subequations}
In \eqs{def_BP_f_b}{def_GP_f_b}, we used that $\cP_\perp^\rho$
can be expressed as $\img\frac{\partial}{\partial b_{\perp\rho}}$ in position space.
The corresponding antiquark correlators are defined as
\begin{subequations} \label{eq:def_BP_fbar}
\begin{align} \label{eq:def_BP_fbar_a}
 \hat B_{\cP\, \bar f/N}^{\alpha'\alpha}(x,\bt) &
 \equiv \frac{-\theta(\w_a)}{2Q} \biggl\{
   \MAe{N}{\chi_{\na}^{\alpha'}(b_\perp) \, \left[ \bar\chi_{\na,\w_a}(0)\,\slashed\bn\,\slashed\cP_\perp^\dagger\right]^{\alpha}}{N}
 \nn\\&\hspace{1.8cm}
   + \MAe{N}{\left[\slashed\cP_\perp\,\slashed\bn\,\chi_{\na}(b_\perp)\right]^{\alpha'} \,
  \chi_{\na,\w_a}^{\alpha}(0)}{N} \biggr\}
\\& \label{eq:def_BP_fbar_b}
  = \frac{-\img}{2Q}\frac{\partial}{\partial b_\perp^\rho}
    \left[ \gamma_\perp^\rho\, \slashed\bn\,,\,\hat B_{\bar f/N}(x,\bt) \right]^{\alpha'\alpha}
\,,\end{align}
\end{subequations}
and
\begin{subequations} \label{eq:def_GP_fbar}
  \begin{align} \label{eq:def_GP_fbar_a}
    \hat\cG_{\cP\, h/\bar f}^{\beta\beta'}(z,\bt) &
    \equiv \frac{-\theta(\w_b)}{4z N_c Q} \SumInt_{X_\bn} \biggl\{
      \MAe{0}{\left[\bar\chi_{\nb}(b_\perp)\,\slashed n\,\slashed\cP_\perp^\dagger\right]^{\beta'} }{h,X_\bn}
      \MAe{h,X_\bn}{\chi_{\nb,-\w_b}^{\beta}(0)}{0}
    \nn\\&\hspace{2cm}
      + \MAe{0}{\bar\chi_{\nb}^{\beta'}(b_\perp)}{h,X_\bn}
        \MAe{h,X_\bn}{\left[\slashed\cP_\perp\slashed n\, \chi_{\nb,-\w_b}(0)\right]^\beta}{0}
    \biggr\}
    \\&  \label{eq:def_GP_fbar_b}
    = \frac{\img}{2Q}\frac{\partial}{\partial b_\perp^\rho}
      \left[ \gamma_\perp^\rho\, \slashed n\,,\,\hat\cG_{h/\bar f}(x,\bt) \right]^{\beta\beta'}
  \,.\end{align}
  \end{subequations}
Notice that the sign difference between \eqs{def_BP_f_a}{def_BP_fbar_a} arises
because the prefactors $1/(2\w_1)$ and $1/(2\w_2)$ in \eq{op_P1}
pick different signs in the quark and antiquark contributions.
There is no such sign difference in \eqs{def_BP_f_b}{def_BP_fbar_b}
because $\cP_\perp$ and $\cP_\perp^\dagger$ act on the opposite fields
when going from \eq{def_BP_f_a} to \eq{def_BP_fbar_a}, which results
in an additional minus sign.

In \eq{WP_hat}, we have not absorbed the soft function into the hadronic  matrix elements.
To do so, we define
\begin{align} \label{eq:WP}
 W^{(1)\mu\nu}_\cP &
 = 4z \int\frac{\df^2\bt}{(2\pi)^2} e^{-\img \bt \cdot \qt} \sum_f \cH^{(0)}_f(q^+q^-)
 \\\nn& \ \times
   \Big\{\Tr\left[ B_{\cP\, f/N}(x,\bt)\, \gamma^\mu \, \cG_{h/f}(z,\bt)\, \gamma^\nu+ B_{f/N}(x,\bt)\, \gamma^\mu \, \cG_{\cP\, h/f}(z,\bt)\, \gamma^\nu\right]
 \nn\\
 &\ \quad +(f\to \bar f,\, \mu\leftrightarrow \nu) \Big\}
.\nn
\end{align}
The soft-subtracted correlators are defined analogous to
\eqs{def_BP_f_b}{def_GP_f_b} as
\begin{align} \label{eq:def_BPf}
 B_{\cP\, f/N}(x,\bt) &
 \equiv \frac{-\img}{2Q} \frac{\partial}{\partial b_\perp^\rho}
        \left[ \gamma_\perp^\rho\, \slashed\bn\,,\, B_{f/N}(x,\bt) \right]
\,,\nn\\
 \cG_{\cP\, h/f}(z,\bt) &
 = \frac{\img}{2Q}\frac{\partial}{\partial b_\perp^\rho}
   \left[ \gamma_\perp^\rho\, \slashed n\,,\,\cG_{h/f}(x,\bt) \right]
\,.\end{align}
Comparing \eqs{def_BP_f}{def_BPf},
\begin{align}  \label{eq:diff_BP}
 B_{\cP\, f/N}(x,\bt) &
 = \hat B_{\cP\, f/N}(x,\bt) \sqrt{S(b_T)}
 - \frac{\img}{2Q}
   \left[ \gamma_\perp^\rho\, \slashed\bn\,,\, \hat B_{f/N}(x,\bt) \right] \frac{\partial}{\partial b_\perp^\rho} \sqrt{S(b_T)}
\,,\nn\\
 \cG_{\cP\, h/f}(x,\bt) &
 = \hat \cG_{\cP\, h/f}(x,\bt) \sqrt{S(b_T)}
 + \frac{\img}{2Q}
   \left[ \gamma_\perp^\rho\, \slashed n\,,\, \hat \cG_{h/f}(x,\bt) \right] \frac{\partial}{\partial b_\perp^\rho} \sqrt{S(b_T)}
\,,\end{align}
we see that one can not absorb the soft function in the same fashion as at LP
due to the additional derivative acting on $S(b_T)$ itself.
This implies that $\hat W^{(1)\mu\nu}_\cP$ and $W^{(1)\mu\nu}_\cP$
also differ by such terms. Fortunately, these turn out to be proportional
to either $n^\mu - \bn^\mu = 2 n_z^\mu$ or $n^\nu - \bn^\nu = 2 n_z^\nu$,
whose contractions with the leading-power projectors
$P^{(0)}_{\mu\nu}$ vanish by current conservation.
Explicitly, we have
\begin{align} \label{eq:W_vs_What}
 W^{(1)\mu\nu}_\cP - \hat W^{(1)\mu\nu}_\cP &
 = 4z \int\frac{\df^2\bt}{(2\pi)^2} e^{-\img \bt \cdot \qt} \sum_f
   \bigl[ \cH^{(0)}_f(q^2) \Delta^{\mu\nu}_{\cP\,f} + (f \to \bar f,\, \mu \leftrightarrow \nu) \bigr]
\end{align}
where the difference of the trace terms evaluates to
\begin{align}
 \Delta^{\mu\nu}_{\cP\,f} &
 = \frac{-\img}{4 Q} \frac{\partial S(b_T)}{\partial b_\perp^\rho} \biggl\{
   \Tr\left[ \hat B_{f/N}(x,\bt)\, \gamma^\mu \, \hat\cG_{h/f}(z,\bt) \bigl(\gamma^\nu \gamma_\perp^\rho \slashed\bn + \gamma_\perp^\rho \slashed n\, \gamma^\nu \bigr) \right]
  \nn\\&\hspace{1.5cm}
  \qquad -\Tr\left[ \hat B_{f/N}(x,\bt) \bigl( \gamma_\perp^\rho \slashed\bn \gamma^\mu + \gamma^\mu \gamma_\perp^\rho \slashed n \bigr) \hat\cG_{h/f}(z,\bt)\, \gamma^\nu\right] \biggr\}
 \nn\\&
 = \frac{-\img}{2 Q} \frac{\partial S(b_T)}{\partial b_\perp^\rho} \biggl\{
   (n^\nu - \bn^\nu)\, \Tr\left[ \hat B_{f/N}(x,\bt)\, \gamma^\mu \, \hat\cG_{h/f}(z,\bt) \gamma_\perp^\rho\right]
  \nn\\&\hspace{1.5cm}
  \qquad
  + (n^\mu - \bn^\mu)\, \Tr\left[ \hat B_{f/N}(x,\bt) \gamma_\perp^\rho \hat\cG_{h/f}(z,\bt)\, \gamma^\nu\right] \biggr\}
\,.\end{align}
Here, we combined the soft factors coming from $B_{\cP\, f/N}, \cG_{h/f}$ and $B_{f/N}, \cG_{\cP\, h/f}$
as $(\partial_{b_\perp^\rho} \sqrt{S}) \sqrt{S} = \frac12 \partial_{b_\perp^\rho} S$,
and in the second step used that the LP correlators obey
$\slashed n \hat B_{f/N} = \slashed{\bn} \hat \cG_{h/f} = 0$.
In conclusion, we find that $W^{(1)\mu\nu}_\cP$ and $\hat W^{(1)\mu\nu}_\cP$ agree up
to subleading (i.e.~at least NNLP) corrections. 

Just as the soft function for the $\cP_\perp$ operators was the same as at leading power, we also can anticipate that the soft subtractions will be the same with various choices of rapidity regulator.  In the soft limit the interactions of the collinear fields in these subleading operators behave like those of the soft Wilson lines in the soft function, so the only difference for the soft subtraction calculation is (potentially) the appearance of a different rapidity regulator. (In the soft subtraction the rapidity regulator must be the same as that used in the unsubtracted collinear matrix elements, while with some choices of rapidity regulator it differs for the soft function itself.) 
In our SCET matrix elements in eqs.~(\ref{eq:def_BP_f}--\ref{eq:def_GP_fbar}), the soft subtractions are encoded as part of the collinear propagators, which act as distributions that induce the subtractions, see~\cite{Manohar:2006nz}. If the soft subtractions are written as distinct vacuum matrix elements, following the method in~\refcite{Lee:2006nr}, then the same argument given above for $S$ will enable us to commute this matrix element with the $\partial/\partial b_\perp^\rho$ derivative.%
\footnote{The argument proceeds as follows. We consider the matrix element in \eq{def_BP_f_a}, but using instead collinear fields without restrictions (which include couplings to the zero-bin collinear gluons), giving a result which we label with an extra prime, $\hat B_{\cP\, f/N}^{\prime\beta\beta'}$. We then make field redefinitions to remove the zero-bin contributions and obtain the collinear fields that we used here~\cite{Lee:2006nr}: one on the $n\cdot A_n'$ collinear gluons and a second one on the Wilson lines $W_n'$. Since $\cP_\perp$ is a linear operator, the same steps used to derive \eq{def_BP_f_b} go through, but now we have $\hat B_{f/N}^\prime = \hat B_{f/N} S^{\rm{0bin}}$ as the argument in the second slot of the commutator. This $S^{\rm{0bin}}$ is the same as subtraction as at leading power, and also the same for TMD PDFs and FFs. Therefore, the same argument leading to our \eq{W_vs_What} implies that the $\partial/\partial b_\perp^\rho$ derivative on $S^{\rm{0bin}}$ can be dropped, implying that $S^{\rm{0bin}}$ can be pulled outside of the commutator and derivative. Together with our result in \eq{def_BP_f} this then implies 
$\hat B_{\cP\, f/N}^{\beta\beta'}=\hat B_{\cP\,f/N}^{\prime\beta\beta'}/ S^{\rm{0bin}}$. Thus the soft subtractions take the same multiplicative form as at leading power, and can be grouped and manipulated along with the soft function.}  
Hence the subtractions do not change any of the manipulations above.

Thus, we have demonstrated the non-trivial fact that the full contributions from $\cP_\perp$ operators can be entirely expressed in terms of the leading power TMDs, which absorb the leading power soft function, and also have the same leading power hard function. Our final result for $W_{\cP}^{(1)\mu\nu}$ is thus given by \eq{WP} with \eq{def_BPf}.  We also remark, that although we have neglected transverse Wilson lines at light-cone infinity by working only in covariant gauges here, the generalization to fully account for these transverse gauge links in the $\cP_\perp$ operators is carried out in \app{Twilsonlines}, leading \eqs{def_BP_f}{def_GP_f} to be modified, and giving the results in \eqs{def_P_hat_trans}{def_GP_f_trans}. It is shown there that the relations in \eqs{def_BP_f_b}{def_GP_f_b} remain valid when transverse gauge links are included.

It is straightforward to use the projection relations $\slashed n\,\hat B=\slashed \bn\,\hat\cG=0$ to show that \eq{WP} is composed of terms with one Lorentz index transverse, and the other one proportional to $n$ or $\bn$ (this point will also be clear from calculations in \sec{results_NLP_P}). 
Therefore, when contracting with the leading power projectors $P_{i\mu\nu}^{(0)}$ (given by \eq{def_P_i} with $\tilde n_{x,y,t}\to n_{x,y,t}$), \eq{WP} has contributions to the structure functions whose corresponding projectors have one transverse index and one $n_t$, namely $W_{1,2,5,6}^X$. In contrast, \eq{WP} does not contribute to the structure functions $W_{-1,3,4,7}^X$, at NLP because their corresponding $P_{i\mu\nu}^{(0)}$ projectors have two transverse indices.

\subsection{NLP Contributions from the Collinear $\cB_{n_i\perp}$ Operators}
\label{sec:factorization_B}

It remains to consider contributions from the operators with an additional $\cB_{n_1\perp}$ or $\cB_{\bn_1\perp}$ field,
given in \eq{Bf}. As usual, we obtain the corresponding contribution to the hadronic tensor
by inserting the currents
\begin{align} \label{eq:WB_sim}
 \Bigl(J^{(1)\dagger\mu}_{\cB 1}+J^{(1)\dagger\mu}_{\cB 2}\Bigr) J^{(0)\nu}
 \quad{\rm and}\quad
 J^{(0)\dagger\mu} \Bigl(J^{(1)\nu}_{\cB 1}+J^{(1)\nu}_{\cB 2}\Bigr)
\end{align}
into \eq{W}. We start by studying the contribution from $J^{(0)\mu\dagger} J^{(1)\nu}_{\cB1}$,
for which the relevant currents are given in \eqs{crrnt_lp}{matched_1gluon_barn1}.
Inserting these into \eq{W} and performing a similar algebra as in \sec{factorization_LP},
we arrive at
\begin{align}  \label{eq:W_J0J1B_1}
 W^{(1)\,\mu\nu}_{J^{(0)} J^{(1)}_{\cB1}} &
 = \SumInt_X \, \delta^4(q + \Pa  - \Pb - \PX) \,
   \braket{N | J^{(0)\mu\dagger}(0) | h,X} \braket{h,X | J^{(1)\nu}_{\cB1}(0) | N}
\nn\\&
 = -2 (\gamma_\perp^\mu)^{\beta'\alpha'} (\gamma_{\perp \rho})^{\alpha\beta}
   \, \sum_f  \int\! \frac{\df^2\bt}{(2\pi)^2}\, e^{-\img \vec{q}_T\cdot\vec{b}_T}
   \int\!\df\xi \, C^{(0)}(-\w_a\w_b)\, C^{(1)}(-\w_a\w_b,\,\xi)
   \nn\\&\quad\times \biggl[
    \frac{\w_a \na^\nu+\w_b \nb^\nu}{\w_b} \hat B_{\cB\,f/N}^{\rho\, \beta \beta'\, \bar b' b}(\w_a, \xi, \bt)
    \, \hat\cG_{h/f}^{\alpha'\alpha\,a'\bar a}(\w_b, \bt)
    \,  S_f^{b' \bar a' a \bar b}(b_T)
  \nn\\& \qquad
   - \frac{\w_a \na^\nu+\w_b \nb^\nu}{\w_a} \hat B_{\bar f/N}^{\alpha'\alpha\,a'\bar a}(\w_a, \bt)
   \, \hat \cG_{\cB\,h/\bar f}^{\dagger\rho\,\beta\beta' \, \bar b' b}(\w_b, \xi, \bt)
   \, S_{\bar f}^{b' \bar a' a \bar b}(b_T)
   \biggr]
\,.\end{align}
Here, the first term in square brackets is the quark contribution,
while the second term is the antiquark contribution.
The relative sign between these terms arises because the kinematic prefactor
$(\lp_1^\nu-\lp_2^\nu+\lp_3^\nu)$ in the Wilson coefficient, \eq{C11C12},
has opposite sign for the two contributions. Since there is only a single
subleading operator insertion, each of the two terms in \eq{W_J0J1B_1}
contains two leading-power correlators, as defined in
\eqs{quark_correlators_LP}{antiquark_correlators_LP},
as well as a single NLP quark-gluon-quark correlator. We define the latter as 
\begin{align}
 \hat B_{\cB\,f/N}^{\rho\, \beta \beta'\, \bar b' b}(\w_a, \xi, \bt) &
 = \theta(\w_a)
   \MAe{N}{ \bar\chi_{\na}^{\beta' \bar b'}(b_\perp) \, g{\cB}_{\na\perp,\,-\xi\w_a}^{\rho\, b\bar c}(0)
            \,\chi_{\na,\,(1-\xi)\w_a}^{\beta c}(0)}{N}
\,,\nn\\
 \hat \cG_{\cB \,h/\bar f}^{\dagger\,\rho\,\beta\beta' \, \bar b' b}(\w_b, \xi, \bt) &
 = \theta(\w_b)
   \SumInt_{X_\bn} \MAe{0}{\bar\chi_{\nb}^{\beta' \bar b'}(b_\perp)}{h,X_\bn}
   \nn\\&\qquad\times
   \MAe{h,X_\bn}{g{\cB}_{\nb\perp,\, \xi\w_b}^{\rho\,b\bar c}(0) \, \chi_{\nb,\,-(1-\xi)\w_b}^{\beta c}(0)}{0}
\,.\end{align}
In both cases, $\rho$ is a Lorentz index, while $\beta\beta'$ are Dirac indices,
and $\xi$ is the momentum fraction of the total momentum $\w_{a,b}$
that is carried away by the collinear gluon. Note that the gluon momenta have different signs,
as in $\hat B_{\cB\, f/N}$ the gluon is incoming ($-\xi \w_a < 0$), while in $\hat\cG_{\cB\, h/f}$
the gluon is outgoing ($\xi \w_b > 0$). The color index $c$ is summed over,
while $\bar b'$ and $b$ are held fixed.
Note that these correlators differ from the quark-gluon-quark correlators
defined in \refcite{Bacchetta:2006tn}, where only the total momentum the quark-gluon
fields that appear in the same collinear direction is constrained, such that there is no analog of our splitting variable $\xi$ there.
We will discuss this further in \sec{tree_level}.

Similar to LP, the collinear matrix elements are diagonal in color,
allowing us to define color-traced objects analogous to \eq{def_B_hat},
\begin{align} \label{eq:def_Bb_hat}
 \hat B_{\cB\,f/N}^{\rho \beta \beta'}(x, \xi, \bt) &
 = \theta(\w_a)
   \MAe{N}{ \bar\chi_{\na}^{\beta'}(b_\perp) \, g{\cB}_{\na\perp,\,-\xi\w_a}^{\rho}(0)
            \,\chi_{\na,\,(1-\xi)\w_a}^{\beta}(0)}{N}
\,,\nn\\
 \hat \cG_{\cB \,h/\bar f}^{\dagger\,\rho\beta\beta'}(z, \xi, \bt) &
 = \frac{-1}{2 z N_c} \delta_{\bar bb'}
  \hat \cG_{\cB \,h/\bar f}^{\dagger\,\rho\,\beta\beta' \, \bar b' b}(\w_b, \xi, \bt)
  \nn\\
 &= \frac{-1}{2 z N_c} \theta(\w_b)
   \SumInt_{X_\bn} \tr  \MAe{0}{\bar\chi_{\nb}^{\beta'}(b_\perp)}{h,X_\bn}
   \nn\\&\qquad\times
   \MAe{h,X_\bn}{g{\cB}_{\nb\perp,\, \xi\w_b}^{\rho}(0) \, \chi_{\nb,\,-(1-\xi)\w_b}^{\beta}(0)}{0}
\,.\end{align}
Here, $x = \w_a / P_N^-$ and $z = P_h^+/\w_b$ are the same collinear momentum fractions as at LP.
We include a minus sign in $\hat\cG_{\cB \bar f}$ to remove the relative sign between the two terms in \eq{W_J0J1B_1}.
Applying \eq{def_Bb_hat} together with \eq{def_B_hat} to \eq{W_J0J1B_1}, we obtain
\begin{align}  \label{eq:W_J0J1B_2}
 W^{(1)\,\mu\nu}_{J^{(0)} J^{(1)}_{\cB1}} &
 = -4z (\gamma_\perp^\mu)^{\beta'\alpha'} (\gamma_{\perp \rho})^{\alpha\beta}
   \, \sum_f  \int\! \frac{\df^2\bt}{(2\pi)^2}\, e^{-\img \vec{q}_T\cdot\vec{b}_T}
   \int\!\df\xi \, \cH_f^{(1)}(-\w_a\w_b,\xi)
   \nn\\&\quad\times \biggl[
    \frac{\w_a \na^\nu+\w_b \nb^\nu}{\w_b}
    \hat B_{\cB\,f/N}^{\rho \beta \beta'}(x, \xi, \bt)
    \, \hat \cG_{h/f}^{\alpha'\alpha}(z, \bt)
    \, S(b_T)
  \nn\\& \qquad
   + \frac{\w_a \na^\nu+\w_b \nb^\nu}{\w_a}
      \hat B_{\bar f/N}^{\alpha'\alpha}(x, \bt)
      \hat \cG_{\cB\,h/\bar f}^{\dagger\,\rho\beta\beta'}(z, \xi, \bt)
   \, S(b_T)
   \biggr]
\,.\end{align}
We also defined the new hard function
\begin{align}
 \cH_f^{(1)}(\tilde q^2, \xi) = C_f^{(0)}(\tilde q^2)  C_f^{(1)}(\tilde q^2,\xi)
\,,\end{align}
where as usual in SIDIS kinematics $\tilde q^2 = -\w_a \w_b < 0$.
To simplify our result, in the following we employ that in the factorization frame
we have $\w_a = q^+ = Q$ and $\w_b = -q^- = Q - q_T^2/Q$, see \eq{q_ff},
and neglet the power suppressed $q_T^2$ correction. With this choice, the factors
of $\w_{a,b}$ cancel in the square brackets in \eq{W_J0J1B_2}.

We next consider the contribution from $J^{(0)\mu\dagger} J^{(1)\nu}_{\cB2}$.
From \eq{Bf}, we see that compared to the $J_{\cB1}$ current, the gluon
is now collinear to the opposite quark. This leads to the following new color-traced correlators:
\begin{align} \label{eq:BBfbar_hat}
 \hat B_{\cB \, \bar f/N}^{\dagger\,\rho\,\alpha'\alpha}(x,\xi,\bt) &
 = -\theta(\w_a)
    \tr \MAe{N}{\chi_{\na}^{\alpha'}(b_\perp)\, \bar\chi_{\na,\,(1-\xi)\w_a}^{\alpha}(0)
      \,g {\cB}_{\na\perp,\,-\xi\w_a}^{\rho}(0)}{N}
\,,\nn\\ 
 \hat \cG_{\cB \, h/f}^{\rho\,\alpha'\alpha}(z,\xi,\bt) &
 = \frac{1}{2z N_c} \theta(\w_b) 
   \SumInt_{X_\bn} \tr \MAe{0}{\chi_{\nb}^{\alpha'}(b_\perp)}{h,X_\bn}
   \nn\\&\qquad\times
   \MAe{h,X_\bn}{\bar\chi_{\nb,\,-(1-\xi)\w_b}^{\alpha}(0)\,g{\cB}_{\nb\perp,\,\xi\w_b}^{\rho}(0)}{0}
\,.\end{align}
Note in \eqs{def_Bb_hat}{BBfbar_hat}, we have written $\hat \cG_{\cB \,h/\bar f}$ and
$\hat B_{\cB \, \bar f/N}$ as daggered objects, defined as
\begin{align}
 \hat B_{\cB \, \bar f/N}^{\dagger\,\rho\,\alpha'\alpha}(x,\xi,\bt) &
 = \bigl[ \hat B_{\cB \, \bar f/N}^{\rho\,\alpha'\alpha}(x,\xi,-\bt) \bigr]^\dagger
\,,\nn\\
 \hat \cG_{\cB \,h/\bar f}^{\dagger\,\rho\,\beta\beta'}(z, \xi, \bt) &
 = \bigl[\hat \cG_{\cB \,h/\bar f}^{\rho\beta\beta'}(z, \xi, -\bt)\bigr]^\dagger
\,.\end{align}
The sign in $-\bt$ arises because by Lorentz invariance the correlator
only depends on the difference of the field positions, which under hermitian conjugation
flips sign, $[\chi_n(b_\perp) \bar\chi_n(0)]^\dagger = \bar\chi_n(0) \chi_n(b_\perp)$.

In our analysis here we are working in a general covariant gauge where transverse Wilson lines at light-cone infinity can be ignored. To obtain full gauge invariant results these lines must be included, and we do so for the quark-gluon-quark TMD PDF and TMD FF correlators in \app{Twilsonlines}. This results in the definitions in \eqs{def_Bb_hat}{BBfbar_hat} being generalized in the manner given in \eq{def_Bb_hat_trans}.

Since the $\cB_{n_i\perp}$ operators do not induce new soft Wilson lines,
as discussed in more detail in \sec{operators_NLP_B}, the factorized hadronic tensor
contains the LP soft function, which can be absorbed into these correlators
in the standard fashion,
\begin{align}\label{eq:def_BB}
 \tilde B_{\cB \, f/N}^{\rho\,\alpha'\alpha}(x,\xi,\bt) &
 = \hat{B}_{\cB \, f/N}^{\rho\,\alpha'\alpha}(x,\xi,\bt)\sqrt{S(b_T)}
\,,\nn\\ 
 \tilde \cG_{\cB \, h/f}^{\rho\,\beta'\beta}(z,\xi,\bt) &
 = \hat{\cG}_{\cB \, h/f}^{\rho\,\beta'\beta}(z,\xi,\bt) \sqrt{S(b_T)}
\,,\end{align}
and likewise for the antiquark correlators. Here and in the following,
we will denote the soft-subtracted quark-gluon-quark correlators with a tilde.
Once again we remark that in SCET the procedure for carrying out soft subtractions at subleading power is well defined, and is contained in the matrix elements of collinear fields in \eqs{def_Bb_hat}{BBfbar_hat}, see~\cite{Manohar:2006nz}.

Also taking the complex conjugate
contributions into account, we arrive at our final factorized expression for the $\cB_{n_i\perp}^\mu$ contributions at NLP
\begin{align} \label{eq:WB}
 W_\cB^{(1)\mu\nu} &
 = -4z \sum_f \int\!\df^2\bt \, e^{-\img \qt \cdot \bt}
   \int\df\xi \, \cH^{(1)}_f(q^+q^-, \xi)\, (n^\nu+\bn^\nu)
   \nn\\&\qquad\times
    \Tr\left[ \tilde B_{\cB \, f/N}^\rho(x,\xi,\bt) \gamma_{\perp}^\mu \cG_{h/f}(z,\bt) \gamma_{\perp\rho}
    +  B_{f/N}(x,\bt) \gamma_\perp^\mu \tilde \cG_{\cB\,h/f}^\rho(z,\xi,\bt) \gamma_{\perp\rho} \right]
    \nn\\
    &\qquad+(f\to\bar f,\,\mu\leftrightarrow\nu,\, \cH^{(1)}_f\to \cH^{(1)*}_f)+\text{h.c.}
\,.\end{align}
In \eq{WB}, h.c.~includes exchanging $\mu \leftrightarrow \nu$ as well as
flipping $\bt \to -\bt$ in the correlators. 
Since the $\mu\nu$ indices involve one transverse index and the other one proportional to $n$ or $\bn$, the situation here is the same as for \eq{WP}: when contracting with the leading power projectors,  $P_{i\mu\nu}^{(0)}$ (given by \eq{def_P_i} with $\tilde n_{x,y,t}\to n_{x,y,t}$), \eq{WB} has contributions to $W_{1,2,5,6}^X$. It does not contribute to $W_{-1,3,4,7}^X$ at NLP, because their corresponding projectors have two transverse indices.
Note that \eq{WB} provides a bare factorization theorem for the $qgq$ contributions. While we do not consider its renormalization here, the terms resulting from renormalization are anticipated to have a similar form.

We can also provide evidence that the convolution over $\xi$ in \eq{WB} will be convergent to all orders in perturbation theory.  This follows from the fact that in SCET divergent convolutions from a collinear momentum-fraction $\xi\to 0$ are in one-to-one correspondence with the presence of a non-trivial soft contributions~\cite{Manohar:2006nz}. These soft contributions correctly describe the small $\xi\sim 0$ momentum region, and the divergent convolution arises from not properly defining the separation between the collinear and soft contributions in the renormalized factorization theorem. (For an SCET example where such endpoint divergences have been treated, see Refs.~\cite{Liu:2020tzd,Liu:2020wbn}.) Since our analysis implies that there are no non-trivial subleading power soft contributions for SIDIS at NLP, we anticipate that the integrals over $\xi$ will correspondingly all be converge.  

\section{Results}
\label{sec:results}
In this section, we explicitly calculate the factorization formulae for different structure functions. To this end, we calculate the spinor traces in the factorized hadronic tensor with substitutions of the spinor decompositions of the several correlators defined in the previous section. Then we contract the hadronic tensor with projectors to give rise to our final formulae of factorized structure functions.

We first show the leading power results in \sec{results_LP}, which are in full agreement with literature (e.g. \refcite{Bacchetta:2006tn}).  The analysis of \sec{factorization} demonstrates that many potential contributions are absent at NLP, including in particular subleading soft corrections and contributions from time-ordered products with subleading power Lagrangian insertions. In \sec{results_NLP}, we include the three non-zero NLP contributions (kinematic, $\cP_\perp$, and $\cB_{n_i\perp}$) to derive factorization for the polarized and unpolarized structure functions $W_{1,2,5,6}^X$ which start at NLP. We then discuss what is new in our results in \sec{results_discussion}, and give an extensive comparison with previous literature in \sec{literature}.

\subsection{Leading Power}
\label{sec:results_LP}

To decompose our hadronic matrix elements $B_{f/N}(x, \bt)$ and $\cG_{h/f}(z,\pt)$
into independent structures, we mostly follow the conventions in \refcite{Bacchetta:2006tn}.
There, the correlators are decomposed in momentum space, so we first require the Fourier-transformed correlators
\begin{align} \label{eq:FT_comparison}
 B_{f/N}(x,\kt) &
 = \int\frac{\df^2\bt}{(2\pi)^2} e^{+\img \bt\cdot\kt} B_{f/N}(x,\bt)
\,,\qquad
 \cG_{h/f}(z,\pt)
 = \int\frac{\df^2\bt}{(2\pi)^2} e^{-\img \bt\cdot\pt} \cG_{h/f}(z,\bt)
\,.\end{align}
Note that with some abuse of notation, we use the same symbols $B_{f/N}$ and $\cG_{h/f}$ for Fourier and momentum space,
as the distinction between the two is clear from the arguments.
The different conventions for the Fourier signs can be understood as follows:
inserting the inverse \eq{FT_comparison} into the LP hadronic tensor in \eq{hadronicts_LP} yields
\begin{align}
 W^{(0) \mu\nu} &
 \sim H(q^2) \int\frac{\df^2\bt}{(2\pi)^2} e^{-\img \bt \cdot \qt} B(x, \bt) \cG(z, \bt)
\nn\\&
 = H(q^2) \int \df^2\kt \, \df^2\pt \, \delta^{(2)}(\qt + \kt - \pt) B(x, \kt) \cG(z, \pt)
\,.\end{align}
Since $q$ and $N$ are incoming, the sum of their transverse momenta must equal
that of the outgoing hadron, as also manifest in \eq{delta_expanded}.

Following the decomposition of the quark-quark correlators in \refcite{Bacchetta:2006tn}, we write
\begin{align} \label{eq:Phi_msp}
 B_{f/N}(x, \kt) &
 = \biggl[ f_{1}
      - f_{1 T}^{\perp} \frac{\eps_\perp^{\rho \sigma} k_{\perp \rho} S_{\perp \sigma}}{M_N}
      + g_{1L} \, S_L \gamma_5
      - g_{1T} \frac{k_\perp \cdot S_\perp}{M_N} \gamma_5
      + h_1 \, \gamma_5 \slashed{S}_\perp
   \nn\\&\qquad
      + h_{1L}^\perp \, S_L\frac{ \gamma_5\,\slashed k_{\perp}}{M_N}
      - h_{1T}^\perp \,\frac{k_\perp^2}{M_N^2}  \Bigl( \frac12 g_\perp^{\rho\sigma} - \frac{k_\perp^\rho  k_\perp^\sigma}{k_\perp^2}  \Bigr) S_{\perp\,\rho} \gamma_\sigma \gamma_5
      +\img h_{1}^{\perp} \frac{\slashed k_{\perp}}{M_N} \biggr] \frac{\slashed{n}}{4}
\,,\\ \label{eq:Delta_msp}
 \cG_{h/f}(z, \pt) &
 = \Bigl( D_1 + \img H_1^\perp \frac{\slashed p_\perp }{M_h} \Bigr) \frac{\slashed{\bn}}{4}
\,.\end{align}
For more details on how to relate our definitions to those in \refcite{Bacchetta:2006tn}
and obtain this decomposition from theirs, see \app{notation_relation}.
Note that our correlators use SCET and hence are defined in terms of good fermion field components only.
Therefore only leading-twist terms contribute to \eqs{Phi_msp}{Delta_msp}.
For the fragmentation function we do not give polarization-dependent pieces, as we only consider the target hadron to be polarized.
On the right-hand side of both \eqs{Phi_msp}{Delta_msp}, we keep the quark flavor $f$ implicit for each individual TMD PDF and TMD FF and suppress their arguments for brevity (which also include $\mu$ and the Collins-Soper scales $\zeta_{a,b})$.
Finally, we remark that the decompositions in \eqs{Phi_msp}{Delta_msp}
apply to the soft-subtracted TMDs, as defined in \eq{BG_soft}, while this soft
factor was not taken into account \refcite{Bacchetta:2006tn}.

For the antiquark distributions, one obtains similar decompositions as in \eqs{Phi_msp}{Delta_msp}
upon taking into account that by charge conjugation~\cite{Mulders:1995dh}
\begin{align} \label{eq:relation_q_barq}
 \Tr\bigl[\gamma^\mu B_{\bar f/N}(x, \kt)\bigr] &
 = \Tr\bigl[\gamma^\mu B_{f/N}^c(x, \kt)\bigr]
\,,\nn\\
 \Tr\bigl[\gamma^\mu \gamma_5 B_{\bar f/N}(x, \kt)\bigr] &
 = -\Tr\bigl[\gamma^\mu \gamma_5 B_{f/N}^c(x, \kt)\bigr]
\,,\nn\\
 \Tr\bigl[\img \sigma^{\mu\nu} \gamma_5 B_{\bar f/N}(x, \kt)\bigr] &
 = \Tr\bigl[\img \sigma^{\mu\nu} \gamma_5 B_{f/N}^c(x, \kt)\bigr]
\,.\end{align}
Here, $B^c_{f/N}$ is the quark distribution evaluated with charge-conjugated fields $\chi^c = C \bar \chi^T$.
\Eq{relation_q_barq} implies that the distributions corresponding to unpolarized and transversely polarized quarks
have identical decompositions for quarks and antiquarks, while those for longitudinally polarized quarks receive a sign flip.
In our case, this only affects $g_{1L}$ and $g_{1T}$.
For the spin-averaged TMDFF, it follows that the we have identical decomposition for quarks and antiquarks.

Using \eqs{app:FT_1}{app:FT_2}, we obtain the Fourier transform of \eqs{Phi_msp}{Delta_msp},
\begin{align} \label{eq:Phi_bsp_new}
 B_{f/N}(x, \bt) &
 = \biggl[ f_{1}
      + \img f_{1 T}^{\perp(1)} M_N b_{\perp\,\rho}\eps_\perp^{\rho \sigma} S_{\perp \sigma}
      + g_{1L} \, S_L \gamma_5
      + \img g_{1T}^{(1)} M_N b_\perp \cdot S_\perp \gamma_5
      + h_1 \, \gamma_5 \slashed{S}_\perp
   \nn\\&\qquad
      - \img  h_{1L}^{\perp(1)} \, S_L M_N  \gamma_5 {\slashed b}_{\perp}
      + h_{1}^{\perp(1)} M_N \slashed b_{\perp}
   \nn\\&\qquad
      + \frac12 h_{1T}^{\perp(2)} \Ma^2 b_\perp^2 \left(  \frac12 g^{\rho\sigma}_\perp - \frac{b_\perp^\rho b_\perp^\sigma}{b_\perp^2} \right) S_{\perp\,\rho} \gamma_\sigma \gamma_5
   \biggr] \frac{\slashed{n}}{4}
\,,\\ \label{eq:Delta_msp_new}
 \cG_{h/f}(z, \bt) &
 = \Bigl( D_1 - H_1^{\perp(1)} M_h \slashed b_\perp \Bigr) \frac{\slashed{\bn}}{4}
\,.\end{align}
Here, we used the abbreviations~\cite{Boer:2011xd}%
\footnote{\Refcite{Boer:2011xd} includes factors of $z$ for the Fourier transform of the TMDFF
due to defining it with respect to the hadron momentum $\vec P_{T,h}$ in a frame where the fragmenting
quark has no transverse momentum, while we work in a frame with vanishing $\vec P_{T,h}$ and thus
the quark has nonvanishing transverse momentum $\pt$.}
\begin{align} \label{eq:def_fn}
 f(x, b_T) &= 2 \pi \int_0^\infty \df p_T \, p_T J_0(b_T p_T) f(x, p_T)
\,,\nn\\
 f^{(n)}(b_T) &
 = \frac{2\pi\, n!}{(b_T M)^n} \int_0^\infty \df p_T \, p_T \left(\frac{p_T}{M}\right)^n J_n(b_T p_T) \, f(p_T)
\,.\end{align}
We next calculate the Dirac trace part required in \eq{hadronicts_LP},
\begin{align}\label{eq:trace_photon}
 & \Tr\bigl[ B_{f/N} (x,\bt) \gamma_\perp^\mu \cG_{h/f}(z,\bt)\gamma_\perp^\nu\bigr]
 \\\nn&
 = \frac{1}{2} \biggl\{
    -\Bigl[f_1  + \img  f_{1 T}^{\perp(1)} \bigl(\Ma b_{\perp \rho} S_{\perp \sigma} \epsilon_{\perp}^{\rho \sigma}\bigr) \Bigr] D_1\, g_\perp^{\mu\nu}
   + \Bigl[ \img g_{1L} \, S_L - \Ma\, b_\perp \cdot S_\perp \,g_{1T}^{(1)} \Bigr] D_1\,\eps_\perp^{\mu\nu}
 \\\nn&\quad
    + 2 h_1^{\perp (1)} H_1^{\perp (1)} \Ma\Mb  b_\perp^2\, \left(\frac12 g_\perp^{\mu\nu} - \frac{b_\perp^\mu b_\perp^\nu}{b_\perp^2}\right)
 \\\nn&\quad
   -\img h_1 H_1^{\perp(1)} \Mb  \left(b_\perp^\mu\eps_\perp^{\nu\rho}S_{\perp\rho}+S_\perp^\nu\eps_\perp^{\mu\rho}b_{\perp\rho}\right)
 \\\nn&\quad
   + h^{\perp(1)}_{1L} H_1^{\perp(1)} \,\Ma\Mb S_L\, \left(b_\perp^\mu\eps_\perp^{\nu\rho}b_{\perp\rho}+b_\perp^\nu\eps_\perp^{\mu\rho}b_{\perp\rho}\right)
 \\\nn&\quad
   + \frac{\img}{2} h_{1T}^{\perp(2)} H_1^{\perp(1)} \Ma^2 \Mb
      \Bigl[ \frac12 b_\perp^2 \bigl(b_\perp^\mu \eps_\perp^{\nu\rho} S_{\perp\rho}  +  S_\perp^\nu  \eps_\perp^{\mu\rho}b_{\perp\rho} \bigr)
  - (b_\perp{\cdot}S_\perp) \left(b_\perp^\mu\eps_\perp^{\nu\rho}b_{\perp\rho}+b_\perp^\nu\eps_\perp^{\mu\rho}b_{\perp\rho}\right) \Bigr]
 \biggr\}
 .
\end{align}
To obtain the structure functions, we first insert the trace in \eq{trace_photon}
into the LP hadronic tensor, \eq{hadronicts_LP}, and contract the result with the LP projectors $P_{i\mu\nu}^{(0)}$ obtained from \eq{def_P_i} by taking $\tilde n_{x,y,t}\to n_{x,y,t}$.
By collecting the resulting expression with respect to $S_L$ and $S_T$ as given in \eq{def_W_i_pol},
one then obtains the polarized structure functions defined in \eq{W_proj_3}.
The structure functions as classified by the angular coefficients are then
constructed following the middle column of \eq{def_W_polarized}.
To express the resulting inverse Fourier transform in a compact fashion, we define
\begin{align}\label{eq:cF}
 \cF[\cH\, g^{(n)} \, D^{(m)}] &
 = 2 z \sum_f \cH_f(q^+ q^-) \int\frac{\df^2\bt}{(2\pi)^2} e^{-\img \qt \cdot \bt}
   (-\img\Ma b_T)^n (\img\Mb b_T)^m \cos[(n+m) \varphi]
   \nn\\&\hspace{3.5cm}\times
   \, g_f^{(n)}(x,b_T) \, D_f^{(m)}(z,b_T) + (f\to\bar f)
 \nn\\&
 = 2 z \sum_{f} \cH_f(q^+ q^-)
   \int_0^\infty \frac{\df b_T \, b_T}{2\pi} (\Ma b_T)^n (-\Mb b_T)^m J_{n+m}(b_T q_T) \,
   \nn\\&\hspace{3.5cm}\times
   \, g_f^{(n)}(x,b_T) \, D_f^{(m)}(z,b_T) + (f\to\bar f)
\,.\end{align}
Here, $\cH$ is the hard function, and $g^{(n)}$ and $D^{(m)}$ denote the Fourier-transformed
TMDPDF and TMDFF as defined in \eq{def_fn}, respectively. For brevity, we suppress
their arguments the sum over flavors on the left hand side, and they given explicitly on the right-hand side.
We have parameterized $\bt = b_T (\cos\varphi, \sin\varphi)$,
such that with our convention for $\qt = -q_T(1,0)$, we have $\qt \cdot \bt = -q_T b_T \cos\varphi$.
In the second step in \eq{cF}, we analytically evaluated the integral over $\varphi$.
We have also absorbed the common factors $(-\img \Ma b_T)$ and $(+\img \Mb b_T)$
that arise in the Fourier transforms of individual TMDs, compare \eqs{Phi_msp}{Phi_bsp_new}.
The nonvanishing LP structure functions then read
\begin{align} \label{eq:W_LP}
  W_{UU,T} &= 
  \phantom{-} \cF \left[ \cH^{(0)}\, f_1 D_1\right]
\,,\nn\\
  W_{UU}^{\cos 2\phi_h} &= 
  -\cF \left[\cH^{(0)}\, h_1^{\perp(1)} H_1^{\perp(1)}\right]
\,,\nn\\
  W_{UL}^{\sin2\phi_h} &= 
  -\cF \left[\cH^{(0)}\, h_{1L}^{\perp(1)}H_1^{\perp(1)}\right]
\,,\nn\\
  W_{LL} &= 
  \phantom{-} \cF \left[\cH^{(0)}\, g_{1L} D_1\right]
\,,\nn\\
  W_{UT,T}^{\sin\left(\phi_h-\phi_S\right)} &= 
  -\cF \left[\cH^{(0)}\, f_{1T}^{\perp(1)} D_1\right]
\,,\nn\\
  W_{UT}^{\sin\left(\phi_h+\phi_S\right)} &=
  -\cF \left[\cH^{(0)}\, h_1 H_1^{\perp(1)}\right]
\,,\nn\\
  W_{UT}^{\sin\left(3\phi_h-\phi_S\right)} &=  -\frac14 \cF \left[\cH^{(0)}\, h_{1T}^{\perp(2)} H_1^{\perp(1)}\right]
\,,\nn\\
  W_{LT}^{\cos\left(\phi_h-\phi_S\right)} &=  \phantom{-} \cF \left[ \cH^{(0)}\, g_{1T}^{(1)} D_1\right]
\,.\end{align}
Here, we suppress that these functions receive corrections in $\lambda$ themselves,
which we will explicitly give in \sec{results_NLP}.
These results agree with Eqs.~(2.23) -- (2.30) in \refcite{Boer:2011xd},
up to a minus sign for all structures involving $H_1^{\perp\,(1)}$ due
to the different sign in the definition of the Fourier transform for TMDFFs.

Note that in \eq{W_LP}, the structure functions $W_{LL}$ and $W_{LT}^{\cos(\phi_h-\phi_S)}$
arise from the antisymmetric projector $P_4^{\mu\nu}$, where exchanging $\mu\leftrightarrow\nu$
for the antiquark contribution in $W^{\mu\nu}$ yields a relative minus sign. Since corresponding
antiquark distributions $g_{1L}$ and $g_{1T}^{(1)}$ receive a minus sign on their own,
such that summing over all quarks and antiquarks as stated in \eq{cF} is correct.
See also the comments below \eq{hadronicts_LP} and \eq{relation_q_barq}.

\subsection*{Result in Momentum Space}
Traditionally, the above structure functions were expressed in momentum space
using convolutions of the form
\begin{align}\label{eq:cF_tilde}
 \tilde\cF[\w\, \cH\, g\, D] &
 = 2z \sum_f \cH_f(q^+q^-) \int\!\df^2\kt\, \df^2\pt \, \delta^{(2)}\bigl(\qt+\kt-\pt\bigr)
  \nn\\ &\qquad\quad
  \times\w(\vec k_T, \vec p_T)\, g_f(x, k_T) \, D_f(z,p_T)+ (f\to\bar f)
\,,\end{align}
where $w(\kt, \pt)$ is weight factor dependent on $\kt$ and $\pt$.
Explicit results for evaluating $\cF[\w \, \cH \, g^{(n)} \, D^{(m)}]$
in terms of \eq{cF_tilde} are given in \app{fourier}.
Applying this to \eq{W_LP}, we obtain
\begin{align}
  W_{UU,T} &= 
  \tilde \cF \left[ \cH^{(0)}\, f_1 D_1\right]
\,,\nn\\
  W_{UU}^{\cos 2\phi_h} &= 
  \tilde\cF \left[\frac{-2 k_{Tx} p_{Tx} + \kt \cdot \pt}{\Ma \Mb} \cH^{(0)}\, h_1^{\perp} H_1^{\perp}\right]
\,,\nn\\
  W_{UL}^{\sin2\phi_h} &= 
  \tilde\cF \left[\frac{-2 k_{Tx} p_{Tx} + \kt \cdot \pt}{\Ma \Mb} \cH^{(0)}\, h_{1L}^{\perp}H_1^{\perp}\right]
\,,\nn\\
  W_{LL} &= 
  \tilde\cF \left[\cH^{(0)}\, g_{1L} D_1\right]
\,,\nn\\
  W_{UT,T}^{\sin\left(\phi_h-\phi_S\right)} &= 
  \tilde\cF \left[ -\frac{k_{Tx}}{\Ma} \cH^{(0)}\, f_{1T}^{\perp} D_1\right]
\,,\nn\\
  W_{UT}^{\sin\left(\phi_h+\phi_S\right)} &=
  \tilde\cF \left[ -\frac{p_{Tx}}{\Mb} \cH^{(0)}\, h_1 H_1^{\perp}\right]
\,,\nn\\
  W_{UT}^{\sin\left(3\phi_h-\phi_S\right)} &=  \tilde\cF \left[\frac{2 k_{Tx}\, (\vec k_T\cdot \vec p_T) + k_T^2\, p_{Tx} - 4k_{Tx}^2\, p_{Tx}}{2 \Ma^2\Mb}\, \cH^{(0)} h_{1T}^\perp H_1^\perp\right]
\,,\nn\\
  W_{LT}^{\cos\left(\phi_h-\phi_S\right)} &= \tilde\cF \left[ \frac{k_{Tx}}{\Ma} \cH^{(0)}\, g_{1T} D_1\right]
\,,\end{align}
where we work in a frame where $\qt = (-q_T, 0)$.

\subsection{Next-to-Leading Power}
\label{sec:results_NLP}

In this section, we derive factorization formulas for the SIDIS structure functions at NLP. 
We again reiterate that in this work we have neglected the impact of leading power Glauber exchange, which spoils factorization. Effectively, if these contributions can be shown to cancel like they do at LP, then we have derived in this work the complete form of the NLP factorization theorems for SIDIS.
We have considered all possible sources of power corrections, namely  kinematic corrections, subleading hard scattering currents generated at both the hard and hard-collinear scales, and subleading \SCETii Lagrangian insertions. The required operators were discussed and derived in \sec{operators}.

Of these NLP effects, all terms that involve time-ordered products in \SCETii were shown to vanish in \sec{fact_vanishing_Lag}, and all other terms that could generate subleading power soft functions were explicitly demonstrated to vanish in \sec{fact_vanishing_soft}.  

The non-vanishing effects identified at NLP include kinematic corrections, given in \sec{factorization_kinematic_corr}, and contributions from $\cP_\perp$ and $\cB_{n_i\perp}$ operators, given in \sec{factorization_P} and \sec{factorization_B}, respectively. We setup the general Lorentz decompositions of the subleading power matrix elements and the contractions needed to obtain the NLP structure functions in the following sections:
kinematic corrections from expanding projectors in \sec{results_NLP_kin},
corrections from $\cP_\perp$ operators in \sec{results_NLP_P}, and 
corrections from operators with collinear $\cB_{n_i\perp}$ gluon insertions in \sec{results_NLP_B}.
The combined final NLP structure function results for $W_{1,2,5,6}^X$ with various spin-polarizations $X$ are then presented in \sec{results_NLP_full} in Fourier space and in \sec{results_NLP_full_msp} in momentum space. The translations between different notations for these structure functions was given above in \eq{def_W_polarized}. 
In this subsection, we will derive NLP factorization formulae for these structure functions, using similar methods as in the leading-power case.

Note that we have not fully determined (or eliminated) NLP corrections to structure functions start off at LP in \eq{W_LP}, namely $W_{-1,3,4,7}^X$. We have demonstrated that most sources of NLP power corrections 
vanish for these structure functions, including: the kinematic corrections, the operators involving insertions of soft gluon or soft derivatives ($\cB_{s\perp}^{(n_i)\mu}$, $\bn\cdot \cB_{s}^{(n)}$, $n\cdot\cB_s^{(\bn)}$, $in\cdot \partial_s$, and $i\bn\cdot \partial_s$), and the operators involving insertions of a $\cP_\perp^\mu$ or a collinear gluon field $\cB_{n_i\perp}^\mu$. However we have not demonstrated that contributions from \SCETii Lagrangian insertions vanish for $W_{-1,3,4,7}^X$, nor have we shown that hard-collinear time-ordered products that arise from the \SCETi operator ${\cal O}_I^{(0)}$ vanish. 
Since the corresponding angular coefficients are dominated by the LP structure functions, potential NLP contributions here are of less phenomenological importance
than for those structure functions that start at NLP, which we do fully treat.
For the inclusive structure function, $W_{-1}^U$, we expect that no NLP corrections arise due to its azimuthal symmetry in $\phi_h$, which is akin to the discussion for unpolarized Drell-Yan
(see for example the discussion in \refcite{Ebert:2020dfc}).

\subsubsection{Kinematic Corrections}
\label{sec:results_NLP_kin}

As discussed in \sec{factorization_kinematic_corr}, only three of the projectors $P_i^{\mu\nu}$
defined in \eq{def_P_i} yield $\cO(\lambda)$ corrections when contracted with
the LP hadronic tensor $W_{\mu\nu}^{(0)}$, namely $P_1^{\mu\nu}, P_5^{\mu\nu}$ and $P_6^{\mu\nu}$
as given in \eq{P_pcs}. By contracting these with the Dirac structures appearing in \eq{trace_photon},
we obtain the induced power corrections. Here, we only calculate the corresponding Dirac contractions, leaving the full summary of the results for structure functions to \secs{results_NLP_full}{results_NLP_full_msp} below.
\Eq{trace_photon} contains the Dirac structures
\begin{align}
 \Bigl\{ g_\perp^{\mu\nu} \,,\quad \eps_
    \perp^{\mu\nu} \,,\quad
    \frac{g_\perp^{\mu\nu}}{2} -\frac{b_\perp^\mu b_\perp^\nu}{b_\perp^2} \,,\quad
    b_\perp^\mu\eps_\perp^{\nu\rho}b_{\perp\rho}+b_\perp^\nu\eps_\perp^{\mu\rho}b_{\perp\rho}  \,,\quad
    b_\perp^\mu\eps_\perp^{\nu\rho}S_{\perp\rho}+S_\perp^\nu\eps_\perp^{\mu\rho}b_{\perp\rho}
 \Bigr\}
\,.\end{align}
The nonvanishing contractions with \eq{P_pcs} are given by
\begin{subequations} \label{eq:subleading_P_traces}
\begin{align}
 P_{1\,\mu\nu} g_\perp^{\mu\nu}
 &= \frac{q_T}{Q}
\,,\\
 P_{6\,\mu\nu} \eps_\perp^{\mu\nu}
 &= \img\frac{q_T}{Q}
\,,\\ \label{eq:subleading_P_traces_c}
 P_{1\,\mu\nu} \left( \frac{g_\perp^{\mu\nu}}{2} -\frac{b_\perp^\mu b_\perp^\nu}{b_\perp^2}\right)
 &= -\frac{q_T}{2Q} \cos(2\varphi)
\,,\nn\\
 P_{5\,\mu\nu} \left( \frac{g_\perp^{\mu\nu}}{2} -\frac{b_\perp^\mu b_\perp^\nu}{b_\perp^2}\right)
 &= -\frac{q_T}{2Q} \sin(2\varphi)
\,,\\
 P_{1\,\mu\nu}
 \left(b_\perp^\mu\eps_\perp^{\nu\rho}b_{\perp\rho}+b_\perp^\nu\eps_\perp^{\mu\rho}b_{\perp\rho}\right)
 &= -b_\perp^2 \frac{q_T}{Q} \sin(2\varphi)
\,,\nn\\
 P_{5\,\mu\nu}
 \left(b_\perp^\mu\eps_\perp^{\nu\rho}b_{\perp\rho}+b_\perp^\nu\eps_\perp^{\mu\rho}b_{\perp\rho}\right)
 &= +b_\perp^2 \frac{q_T}{Q} \cos(2\varphi)
\,,\\
 P_{1\,\mu\nu}
  \left(b_\perp^\mu\eps_\perp^{\nu\rho}S_{\perp\rho}+S_\perp^\nu\eps_\perp^{\mu\rho}b_{\perp\rho}\right)
 &= \frac{q_T b_T}{Q} \bigl(S_{Tx} \sin\varphi + S_{Ty}  \cos\varphi\bigr)
\,,\\
 P_{5\,\mu\nu}
  \left(b_\perp^\mu\eps_\perp^{\nu\rho}S_{\perp\rho}+S_\perp^\nu\eps_\perp^{\mu\rho}b_{\perp\rho}\right)
 &= \frac{q_T b_T}{Q} \bigl(-S_{Tx} \cos\varphi + S_{Ty}  \sin\varphi\bigr)
\,.\end{align}
\end{subequations}
The contraction given by the fourth equality (second line of \eq{subleading_P_traces_c}) does not appear in our final results,
as in \eq{trace_photon} the traceless tensor only multiplies spin-independent terms,
but all structure functions in \eq{def_W_polarized} defined from $P_5$ involve spin structures.

\subsubsection{Contributions from the $\cP_\perp$ Operators}
\label{sec:results_NLP_P}

As discussed in \sec{factorization_P}, the $\cP_\perp$ operators introduce new
subleading correlators, which at NLP accuracy are defined in \eq{def_BPf}. At NLP they were proven to be completely determined by the LP TMD PDFs and TMD FFs. A key step in this proof was demonstrating that the same LP soft function could be absorbed into these correlators, since doing so only induces a NNLP correction that we then can neglect.
The corresponding expressions in momentum space can be obtained
by applying \eq{FT_1} and integrating by parts,
\begin{align} \label{eq:BPf_msp_0}
 B_{\cP\, f/N}(x,\kt) &
 = \int\frac{\df^2\bt}{(2\pi)^2} e^{+\img \bt \cdot \kt}
   \frac{-\img}{2Q} \frac{\partial}{\partial b_\perp^\rho}
   \left[ \gamma_\perp^\rho\, \slashed\bn\,,\, B_{f/N}(x,\bt) \right]
 = \frac{1}{2Q} \left[ B_{f/N}(x,\kt) \,,\, \slashed\bn \slashed{k}_\perp \right]
\,,\nn\\
 \cG_{\cP\, h/f}(z,\pt) &
 = \int\frac{\df^2\bt}{(2\pi)^2} e^{-\img \bt \cdot \pt}
   \frac{\img}{2Q}\frac{\partial}{\partial b_\perp^\rho}
   \left[ \gamma_\perp^\rho\, \slashed n\,,\,\cG_{h/f}(x,\bt) \right]
 = \frac{1}{2Q} \left[ \cG_{h/f}(x,\kt) \,,\, \slashed{n} \slashed{p}_\perp \right]
\,,\end{align}
where we assumed that the correlator vanishes sufficiently fast as $b_T \to \infty$
as stated by the Riemann-Lebesgue Lemma.

One advantage of performing this Fourier transform is that it allows us to
use integration by parts to avoid derivatives of the LP correlators.
In $b_T$ space, this can also be done, but only at the level of the hadronic tensor.
Furthermore, it allows us to compare our results for $B_{\cP\, f/N}$ and $\cG_{\cP\, h/f}$
to the NLP terms in the quark-quark correlators in the literature.
By inserting \eqs{Phi_msp}{Delta_msp} into \eq{BPf_msp_0}, we obtain
\begin{align} \label{eq:BPf_msp}
 B_{\cP\, f/N}(x,\kt) &
 = \frac{1}{2Q} \biggl\{
     f_1 \slashed{k}_\perp
    - f_{1 T}^{\perp} \frac{\eps_\perp^{\rho \sigma} k_{\perp \rho} S_{\perp \sigma}}{M_N} \slashed{k}_\perp
    - g_{1L} \, S_L \slashed{k}_\perp \gamma_5
    + g_{1T} \frac{k_\perp \cdot S_\perp}{M_N} \slashed{k}_\perp \gamma_5
    \nn\\&\hspace{1cm}
    + \frac12 h_1 \Bigl( [\slashed{S}_\perp, \slashed{k}_\perp] + \frac{1}{2} k_\perp \cdot S_\perp [\slashed n , \slashed{\bn}] \Bigr) \gamma_5
    + h_{1L}^\perp \, S_L \frac{k_\perp^2}{4 \Ma} [\slashed{n}, \slashed{\bn}] \gamma_5
    \nn\\&\hspace{1cm}
    + h_{1T}^\perp \frac{k_\perp^2}{4M_N^2}  \Bigl( [\slashed{S}_\perp , \slashed{k}_\perp]  - \frac12 k_\perp \cdot S_\perp [\slashed n , \slashed{\bn}] \Bigr) \gamma_5
   +\img h_{1}^{\perp} \frac{k_\perp^2}{4\Ma} [\slashed{n}, \slashed{\bn}]
   \biggr\}
\,,\nn\\
 \cG_{\cP\, h/f}(z,\pt) &
 = \frac{1}{2Q} \biggl\{ D_1 \slashed p_\perp
   - \img H_1^\perp \frac{ p_\perp^2}{4M_h} \bigl[ \slashed n, \slashed \bn \bigr] \biggr\}
\,.\end{align}
This produces the $\kt$-dependent Dirac structure of the quark-quark correlators at twist-3 in Ref.~\cite{Bacchetta:2006tn}.
Fourier transforming \eq{BPf_msp} yields the corresponding results in $\bt$ space,
\begin{align} \label{eq:BPf_bsp}
 B_{\cP\, f/N}(x,\bt) &
 = \frac{\Ma}{2Q} \biggl\{
    -\img \Ma f_1^{(1)} \slashed{b}_\perp
    + \Big[ \frac{\Ma^2}{2} f_{1 T}^{\perp(2)} (b_{\perp \rho} \eps_\perp^{\rho \sigma} S_{\perp \sigma}) \slashed{b}_\perp
    + f_{1 T}^{\perp(1)} \gamma_\rho \eps_\perp^{\rho \sigma} S_{\perp \sigma} \Bigr]
    \nn\\&\hspace{1.5cm}
    + \img \Ma g_{1L}^{(1)} \, S_L \slashed{b}_\perp \gamma_5
    - \Bigl[\frac{\Ma^2}{2} g_{1T}^{(2)} (b_\perp {\cdot} S_\perp) \slashed{b}_\perp
             + g_{1T}^{(1)} \slashed{S}_\perp  \Bigr] \gamma_5
    \nn\\&\hspace{1.5cm}
    - \frac{\img}{2} \Ma h_1^{(1)}  \Bigl( [\slashed{S}_\perp, \slashed{b}_\perp] + \frac{1}{2} b_\perp \cdot S_\perp [\slashed n , \slashed{\bn}] \Bigr) \gamma_5
    \nn\\&\hspace{1.5cm}
    -  \frac{S_L}{4} h_{1L}^{\perp\,(0')} [\slashed{n}, \slashed{\bn}] \gamma_5
    - \frac{\img}{4} h_{1}^{\perp\,(0')} [\slashed{n}, \slashed{\bn}]
    \nn\\&\hspace{1.5cm}
    + \frac{\img}{4} h_{1T}^{\perp\,(1')} \Ma
      \Bigl( [\slashed{S}_\perp , \slashed{b}_\perp]  - \frac12 b_\perp {\cdot} S_\perp [\slashed n , \slashed{\bn}] \Bigr) \gamma_5
   \biggr\}
\,,\nn\\
 \cG_{\cP\, h/f}(z,\bt) &
 = \frac{\Mb}{2Q} \biggl\{ \img  D_1^{(1)} \Mb \slashed b_\perp
   + \frac{\img}{4} H_1^{\perp\,(0')} \bigl[ \slashed n, \slashed \bn \bigr] \biggr\}
\,.\end{align}
Here, we made use of the abbreviations
\begin{align} \label{eq:0p_1p}
 f^{(0')}(b_T) &= \int\df^2\pt \,  e^{-\img \bt \cdot \pt} \frac{p_T^2}{\Ma^2} f(p_T)
\,,\nn\\
 -\img \Ma b_\perp^\mu f^{(1')}(b_T) &
 = \int\df^2\pt \,  e^{-\img \bt \cdot \pt} \frac{p_\perp^\mu p_T^2}{\Ma^3} f(p_T)
\,.\end{align}
see \eq{app:FT_3} for explicit expressions in terms of the standard $f^{(n)}$.

To obtain the NLP contribution from $\cP_\perp$ operators to the hadronic tensor,
we insert \eq{BPf_bsp} into \eq{WP}. Here, we only give the result for the resulting traces,
while the results for the structure functions will be given below in \secs{results_NLP_full}{results_NLP_full_msp}.
We obtain
\begin{align} \label{eq:WP_trace_a}
  &\Tr\left[ B_{\cP\, f/N}(x,\bt)\, \gamma^\mu \, \cG_{h/f}(z,\bt)\, \gamma^\nu\right]
  \\ \nn&
  =\frac{\Ma}{2Q} \biggl\{
   \Bigl( -\img f_1^{(1)} + \frac{\Ma}{2}f_{1 T}^{\perp(2)} \eps_\perp^{\rho \sigma} b_{\perp \rho} S_{\perp \sigma} \Bigr)
   D_1  \, \Ma \left(b_\perp^\mu\bn^\nu+\bn^\mu b_\perp^\nu\right)
  \\ \nn&\hspace{1.2cm}
  + f_{1 T}^{\perp(1)} D_1 \left(\bn^\mu\eps_{\perp}^{\nu\rho}S_{\perp\rho} + \bn^\nu\eps_{\perp}^{\mu\rho}S_{\perp\rho}\right)
  \\ \nn&\hspace{1.3cm}
  - \left(S_L\,g_{1 L}^{(1)} + \frac{\img\Ma}{2}\, b_\perp \cdot S_\perp\,g_{1T}^{(2)} \right) D_1
     \, \Ma \eps^{\mu\nu\rho\sigma}\bn_\rho b_{\perp\sigma}
  - \img g_{1T}^{(1)}  D_1\, \eps^{\mu\nu\rho\sigma}\bn_\rho S_{\perp\sigma}
  \\ \nn&\hspace{1.2cm}
  -\img\Mb h_1^{\perp(0')} H_1^{\perp (1)}\left(b_\perp^\mu\bn^\nu+\bn^\mu b_\perp^\nu\right)
  -\img\Mb S_L h_{1L}^{\perp(0')} H_1^{\perp(1)} \left(\bn^\mu\eps_{\perp}^{\nu\rho}b_{\perp\rho}+\bn^\nu\eps_{\perp}^{\mu\rho}b_{\perp\rho}\right)
  \\ \nn&\hspace{1.2cm}
  +\frac12\Ma\Mb\, h_{1T}^{\perp(1')} H_1^{\perp(1)}
  \bigl[ 
  b_\perp{\cdot}S_\perp \left(\bn^\mu\eps_{\perp}^{\nu\rho}b_{\perp\rho} + \bn^\nu\eps_{\perp}^{\mu\rho}b_{\perp\rho}\right)
  -\eps_\perp^{\rho\sigma}b_{\perp\rho}S_{\perp\sigma}\left(\bn^\mu b_\perp^\nu+\bn^\nu b_\perp^\mu\right) \bigr]
  \\ \nn&\hspace{1.2cm}
  +\Ma\Mb\, h_1^{(1)}H_1^{\perp(1)} \bigl[b_\perp {\cdot}  S_\perp\left(\bn^\mu\eps_{\perp}^{\nu\rho}b_{\perp\rho} + \bn^\nu\eps_{\perp}^{\mu\rho}b_{\perp\rho}\right)
  +\eps_\perp^{\rho\sigma}b_{\perp\rho}S_{\perp\sigma}\left(\bn^\mu b_\perp^\nu+\bn^\nu b_\perp^\mu\right) 
  \bigr]
  \biggr\}
,\\ \label{eq:WP_trace_b}
 &\Tr\left[ B_{f/N}(x,\bt)\, \gamma^\mu \, \cG_{\cP\, h/f}(z,\bt)\, \gamma^\nu\right]
 \\ \nn&
 = \frac{\Mb}{2Q} \biggl\{
   \left(\img f_1  - \Ma f_{1 T}^{\perp(1)}\, \eps_\perp^{\rho \sigma} b_{\perp \rho} S_{\perp \sigma}\right)
    D_1^{(1)} \Mb \left(b_\perp^\mu n^\nu+ n^\mu b_\perp^\nu\right)
  \\ \nn&\hspace{1.2cm}
  - \left(S_L\,g_{1L} +\img\Ma\, b_\perp \cdot S_\perp\,g_{1T}^{(1)}\right)D_1^{(1)}\, \Mb n_\rho \eps^{\mu\nu\rho\sigma}  b_{\perp\sigma}
  \\ \nn&\hspace{1.2cm}
  +\img \Ma\, h_1^{\perp (1)} H_{1}^{\perp(0')} \left(b_\perp^\mu n^\nu+ n^\mu b_\perp^\nu\right)
  + \img\Ma\,S_L h_{1L}^{\perp(1)} H_{1}^{\perp(0')} \left(n^\mu\eps_{\perp}^{\nu\rho}b_{\perp\rho}+n^\nu\eps_{\perp}^{\mu\rho}b_{\perp\rho}\right)
  \\ \nn&\hspace{1.2cm}
  +\frac{\Ma^2}{2}\,h_{1T}^{\perp(2)} H_{1}^{\perp(0')} \biggl[
   \frac12 b_\perp^2 \bigl(n^\mu \eps_\perp^{\nu\rho} S_{\perp\rho}  + n^\nu \eps_\perp^{\mu\rho} S_{\perp\rho} \bigr)
  - (b_\perp {\cdot} S_\perp) \left(n^\mu\eps_\perp^{\nu\rho}b_{\perp\rho}+n^\nu\eps_\perp^{\mu\rho}b_{\perp\rho}\right) \biggr]
  \\ \nn&\hspace{1.2cm}
  - h_1 H_{1}^{\perp(0')} \left(n^\mu\eps_{\perp}^{\nu\rho}S_{\perp\rho}+n^\nu\eps_{\perp}^{\mu\rho}S_{\perp\rho}\right)
  \biggr\}
\,.\end{align}

\subsubsection{Contributions from the Collinear $\cB_{n_i\perp}$ Operators}
\label{sec:results_NLP_B}

The operators containing an insertion of a collinear gluon field give rise to the quark-gluon-quark correlators
in \eq{def_Bb_hat}. A key feature of this correlators is that the quark and gluon fields at the same transverse position
can freely exchange longitudinal momentum, and the correlator is sensitive to two momentum fractions.
This is distinct from similar quark-gluon-quark correlators studied already a long time ago in \refscite{Boer:2003cm, Bacchetta:2006tn},
which only depend on the total momentum of the quark-gluon system.%
\footnote{Correlators with a distinct dependence on both momenta were found independently
in the recent work of \refcite{Vladimirov:2021hdn}.}
Despite this difference, it will still be useful to employ the same Lorentz index decomposition of the quark-gluon-quark correlator
as in \refcite{Bacchetta:2006tn}, up to including this additional momentum dependence.%
\footnote{In \app{qgq_relation}, we provide more details on how to relate our $\hat B_{\cB}^{\rho}$
and $\hat \cG_{\cB}^{\rho}$ to their $\tilde \Phi_A^\rho$ and $\tilde \Delta_A^\rho$, respectively.
The key results are given in \eqs{Bb_hat_integrated_2}{BBfbar_hat_integrated_2},
which relate our correlators \emph{integrated over $\xi$ and before the soft subtraction}
to their purely hadronic correlators.}

We define our correlators in momentum space as
\begin{align} \label{eq:BB_decomposition}
 \tilde B_{\cB\,f/N}^{\rho}(x, \xi, \kt) 
 &
 = \frac{\Ma}{4 \Pa^-}  \biggl\{
  \biggl[
  \munderbarf\tilde{f}^{\perp}
          \frac{k_{\perp \sigma}}{\Ma}
  -\munderbarf\tilde{f}_T
       \,\eps_{\perp \sigma\delta} S_\perp^{\delta}
  -S_L\,\munderbarf\tilde{f}_L^{\perp}
       \frac{\eps_{\perp \sigma \delta}\, k_{\perp}^{\delta}}{\Ma} 
  \nn\\*
  &\hspace{1.8cm}
   -\frac{\munderbarf\tilde{f}_T^\perp}{\Ma^2} \left({\frac{k_\perp^2}{2}\eps_{\perp\sigma\delta}S_\perp^\delta}-k_\perp\cdot S_\perp\eps_{\perp\sigma\delta}k_\perp^\delta\right)
   \biggr]
  \bigl(g_\perp^{\rho \sigma} - \img \eps_\perp^{\rho\sigma} \gamma_5\bigr)
  \nn\\*
  &\hspace{1.8cm}
 - \Bigl[ S_L\, \munderbarh\tilde{h}_L-\frac{k_\perp\cdot S_\perp}{\Ma}\munderbarh\tilde{h}_T \Bigr]
         \gamma_\perp^{\rho}\,\gamma_5
         +\Bigl[\munderbarh\tilde{h} 
   +\munderbarh\tilde{h}_T^{\perp}\,
    \frac{\eps_\perp^{\sigma \delta} k_{\perp\sigma}\, S_{\perp\delta}}{\Ma}
  \,\Bigr]
   \img \gamma_\perp^{\rho}
   \nn\\*
  &\hspace{1.8cm}
 + \ldots \bigl(g_\perp^{\rho \sigma}
                + \img \eps_\perp^{\rho\sigma} \gamma_5\bigr)
 \biggr\}  \frac{\slashed{n}}{2}
 \nn\\
 &
 = \frac{\Ma}{4 \Pa^-}  \biggl\{
  \biggl[
  \bigl(\tilde{f}^{\perp}-\img \tilde{g}^{\perp} \bigr)
          \frac{k_{\perp \sigma}}{\Ma}
  -\bigl(\tilde{f}_T+ \img \tilde{g}_T\bigr)
       \,\eps_{\perp \sigma\delta} S_\perp^{\delta}
  -S_L\bigl(\tilde{f}_L^{\perp}+\img\,\tilde{g}_L^{\perp}\bigr)
       \frac{\eps_{\perp \sigma \delta}\, k_{\perp}^{\delta}}{\Ma}
  \nn\\*
  &\hspace{1.8cm}
  -\frac{\tilde f_T^\perp+\img \tilde g_T^\perp}{\Ma^2} \left({\frac{k_\perp^2}{2}\eps_{\perp\sigma\delta}S_\perp^\delta}-k_\perp\cdot S_\perp\eps_{\perp\sigma\delta}k_\perp^\delta\right)
   \biggr]
  \bigl(g_\perp^{\rho \sigma} - \img \eps_\perp^{\rho\sigma} \gamma_5\bigr)
  \nn\\*
  &\hspace{1.8cm}
 - \Bigl[ S_L \bigl(\tilde{h}_L + \img\,\tilde{e}_L\bigr)-\frac{k_\perp\cdot S_\perp}{\Ma}\bigl(\tilde{h}_T + \img\,\tilde{e}_T\bigr)\Bigr]
         \gamma_\perp^{\rho}\,\gamma_5
   \nn\\*
   &\hspace{1.8cm}
         +\Bigl[\bigl(\tilde{h} + \img\,\tilde{e}\bigr)
   +\bigl( \tilde{h}_T^{\perp}
         - \img\,\tilde{e}_T^{\perp}\bigr)\,
    \frac{\eps_\perp^{\sigma \delta} k_{\perp\sigma}\, S_{\perp\delta}}{\Ma}
  \,\Bigr]
   \img \gamma_\perp^{\rho}
   \nn\\*
  &\hspace{1.8cm}
 + \ldots \bigl(g_\perp^{\rho \sigma}
                + \img \eps_\perp^{\rho\sigma} \gamma_5\bigr)
 \biggr\}  \frac{\slashed{n}}{2}
\,, \\
 \label{eq:GB}
 \tilde \cG_{\cB\, h/f}^{\rho}(z, \xi, \pt) &
 = \frac{M_h}{4 \Pb^+}\,
  \biggl\{
  \munderbarD\tilde{D}^\perp\,
     \frac{p_{\perp \sigma}}{M_h}\,
     \bigl(g_\perp^{\rho \sigma} - \img \eps_\perp^{\rho\sigma} \gamma_5\bigr)
  + \munderbarD\tilde{H} \, \img \gamma_\perp^{\rho}
  + \ldots
    \bigl(g_\perp^{\rho \sigma} + \img \eps_\perp^{\rho\sigma} \gamma_5\bigr)
  \biggr\} \frac{\slashed \bn}{2}
  \nn\\
  & = 
  \frac{M_h}{4 \Pb^+}\,
  \biggl\{
  \bigl(\tilde{D}^\perp+\img\, \tilde{G}^{\perp} \bigr)\,
     \frac{p_{\perp \sigma}}{M_h}\,
     \bigl(g_\perp^{\rho \sigma} - \img \eps_\perp^{\rho\sigma} \gamma_5\bigr)
   \nn\\&\hspace{1.8cm}
  + \bigl(\tilde{H} - \img\,\tilde{E}\bigr)\, \img \gamma_\perp^{\rho}
  + \ldots
    \bigl(g_\perp^{\rho \sigma} + \img \eps_\perp^{\rho\sigma} \gamma_5\bigr)
  \biggr\} \frac{\slashed \bn}{2}
  \,.\end{align}
We have defined the complex function $\munderbarf\tilde{f}^\perp\equiv \tilde{f}^\perp-\img\tilde{g}^\perp$, and similarly for other cases  as indicated above.
For brevity, we suppress the arguments on the right-hand side,
i.e.~we have the arguments 
$\tilde f^\perp \equiv \tilde f^\perp_{f/N}(x, \xi, k_T,\mu,\zeta_a)$ and 
$\tilde D^\perp\equiv \tilde D^\perp_{h/f}(z,\xi,p_T,\mu,\zeta_b)$,
and likewise for all other TMDs. We have also suppressed the dependence on the renormalization scale $\mu$ and Collins-Soper parameters $\zeta_{a,b}$ on the left hand side.
The terms in ellipses in \eqs{BB_decomposition}{GB}
do not contribute to the SIDIS structure functions~\cite{Bacchetta:2006tn}.%
\footnote{Compared to the notation for the Lorentz decomposition in~\cite{Bacchetta:2006tn}, we have already made use in our \eq{BB_decomposition} of the relations
\begin{align}
 \tilde f_T = \tilde f'_T - \frac{k_\perp^2}{2\Ma^2} \tilde f_T^\perp
\,,\quad
 \tilde g_T = \tilde g'_T - \frac{k_\perp^2}{2\Ma^2} \tilde g_T^\perp
\,,\end{align}
and hence do not define $\tilde f'_T$ or $\tilde g'_T$ here.
}

We stress that while the structure of \eqs{BB_decomposition}{GB}
is consistent with that obtained by applying \eqs{Bb_hat_integrated_2}{BBfbar_hat_integrated_2}
to decomposition in \refcite{Bacchetta:2006tn}, a key difference is that here all TMDs here are defined after absorbing the LP soft function as in \eq{def_BB},
and with an extra longitudinal momentum fraction $\xi$ dependence that was not present in \refcite{Bacchetta:2006tn}.

For the antiquark contribution, analogous to \eq{relation_q_barq}, our quark-gluon-quark correlator satisfies the following relations,
\begin{align} \label{eq:relation_qgq_q_barq}
  \Tr\bigl[\sigma_{\rho+}\, \tilde B_{\cB\, \bar f/N}^{\rho} (x,\xi, \kt)\bigr] &
  = \Tr\bigl[\sigma_{\rho+}\, \tilde B_{\cB\, f/N}^{\rho\, c} (x,\xi, \kt)\bigr]
 \,,\nn\\
 \Tr\bigl[\img\sigma_{\rho+}\gamma_5\, \tilde B_{\cB\, \bar f/N}^{\rho} (x,\xi, \kt)\bigr] &
 = \Tr\bigl[\img\sigma_{\rho+}\gamma_5\, \tilde B_{\cB\, f/N}^{\rho\, c} (x,\xi, \kt)\bigr]
 \,,\nn\\
 \Tr\bigl[\bigl(g_\perp^{\sigma\rho} + \img \eps_\perp^{\sigma\rho} \gamma_5\bigr) \gamma^+\, \tilde B_{\cB\, \bar f/N,\,\rho} (x,\xi, \kt)\bigr] &
 = \Tr\bigl[\bigl(g_\perp^{\sigma\rho} - \img \eps_\perp^{\sigma\rho} \gamma_5\bigr) \gamma^+\, \tilde B_{\cB\, f/N,\,\rho}^{c} (x,\xi, \kt)\bigr]
 \,.\end{align}
 Here, $\tilde B_{\cB\, f/N}^{\rho\, c}$ is $\tilde B_{\cB\, f/N}^{\rho}$ evaluated with charge-conjugated fields $\chi^c = C \bar \chi^T$ and $\cB_{n\perp\, \rho}^c= -\cB_{n\perp\, \rho}^T $.
Notice that $\tilde B_{\cB\, f/N}^{\rho\, c}$ then (roughly) corresponds to the hermitian conjugate of $\tilde B_{\cB\, f/N}^{\rho}$, namely, flipping sign of every $\img$ in \eq{BB_decomposition} except the one in $\img\gamma_\perp^\rho$. Therefore, the decomposition of $\tilde B_{\cB\, \bar f/N}^{\rho}$ is identical to \eq{BB_decomposition}, except that $\tilde g^\perp$, $\tilde g_T$, $\tilde g^\perp_L$, $\tilde g^\perp_T$, $\tilde e_L$, $\tilde e_T$, $\tilde e$ and $\tilde e^\perp_T$ would receive a sign flip. Similarly, $\tilde \cG_{\cB\, h/\bar f}^{\rho}(z, \xi, \pt)$ has the same decomposition as \eq{GB} except that $\tilde G$ and $\tilde E$ have a sign flip. 

The Fourier transforms of \eqs{BB_decomposition}{GB} can be obtained
in the usual fashion using the results in \app{fourier}:
\begin{subequations}  \label{eq:qgqdecomp}
\begin{align}
 \tilde B_{\cB\,f/N}^\rho(x,\xi,\bt)
 &
   = \frac{\Ma}{4 \Pa^-} \biggl\{
       \biggl[ -\img\Ma \munderbarf\tilde{f}^{\perp(1)}\, b_{\perp \sigma}
               - \munderbarf\tilde{f}_T \,\eps_{\perp \sigma\delta} S_\perp^{\delta}
               +\img\Ma S_L \, \munderbarf\tilde{f}_L^{\perp(1)}\, \eps_{\perp \sigma \delta} b_{\perp}^{\delta}
               \nn\\&\qquad
               + \frac12 \Ma^2 \munderbarf\tilde{f}_T^{\perp(2)}\, \eps_{\perp\sigma\delta}
                 \Bigl(\frac12 b_\perp^2 S_\perp^\delta -b_\perp\cdot S_\perp b_\perp^\delta \Bigr)
       \biggr] \bigl(g_\perp^{\rho \sigma} - \img \eps_\perp^{\rho\sigma} \gamma_5\bigr)
      \nn\\&\qquad
      -\left[S_L\, \munderbarh\tilde{h}_L +\img\Ma \,b_\perp\cdot S_\perp\, \munderbarh\tilde{h}_T^{(1)}  \right]
              \gamma_\perp^{\rho}\,\gamma_5
      +\Bigl[\munderbarh\tilde{h} 
        -\img\Ma\, \munderbarh\tilde{h}_T^{\perp(1)}\,
         \eps_\perp^{\sigma \delta} b_{\perp\sigma}\, S_{\perp\delta}
       \,\Bigr]
        \img \gamma_\perp^{\rho}
      \nn\\&\hspace{1.8cm}
      + \ldots \bigl(g_\perp^{\rho \sigma}
                     + \img \eps_\perp^{\rho\sigma} \gamma_5\bigr)
      \biggr\} \frac{\slashed n}{2}
      \nn\\ &
 = \frac{\Ma}{4 \Pa^-} \biggl\{
    \biggl[ -\img\Ma\bigl(\tilde{f}^{\perp(1)}-\img \tilde{g}^{\perp(1)}\bigr) b_{\perp \sigma}
            - \bigl(\tilde{f}_T+ \img \tilde{g}_T \bigr) \eps_{\perp \sigma\delta} S_\perp^{\delta}
            \nn\\&\hspace{2cm}
            +\img\Ma S_L\bigl(\tilde{f}_L^{\perp(1)}+\img\,\tilde{g}_L^{\perp(1)}\bigr) \eps_{\perp \sigma \delta} b_{\perp}^{\delta}
            \nn\\&\hspace{2cm}
            + \frac12 \Ma^2\bigl(\tilde{f}_T^{\perp(2)}+\img\,\tilde{g}_T^{\perp(2)}\bigr) \eps_{\perp\sigma\delta}
              \Bigl(\frac12 b_\perp^2 S_\perp^\delta -b_\perp\cdot S_\perp b_\perp^\delta \Bigr)
    \biggr] \bigl(g_\perp^{\rho \sigma} - \img \eps_\perp^{\rho\sigma} \gamma_5\bigr)
   \nn\\&\hspace{1.8cm}
   -\left[S_L\bigl(\tilde{h}_L + \img\,\tilde{e}_L\bigr)+\img\Ma \,b_\perp\cdot S_\perp \bigl(\tilde{h}_T^{(1)} + \img\,\tilde{e}_T^{(1)}\bigr) \right]
           \gamma_\perp^{\rho}\,\gamma_5
   \nn\\&\hspace{1.8cm}
   +\Bigl[\bigl(\tilde{h} + \img\,\tilde{e}\bigr)
     -\img\Ma\bigl( \tilde{h}_T^{\perp(1)}
           - \img\,\tilde{e}_T^{\perp(1)}\bigr)\,
      \eps_\perp^{\sigma \delta} b_{\perp\sigma}\, S_{\perp\delta}
    \,\Bigr]
     \img \gamma_\perp^{\rho}
   \nn\\&\hspace{1.8cm}
   + \ldots \bigl(g_\perp^{\rho \sigma}
                  + \img \eps_\perp^{\rho\sigma} \gamma_5\bigr)
   \biggr\} \frac{\slashed n}{2}
\,,\\ \label{eq:deltaA}
 \tilde \cG_{\cB\, h/f}^{\rho}(z, \xi, \bt) 
 &
 = \frac{M_h}{4 \Pb^+} \Bigl\{ 
       \img\Mb\, \munderbarD\tilde{D}^{\perp(1)} 
       b_{\perp \sigma} \bigl(g_\perp^{\rho \sigma} - \img \eps_\perp^{\rho\sigma} \gamma_5\bigr)
      + \munderbarD\tilde{H} \, \img \gamma_\perp^{\rho}
      + \ldots
        \bigl(g_\perp^{\rho \sigma} + \img \eps_\perp^{\alpha\sigma} \gamma_5\bigr)
      \Bigr\} \frac{\slashed \bn}{2}
    \nn\\
      &
 = \frac{M_h}{4 \Pb^+} \Bigl\{
   \img\Mb\bigl(\tilde{D}^{\perp(1)} + \img\, \tilde{G}^{\perp(1)} \bigr)
   b_{\perp \sigma} \bigl(g_\perp^{\rho \sigma} - \img \eps_\perp^{\rho\sigma} \gamma_5\bigr)
 \nn\\&\hspace{1.5cm}
  + \bigl(\tilde{H} - \img\,\tilde{E}\bigr)\, \img \gamma_\perp^{\rho}
  + \ldots
    \bigl(g_\perp^{\rho \sigma} + \img \eps_\perp^{\alpha\sigma} \gamma_5\bigr)
  \Bigr\} \frac{\slashed \bn}{2}
\,.\end{align}
\end{subequations}
Here, the $f^{(n)}$ are defined in \eq{def_fn} as usual.

To obtain the contributions from these correlators to the NLP hadronic tensor
as given by \eq{WB}, we require the traces
\begin{align} \label{eq:WB_trace}
  & \Tr\left[ \tilde B_{\cB \, f/N}^\rho \gamma_{\perp}^\mu \cG_{h/f} \gamma_{\perp\rho}
    +  B_{f/N} \gamma_\perp^\mu \tilde \cG_{\cB\,h/f}^\rho \gamma_{\perp\rho} \right]
 \nn\\&
 = - \frac{\Ma}{2 \Pa^-} \, \bigg\{
    - \img\Ma \left(\tilde{f}^{\perp(1)}-\img\,\tilde{g}^{\perp(1)}\right)D_1\, b_\perp^\mu
    - \bigl(\tilde{f}_T+ \img\, \tilde{g}_T \bigr) D_1 \eps_\perp^{\mu\sigma}S_{\perp\sigma}
    \nn\\&\hspace{2cm}
    + \img\Ma S_L\bigl(\tilde{f}_L^{\perp(1)}+\img\,\tilde{g}_L^{\perp(1)}\bigr) D_1\, \eps_\perp^{\mu\sigma}\,b_{\perp\sigma}
    \nn\\&\hspace{2cm}
    + \frac{\Ma^2}{2}\, \bigl(\tilde{f}_T^{\perp(2)}+\img\,\tilde{g}_T^{\perp(2)}\bigr) D_1 \eps_\perp^{\mu\sigma}
       \Bigl[ \frac12 b_\perp^2 S_{\perp\sigma} -(b_\perp\cdot S_\perp)\,b_{\perp\sigma} \Bigr]
    \nn\\&\hspace{2cm}
    + \img\Mb\left[S_L\bigl(\tilde{h}_L+\img\,\tilde{e}_L\bigr)+ \img\Ma\, b_\perp\cdot S_\perp \bigl(\tilde{h}_T^{(1)}+\img\,\tilde{e}_T^{(1)}\bigr)\right]
      H_1^{\perp(1)}\eps_\perp^{\mu\sigma}b_{\perp\sigma}
    \nn\\&\hspace{2cm}
   + \img\Mb \left[ \left(\tilde{h} + \img\,\tilde{e}\right) -\img\Ma\left( \tilde{h}_T^{\perp(1)}
                    - \img\,\tilde{e}_T^{\perp(1)}\right)\, \eps_\perp^{\sigma \delta} b_{\perp\sigma}\, S_{\perp\delta}
             \right]H_1^{\perp(1)} b_{\perp}^\mu
  \bigg\}
  \nn\\&
  - \frac{\Mb}{2 \Pb^+}\, \bigg\{ \img\Mb \left(f_1+\img\Ma f_{1T}^{\perp(1)} \eps_\perp^{\sigma\delta}b_{\perp\sigma}S_{\perp\delta}\right) \left(\tilde{D}^{\perp(1)}+\img\, \tilde{G}^{\perp(1)}\right)b_\perp^\mu
    \nn\\&\hspace{2cm}
  +\img\Mb\left(S_L g_{1L}+ \img\Ma\, b_\perp\cdot S_\perp\, g_{1T}^{(1)}\right)\left(\tilde{G}^{\perp(1)}-\img\,\tilde{D}^{\perp(1)}\right)\eps_\perp^{\mu\sigma}\,b_{\perp\sigma}
    \nn\\&\hspace{2cm}
  + \Bigl[ - \img \Ma\, h_1^{\perp(1)} b_\perp^\mu
           - \img \Ma S_L h_{1L}^{\perp(1)} \eps_\perp^{\mu\sigma}\,b_{\perp\sigma}
           + h_1\eps_\perp^{\mu\sigma}S_{\perp\sigma} \Bigr] \bigl(\tilde{H} - \img\,\tilde{E}\bigr)
    \nn\\&\hspace{2cm}
  -\frac{\Ma^2}{2} h_{1T}^{\perp(2)}\bigl(\tilde{H} - \img\,\tilde{E}\bigr)\,
  \eps_\perp^{\mu\sigma}
  \Bigl[ \frac12 b_\perp^2 S_{\perp\sigma} -(b_\perp{\cdot}S_\perp)\,b_{\perp\sigma} \Bigr]
  \biggr\}
\,.\end{align}
This can be inserted into \eq{WB} to obtain the NLP contributions
to the hadronic tensor arising from an additional gluon field $\cB_{n_i\perp}$.

\subsubsection{Combined Results in Fourier Space}
\label{sec:results_NLP_full}

In the previous sections, we provided all the ingredients required to calculate
the individual contributions to the subleading structure functions from
\begin{itemize}
 \item \textbf{\colkin{Kinematic corrections}} in \sec{results_NLP_kin}:
      contract the leading-power hadronic tensor in \eq{hadronicts_LP}
      with the subleading projectors in \eq{P_pcs} to obtain $P^{(1)}_{i\,\mu\nu} W^{(0)\,\mu\nu}$.
      The contractions to take into account are given in \eq{subleading_P_traces}.
 \item \textbf{\colp{$\cP_\perp$ operators}}  in \sec{results_NLP_P}:
      insert \eqs{WP_trace_a}{WP_trace_b} into \eq{WP} to obtain $W_\cP^{(1)\,\mu\nu}$,
      and contract with the LP projectors to obtain $P^{(0)}_{i\,\mu\nu} W_\cP^{(1)\,\mu\nu}$
 \item\textbf{\colbn{$\cB_{n_i\perp}$ operators}}  in \sec{results_NLP_B}:
      insert \eq{WB_trace} into \eq{WB} to  obtain $W_\cB^{(1)\,\mu\nu}$,
      and contract with the LP projectors to obtain $P^{(0)}_{i\,\mu\nu} W_\cB^{(1)\,\mu\nu}$
\,.\end{itemize}
By adding these contributions, we obtain the subleading structure functions as
\begin{align}
 W_i = \colkin{P^{(1)}_{i\,\mu\nu} W^{(0)\,\mu\nu}}
     + \colp{P^{(0)}_{i\,\mu\nu} W_\cP^{(1)\,\mu\nu}}
     + \colbn{P^{(0)}_{i\,\mu\nu} W_\cB^{(1)\,\mu\nu}}
\,.\end{align}
By separating $W_i$ according to \eq{def_W_i_pol} into the different spin structures,
we obtain the individual $W_i^U$, $W_i^L$, $W_i^{Tx}$ and $W_i^{Ty}$.
Combining these following \eq{def_W_polarized} then yields the polarized structure functions at NLP.
In the following results, we will always use the above color scheme
to indicate the different sources of power corrections,
except when we combine the individual pieces to obtain more compact final results.

Interestingly, $W_{LU}^{\sin\phi_h}$ is the only structure function that does not receive kinematic corrections
and contributions from $\cP_\perp$ operators, and hence only has $\cB_{n_i\perp}$ contributions. All the other structure functions receive contributions from all three sources of power corrections.

At leading power, we expressed all structure functions in terms of the Fourier transform in \eq{cF}.
While we still encounter the same Fourier transform for (some of) the kinematic and $\cP_\perp$ corrections,
the $\cB_{n_i\perp}$ operators give rise an additional convolution in $\xi$.
Thus, we have to slightly adapt $\cF$ at NLP, and define three versions of it, which varies depending on which functions it acts:
\begin{align} \label{eq:cF_NLP}
 \cF[\cH\, g^{(n)} D^{(m)}] &
 = 2 z \sum_{f} \cH_f(q^+ q^-)
   \int_0^\infty \frac{\df b_T \, b_T}{2\pi} (\Ma b_T)^n (-\Mb b_T)^m J_{n+m}(b_T q_T) \,
   \nn\\&\hspace{3.5cm}\times
   \, g_f^{(n)}(x,b_T) \, D_f^{(m)}(z,b_T) + (f\to\bar f)
\,,\nn\\
 \cF[\cH\, \tilde g^{(n)} D^{(m)}] &
 = 2 z \sum_{f} \int\!\df \xi \, \cH_f(q^+ q^-, \xi)
   \int_0^\infty \frac{\df b_T \, b_T}{2\pi} (\Ma b_T)^n (-\Mb b_T)^m J_{n+m}(b_T q_T) \,
   \nn\\&\hspace{3.5cm}\times
   \, \tilde g_f^{(n)}(x, \xi, b_T) \, D_f^{(m)}(z,b_T) + (f\to\bar f)
\,,\nn\\
 \cF[\cH\, g^{(n)} \tilde D^{(m)}] &
 = 2 z \sum_{f} \int\!\df \xi \, \cH_f(q^+ q^-, \xi)
   \int_0^\infty \frac{\df b_T \, b_T}{2\pi} (\Ma b_T)^n (-\Mb b_T)^m J_{n+m}(b_T q_T) \,
   \nn\\&\hspace{3.5cm}\times
   \, g_{f}^{(n)}(x, b_T) \, \tilde D_{f}^{(m)}(z, \xi, b_T) + (f\to\bar f)
\,.\end{align}
Here, the $f^{(n)}$ are as usual defined as the Fourier transforms in \eq{FT_derivative}.
The definition in the first line of \eq{cF_NLP} is identical to \eq{cF}.
In the second (third) line, there is a convolution over $\xi$ between the hard function $\cH$
and the quark-gluon-quark TMDPDF $\tilde g$ (quark-gluon-quark TMDFF $\tilde D$), as indicated by the explicit arguments for these TMDs.
Since the distinction is unique from the arguments, we use the same symbol $\cF$ in all cases.
The functions $f^{(0')}$ and $f^{(1')}$ defined in \eq{0p_1p} are treated as $n=0$ and $n=1$, respectively.  
Note that on the RHS of \eq{cF_NLP} we have once again suppressed important dependence on the renormalization scale $\mu$, and Collins-Soper scales $\zeta_{a,b}$, which can be restored by writing $\cH_f(q^+q^-,\mu)$, $\cH_f(q^+q^-,\xi,\mu)$, $g_f^{(n)}(x,b_T,\mu,\zeta_a)$,
$\tilde g_f^{(n)}(x,\xi,b_T,\mu,\zeta_a)$, 
$D_f^{(m)}(z,b_T,\mu,\zeta_b)$, and $\tilde D_f^{(m)}(z,\xi,b_T,\mu,\zeta_b)$.

At LP, the above Fourier transform was sufficient, as all powers of $b_T$
were compensated by a corresponding factor of $\cos\varphi$, where $\cos\varphi = -\bt \cdot \qt / (b_T q_T)$, 
leading to the particularly simple form of \eq{cF_NLP} as an integral over $J_{n+m}$.
At NLP, this is not true anymore, and we encounter powers of $b_T$
without the corresponding $\cos\varphi$. To handle these, we also define the modified Fourier transform
\begin{align} \label{eq:cFp}
  \cF'[\cH\, g^{(n)} \, D^{(m)}] &
  = 2 z \sum_f \cH_f(q^+ q^-) \int\frac{\df^2\bt}{(2\pi)^2} e^{-\img \qt \cdot \bt}
    (-\img\Ma b_T)^n (\img\Mb b_T)^m
    \nn\\&\hspace{3.5cm}\times
    \, g_f^{(n)}(x,b_T) \, D_f^{(m)}(z,b_T) + (f\to\bar f)
  \nn\\&
  = 2 z \sum_{f} \cH_f(q^+ q^-)
    \int_0^\infty \frac{\df b_T \, b_T}{2\pi} (-\img\Ma b_T)^n (\img\Mb b_T)^m J_{0}(b_T q_T) \,
    \nn\\&\hspace{3.5cm}\times
    \, g_f^{(n)}(x,b_T) \, D_f^{(m)}(z,b_T) + (f\to\bar f)
\,, \nn\\
     \cF'[\cH\, \tilde g^{(n)} D^{(m)}] &
     = 2 z \sum_{f} \int\!\df \xi \, \cH_f(q^+ q^-, \xi)
       \int_0^\infty \frac{\df b_T \, b_T}{2\pi} (-\img\Ma b_T)^n (\img\Mb b_T)^m J_0(b_T q_T) \,
       \nn\\&\hspace{3.5cm}\times
       \, \tilde g_f^{(n)}(x, \xi, b_T) \, D_f^{(m)}(z,b_T) + (f\to\bar f)
    \,,\nn\\
     \cF'[\cH\, g^{(n)} \tilde D^{(m)}] &
     = 2 z \sum_{f} \int\!\df \xi \, \cH_f(q^+ q^-, \xi)
       \int_0^\infty \frac{\df b_T \, b_T}{2\pi} (-\img\Ma b_T)^n (\img\Mb b_T)^m J_0(b_T q_T) \,
       \nn\\&\hspace{3.5cm}\times
       \, g_{f}^{(n)}(x, b_T) \, \tilde D_{f}^{(m)}(z, \xi, b_T) + (f\to\bar f)
 \,.
\end{align}
For $n=m=0$, $\cF'$ and $\cF$ coincide.

In the following we provide all structure functions that \emph{start} at NLP,
since as previously mentioned we do not consider NLP corrections to structure functions already contributing at LP.
Here, we only provide expressions in position space.
Explicit results in momentum space will be presented in \sec{results_NLP_full_msp}.

Note that the structure functions $W_{LU}^{\sin\phi_h}$, $ W_{LL}^{\cos\phi_h}$, $W_{LT}^{\cos\phi_S}$ and $W_{LT}^{\cos\left(2\phi_h -\phi_S\right)}$
arise from the antisymmetric projector $P_2^{\mu\nu}$ and $P_6^{\mu\nu}$, where exchanging $\mu\leftrightarrow\nu$
for the antiquark contribution in $W^{\mu\nu}$ yields a relative minus sign. Since the corresponding
antiquark distributions $g_{1L}$, $g_{1T}$, and all the functions mentioned below \eq{relation_qgq_q_barq} also receive a minus sign of their own,
simply summing over all quarks and antiquarks as stated in \eqs{cF_NLP}{cFp} is correct.
Below we use the notation $\Re$ and $\Im$ for the real and imaginary parts respectively.

\paragraph{Unpolarized structure functions.}
\begin{align}
 W_{UU}^{\cos\phi_h} &
 = \cF\biggl\{ \colkin{\frac{q_{T}}{Q}\,\cH^{(0)}\left[-f_1D_1+ h_1^{\perp(1)} H_1^{\perp(1)}\right]}
  \nn\\
  &\qquad \colp{
  +\cH^{(0)}\left[-\frac{\Ma}{Q}f_1^{(1)} D_1-\frac{\Mb}{Q}f_1 D_1^{(1)}+\frac{\Ma}{Q} h_1^{\perp (0')} H_1^{\perp (1)} +  \frac{\Mb}{Q} h_1^{\perp (1)} H_{1}^{\perp(0')}\right]}\biggl\}
  \nn\\
  &\quad - \Re\biggl[ \colbn{
  \cH^{(1)}\left[\frac{2x\Ma}{Q}\left(\munderbarf{\tilde{f}}^{\perp(1)} D_1+\munderbarh{\tilde{h}}\,H_1^{\perp(1)}\right)+\frac{2\Mb}{zQ}\left(f_1 \munderbarD{\tilde{D}}^{\perp(1)}+ h_1^{\perp(1)} \munderbarD{\tilde{H}}\right)\right]}\biggl]\biggr\}
\,,\\
 W_{LU}^{\sin\phi_h} &
 = \cF\biggl\{ -\Im\biggl[ 
    \colbn{\cH^{(1)}\left[\frac{2x\Ma}{Q}\left(\munderbarf\tilde{f}^{\perp(1)} D_1+ \munderbarh{\tilde{h}}\,H_1^{\perp(1)}\right)+ \frac{2\Mb}{zQ}\left( f_1\munderbarD{\tilde{D}}^{\perp(1)}+ h_1^{\perp(1)} \munderbarD{\tilde{H}}\right)\right]}\biggr]\biggr\}
\,.\end{align}

\paragraph{Longitudinally polarized structure functions.}
\begin{align}
 W_{UL}^{\sin\phi_h} &
 = \cF\biggl\{ \colkin{ \frac{q_{T}}{Q}\,\cH^{(0)}\,  h_{1L}^{\perp(1)} H_1^{\perp(1)}}
   \colp{ +\cH^{(0)} \left(\frac{\Ma}{Q} h_{1L}^{\perp (0')} H_1^{\perp (1)}+ \frac{\Mb}{Q} h_{1L}^{\perp (1)} H_{1}^{\perp(0')}\right)}
 \nn\\&\qquad
   + \Re\biggl[\colbn{\cH^{(1)}\left[\frac{2x\Ma}{Q}\left(\munderbarf\tilde{f}^{\perp(1)}_{L} D_1- \munderbarh\tilde{h}_{L}H_1^{\perp(1)}  \right)-\frac{2\Mb}{zQ}\left( -\img g_{1L} \munderbarD\tilde{D}^{\perp(1)}+ h_{1L}^{\perp(1)} \munderbarD\tilde{H}\right) \right]}\biggr] \biggr\}
\,,\\
 W_{LL}^{\cos\phi_h}
 &= \cF\biggl\{\colkin{-\frac{q_T}{Q}\cH^{(0)}g_{1L}D_1}
  -\colp{\cH^{(0)} \left(\frac{\Ma}{Q}\, g_{1L}^{(1)}D_1 + \frac{\Mb}{Q}\, g_{1L}D_1^{(1)}\right)}
  \nn\\ &\qquad - \Im\biggl[ \colbn{
   \cH^{(1)}\left[\frac{2x\Ma}{Q}\left(\munderbarf\tilde{f}^{\perp(1)}_{L} D_1-\munderbarh\tilde{h}_L H_1^{\perp(1)}\right)-\frac{2\Mb}{zQ}\left(-\img g_{1L} \munderbarD\tilde{D}^{\perp(1)}+ h_{1L}^{\perp(1)} \munderbarD\tilde{H}\right)\right]} \biggr]\biggr\}
\,.\end{align}

\paragraph{Transversely polarized structure functions.}
\begin{align}
 W_{UT}^{\sin\phi_S} &
 = \cF\biggl\{ \colkin{ -\frac{q_T}{2Q}\,\cH^{(0)}\left( f_{1T}^{\perp(1)} D_1 - 2 h_1 H_1^{\perp(1)}\right)}\biggr\}
  \nn\\&
  +\cF' \biggl\{\colp{
  \cH^{(0)}\left(-\frac{\Ma}{2Q} f_{1T}^{\perp(0')} D_1-\frac{\Mb}{2Q} f_{1T}^{\perp(1)} D_1^{(1)}  +\frac{\Mb}{Q} h_{1} H_1^{\perp(0')}
  +\frac{\Ma}{Q} \,h_{1}^{(1)} H_1^{\perp(1)}\right)}
  \nn\\&\qquad
  +  \Re\biggl[
  \colbn{ \cH^{(1)}\biggl[\frac{2x\Ma}{Q}\munderbarf\tilde{f}_T D_1  -\frac{2\Mb}{zQ}h_1 \munderbarD\tilde{H}
          - \frac{x\Ma}{Q}   \Bigl(\munderbarh\tilde{h}_T^{(1)} -\munderbarh\tilde{h}_T^{\perp(1)}\Bigr) H_1^{\perp(1)} }
  \nn\\&\hspace{2cm}
  \colbn{  - \frac{\Mb}{zQ} \left(-\img g_{1T}^{(1)} \, \munderbarD\tilde{D}^{\perp(1)} + f_{1T}^{\perp(1)} \munderbarD\tilde{D}^{\perp(1)}\right) \biggr] } \biggr]\biggr\}
\,,\\
 W_{UT}^{\sin\left(2\phi_h -\phi_S\right)} &
 = \cF\biggl\{
   \colkin{ \frac{q_T}{2Q}\,\cH^{(0)}\left[f_{1T}^{\perp(1)} D_1 +\frac12\, h_{1T}^{\perp(2)} H_1^{\perp(1)}\right]}
  \nn\\ &\qquad
   \colp{+ \frac{1}{2Q} \cH^{(0)} \biggl[\frac12\Ma f_{1T}^{\perp(2)}D_1+ \Mb f_{1T}^{\perp(1)}D_1^{(1)}
         + \Ma h_{1T}^{\perp(1')}H_1^{\perp(1)}}
  \nn\\&\hspace{2.5cm}
   \colp{+ \frac12 \Mb\, h_{1T}^{\perp(2)}H_1^{\perp(0')} \biggr] }
  \nn\\&\qquad + 
  \Re\biggl[
  \colbn{\cH^{(1)} \biggl[\frac{x\Ma}{Q}\left(\frac12 \munderbarf\tilde{f}_T^{\perp(2)} D_1
                             - \bigl(\munderbarh\tilde{h}_T^{(1)} +\munderbarh\tilde{h}_T^{\perp(1)}\bigr) H_1^{\perp(1)}\right)}
  \nn\\&\hspace{2cm}
  \colbn{                    - \frac{\Mb}{zQ}\left(\frac12 h_{1T}^{\perp(2)}\munderbarD\tilde{H}- \img g_{1T}^{(1)} \munderbarD\tilde{D}^{\perp(1)} -f_{1T}^{\perp(1)} \munderbarD\tilde{D}^{\perp(1)}\right)
                      \biggr]}\biggr]\biggr\}
\,,\\
 W_{LT}^{\cos\phi_S} &
 = \cF\biggl\{ \colkin{ -\frac{q_T}{2Q}\cH^{(0)}\, g_{1T}^{(1)} D_1 }\biggr\}
 \nn\\&
   + \cF' \biggl\{
         \colp{ \cH^{(0)}\left[- \frac{\Ma}{2Q} g_{1T}^{(0')} D_1 -\frac{\Mb}{2Q} g_{1T}^{(1)} D_1^{(1)} \right]}
  \nn\\&\qquad + \Im\biggl[
         \colbn{\cH^{(1)}\biggl[ \frac{x\Ma}{Q}\left(-2\munderbarf\tilde{f}_T D_1+\bigl(\munderbarh\tilde{h}_T^{(1)} -\munderbarh\tilde{h}_T^{\perp(1)}\bigr) H_1^{\perp(1)}\right)}
  \nn\\&\hspace{2cm}
         \colbn{+\frac{\Mb}{zQ}\left(2h_1 \munderbarD\tilde{H}- \img g_{1T}^{(1)}\, \munderbarD\tilde{D}^{\perp(1)} +f_{1T}^{\perp(1)} \munderbarD\tilde{D}^{\perp(1)}\right)\biggr] } \biggr] \biggr\}
\,,\\
 W_{LT}^{\cos\left(2\phi_h -\phi_S\right)} &
 = \cF\biggl\{
   \colkin{ -\frac{q_T}{2Q}\cH^{(0)} g_{1T}^{(1)} D_1}
   ~\colp{-~\cH^{(0)}\left[ \frac{\Ma}{4Q} g_{1T}^{(2)}D_1 + \frac{\Mb}{2Q} g_{1T}^{(1)}D_1^{(1)} \right]}
  \nn\\&\qquad + \Im\biggl[
   \colbn{ \cH^{(1)} \biggl[\frac{x\Ma}{Q} \left(-\frac12 \munderbarf\tilde{f}_T^{\perp(2)} D_1 +\bigl(\munderbarh\tilde{h}_T^{(1)} +\munderbarh\tilde{h}_T^{\perp(1)}\bigr) H_1^{\perp(1)}\right) }
  \nn\\&\hspace{2cm}
   \colbn{+\frac{\Mb}{zQ} \left(\frac12 h_{1T}^{\perp(2)} \munderbarD\tilde{H}- \img g_{1T}^{(1)} \munderbarD\tilde{D}^{\perp(1)} -f_{1T}^{\perp(1)} \munderbarD\tilde{D}^{\perp(1)}\right)\biggr]}\biggr] \biggr\} 
\,.\end{align}

\subsubsection{Combined Results in Momentum Space}
\label{sec:results_NLP_full_msp}

In \sec{results_NLP_full}, we provided the full results in position space,
which leads to particularly compact formulas. Historically, it was more common
to express the factorization as a convolution in momentum space.
In particular, \refcite{Bacchetta:2006tn} only provided results for the subleading
structure functions in SIDIS in momentum space and we will make a comparison to their results below in \sec{tree_level}.

In the following we therefore repeat our results
in momentum space. To so, we only need write the Fourier transforms $\cF$ and $\cF'$
defined in \eqs{cF}{cFp} in terms of explicit momentum space convolutions $\tilde\cF$
as defined in \eq{cF_tilde}. As before, we have to modify the meaning of $\tilde\cF$
to include the integral over $\xi$ for the subleading TMDs, and define
\begin{align}\label{eq:cF_tilde_NLP}
 \tilde\cF[\w\, \cH\, g\, D] &
 = 2z \sum_f \cH_f(q^+q^-) \int\!\df^2\kt\, \df^2\pt \, \delta^{(2)}\bigl(\qt+\kt-\pt\bigr)
  \nn\\ &\qquad\quad
  \times\w(\vec k_T, \vec p_T)\, g_f(x, k_T) \, D_f(z,p_T)+ (f\to\bar f)
\,,\nn\\
 \tilde\cF[\w\, \cH\, \tilde g\, D] &
 = 2z \sum_f \int\!\df\xi \, \cH_f(q^+q^-, \xi) \int\!\df^2\kt\, \df^2\pt \, \delta^{(2)}\bigl(\qt+\kt-\pt\bigr)
  \nn\\ &\qquad\quad
  \times\w(\vec k_T, \vec p_T)\, \tilde g_f(x, \xi, k_T) \, D_f(z,p_T)+ (f\to\bar f)
\,,\nn\\
 \tilde\cF[\w\, \cH\, g \, \tilde D] &
 = 2z \sum_f \int\!\df\xi \, \cH_f(q^+q^-, \xi) \int\!\df^2\kt\, \df^2\pt \, \delta^{(2)}\bigl(\qt+\kt-\pt\bigr)
  \nn\\ &\qquad\quad
  \times\w(\vec k_T, \vec p_T)\, g_f(x, k_T) \, \tilde D_f(z, \xi, p_T)+ (f\to\bar f)
\,,\end{align}
where the $\tilde g$ and $\tilde D$ are the subleading TMDs, respectively.
Explicit results for expressing all appearing Fourier transforms $\cF[ \cH g^{(n)} D^{(m)}]$,
$\cF[ \cH \tilde g^{(n)} D^{(m)}]$ and $\cF[ \cH g^{(n)} \tilde D^{(m)}]$
in terms of the convolutions in \eq{cF_tilde_NLP} are provided in \app{fourier}.
Again on the RHS of \eq{cF_tilde_NLP} we have suppressed dependence of the various functions on the renormalization scale $\mu$ and Collins-Soper parameters $\zeta_{a,b}$, which can be restored in the same manner that we discussed above for the Fourier space results.

In the following, for each NLP structure function we first show the direct
Fourier transform from the results in \sec{results_NLP_full}, where we
still use color coding to separate contributions from \colkin{kinematic corrections},
\colp{$\cP_\perp$ operators}, and \colbn{$\cB_{n_i\perp}$ operators}.
Secondly, we provide simplified results by using that in our frame choice,
$\kt + \qt = \pt$ implies
\begin{align}  \label{eq:Tframechoice}
 q_T = k_{Tx} - p_{Tx}
 \,,\qquad
 0 = k_{Ty} - p_{Ty}
\,.\end{align}
(Recall that $\qt = q_T(-1,0)$.)
Using \eq{Tframechoice} gives us results that are written in a form similar to those in \refcite{Bacchetta:2006tn}.

\paragraph{Unpolarized structure functions.}

\begin{align} \label{eq:W_UU_cosphi_msp}
 W_{UU}^{\cos\phi_h} &
 = \tilde \cF\biggl\{ \colkin{\frac{q_{T}}{Q}\,\cH^{(0)} \biggl[-f_1D_1 + \frac{2 k_{Tx} p_{Tx} - \kt \cdot \pt}{\Ma \Mb}h_1^{\perp} H_1^{\perp} \biggr]}
  \\ \nn
  &\qquad \colp{
  + \cH^{(0)} \biggl[- \frac{k_{Tx} + p_{Tx}}{Q} f_1 D_1
                    + \frac{k_T^2 p_{Tx} + p_T^2 k_{Tx}}{\Ma \Mb Q} h_1^{\perp} H_1^{\perp} \biggr]}
  \\ \nn
  &\qquad - \Re\biggl[\colbn{
  \cH^{(1)} \left[ \frac{2x\Ma}{Q} \left(\frac{k_{Tx}}{\Ma} \munderbarf\tilde{f}^{\perp} D_1 + \frac{p_{Tx}}{\Mb} \munderbarh\tilde{h}\,H_1^{\perp}\right)
                    +\frac{2\Mb}{zQ} \left(\frac{p_{Tx}}{\Mb} f_1\munderbarD\tilde{D}^{\perp} + \frac{k_{Tx}}{\Ma} h_1^{\perp} \munderbarD\tilde{H}\right)\right]}\biggr]\biggl\}
\\ \nn&
 = \tilde \cF\biggl\{
   \cH^{(0)} \biggl[ - \frac{2 k_{Tx}}{Q} f_1 D_1
                     + \frac{2 k_T^2 p_{Tx}}{\Ma \Mb Q} h_1^{\perp} H_1^{\perp} \biggr]
  \\ \nn
  &\qquad  - \Re\biggl[ \cH^{(1)} \left[ \frac{2x\Ma}{Q} \left(\frac{k_{Tx}}{\Ma} \munderbarf\tilde{f}^{\perp} D_1 + \frac{p_{Tx}}{\Mb} \munderbarh\tilde{h}\,H_1^{\perp}\right)
                    +\frac{2\Mb}{zQ} \left(\frac{p_{Tx}}{\Mb} f_1\munderbarD\tilde{D}^{\perp} + \frac{k_{Tx}}{\Ma} h_1^{\perp} \munderbarD\tilde{H}\right)\right]\biggr]  \biggl\}
\,,\end{align}
\begin{align}
 W_{LU}^{\sin\phi_h} &
 = \tilde \cF\biggl\{
    \Im\biggl[\colbn{- \cH^{(1)} \biggl[
       \frac{2x\Ma}{Q} \left( \frac{k_{Tx}}{\Ma} \munderbarf\tilde{f}^{\perp} D_1 + \frac{p_{Tx}}{\Mb}  \munderbarh\tilde{h}\,H_1^{\perp}\right)}
    \colbn{ + \frac{2\Mb}{zQ}\left( \frac{p_{Tx}}{\Mb} f_1\munderbarD\tilde{D}^{\perp} + \frac{k_{Tx}}{\Ma}  h_1^{\perp} \munderbarD\tilde{H}\right)
    \biggr]}\biggr] \biggr\}
\,.\end{align}

\paragraph{Longitudinally polarized structure functions.}
\begin{align}
 W_{UL}^{\sin\phi_h} &
 = \tilde\cF\biggl\{
   \colkin{ \frac{q_{T}}{Q} \frac{2 k_{Tx} p_{Tx} - \kt \cdot \pt}{\Ma \Mb} \,\cH^{(0)}\,  h_{1L}^{\perp} H_1^{\perp}}
   ~+~\colp{\cH^{(0)} \frac{k_T^2 p_{Tx} + k_{Tx} p_T^2}{\Ma \Mb Q} h_{1L}^{\perp} H_1^{\perp} }
   \nn\\&\qquad
   +\Re\biggl[\colbn{ \cH^{(1)} \biggl[
      \frac{2x\Ma}{Q} \left( \frac{k_{Tx}}{\Ma} \munderbarf\tilde{f}^{\perp}_{L} D_1 - \frac{p_{Tx}}{\Mb} \munderbarh\tilde{h}_{L}H_1^{\perp} \right)}
   \nn\\&\hspace{2cm}
   \colbn{-\frac{2\Mb}{zQ} \left(-\frac{p_{Tx}}{\Mb}  \img g_{1L}\munderbarD\tilde{D}^{\perp} + \frac{k_{Tx}}{\Ma} h_{1L}^{\perp} \munderbarD\tilde{H} \right)
   \biggr]} \biggr]\biggr\}
\nn\\&
 = \tilde\cF\biggl\{ \frac{2 k_T^2 p_{Tx}}{\Ma \Mb Q} \,\cH^{(0)}\,  h_{1L}^{\perp} H_1^{\perp}
   + \Re\biggl[ \cH^{(1)} \biggl[
      \frac{2x\Ma}{Q} \left( \frac{k_{Tx}}{\Ma} \munderbarf\tilde{f}^{\perp}_{L} D_1 - \frac{p_{Tx}}{\Mb} \munderbarh\tilde{h}_{L}H_1^{\perp} \right)
   \nn\\&\qquad
   - \frac{2\Mb}{zQ} \left(-\frac{p_{Tx}}{\Mb}  \img g_{1L}\munderbarD\tilde{D}^{\perp} + \frac{k_{Tx}}{\Ma} h_{1L}^{\perp} \munderbarD\tilde{H} \right)
   \biggr]\biggr] \biggr\}
\,,\end{align}
\begin{align}
 W_{LL}^{\cos\phi_h} &
 = \tilde\cF\biggl\{ \colkin{-\frac{q_T}{Q}\cH^{(0)}g_{1L}D_1}
  -\colp{\cH^{(0)} \frac{k_{Tx} + p_{Tx}}{Q}  g_{1L} D_1}
  \nn\\ &\quad \colbn{
   - \Im\biggl[ \cH^{(1)}\left[\frac{2x\Ma}{Q} \left( \frac{k_{Tx}}{\Ma} \munderbarf\tilde{f}^{\perp}_{L} D_1  - \frac{p_{Tx}}{\Mb}  \munderbarh\tilde{h}_L H_1^{\perp}\right)
       - \frac{2\Mb}{zQ} \left( -\frac{p_{Tx}}{\Mb} \img g_{1L} \munderbarD\tilde{D}^{\perp} + \frac{k_{Tx}}{\Ma} h_{1L}^{\perp} \munderbarD\tilde{H}\right)\right]}\biggr] \biggr\}
 \nn\\&
 = \tilde\cF\biggl\{ -\frac{2 k_{Tx}}{Q} \cH^{(0)}g_{1L}D_1
   -  \Im\biggl[\cH^{(1)} \biggl[\frac{2x\Ma}{Q} \left( \frac{k_{Tx}}{\Ma} \munderbarf\tilde{f}^{\perp}_{L} D_1  - \frac{p_{Tx}}{\Mb}  \munderbarh\tilde{h}_L H_1^{\perp}\right)
   \nn\\&\quad
       - \frac{2\Mb}{zQ} \left( -\frac{p_{Tx}}{\Mb} \img g_{1L} \munderbarD\tilde{D}^{\perp} + \frac{k_{Tx}}{\Ma} h_{1L}^{\perp} \munderbarD\tilde{H}\right)\biggr]\biggr] \biggr\}
\,.\end{align}

\paragraph{Transversely polarized structure functions.}
\begin{align}
 W_{UT}^{\sin\phi_S} &
 = \tilde \cF\biggl\{ \colkin{ -\frac{q_T}{2Q}\,\cH^{(0)}\left( \frac{k_{Tx}}{\Ma} f_{1T}^{\perp} D_1 - \frac{2 p_{Tx}}{\Mb} h_1 H_1^{\perp}\right)}
  \nn\\&\qquad
  + \colp{
  \cH^{(0)} \left(
   - \frac{k_T^2 + \kt \cdot \pt}{2 \Ma Q} f_{1T}^{\perp} D_1
   + \frac{p_T^2 + \kt \cdot \pt}{\Mb Q}  h_{1} H_1^{\perp} \right)}
  \nn\\&\qquad +  \Re\biggl[
  \colbn{\cH^{(1)} \biggl[
         \frac{x\Ma}{Q} \biggl( 2\munderbarf\tilde{f}_T D_1 - \frac{\kt \cdot \pt}{\Ma \Mb}  \Bigl(\munderbarh\tilde{h}_T -\munderbarh\tilde{h}_T^{\perp}\Bigr) H_1^{\perp} \biggr)}
  \nn\\&\hspace{2cm}
  \colbn{  - \frac{\Mb}{zQ} \biggl( 2 h_1 \munderbarD\tilde{H} + \frac{\kt \cdot \pt}{\Ma \Mb} \left(-\img g_{1T} \munderbarD\tilde{D}^{\perp} + f_{1T}^{\perp} \munderbarD\tilde{D}^{\perp}\right) \biggr) \biggr] } \biggr] \biggr\}
 \nn\\&
 = \tilde \cF\biggl\{
   \cH^{(0)} \biggl[ - \frac{k_T^2}{\Ma Q} f_{1T}^{\perp} D_1
                     + \frac{2 \kt \cdot \pt}{\Mb Q} h_1 H_1^{\perp} \biggr]
  \nn\\&\qquad
  +  \Re\biggl[\cH^{(1)} \biggl[
         \frac{x\Ma}{Q} \biggl( 2\munderbarf\tilde{f}_T D_1 - \frac{\kt \cdot \pt}{\Ma \Mb}  \Bigl(\munderbarh\tilde{h}_T -\munderbarh\tilde{h}_T^{\perp}\Bigr) H_1^{\perp} \biggr)
  \nn\\&\hspace{2cm}
    - \frac{\Mb}{zQ} \biggl( 2 h_1 \munderbarD\tilde{H} + \frac{\kt \cdot \pt}{\Ma \Mb} \left(-\img g_{1T} \munderbarD\tilde{D}^{\perp} + f_{1T}^{\perp} \munderbarD\tilde{D}^{\perp}\right) \biggr) \biggr] \biggr] \biggr\}
\,.\end{align}
\begin{align}
 W_{UT}^{\sin\left(2\phi_h -\phi_S\right)} &
 = \tilde\cF\biggl\{
   \colkin{ \frac{q_T}{2Q}\,\cH^{(0)} \biggl[
         \frac{k_{Tx}}{\Ma} f_{1T}^{\perp} D_1
         + \frac{4 k_{Tx}^2\, p_{Tx} - 2 k_{Tx}\, (\vec k_T\cdot \vec p_T) - k_T^2\, p_{Tx}}{\Ma^2\Mb} h_{1T}^{\perp} H_1^{\perp} \biggr] }
  \nn\\ &\qquad
   \colp{+ \cH^{(0)} \biggl[
         + \frac{k_{Tx}^2 - k_{Ty}^2 + 2 k_{Tx} p_{Tx} - \kt \cdot \pt}{2 \Ma Q}  f_{1T}^{\perp} D_1 }
  \nn\\&\hspace{-0.5cm}\hspace{2cm}
   \colp{+ \frac{k_T^2 (2 p_{Tx} k_{Tx} - \kt \cdot \pt) + p_T^2 (k_{Tx}^2 - k_{Ty}^2)}{2 \Ma^2 \Mb Q} \biggr] h_{1T}^{\perp}H_1^{\perp} }
  \nn\\&\qquad + \Re\biggl[
  \colbn{ \cH^{(1)} \biggl[ \frac{x\Ma}{Q} \biggl( \frac{k_{Tx}^2 - k_{Ty}^2}{\Ma^2} \munderbarf\tilde{f}_T^{\perp} D_1
   - \frac{2 k_{Tx} p_{Tx} - \kt \cdot \pt}{\Ma \Mb} \bigl(\munderbarh\tilde{h}_T +\munderbarh\tilde{h}_T^{\perp}\bigr) H_1^{\perp}\biggr)}
  \nn\\&\hspace{-0.5cm}\hspace{2cm}
  \colbn{ - \frac{\Mb}{zQ}\biggl( \frac{k_{Tx}^2 - k_{Ty}^2}{\Ma^2} h_{1T}^{\perp}\munderbarD\tilde{H}
   + \frac{2 k_{Tx} p_{Tx} - \kt \cdot \pt}{\Ma \Mb} \bigl(-\img g_{1T} \munderbarD\tilde{D}^{\perp} -f_{1T}^{\perp} \munderbarD\tilde{D}^{\perp}\bigr) \biggr)
                      \biggr]}\biggr]\biggr\}
\nn\\&\hspace{-0.8cm}
 = \tilde\cF\biggl\{
    \,\cH^{(0)} \biggl[
         \frac{k_{Tx}^2 - k_{Ty}^2}{\Ma Q} f_{1T}^{\perp} D_1
         + \frac{k_T^2 (2 k_{Tx} p_{Tx} - \pt \cdot \kt)}{\Ma^2 \Mb Q} h_{1T}^{\perp} H_1^{\perp} \biggr]
  \nn\\&\hspace{-0.8cm}\qquad
   + \Re\biggl[ \cH^{(1)} \biggl[ \frac{x\Ma}{Q} \biggl( \frac{k_{Tx}^2 - k_{Ty}^2}{\Ma^2} \munderbarf\tilde{f}_T^{\perp} D_1
   - \frac{2 k_{Tx} p_{Tx} - \kt \cdot \pt}{\Ma \Mb} \bigl(\munderbarh\tilde{h}_T +\munderbarh\tilde{h}_T^{\perp}\bigr) H_1^{\perp}\biggr)
  \nn\\&\hspace{-0.8cm}\hspace{2cm}
   - \frac{\Mb}{zQ}\biggl( \frac{k_{Tx}^2 - k_{Ty}^2}{\Ma^2} h_{1T}^{\perp}\munderbarD\tilde{H}
     + \frac{2 k_{Tx} p_{Tx} - \kt \cdot \pt}{\Ma \Mb} \bigl(-\img g_{1T} \munderbarD\tilde{D}^{\perp} -f_{1T}^{\perp} \munderbarD\tilde{D}^{\perp}\bigr) \biggr)
   \biggr]\biggr]\biggr\}
 .
\end{align}
\begin{align}
 W_{LT}^{\cos\phi_S} &
 = \tilde\cF\biggl\{
   \colkin{ -\frac{q_T}{2Q} \frac{k_{Tx}}{\Ma} \cH^{(0)} g_{1T} D_1 }
  -  \colp{  \frac{k_T^2 + \kt \cdot \pt}{2 \Ma Q} \cH^{(0)} g_{1T} D_1 }
  \nn\\&\qquad
  + \Im\biggl[\colbn{\cH^{(1)}\biggl[ \frac{x\Ma}{Q} \Bigl( -2\munderbarf\tilde{f}_T D_1
  + \frac{\kt \cdot \pt}{\Ma \Mb} \bigl(\munderbarh\tilde{h}_T -\munderbarh\tilde{h}_T^{\perp}\bigr) H_1^{\perp} \Bigr)}
  \nn\\&\hspace{2cm}
         \colbn{+\frac{\Mb}{zQ} \Bigl( 2h_1 \munderbarD\tilde{H} + \frac{\kt \cdot \pt}{\Ma \Mb} \bigl( -\img g_{1T} \munderbarD\tilde{D}^{\perp} +f_{1T}^{\perp} \munderbarD\tilde{D}^{\perp}\bigr) \Bigr)\biggr] }\biggr\}
\nn\\&
 = \tilde\cF\biggl\{
    - \frac{k_T^2}{\Ma Q} \cH^{(0)} g_{1T} D_1
  + \Im\biggl[\cH^{(1)}\biggl[ \frac{x\Ma}{Q} \Bigl( -2\munderbarf\tilde{f}_T D_1
    + \frac{\kt \cdot \pt}{\Ma \Mb} \bigl(\munderbarh\tilde{h}_T -\munderbarh\tilde{h}_T^{\perp}\bigr) H_1^{\perp} \Bigr)
  \nn\\&\qquad\quad
    +\frac{\Mb}{zQ} \Bigl( 2h_1 \munderbarD\tilde{H} + \frac{\kt \cdot \pt}{\Ma \Mb} \bigl( -\img g_{1T} \munderbarD\tilde{D}^{\perp} +f_{1T}^{\perp} \munderbarD\tilde{D}^{\perp}\bigr) \Bigr)\biggr] \biggr]\biggr\}
\,.\end{align}
\begin{align}
 W_{LT}^{\cos\left(2\phi_h -\phi_S\right)} &
 = \tilde\cF\biggl\{
   \colkin{ -\frac{q_T}{2Q} \frac{k_{Tx}}{\Ma} \cH^{(0)} g_{1T} D_1}
   - \colp{\cH^{(0)}\biggl[ \frac{k_{Tx}^2 - k_{Ty}^2}{2 \Ma Q} + \frac{2 k_{Tx} p_{Tx} - \kt \cdot \pt}{2 \Ma Q} \biggr] g_{1T} D_1}
  \nn\\&\qquad + \Im \biggl[
   \colbn{ \cH^{(1)} \biggl[ \frac{x\Ma}{Q} \biggl( -\frac{k_{Tx}^2 - k_{Ty}^2}{\Ma^2} \munderbarf\tilde{f}_T^{\perp} D_1
             + \frac{2 k_{Tx} p_{Tx} - \kt \cdot \pt}{\Ma \Mb} \bigl(\munderbarh\tilde{h}_T +\munderbarh\tilde{h}_T^{\perp}\bigr) H_1^{\perp}\biggr) }
  \nn\\&\hspace{-0.6cm}\hspace{2cm}
   \colbn{+\frac{\Mb}{zQ} \biggl( \frac{k_{Tx}^2 - k_{Ty}^2}{\Ma^2} h_{1T}^{\perp} \munderbarD\tilde{H}
           + \frac{2 k_{Tx} p_{Tx}-\kt \cdot \pt}{\Ma \Mb}
             \bigl(-\img g_{1T} \munderbarD\tilde{D}^{\perp} -f_{1T}^{\perp} \munderbarD\tilde{D}^{\perp}\bigr) \biggr)\biggr]\biggr]}\biggr\}
\nn\\&
 = \tilde\cF\biggl\{-
   \frac{k_{Tx}^2 - k_{Ty}^2}{\Ma Q} \cH^{(0)} g_{1T} D_1
  \nn\\&\qquad
   { + \Im\biggl[\cH^{(1)} \biggl[ \frac{x\Ma}{Q} \biggl(- \frac{k_{Tx}^2 - k_{Ty}^2}{\Ma^2} \munderbarf\tilde{f}_T^{\perp} D_1
             + \frac{2 k_{Tx} p_{Tx} - \kt \cdot \pt}{\Ma \Mb} \bigl(\munderbarh\tilde{h}_T +\munderbarh\tilde{h}_T^{\perp}\bigr) H_1^{\perp}\biggr) }
  \nn\\&\hspace{-0.6cm}\hspace{2cm}
   {+\frac{\Mb}{zQ} \biggl( \frac{k_{Tx}^2 - k_{Ty}^2}{\Ma^2} h_{1T}^{\perp} \munderbarD\tilde{H}
           + \frac{2 k_{Tx} p_{Tx}-\kt \cdot \pt}{\Ma \Mb}
             \bigl(-\img g_{1T} \munderbarD\tilde{D}^{\perp} -f_{1T}^{\perp} \munderbarD\tilde{D}^{\perp}\bigr) \biggr)\biggr]\biggr]}\biggr\}
\,.\end{align}

\subsection{Discussion of Results}
\label{sec:results_discussion}

In this section, we discuss the key features of the factorization formulae we have derived for the NLP structure functions.
First of all we note that the basic structure and number of terms in our momentum space results agrees with that in the literature, where results 
have been derived based on the parton model and tree level analysis in QCD,
see e.g.~\refcite{Bacchetta:2006tn}. From our derivation of the factorization formulae here we obtain a number of important new observations and features, which we will discuss in this section.
A detailed comparison to the literature is then given in \sec{literature}.

We recall that we obtain three non-zero sources of power corrections at NLP,
namely kinematic corrections from expanding the projectors,
insertions of $\cP_\perp$ operators, and insertions of $\cB_{n_i\perp}$ operators.
While the first two sources induce new kinematic structures, they are entirely given
in terms of the leading power hard function $\cH^{(0)}(q^+q^-,\mu)$ and leading power TMD PDFs and FFs.
For the $\cP_\perp$ operators, the fact that the same $\cH^{(0)}(q^+q^-,\mu)$ appears to all orders in $\alpha_s$ is non-trivial,
but follows from the reparameterization invariance of SCET. The fact that only LP TMDs appear follows from the commutator result in \eq{def_BPf}.

The $\cB_{n_i\perp}$ operators involve quark-gluon-quark correlators which result in truly new TMD PDFs and TMD FFs,
and in the previous formulas were denoted by functions with tildes, $\tilde f^\perp$, $\tilde e$, $\tilde D^\perp$, $\tilde H$, etc. We have demonstrated that all these terms involve only a single new hard coefficient function,  $\cH^{(1)}(q^+q^-, \xi,\mu)$, whose allowed functional form is determined to all orders in $\alpha_s$ by our analysis. Crucially, these correlators
not only depend on the total lightcone momentum $\omega$ of the quark-gluon system (parameterized by the $x$ and $z$ dependence),
but allow for a splitting into a large longitudinal momentum $\xi \w$ carried by the gluon
and $(1-\xi)\w$ carried by the quark. Likewise, $\cH^{(1)}(q^+q^-, \xi,\mu)$ encodes perturbative loop corrections at the hard scale which depend in a non-trivial way on
the momentum fraction $\xi$. In general the $\cB_{n_i\perp}$ TMD correlators are integrated against
this hard function in the variable $\xi$. While this convolution is absent at tree level where $\cH^{(1)}$ is $\xi$-independent, at 
NLO, calculations in SCET for the Wilson coefficient $C^{(1)}$ that enters $\cH^{(1)}$
have explicitly demonstrated non-trivial $\xi$ dependence~\cite{Beneke:2017ztn, Beneke:2018rbh}.

The soft sector plays a special role for TMDs due to the appearance of rapidity divergences.
As discussed in \sec{fact_vanishing_soft}, although there are various sources for NLP soft operators, they all can be shown to lead to vanishing contributions
to the structure functions at NLP, to all orders in $\alpha_s$. This type of contribution only starts at NNLP. In every NLP contribution that we obtained, the same soft function $S(b_T)$ that appeared at LP, also appeared.
Even the $\cB_{n_i\perp}$ operators do not induce new soft functions, since from the point of view of the soft scale the quark-gluon fields in the same collinear direction act like a single color source in the fundamental representation, despite the fact that they act as distinct partons from the point of view of the hard interactions. 
Furthermore, we showed that the LP soft function could always be absorbed into the definitions of the various LP and NLP TMDs that appeared, just like what is done for the TMDs appearing in the LP factorization theorem. 
In this case, the fact that the same LP soft function could be absorbed for the $\cP_\perp^\mu=i\partial/\partial b_{\perp\mu}$ operators was non-trivial, since it involved commuting $\cP_\perp$ and $S(b_T)$, and the argument for this was given in \sec{factorization_P}.

The factorization formulae that we have derived also have interesting implications for the structure of the renormalization group evolution, both for the standard invariant mass evolution in $\mu$, and for the rapidity evolution, which is given by differential equations in $\zeta_{a,b}$. The appearance of the same leading power hard function $\cH^{(0)}(q^+q^-,\mu) = [C^{(0)}(q^+q^-,\mu)]^2$ for the kinematic and $\cP_\perp$ contributions, implies that this term has the same $\mu$-evolution equations as at LP. Thus the known next-to-next-to-next-to-leading logarithmic (N$^3$LL) order resummation for $C^{(0)}$ can be used for these terms. For the terms with quark-gluon-quark correlators we have $\cH^{(1)}(q^+q^-,\xi,\mu)=C^{(1)}(q^+q^-,\xi,\mu) C^{(0)}(q^+q^-,\mu)$.  It is known from the general properties of SCET that the evolution equation for $C^{(1)}$ will involve a convolution in $\xi$, much like DGLAP evolution for standard longitudinal PDFs. This RGE takes the form
\begin{align} \label{eq:C1murge}
   \mu\frac{\df}{\df\mu}C^{(1)}(q^+q^-,\xi,\mu)
   = \int\! \frac{\df\xi'}{\xi'}\,
  \gamma^{(1)}\big(\xi,\xi',q^+q^-,\mu\big)\: C^{(1)}(\xi',q^+q^-,\mu)
  \,.
\end{align}
In fact, the required anomalous dimension $\gamma^{(1)}(\xi,\xi',q^+q^-,\mu)$ has been considered in the context of subleading power $N$-jet SCET operators, and calculated to one loop order~\cite{Beneke:2017ztn, Beneke:2018rbh}. The authors found non-trivial dependence on the variables $\xi$, $\xi'$ appearing in the non-cusp part of the anomalous dimension, which will start to play an important role for NLP resummation of the large $\ln(q_T/Q)$ logarithms at next-to-leading-logarithmic (NLL) order. 
By RGE consistency the $\mu$-anomalous dimensions for $\tilde{B}_{\cB \, f/N}^{\rho}(x,\xi,\vec b_T,\mu,\zeta_a)$ and $\tilde{\cG}_{\cB \, h/f}^{\rho}(z,\xi,\vec b_T,\mu,\zeta_b)$ will also involve the same function, $-\gamma^{(1)}(\xi',\xi,\zeta_{a},\mu)$ and $-\gamma^{(1)}(\xi',\xi,\zeta_{b},\mu)$, as well as other terms associated with the anomalous dimension for $C^{(0)}$. 

Another implication of our NLP factorization formula is that the new quark-gluon-quark correlators have the same rapidity anomalous dimension as at LP, and thus satisfy the same LP Collins-Soper evolution equation~\cite{Collins:1981uk,Collins:1981va,Collins:1984kg}. 
To demonstrate this there is a subtle complication that we have to rule out: the rapidity evolution of subleading power operators can cause them to mix into other operators, without violating the form of the factorization theorem. Essentially, because subleading power factorization theorems involve a sum of terms, the RGE evolution is allowed to cause new terms to appear as long as they cancel out in the sum.  The existence and necessity of this type of mixing for power suppressed soft and collinear functions was first discussed for $\mu$ evolution in \SCETi in \refcite{Moult:2018jjd}. Furthermore, in \SCETii it is known that the rapidity evolution of power suppressed terms causes this type of mixing, since this has been demonstrated explicitly for the NNLP corrections to the EEC observable in \refcite{Moult:2019vou} (where also leading-logarithmic resummation was carried out).  This mixing effect occurs for a case where the leading order perturbative result for the power correction involves a rapidity logarithm that is factorized into rapidity divergent subleading power soft and collinear functions.  In \refcite{Moult:2019vou} it was shown that such effects involve mixing between a triplet of subleading power functions, one soft, and two collinear (one in $\bn$, and one in $n$). For our case, we are at NLP rather than NNLP, and have demonstrated by explicit constrution that there is no subleading power soft function at this order.  Therefore we will not have this type of rapidity evolution mixing effect. Our kinematic and $\cP_\perp$ NLP corrections involved only LP TMDs, and hence obviously have the same rapidity evolution equations in $b_T$ space where the evolution equation is multiplicative. The only genuinely new subleading power functions are the quark-gluon-quark correlators for the TMD PDFs and TMD FFs, denoted with tildes in the decomposition in \eq{qgqdecomp}.  However, if there is no mixing then the rapidity evolution equations for these correlators must also be multiplicative. This follows from the fact that the evolution in $\zeta_{a,b}$ is related to the cancellation of rapidity divergences between collinear and soft matrix elements, and for the q-g-q correlators we have the same LP soft function. This can be seen quite explicitly when we follow the usual SCET approach where the collinear and soft functions are separately UV and rapidity renormalized, giving rise to correlators depending on $\mu$ and the rapidity renormalization scale $\nu$ (see the discussion around \eq{soft_relation_sub} for an introduction to $\nu$, and further references). The complete correlator is then obtained from a product of renormalized functions.  
For the quark-gluon-correlator defined in \eq{def_BB}, using the $\eta$ rapidity regulator~\cite{Chiu:2012ir} and carrying out the renormalization, the result for the NLP TMD PDFs reads
\begin{align} \label{eq:tildeB_absorbed}
 \tilde{B}_{\cB \, f/N}^{\rho}(x,\xi,\bt,\mu,\zeta_a)
 = \hat{B}_{\cB \, f/N}^{\rho}(x,\xi,\bt,\mu,\nu^2/\zeta_a) \sqrt{S(b_T,\mu,\nu)}
\,.
\end{align}
By combining these functions as shown, the $\nu$ dependence cancels,
and one is left with a result that is only a function of the Collins-Soper scale $\zeta_a$. However, this same type of formula is also present at LP
\begin{align} \label{eq:LPbeamsoftTMD}
 {B}_{f/N}(x,\bt,\mu,\zeta_a)
 = \hat{B}_{f/N}^{\rho}(x,\bt,\mu,\nu^2/\zeta_a) \sqrt{S(b_T,\mu,\nu)}
\,,
\end{align}
with the same soft function. The same results are also true for the LP and NLP TMD fragmentation functions, $\cG_{h/f}(z,\vec b_T,\mu,\zeta_b)$ and $\tilde{\cG}_{\cB\, h/f}^\rho(x,\xi,\vec b_T,\mu,\zeta_b)$.
Consequently, the rapidity anomalous dimension of subleading quark-gluon-quark TMD in $b_T$ space is related to
that of the LP soft function, and to the leading power TMDs
\begin{equation}
 \frac{\df\ln \tilde{B}_{\cB \, f/N}^{\rho}}{\df\ln \zeta}
 = \frac{\df\ln \tilde {\cG}_{\cB \, h/f}^{\rho}}{\df\ln \zeta}
 =  \frac{\df\ln {B}_{f/N}}{\df\ln \zeta}
 =  \frac{\df\ln {\cG}_{h/f}}{\df\ln \zeta}
 = \frac14 \frac{\df\ln S}{\df\ln\nu}
 =\frac14 \gamma_\nu(\mu,b_T) 
\,.\end{equation}
Note that other notations are in common use for this anomalous dimension in the literature, and the relations are: $\gamma_\nu(\mu,b_T) = 2\gamma_\zeta(\mu,b_T) = \tilde K(b_T,\mu)=-2{\cal D}^q(\mu,b_T)$.
Hence, the NLP quark-gluon-quark TMD PDFs and FFs have the same rapidity evolution equation as at LP, which is known perturbatively to three-loop order~\cite{Li:2016ctv,Vladimirov:2017ksc}.
This gives us knowledge about an important class of logarithms, and complements the presence
of the more complicated $\mu$ evolution in \eq{C1murge} which is only known to NLL. It also implies that the known perturbative structure of $\gamma_\nu(\mu,b_T)$ to all orders in $\alpha_s$ constrains the form of the $\mu$-anomalous dimensions for the $\tilde B_{\cB\, f/N}^\rho$ and ${\tilde \cG}_{\cB\, h/f}^\rho$ functions to involve a term with the cusp-anomalous dimension, of the form $\Gamma_{\rm cusp}[\alpha_s(\mu)]\ln\mu^2/\zeta$.

\subsection{Comparison to Literature}
\label{sec:literature}

The factorization of SIDIS at small transverse momentum has been studied in the literature since the mid-90's.
Building on the tree-level analysis in \refcite{Mulders:1995dh},
the structure of the expected quark-quark and quark-gluon-quark correlators
was studied extensively~\cite{Boer:1997nt, Boer:1999uu, Collins:2002kn, Bomhof:2004aw, Collins:2004nx, Bacchetta:2005rm, Boer:2003cm, Bacchetta:2004zf,Goeke:2005hb}.
A useful summary of these developments is provided in \refcite{Bacchetta:2006tn}. 
Note that this work is not based on deriving factorization formulae as we have done here, but instead works out the structure of power corrections based on low order QCD diagrams and the parton model.
They therefore did not explicitly include soft factors,
but mentioned that they should contribute based on the LP analyses in \refscite{Collins:2004nx,Ji:2004wu,Ji:2004xq}.
Furthermore, the quark-gluon-quark correlators were defined in terms of a single longitudinal momentum fraction only,
as opposed to having the additional momentum fraction $\xi$ that was found in the all orders factorization analysis in our work.

\Refcite{Bacchetta:2008xw} performed a first validation of the tree-level factorization
suggested in these early works by comparing at NLO its predictions to those of collinear factorization
which holds at large $P_{hT} \sim Q$, in the cases where the TMD to collinear matching starts at most at twist-3.
Agreement was only found for LP structure functions, while the azimuthal asymmetries
starting at NLP did not match between the two approaches, see table 2 in \refcite{Bacchetta:2008xw}
for an overview. 

While \refcite{Bacchetta:2008xw} included a soft factor in their factorization,
it was argued to have no effect in the large-$q_T$ region, and thus excluded in the perturbative comparison.
This was clarified in \refcite{Bacchetta:2019qkv}, where the LP soft function was reinstated at NLP,
and agreement with collinear factorization was found at NLO.
Furthermore, since the earlier works were based on a tree level analysis
without virtual corrections, it was conjectured that a hard function is required in the factorization.
However, their perturbative validation was performed at NLO, where for finite $P_{hT}$
the hard function only enters at tree level, and thus its presence could not be validated, nor could its form be constrained for various different NLP corrections.
For the same reason, it was again not yet noticed that there is a convolution over the additional momentum fraction $\xi$.
In \sec{tree_level}, we will discuss in more detail how our results reduce to these earlier works
at tree level, in which case the convolutions in $\xi$ are still absent.

Very recently, \refcite{Vladimirov:2021hdn} presented a systematic framework
for studying the power expansion of TMDs. They did not provide fatorization formulae 
for the structure functions themselves, but from their setup it is clear
that one receives contributions from additional momentum fractions.
We will compare to their methodology in \sec{VMS}.

\subsubsection{Comparison at Tree Level}
\label{sec:tree_level}

In the following, we compare in more detail the ingredients in our work,
and those encountered in the previous works, for which we use \refcite{Bacchetta:2006tn} as a reference.
As discussed above, their work is valid at tree level, and where necessary
we will employ the tree-level matching to allow for a direct comparison.
Note that whenever we compare explicit definitions, we will limit ourselves
to the case of the TMD PDF for brevity, as the conclusions obtained immediately carry over the TMD FFs.

\subsubsection*{Hard Function}

Hard functions encode the virtual corrections to the underlying process,
and are composed of the Wilson coefficients $C_f$ that arise from matching
the full theory onto the hard-scattering operators.
At leading order, both leading and subleading-power Wilson coefficients reduce to the quark charge $Q_f$
by virtue of the normalization of the leading subleading-power currents,
\begin{align}
  C_f^{(0)}(q^+ q^-, \mu) = Q_f + \cO(\as)
\,,\qquad
  C_f^{(1)}(q^+ q^-, \xi, \mu) = Q_f + \cO(\as)
\,.\end{align}
The LP and NLP hard functions are then given by
\begin{align} \label{eq:H01tree}
  \cH_f^{(0)}(q^+ q^-, \mu) = Q_f^2 + \cO(\as)
\,,\qquad
  \cH_f^{(1)}(q^+ q^-, \xi, \mu) = Q_f^2 + \cO(\as)
\,.\end{align}
At tree level, they become constants and identical to each other.
In particular, this implies that $\cH_f^{(1)}$ is independent of $\xi$
and the TMD PDFs and FFs can be integrated over $\xi$.

In \refcite{Bacchetta:2006tn}, hard functions are not included, but their momentum-space
convolutions involve sums over flavors that are weighted by the quark charges, corresponding
to exactly the above tree-level results for our hard functions.

\subsection*{Quark-Quark Correlators}

Our quark-quark correlator, in the SCET jargon called a beam function,
is defined in \eq{def_B_hat} as
\begin{align} \label{eq:def_Bhat_rep}
 \hat B^{\beta\beta'}(x = \w_a/P_N^-, \bt) &
 = \theta(\w_a) \, \braket{N | \bar\chi_{n}^{\beta'}(b_\perp) \chi_{n,\w_a}^{\beta}(0) | N}
\,.\end{align}
Here, we have not yet absorbed the soft function.
For comparison, the quark-quark correlator in \refcite{Bacchetta:2006tn},
converted to our notation for light-cone coordinates is defined as
\begin{align} \label{eq:def_Phi}
 \Phi^{\beta\beta'}(x, \bt) &
 = \frac{1}{\sqrt2} \int\frac{\df b^+}{2\pi} e^{-\frac{\img}{2} x P^- b^+}
   \braket{N | \bar\psi^{\beta'}(b) W_n(b) W_n^\dagger(0) \psi^\beta(0) | N}
\,.\end{align}
For more details on the conversion, see \app{qq_relation}.
In \eq{def_Phi}, the $\sqrt2$ arises from different lightcone conventions.
Compared to \eq{def_Bhat_rep}, it contains the full QCD quark fields $\psi$
and explicit Wilson lines, while \eq{def_Bhat_rep} is built from collinear
quark fields that are rendered collinear gauge invariant by including Wilson lines inside $\chi_n$,
see \eq{chiB}. Furthermore, the fields in \eq{def_Bhat_rep} have fixed minus-momenta,
while in \eq{def_Phi} the momentum dependence is encoded by Fourier-transforming
the field at position $b$. Since there is a one-to-one correspondence between the two,
this simply amounts to a matter of taste.

An important difference between \eqs{def_Bhat_rep}{def_Phi} is that the collinear fields $\chi_n$ are
defined purely in terms of the good fermion components. Projecting these out from
the quark fields in \eq{def_Phi}, we obtain the relation derived in \eq{relation_B_phi},
\begin{align} \label{eq:relation_B_phi_rep}
 \hat B(x,\bt) &
 = \frac{(\slashed{n} \slashed{\bn})}{4}
   \, \frac{\Phi(x, \bt)}{\sqrt2} \,
   \frac{(\slashed{\bn} \slashed{n})}{4}
\,.\end{align}
Here, the sandwiching by $\slashed{n}\slashed{\bn}/4$ and $\slashed{\bn}\slashed{n}/4$ projects onto the good components of the fermion fields.

An immediate consequence of \eq{relation_B_phi_rep} is that the subleading-twist
contributions that appear in $\Phi$, instead vanish for the beam function $\hat B$.
Schematically, one can expand
\begin{align} \label{eq:Phi_expansion_schematic}
 \sqrt2 \Phi(x, \kt) &
 = \bigl( f_1 + \cdots \bigr) \frac{\slashed{n}}{2}
 \nn\\&\quad
   + \frac{x \Ma}{2 Q} \Bigl\{ e + \cdots + f^\perp \frac{\slashed{k}_\perp}{\Ma} + \cdots
                            + \frac{\img}{\sqrt 2} h \frac{[\slashed{n}, \slashed{\bn}]}{2} + \cdots \Bigr\}
 \nn\\&\quad
  + \cdots
\,.\end{align}
Here, the first line are typically referred to as leading power (or leading-twist) contributions,
while the second-line denotes the subleading power (subleading-twist) terms suppressed at large momentum,
and we do not show higher order contributions.
Inserting this into \eq{relation_B_phi_rep}, only the leading power contributions remain,
such that the beam function is indeed a pure leading power object.
Of course, the subleading power terms are not absent in SCET, but will instead systematically reappear in the correlators discussed below.

\subsection*{Quark-Quark Correlators with a $\cP_\perp$ Insertion}

In \sec{factorization_P}, we constructed quark-quark correlators with a single $\cP_\perp$ insertion
from the corresponding subleading scattering operator.
For example, we have from \eq{def_BP_f_a}
\begin{align} \label{eq:def_BP_f_a_rep}
 \hat B_{\cP}^{\beta\beta'}(x,\bt)
 = \frac{-\img}{2Q} \frac{\partial}{\partial b_\perp^\rho}
   \left[ \gamma_\perp^\rho\, \slashed\bn\,,\,\hat B_{f/N}(x,\bt) \right]^{\beta\beta'}
\,.\end{align}
It is simply given by a commutator involving the LP beam function,
and thus its spin-decomposition is given by LP TMDs only.
We provide an explicit expressions for these in \eq{BPf_msp}.
Limiting ourselves to showing a few terms here for brevity, we have
\begin{align} \label{eq:BPf_msp_rep}
 \hat B_{\cP}(x,\kt) &
 = \frac{1}{2Q} \biggl\{ f_1 \slashed{k}_\perp + \cdots
    -\img h_{1}^{\perp} \frac{k_T^2}{4\Ma} [\slashed{n}, \slashed{\bn}] \biggr\}
\,.\end{align}
The two terms we display have remarkable similarity with
those shown in \eq{Phi_expansion_schematic}. In fact, they are related
by the equation of motions as~\cite{Mulders:1995dh, Boer:1997bw}
\begin{align} \label{eq:relation_eom}
 f_1 + x \tilde{f}^{\perp} = x f^\perp
\,,\qquad
 x \tilde{h} -  \frac{k_T^2}{\Ma^2} h_1^{\perp} = x h
\,,\end{align}
where the functions with the tilde will arise in the quark-gluon-quark correlators below.\footnote{When the functions with the tilde are dropped, these results are known as Wandzura-Wilczek relations~\cite{Wandzura:1977qf}.}
These relations immediately illustrate that the correlators shown in
\eqs{Phi_expansion_schematic}{BPf_msp_rep} are not independent.
In fact, by comparing \eq{BPf_msp} to the full decomposition of $\Phi(x, \kt)$
in \refcite{Bacchetta:2006tn} one finds that $\cB_{\cP}$ reproduces \emph{all}
subleading-twist Dirac structures that involve the transverse momentum $k_T$ and LP TMDs.

The upshot here is that in the SCET approach the LP beam function is defined by good fermion field components only, and its subleading component only enter via SCET building blocks like $\cP_\perp^\mu$, whose matrix elements are related back to those at LP. The SCET organization 
removes the redundancy between subleading components from the start without ever having to use relations like the ones in \eq{relation_eom}.

\subsection*{Quark-Gluon-Quark Correlators}

We finally turn to the quark-gluon-quark correlators.
As discussed previously, from the results we have derived they in general measure two longitudinal momentum fractions, $(x,\xi)$ for NLP TMD PDFs and $(z,\xi)$ for NLP TMD FFs. 
The new $\xi$ dependence is integrated against that appearing in the new hard function $\cH^{(1)}$ for these terms.
At tree level in the hard matching, the hard function $\cH^{(1)}$ is independent of $\xi$ as in \eq{H01tree},
so one can freely integrate the TMD correlators over $\xi$.
We show explicitly how this is achieved by rewriting the correlators entirely
in Fourier space in \app{qgq_relation}. Here, we only quote the result
in \eq{Bb_hat_integrated},
\begin{align} \label{eq:integrated_correlators}
 \int\!\df\xi\, \hat B_{\cB}^{\rho \beta \beta'}(x, \xi, \bt) &
 = \frac{1}{x \Pa^-} \int\frac{\df b^+}{4\pi} e^{-\frac{\img}{2} b^+ x \Pa^- }
   \MAe{N}{ \bar\chi_{\na}^{\beta'}(b) \,g  {\cB}_{\na\perp}^{\rho}(0) \chi_{\na}^{\beta}(0)}{N}
\,,\end{align}
where $b^\mu = (b^+, 0, b_\perp)$. Again we remind the reader that the function with a hat, $\hat B_{\cB}^{\rho \beta \beta'}$, does not absorb the LP soft function. The fact that the gluon and quark field that come as a pair
sit at the same position implies that these correlators only measure their combined momentum,
which by momentum conservation is identical to that of the remaining quark field, namely $x\Pa^-$.

Quark-gluon-quark correlators similar to \eq{integrated_correlators} are also present
in the tree-level results of \refscite{Mulders:1995dh, Boer:2003cm},
where they are defined as~\cite{Bacchetta:2006tn}
\begin{align} \label{eq:Phi_D}
 (\Phi^\rho_D)^{\beta\beta'}(x, \kt) &
 = \int\frac{\df b^+ \df^2\bt}{\sqrt2 (2\pi)^3} e^{-\img p \cdot b}
   \braket{ P | \bar\psi^{\beta'}(b) W_n(b) W^\dagger_n(0) \img D^\rho(0) \psi^\beta(0) | P}
\,.\end{align}
Here, we have converted their expression to our notation, see  \app{qq_relation} for details.
In \eq{Phi_D}, $D^\rho$ is the covariant derivative, which together with the Wilson lines renders the matrix element gauge invariant.
However, from the identity $W_n^\dagger \img \bn\cdot D \psi = \img \bn\cdot \partial [W_n^\dagger \psi]$,
it follows that $\Phi_D^-(x, \kt) = x \Pa^- \Phi(x, k_T)$, and thus the large light-cone component
$\Phi_D^-$ is not independent from the LP operator $\Phi$. In order to separate out the
independent subleading component, one defines~\cite{Bacchetta:2006tn}
\begin{align} \label{eq:Phi_A}
 \tilde \Phi_A^\rho(x, \kt) = \Phi_D^\rho(x, k_T) - k_T^\rho \Phi(x, k_T)
\,.\end{align}
This difference projects onto terms that must involve a gluon field $A$. 
In our approach using SCET, the identification of terms purely involving a gluon field is built into the formalism from the start, through the use of the gluon building block operators $\cB_{n_i\perp}^\mu$ which include Wilson lines and are collinear gauge invariant. Thus for the matrix elements we consider we immediately obtain quark-gluon-quark correlators from operators containing the collinear gluon field $\cB_{n_i\perp}^\rho$, without need for subtractions like the one in \eq{Phi_A}.

The correlator in \eq{Phi_A} can be directly related to our $\xi$-integrated correlators.
In \app{qgq_relation}, we carefully convert between the notations in our and their work,
leading to the simple relations in \eq{Bb_hat_integrated}
\begin{align}
 \int\!\df\xi\, \hat B_{\cB}^{\rho \beta \beta'}(x, \xi, \bt) &
 = \frac{(\slashed{n} \slashed{\bn})^{\beta\alpha}}{4}
   \frac{(\tilde\Phi_A^\rho)^{\alpha\alpha'}(x, \bt)}{\sqrt2 x \Pa^-}
   \frac{(\slashed{\bn} \slashed{n})^{\alpha'\beta'}}{4}
\,.\end{align}
This relation differs from the similar relation in \eq{relation_B_phi_rep}
for the quark-quark correlator only by the addition of a $1/(x \Pa^-)$ term to compensate for the lack of mass dimension for the gluon building block fields $\cB_{n\perp,\w}^\rho$ with fixed momentum $\w$, defined as in \eq{chiB}.

\subsection*{Illustration for $W_{UU}^{\cos\phi_h}$}

We now explicitly illustrate for the example of the $W_{UU}^{\cos\phi_h}$ structure function
that at tree level, our final results reproduce the same terms as in \refcite{Bacchetta:2006tn}
(once we implement in our result that all TMDs must include a square-root of the leading power soft function in their definitions).
To make this comparison, first note that integrating the quark-gluon-quark correlators over $\xi$
is equivalent to integrating all individual scalar NLP TMDs over $\xi$.
For example, at tree level
\begin{equation} \label{eq:integrator}
 \tilde f^\perp(x,\xi,p_T) \quad\rightarrow\quad
 \tilde f^\perp(x, p_T) \equiv \int\!\df\xi\,\tilde f^\perp(x,\xi,p_T)
\,,\end{equation}
and likewise for all correlators denoted with a tilde.
We also define the convolution operator as in \refcite{Bacchetta:2006tn},
\begin{align} \label{eq:C_integral}
 \cC[\w g D] \equiv x \sum_f Q_f^2 \int\df^2\kt \df^2\pt \, \delta^{(2)}(\qt + \kt - \pt) g_f(x, k_T) D_f(z, p_T) + (f \to \bar f)
\,.\end{align}
When evaluated for functions with a tilde, \eq{integrator} is assumed.
It is related to our $\tilde\cF[\w\, \cH\, g\, D]$ defined in \eq{cF_tilde}  as
\begin{align}
 \cC[\w g D] = \frac{x}{2 z} \tilde\cF[\w\, g\, D]
 \quad\mathrm{with}\quad
 \tilde f(x, \xi, \kt) \to \int\df\xi \, \tilde f(x, \xi, \kt)
\,.\end{align}
Here, the prefactor $x/(2z)$ precisely compensates for the overall difference in normalization,
see \eq{relation_W_F}, such that our expressions with $\tilde\cF$ take the same form as their
expressions with $\cC$. At tree level, our \eq{W_UU_cosphi_msp} can be written as
\begin{align} \label{eq:WUU_ours}
  \frac{x}{2z} W_{UU}^{\cos\phi_h} &
 = \cC\biggl\{ \biggl[ - \frac{2 k_{Tx}}{Q} f_1 D_1
                       + \frac{2 k_T^2 p_{Tx}}{\Ma \Mb Q} h_1^{\perp} H_1^{\perp} \biggr]
  \nn\\&\qquad
              - \biggl[ \frac{2x\Ma}{Q} \left(\frac{k_{Tx}}{\Ma} \tilde{f}^{\perp} D_1 + \frac{p_{Tx}}{\Mb} \tilde{h}\,H_1^{\perp}\right)
                    +\frac{2\Mb}{zQ} \frac{p_{Tx}}{\Mb} f_1\tilde{D}^{\perp} + \frac{2\Mb}{zQ}\frac{k_{Tx}}{\Ma} h_1^{\perp} \tilde{H} \biggr]\biggl\}
 \nn\\&
 = \frac{2 \Ma}{Q} \cC\biggl\{ \frac{-k_{Tx}}{\Ma} \left[  (f_1 + x \tilde{f}^{\perp}) D_1 + \frac{\Mb}{\Ma} x h_1^{\perp} \frac{\tilde{H}}{z} \right]
  \nn \\&\hspace{1.8cm}
              - \frac{p_{Tx}}{\Mb} \biggl[ \biggl(x \tilde{h} -  \frac{k_T^2}{\Ma^2} h_1^{\perp} \biggr) H_1^{\perp} + \frac{\Mb}{\Ma} f_1 \frac{\tilde{D}^{\perp}}{z} \biggr]
         \biggl\}
\,.\end{align}
In the first equation, we give the result as in \eq{W_UU_cosphi_msp},
in which the second square bracket contains the terms that in the full result involve a convolution over $\xi$.
In the second step, we collect by $k_T$ and $p_T$ to simplify the comparison with Eq.~(4.5) in \refcite{Bacchetta:2006tn},
which reads%
\footnote{Compared to their original notation, we have exchanged $\kt \leftrightarrow \pt$,
and used that their unit vector $\hat h = \vec P_{hT} / P_{hT} = (1, 0)$ in our conventions.}
\begin{align} \label{eq:WUU_theirs}
  \frac{x}{2z} W_{UU}^{\cos\phi_h} &
 = \frac{2 \Ma}{Q} \cC\biggl\{
         - \frac{k_{Tx}}{\Mb} \biggl[ x f^\perp D_1 + \frac{\Mb}{\Ma} h_1^\perp \frac{\tilde H}{z} \biggr]
         - \frac{p_{Tx}}{\Ma} \biggl[ x h H_1^\perp + \frac{\Mb}{\Ma} f_1 \frac{\tilde D^\perp}{z}  \biggr]
   \biggl\}
\,.\end{align}
Comparing \eqs{WUU_ours}{WUU_theirs}, we find agreement if
\begin{align}
 f_1 + x \tilde{f}^{\perp} = x f^\perp
\,,\qquad
 x \tilde{h} -  \frac{k_T^2}{\Ma^2} h_1^{\perp} = x h
\,,\end{align}
which is precisely \eq{relation_eom} that results from the QCD equations of motion~\cite{Mulders:1995dh, Boer:1997bw}.
Such relations are widely used to discuss correlators at different twist~\cite{Bacchetta:2006tn},
but are not required in our approach, since starting with a minimal basis of independent SCET operators naturally separates out the independent sources of power corrections.

In a similar fashion, we have checked that at tree-level, all our results in \sec{results_NLP_full_msp}
reduce to those \refcite{Bacchetta:2006tn}, if we take the definitions used there and absorb a leading power soft function into the TMDs at both LP and NLP.

\subsubsection{Comparison to the TMD Operator Expansion}
\label{sec:VMS}

The first systematic study of subleading-power TMD factorization was presented
recently by Vladimirov, Moos and Scimemi in \refcite{Vladimirov:2021hdn}, while the work presented here was in the final stages of preparation.
The focus of \refcite{Vladimirov:2021hdn} was to develop a generic formalism to obtain the power expansion of TMDs
in Drell-Yan, SIDIS and semi-inclusive annhilation (SIA),
and hence did not yet provide explicit results for structure functions.
In the following, we briefly compare their approach to determining
the operators required at subleading power to ours.

The method of \refcite{Vladimirov:2021hdn}, dubbed \emph{TMD operator expansion}, is based on expressing
the hadronic tensor in \eq{W} in terms of a functional integral,
which in the absence of time ordering involves a causal and an anti-causal field~\cite{Keldysh:1964ud}.
One then assumes that the physics associated with the hadrons,
which define the $n$ and $\bn$ directions as in our case,
are dominated by collinear fields with the scalings $p_n^\mu \lesssim Q (\lambda^2, 1, \lambda)$
and $p_\bn^\mu \lesssim Q (1, \lambda^2, \lambda)$, respectively, which are the same infrared regions covered by the collinear fields in SCET.
Following the background field method,
both causal and anti-causal QCD fields are then split into a dynamical field
and a background field, the latter consisting of collinear and anti-collinear modes.
By working in the background field gauge, the gauge transformations of dynamic and background
fields are decoupled. After imposing that the hadronic states are entirely built
from background fields, one can perform the functional integral over the dynamic fields
independent of the hadronic states. This gives rise to an effective operator,
which encodes the interactions between the collinear and anti-collinear sectors
and the hard dynamic fields. After multipole expanding the (anti-)collinear fields
and reducing the resulting operators to a minimal basis, one finally
obtains the desired TMD operators.

Our approach is based on SCET, where one constructs all possible operators
from a set of minimal building blocks in the EFT, where the power counting
of individual building blocks limits the operators allowed at a certain order
in the power counting. Indeed, for parts of our analysis we could make use of the construction of complete basis of subleading power \SCETi operators derived in earlier literature~\cite{Feige:2017zci}. Operators arise from two different sources, namely
hard-scattering operators that couple the collinear and soft sectors to the mediating current,
and the dynamic Lagrangian that encodes interactions within and between the collinear and soft sectors.
This separation into different sectors thus is akin to the different terms in
the action in \refcite{Vladimirov:2021hdn}, cf.~e.g~their Eq.~(2.27).
It is not surprising that both approaches in principle give the same interaction terms;
in fact, one can also formulate SCET in terms of functional integrals~\cite{Bauer:2008qu}.
However, given the vastly different notation,
we have not attempted a comparison of the collinear operators in our and their work.

A key difference between our work and \refcite{Vladimirov:2021hdn} is the treatment of the soft sector.
In SCET, soft modes corresponding to momenta $k^\mu \lesssim Q (\lambda, \lambda, \lambda)$ 
are an important dynamic degree of freedom, which ultimately gives rise to the soft function. The soft momentum region is also an important leading region in the Collins-Soper-Sterman approach to factorization~\cite{Collins:2011zzd}. 
Naturally, the soft modes in SCET have overlap with the collinear sectors encoding momenta
$p_n^\mu \lesssim Q (\lambda^2, 1, \lambda)$ and $p_\bn^\mu \lesssim Q (1, \lambda^2, \lambda)$.
As explained in \sec{SCET_review}, modes in SCET are infrared in origin,
and thus extend down to zero momentum, with appropriate zero-bin subtractions
to avoid double counting in the infrared region, contrary to the impression one might get about SCET from reading \refcite{Vladimirov:2021hdn}.
While there is significant overlap between collinear and soft modes,
the region where both $k^+, k^- \sim \lambda$ is not covered by the overlap
of collinear modes. Thus, soft modes can not be fully absorbed into the collinear sectors,
and arise as dynamical modes on their own.
\Refcite{Vladimirov:2021hdn} explicitly excludes such soft modes from their setup;
instead they only consider the overlap of the collinear sectors,
where $p^\mu \sim Q (\lambda^2, \lambda^2, \lambda)$.
Such scalings are not associated with their own background fields,
but instead only used to construct the LP soft function to subtract overlap
between collinear sectors.
In SCET, such offshell modes are classified as Glauber contributions~\cite{Rothstein:2016bsq} and potentially violate factorization.
Indeed, it is well known that the decoupling of this Glauber momentum region is an important step in the proof of factorization theorems~\cite{Collins:2011zzd}. In \refcite{Vladimirov:2021hdn} it is assumed without comment that interactions from this region, such as those with spectator particles, will not spoil factorization in SIDIS beyond LP. In the SCET approach we can make this assumption precise and hence we have stated it explicitly, we assumed that contributions from the leading power Glauber Lagrangian, ${\cal L}_G^{(0)}$ can either be canceled out in the NLP SIDIS observables, or absorbed into the directions of the soft and collinear Wilson lines, just like what occurs for interactions involving ${\cal L}_G^{(0)}$ at LP. 

The treatment of the soft factor has important consequences for TMDs.
It is well known that (unsubtracted) collinear matrix elements are rapidity divergent,
and at LP these divergences can be canceled by absorbing the soft function.
In SCET-based approaches, it is also common to separately rapidity-renormalize
collinear and soft functions. This reflects that in SCET, collinear and soft modes
are independent degrees of freedom,
and thus arise as independent functions at the level of the factorization theorem.%
\footnote{For many observables that fall into the realm of \SCETi, collinear and soft
modes have distinct virtualities $p_{n,\bn}^2 \sim Q^2 \lambda^2, p_s^2 \sim Q^2 \lambda^4$,
in which case this is the natural approach.}
\footnote{This is independent of the motivation to absorb the soft function
into the collinear matrix elements to reduce the number of independent nonperturbative objects
to account for in nonperturbative extractions.}
Although rapidity divergences are multiplicatively renormalizable at LP~\cite{Vladimirov:2017ksc}, this is no longer the case in general for power suppressed terms, as demonstrated by the observed mixing of operators in the rapidity RGE at NNLP in \refcite{Moult:2019vou}. 
\Refcite{Vladimirov:2021hdn} adopts without explicit proof the viewpoint that soft effects at NLP merely occur through  the rapidity counterterm
to the collinear matrix elements at NLP, and by arguing
that the same multiplicative renormalization occurs at NLP, adopt the same LP
soft function also at NLP. 
It is constructed by splitting the $p^\mu \sim Q (\lambda^2, \lambda^2, \lambda)$
fields from the collinear fields and power expanding the resulting operator,
which as a soft field is then evaluated in a vacuum matrix element.
However, they stress themselves that this multiplicative structure most likely does not hold beyond NLP.  Based on the appearance of the same soft function, Ref.~\cite{Vladimirov:2021hdn} draws the same conclusion that we have arrived at here, that the rapidity anomalous dimensions appearing for the TMDs at NLP are the same as at LP. 

In our analysis we have given a full treatment of dynamical soft effects that can potentially arise at NLP, including through insertions of soft gluon field strengths that arise at NLP, and soft derivative operators. In addition there are dynamical effects involving interactions between soft and collinear fields in the subleading power \SCETii Lagrangian, which involve four or more fields and have potential contributions at both ${\cal O}(\lambda^{1/2})$ and ${\cal O}(\lambda)$ which must be considered at NLP. For SIDIS at NLP we have demonstrated explicitly that all non-trivial dynamic soft effects vanish. Our arguments involved a combination of power counting, discrete symmetries that were used to constrain the form of factorized NLP soft functions, boost invariance along the $z$-axis in the back-to-back frame, and the polarization structure of the indices in $W^{\mu\nu}$ that are relevant for the structure functions that start at NLP. The result of this is that we were able to demonstrate that the only soft effects for our NLP results in SIDIS are those arising from the same soft function as at LP, $S(b_T)$, which is defined by a vacuum matrix element of Wilson lines.
This $S(b_T)$ is generated by the momentum sector $p^\mu \lesssim Q(\lambda,\lambda,\lambda)$, just like for the LP soft effects in the classic CSS proof. Furthermore, we have demonstrated that it is possible to absorb this LP soft function into the 
the collinear functions just like at LP, $f \sim B \sqrt{S}$. Here we have a $\sqrt{S}$ as opposed to the $1/\sqrt{S}$ that is more commonly encountered in the literature. (This is because the precise form of this relation depends on the nature of 
the soft zero-bin subtractions, which in our notation are contained with $B$ itself. If we define $B^{(u)}$ as the correlator without soft zero-bin subtractions, then one has $f \sim B^{(u)} \sqrt{S} / S$ in our setup, for many choices of rapidity regulators.
In SCET, this process of absorbing the (physical) soft function
and subtracting the zero-bin region are commonly performed separately.)
While it seems that our result for the absorption of $S$ agrees with \refcite{Vladimirov:2021hdn}, it should be noted that the soft effects under discussion arise from different momentum regions, and that we have done a much more extensive analysis of all possible NLP soft effects to arrive at this conclusion. 
In particular, \refcite{Vladimirov:2021hdn} argued that a derivative of a soft Wilson line or the addition of an extra soft gluon leads to an additional power suppression, whereas we find that such terms in fact already enter at NLP at the operator level, and could only be shown to vanish when factorized for the SIDIS process at NLP.

Finally we remark that \refcite{Vladimirov:2021hdn} has carried out a complete NLO calculation for their analog of our Wilson coefficient $C^{(1)}$, which is given by the formula for $C_2(x_{1,2})$ in their Eq.~(6.15). This is a very useful result for applications of the NLP factorization formula for SIDIS beyond LO.  Furthermore, the explicit calculations done in \refcite{Vladimirov:2021hdn} confirm at one-loop that the same rapidity anomalous dimension appears as at LP, which is consistent with the appearance of only the LP soft function.

\section{Conclusion}
\label{sec:conclusion}

We have studied the factorization of azimuthal asymmetries in the SIDIS process
that first appear at subleading power, i.e.~are suppressed by $\cO(q_T/Q)$ relative
to the leading processes. Working under the assumption that leading power Glauber interactions
do not spoil factorization at NLP, we have used the Soft Collinear Effective Theory
to derive the form that an all orders in $\alpha_s$ factorization formulae for power suppressed hard scattering effects in SIDIS must take.
In addition to deriving and constraining all necessary operators needed for our analysis, we present final explicit factorization formulae for all
SIDIS structure functions at NLP in \secs{results_NLP_full}{results_NLP_full_msp}, including the full spin dependence of the target nucleon.

A summary of the main results achieved by our analysis is as follows
\begin{itemize}
\item A derivation of a minimal basis of all \SCETii hard-scattering operators that enter for observables with two collinear directions at NLP. This included matching contributions from both the hard and hard-collinear regions of momentum space. 
\item A demonstration that subleading power dynamic \SCETii Lagrangians do not contribute to SIDIS at NLP.
\item At NLP there are subleading hard \SCETii 
operators involving soft gluon fields and soft derivatives.
In all cases, we find that these soft contributions vanish for the NLP
structure functions under consideration. In the most difficult case to rule out, involving the $\cB_{s\perp}^{(n_i)\mu}$ operator, this followed after factorization of terms into an NLP vacuum soft matrix
elements, which could then be argued to vanish based in part on the charge conjugation and  parity invariance of the QCD vacuum. Our analysis
shows that new soft contributions will appear at NNLP and beyond. We note that such soft contributions have not been considered in recent independent approaches
to subleading power factorization of TMDs~\cite{Inglis-Whalen:2021bea, Vladimirov:2021hdn}.
\item At NLP we demonstrated that all soft effects are given by the same leading power soft function, $S(b_T)$, and that in all cases it can be absorbed into the leading and subleading power TMDs. 
\item There are three non-zero sources of power corrections to SIDIS at NLP,
namely kinematic corrections from subleading terms in the projectors $P_i^{\mu\nu}$ which determine structure functions $W_i= P_i^{\mu\nu} W_{\mu\nu}$, and hard scattering operators with $\cP_\perp$ and $\cB_{n_i\perp}$ insertions.
The first two can be fully expressed in terms of the LP TMDs,
while the $\cB_{n_i\perp}$ operators give rise to quark-gluon-quark correlators and thus new TMDs that do not appear at LP. These correlators dependend on variable in the following form: $\tilde B_{\cB\, f/N}^\rho(x,\xi,\vec b_T,\mu,\zeta_a)$ for TMD PDFs and $\tilde \cG_{\cB\, h/f}^\rho(z,\xi,\vec b_T,\mu,\zeta_b)$ for TMD FFs. In both cases they depend on two longitudinal momentum fractions, with the appearance of a new variable $\xi$ determining the split of large momentum between gluon and quark fields that appear in the same collinear direction at the hard scale.  
\item The factorization formulae for all spin polarized NLP structure functions is proven to have only two independent hard functions, the same function $\cH^{(0)}(q^+q^-,\mu)$ that already appears at LP, and a single new function $\cH^{(1)}(q^+q^-,\xi,\mu)$ which appears along with the quark-gluon-quark correlators that depend on this same variable $\xi$. These hard functions encode perturbative $\alpha_s$ corrections, and also determine the structure of large logarithms from the invariant mass renormalization group evolution in $\mu$ for the NLP structure functions.
\item The structure of our factorization formula have important implications for the resummation of large logarithms, $\ln(q_T/Q)$ for SIDIS at NLP. In particular, we have demonstrated that the same rapidity anomalous dimension (Collins-Soper equation) governs the evolution in rapidity for all NLP terms. For the evolution in invariant mass governed by $\mu$, some terms inherit the known resummation from LP since they involve the same $\cH^{(0)}$, while there is a new type of evolution for the terms involving $\cH^{(1)}$. Results from the SCET literature~\cite{Beneke:2017ztn, Beneke:2018rbh} imply that the evolution equation for $\cH^{(1)}$ is known at NLL order.
\item  At tree level $\cH^{(1)}$ is independent of $\xi$, and integrating over $\xi$ in the quark-gluon-quark correlators $\tilde B_{\cB\, f/N}^\rho$ and $\tilde \cG_{\cB\, h/f}^\rho$ gives rise to quark-gluon-quark correlators of this form that were studied in previous literature. Accounting for the required absorption of the LP soft function, we find that at tree level our results for the NLP structure functions reduce to exactly the momentum space results given by the parton based analysis of \refcite{Bacchetta:2006tn}. The non-trivial integral over $\xi$ first starts to contribute at NLO.
\item We have presented complete factorization formula for the structure functions $W_{1,2,5,6}^X$, where $X$ denotes possible spin polarizations. These results are given in transverse position space in \sec{results_NLP_full} and in transverse momentum space in \sec{results_NLP_full_msp}. 
\end{itemize}

An important future direction is to fully demonstrate (or not) whether the leading power Glauber effects from the SCET Lagrangian ${\cal L}_G^{(0)}$, which potentially spoil factorization, can be eliminated for the NLP SIDIS analysis in the manner we have assumed.  At LP proving the absence of factorization violating Glauber effects in SIDIS is much simpler than in Drell-Yan~\cite{Collins:2011zzd}, giving us some hope that these effects will indeed not spoil factorization at NLP. Another indication that points in a positive direction for a complete factorization proof for SIDIS at NLP, is the cancellation of various NLP soft contributions demonstrated here. In the CSS approach the soft and Glauber effects are kept combined until a later step in the analysis than in SCET, which may indicate that the absence of soft effects other than through the LP soft function, also indicates that the Glauber contributions at NLP will follow the same pattern for SIDIS as they do at LP. This may imply that the same type of steps used to demonstrate the absence of factorization violating Glauber effects in SIDIS at LP, will also work at NLP for the terms considered here. We leave such investigations to future work. 
 
Another interesting application of our results is to utilize the factorization formula for the azimuthal asymmetry structure functions, $W_{1,2,5,6}^X$, to carry out phenomenological analyses beyond LO. Combining our results with what is known in the literature, such analyses are now possible at LL and NLL orders. 

It should also be evident that the same techniques used here can be adapted to analyze factorization for NLP corrections for spin-polarized TMD observables in Drell-Yan and in $e^+e^-$ annihilation with fragmentation to almost back-to-back hadrons. 

\section*{Acknowledgments}

We thank Johannes Michel, Gherardo Vita, Leonard Gamberg, Aditya Pathak, and Alexey Vladimirov for useful discussions. This work was supported by the U.S.~Department of Energy, Office of Science, Office of Nuclear Physics, from DE-SC0011090 and within the framework of the TMD Topical Collaboration. I.S.~was also supported in part by the Simons Foundation through the Investigator grant 327942. M.E.~was also supported by the Alexander von Humboldt Foundation through a Feodor Lynen Research Fellowship.

\appendix

\section{Coordinate Systems}
\label{app:frames}

In our treatment of SIDIS, wee have considered various reference frames.
Physical observables are defined in the target rest frame following the Trento conventions~\cite{Bacchetta:2004jz},
while the hadronic tensor decomposition is more naturally addressed in the hadronic frame, see also \fig{trento} for an illustration.
The derivation of the factorization is carried using lightcone coordinates.
Here, we collect explicit expressions of the unit vectors defining these different frames,
as well as explicit parameterizations of all particle momenta in these frames.

\subsection{Rest Frame using Trento Conventions}
\label{app:Trento_frame}

We construct the target rest frame in terms of orthornormal unit vectors $n_i^{\prime\,\mu}$ satisfying
\begin{align} \label{eq:unit_vectors}
 n_t^{\prime\,2} = 1 \,,\qquad n_x^{\prime\,2} = n_y^{\prime\,2} = n_z^{\prime\,2} = -1
\,,\qquad
 n'_i \cdot n'_j = 0 \quad {\rm for}~i \ne j
\,.\end{align}
Following the Trento conventions~\cite{Bacchetta:2004jz}, the lepton plane coincides with the $x-z$ plane,
such that $q_z > 0$ and $\pa$ and $\pb$ have positive $x$-component. This uniquely fixes
\begin{align} \label{eq:vectors_R}
 n_t^{\prime\,\mu} &= \frac{\Pa^\mu}{M_N}
\,,\qquad
 n_z^{\prime\,\mu} = \frac{-2 x \Pa^\mu + \gamma^2 q^\mu}{Q \gamma \sqrt{1+\gamma^2}}
\,,\qquad
 n_y^{\prime\,\mu} = \eps^{\mu\nu\rho\sigma} n_{t\,\nu} n_{x\,\rho} n_{z\,\sigma}
\,,\nn\\
 n_x^{\prime\,\mu} &
 =\sqrt{\frac{2}{\eps}} \biggl[ \sqrt{1-\eps} \frac{\pa^\mu}{Q} - x \sqrt{\frac{1+\eps}{1+\gamma^2}} \frac{\Pa^\mu}{Q}
   - \frac{1}{2} \biggl(\sqrt{1-\eps} + \sqrt{\frac{1+\eps}{1+\gamma^2}} \biggr) \frac{q^\mu}{Q} \biggr]
\,,\end{align}
which holds for arbitrary hadron masses, with mass corrections encoded in
$\gamma = 2 x M_N / Q$ and $\eps$ as defined in \eqs{kappa_gamma}{eps}, as well as
\begin{align} \label{eq:kappa}
 \kappa = \frac{\sqrt{1 - (m_{Th} \gamma)^2 / (Q z)^2}}{\sqrt{1+\gamma^2}}
\,,\qquad
 m_{Th}^2 = M_h^2 + \PTb^2
\,.\end{align}
In this coordinate system, the lepton and hadron momenta are parameterized as
\begin{align} \label{eq:momenta_Trento}
 \pa^\mu &= \frac{Q}{2\gamma} \Bigl(
               \frac{2}{y} \,,\,
               \sqrt{\frac{2 \eps}{1-\eps}} \gamma \,,\,
               0 \,,\,
               \sqrt{\frac{1+\eps}{1-\eps}} + \sqrt{1+\gamma^2} \Bigr)_T
\,,\nn\\
 \pb^\mu &= \frac{Q}{2\gamma} \Bigl(
               \frac{2(1-y)}{y} \,,\,
               \sqrt{\frac{2 \eps}{1-\eps}} \gamma \,,\,
               0 \,,\,
               \sqrt{\frac{1+\eps}{1-\eps}} - \sqrt{1+\gamma^2} \Bigr)_T
\,,\nn\\
 \Pa^\mu &=  M_N \bigl(1, 0, 0, 0\bigr)_T
\,,\nn\\
 \Pb^\mu &= \Bigl( \frac{Q z}{\gamma} \,,\, P_{hT} \cos\phiH \,,\, P_{hT} \sin\phiH \,,\,  \frac{Q z}{\gamma} \kappa \sqrt{1+\gamma^2} \Bigr)_T
\,,\nn\\
 q^\mu &= \frac{Q}{\gamma} \Bigl(1, 0, 0, \sqrt{1+\gamma^2}\Bigr)_T
\,,\end{align}
where the subscript $T$ signals using the rest frame in the Trento conventions.

\subsection{Hadronic Breit Frame}
\label{app:Breit_frame}

The hadronic Breit frame is defined in terms of orthornormal unit vectors $\tilde n_i^\mu$ obeying \eq{unit_vectors}.
As defined in \sec{tensor_decomposition} and illustrated in \fig{trento}, the frame is constructed such that
$q^\mu = (0,0,0,-Q)$ and $\PTb$ is aligned along the $x$ axis. This uniquely fixes
\begin{align} \label{eq:vectors_B}
 \tilde n_t^\mu &= \frac{2x \Pa^\mu + q^\mu}{Q \sqrt{1 + \gamma^2}}
\,,\qquad
 \tilde n_y^\mu = \eps^{\mu\nu\rho\sigma} \tilde n_{t\,\nu} \tilde n_{x\,\rho} \tilde n_{z\,\sigma}
\,,\qquad
 \tilde n_z^\mu = -\frac{q^\mu}{Q}
\,,\nn\\
 \tilde n_x^\mu &= \frac{1}{P_{hT}}  \Bigl( \Pb^\mu - z \kappa q^\mu + 2 x z \frac{\kappa - 1}{\gamma^2} \Pa^\mu \Bigr)
\,,\end{align}
which is valid for arbitrary hadron masses, with $\kappa$ given by \eq{kappa}.
In this coordinate system, the lepton and hadron momenta are parameterized as
\begin{align} \label{eq:momenta_Breit_had}
 \pa^\mu &= \frac{Q}{2} \Bigl( \sqrt{\frac{1+\eps}{1-\eps}} \,,\,
                      \sqrt{\frac{2\eps}{1-\eps}} \cos\phiH \,,\,
                      \sqrt{\frac{2\eps}{1-\eps}} \sin\phiH \,,\,
                      - 1\Bigr)_B
\,,\nn\\
 \pb^\mu &= \frac{Q}{2} \Bigl( \sqrt{\frac{1+\eps}{1-\eps}} \,,\,
                      \sqrt{\frac{2\eps}{1-\eps}} \cos\phiH \,,\,
                      \sqrt{\frac{2\eps}{1-\eps}} \sin\phiH \,,\,
                      + 1\Bigr)_B
\,,\nn\\
 \Pa^\mu &=  \frac{Q}{2x} \Bigl( \sqrt{1 + \gamma^2} \,,\, 0 \,,\, 0 \,, +1 \Bigr)_B
\,,\nn\\
 \Pb^\mu &
 = \bigl( m_{Th} \cosh Y_h \,, P_{hT} \,, 0 \,, m_{Th}  \sinh Y_h \bigr)_B
\,,\nn\\
 q^\mu &= (0, 0, 0, -Q)_B
\,,\end{align}
where the rapidity of the outgoing hadron is defined as
\begin{align}
 \sinh Y_h = \frac{Q z}{m_{Th} \gamma^2} \bigl[1 - (1+\gamma^2) \kappa\bigr]
\,.\end{align}
The rest frame and Breit frame are related through a longitudinal Lorentz boost, a rotation about the $y$ axis
(to revert the direction of the $z$ axis), and a rotation about the $z$ axis, given by
\begin{align} \label{eq:relation_unit_vectors}
 \begin{pmatrix} n'_t \\ n'_z \end{pmatrix}
 &= \frac{1}{\gamma}
 \begin{pmatrix} \sqrt{1 + \gamma^2} & 1 \\
                 -1  & -\sqrt{1 + \gamma^2} \\ \end{pmatrix}
 \begin{pmatrix} \tilde n_t \\ \tilde n_z \end{pmatrix}
\,,\quad
 \begin{pmatrix} n'_x \\  n'_y \end{pmatrix}
 = \begin{pmatrix} \cos\phiH & \phantom{-} \sin\phiH \\
                    \sin\phiH & -\cos\phiH \\ \end{pmatrix}
 \begin{pmatrix} \tilde n_x \\ \tilde n_y \end{pmatrix}
\,.\end{align}
In particular, for $\phiH = 0$, the $x$ axis of both frames coincide,
while the $y$ axis has opposite signs reflecting the reversion of the $z$ axis.

\section{Fourier Transformation}
\label{app:fourier}

Our convention for the Fourier transform and its inverse are
\begin{alignat}{2} \label{eq:FT_1}
 \tilde f(\bt) &= \int\df^2\kt e^{-\img \kt \cdot \bt} f(\kt)
\,,\quad
 ~f(\kt) &&= \int\frac{\df^2\bt}{(2\pi)^2} e^{+\img \kt \cdot \bt} \tilde f(\bt)
\,,\nn\\
 \tilde D(\bt) &= \int\df^2\pt e^{+\img \pt \cdot \bt} D(\kt)
\,,\quad
 D(\pt) &&= \int\frac{\df^2\bt}{(2\pi)^2} e^{-\img \pt \cdot \bt} \tilde D(\bt)
\,.\end{alignat}
Here, $\bt$, $\kt$ and $\pt$ are transverse vectors in Euclidean space,
i.e.~ $b_\perp^2 = -\bt^2 = -b_T^2$, $\bt \cdot \kt = - b_T \cdot k_\perp$, etc.
In this appendix, for clarity we denote the Fourier transform by a tilde,
while in the main body of this paper we will drop the tilde
as it is used for the subleading TMDs, and the distinction is clear by the arguments.
In the first line of \eq{FT_1}, $f(\kt)$ is a TMDPDF, while in the second line $D(\pt)$ is a TMDFF.
As discussed in \sec{results_LP}, they are defined with opposite conventions for their Fourier phases.
In the following, we only show explicit results for the Fourier transform of the TMPDF,
as the corresponding results for the TMDFF can be easily obtained by letting $\img \to -\img$ and $\Ma\to\Mb$.

Often, we encounter functions that only depend on the magnitude of $\bt$ or $\kt$,
i.e.~$f(\kt) = f(k_T)$ with $k_T = |\kt|$. In this case, \eq{FT_1} simplifies to
\begin{align} \label{eq:FT_2}
 \tilde f(b_T) = 2 \pi \int_0^\infty \df k_T \, k_T J_0(b_T k_T) f(k_T)
\,,\quad
 f(k_T) = \frac{1}{2 \pi} \int_0^\infty \df b_T \, b_T J_0(b_T k_T) \tilde f(b_T)
\,,\end{align}
where $J_n(x)$ is the $n$-th order Bessel function of first kind.

In addition, we frequently require Fourier transforms of functions of the form
\begin{align} \label{eq:FT_3}
 \int\df^2\kt e^{-\img \kt \cdot \bt} (k_\perp^\mu \cdots k_\perp^\nu) f(k_T) &
 = (-\img \partial^\mu) \cdots (-\img \partial^\nu) f(b_T)
 \nn\\&
 = (-\img \partial^\mu) \cdots (-\img \partial^\nu) 2 \pi \int_0^\infty \df k_T \, k_T J_0(b_T k_T) f(k_T)
\,.\end{align}
Here, $\partial^\mu = \partial / \partial b_{\perp \mu}$, and it is crucial
to note that $k_\perp^\mu$ and $b_\perp^\mu$ are vectors in Minkowski space,
as opposed to the Euclidean vectors $\kt$ and $\bt$.
Using \eq{FT_3} together with the Bessel function identity
\begin{align}
 \frac{\df}{\df z} z^{-m} J_m(z) = -z^{-m} J_{m+1}(z)
\,,\end{align}
one can easily obtain explicit results for the required Fourier transforms.
For example, one obtains
\begin{alignat}{2} \label{eq:FT_4}
 \int\df^2\kt \, e^{-\img \kt \cdot \bt} \, \frac{k_\perp^\mu}{k_T} f(k_T)
 & = (-\img) \frac{b_T^\mu}{b_T} \ 2\pi\int_0^\infty\!\!\! \df k_T \, k_T \, J_1(b_T k_T) f(k_T)
\,, \\ \nn
 \int\df^2\kt \, e^{-\img \kt \cdot \bt} \biggl(\frac{g_T^{\mu\nu}}{2} + \frac{k_\perp^\mu k_\perp^\nu}{k_T^2} \biggr) f(k_T)
 & = (-\img)^2 \biggl(\frac{g_T^{\mu\nu}}{2}  +  \frac{b_T^\mu b_T^\nu}{b_T^2} \biggr)
      \, 2\pi\! \int_0^\infty\!\!\! \df k_T \, k_T \, J_2(b_T k_T) \, f(k_T)
.
\end{alignat}
These relations match the dimensionless prefactors, which in the second line is also traceless,
onto corresponding prefactors in Fourier space. The integrals themselves have a form very
similar to \eq{FT_2}, up to replacing $J_0$ by $J_1$ and $J_2$, respectively.

In practice, due to the traditional normalization of correlators as $k_\perp^\mu/\Ma$
rather than $k_\perp^\mu/k_T$, we need slightly modified transforms.
To absorb these mass factors, following \refcite{Boer:2011xd}
we define the derivative of the Fourier transform as
\begin{align} \label{eq:FT_derivative}
 \tilde f^{(n)}(b_T) &
= n! \left(\frac{-1}{\Ma^2 b_T} \frac{\partial}{\partial b_T} \right)^n \tilde f(b_T)
\nn \\ &
 = \frac{2\pi\, n!}{(b_T \Ma)^n} \int_0^\infty \df k_T \, k_T \left(\frac{k_T}{\Ma}\right)^n J_n(b_T k_T) \, f(k_T)
\,.\end{align}
Note that $n!$ factor is included for historical reasons for consistency
with moments of TMD distributions with respect to $k_T$~\cite{Boer:2011xd}.
The inverse of \eq{FT_derivative} is given by
\begin{align} \label{eq:FT_derivative_inv}
 f(k_T) &
 = \frac{\Ma^{2n}}{ 2 \pi\, n!} \int_0^\infty \df b_T \, b_T  \left(\frac{b_T}{k_T}\right)^n J_n(b_T k_T) \, \tilde f^{(n)}(b_T)
\,,\end{align}
which follows from the orthogonality relation of Bessel functions,
\begin{align}
 \int_0^\infty \df b_T \, b_T \, J_n(k_T b_T) \, J_n(p'_T b_T) = \frac{1}{k_T} \delta(k_T-p'_T)
\,.\end{align}
Using \eqs{FT_3}{FT_derivative}, the Fourier transforms required in this paper are obtained as
\begin{align} \label{eq:app:FT_1}
 \int\df^2\kt \, e^{-\img \bt \cdot \kt} \frac{k_\perp^\mu}{\Ma} f(k_T)
 & = -\img\Ma b_\perp^\mu \tilde f^{(1)}(b_T)
\,,\nn\\
 \int\df^2\kt \, e^{-\img \bt \cdot \kt} \frac{k_\perp^\mu k_\perp^\nu}{\Ma^2} f(k_T)
 &= \frac{(-\img)^2}{2}\Ma^2\, b_\perp^\mu b_\perp^\nu \, \tilde f^{(2)}(b_T)
    +(-\img)^2 g_\perp^{\mu\nu} \tilde f^{(1)}(b_T)
\,,\nn\\
 \int\df^2\kt \, e^{-\img \bt \cdot \kt} \frac{k_\perp^\mu k_\perp^\nu k_\perp^\rho}{\Ma^3} f(k_T)
 &= \frac{(-\img)^3}{6}\Ma^3\, b_\perp^\mu b_\perp^\nu b_\perp^\rho \, \tilde f^{(3)}(b_T)
 \nn\\&\quad
  +\frac{(-\img)^3}{2} \Ma \left(g_\perp^{\mu\nu}b_\perp^\rho+g_\perp^{\nu\rho}b_\perp^\mu+g_\perp^{\rho\mu}b_\perp^\nu\right) \tilde f^{(2)}(b_T)
\,.\end{align}
The following traceless relation is also useful,
\begin{align} \label{eq:app:FT_2}
 \int\df^2\kt \, e^{-\img \bt \cdot \kt} \frac{k_T^2}{\Ma^2}
              \left(\frac12 g_\perp^{\mu\nu} + \frac{k_\perp^\mu k_\perp^\nu}{k_T^2} \right) f(k_T) &
 = - \frac12 \Ma^2 b_T^2 \left(  \frac12 g_\perp^{\mu\nu} + \frac{b_\perp^\mu b_\perp^\nu}{b_T^2} \right) \tilde f^{(2)}(b_T)
\,.\end{align}
At subleading power, we also encounter Fourier transformations of the form
\begin{align} \label{eq:app:FT_3}
 f^{(0')}(b_T) &\equiv
 \int\df^2\kt \,  e^{-\img \bt \cdot \kt} \frac{k_T^2}{\Ma^2} f(k_T)
 = 2 \pi \int_0^\infty \df k_T \, k_T J_0(b_T k_T) \frac{k_T^2}{\Ma^2} f(k_T)
 \nn\\&
 = \frac{(-\img)^2}{2}\Ma^2\, b_T^2 \tilde f^{(2)}(b_T)
     - 2 (-\img)^2  \tilde f^{(1)}(b_T)
\,,\nn\\
 -\img \Ma b_\perp^\mu f^{(1')}(b_T) &\equiv
 \int\df^2\kt \,  e^{-\img \bt \cdot \kt} \frac{k_\perp^\mu k_T^2}{\Ma^3} f(k_T)
 = -\img \frac{b_\perp^\mu}{b_T}  2\pi \int_0^\infty \df k_T \, k_T J_1(b_T k_T) \frac{k_T^3}{\Ma^3} f(k_T)
 \nn\\&
 = -\img \Ma b_\perp^\mu \Bigl[ \frac{(-\img)^2}{6}\Ma^2\, b_T^2 \, \tilde f^{(3)}(b_T)
  - 2 (-\img)^2 \tilde f^{(2)}(b_T) \Bigr]
\,.\end{align}
The expressions in terms of the $\tilde f^{(n)}$ can easily be obtained from appropriate contractions of \eq{FT_1}.

\subsection*{Fourier Transforms and Convolutions in Momentum Space}

The structure functions require evaluations of the Fourier transform in \eq{cF},
\begin{align} \label{eq:cF_rep}
 \cF[\cH\, g^{(n)} \, D^{(m)}] &
 = 2 z \sum_f \cH_f(q^+ q^-) \int\frac{\df^2\bt}{(2\pi)^2} e^{-\img \qt \cdot \bt}
   (-\img\Ma b_T)^n (\img\Mb b_T)^m \cos[(n+m) \varphi)]
   \nn\\&\hspace{3.5cm}\times
   \, g_f^{(n)}(x,b_T) \, D_f^{(m)}(z,b_T) + (f\to\bar f)
\,.\end{align}
Here, we work in a frame where $\qt = q_T (-1, 0)$ and $\bt = b_T (\cos\varphi, \sin\varphi)$.
Using the covariant vector $\hat h_\perp = q_\perp^\mu/q_T$, we can rewrite the $\cos\varphi$
dependence as
\begin{align}
 \cos(\varphi) &
 = \frac{b_\perp^\mu \hat h_{\perp\,\mu}}{b_T}
\,,\nn\\
 \cos(2\varphi) &
 = \frac{2 b_\perp^\mu b_\perp^\nu}{b_T^2} \Bigl( \frac{g_{\perp\,\mu\nu}}{2} + \hat h_{\perp\,\mu} \hat h_{\perp\,\nu} \Bigr)
\,,\end{align}
and likewise for higher terms.
Using this to replace the $\cos$ term in \eq{cF_rep}, we can reduce the integrand
to the particular combinations in \eq{app:FT_1}, after which one can easily read off
the momentum-space result, which we express using \eq{cF_tilde},
\begin{align} \label{eq:cF_tilde_rep}
 \tilde\cF[\w\, \cH\, g\, D] &
 = 2z \sum_f \cH_f(q^2) \int\!\df^2\kt\, \df^2\pt \, \delta^{(2)}\bigl(\qt+\kt-\pt\bigr)
  \nn\\ &\qquad\quad
  \times\w(\vec k_T, \vec k_T)\, g_f(x, k_T) \, D_f(z,p_T)+ (f\to\bar f)
\,.\end{align}
The expressions required at LP are
\begin{subequations} \label{eq:cF_sol_LP}
\begin{align}
 \cF[\cH\, g \, D] &
 = \tilde\cF[\cH g D]
\,,\\
 \cF[\cH\, g^{(1)} \, D^{(0)}] &
 = \tilde\cF\biggl[\frac{\hat h_\perp \cdot k_\perp}{\Ma} \cH g D\biggr]
 \nn\\&
 = \tilde\cF\biggl[\frac{k_{Tx}}{\Ma} \cH g D\biggr]
\,,\\
 \cF[\cH\, g^{(0)} \, D^{(1)}] &
 = \tilde\cF\biggl[\frac{\hat h_\perp \cdot p_\perp}{\Mb} \cH g D\biggr]
 \nn\\&
 = \tilde\cF\biggl[\frac{p_{Tx}}{\Mb} \cH g D\biggr]
\,,\\
 \cF[\cH\, g^{(1)} \, D^{(1)}] &
 = \tilde\cF\biggl[ \frac{2 (k_\perp{\cdot}\hat h_\perp)(p_\perp{\cdot}\hat h_\perp) + k_\perp \cdot p_\perp}{\Ma \Mb} \cH g D \biggr]
 \nn\\&
 = \tilde\cF\biggl[ \frac{2 k_{Tx} p_{Tx} - \kt \cdot \pt}{\Ma \Mb} \cH g D \biggr]
\,,\\
 \cF[\cH \, g^{(2)} D^{(1)}] &
 = \tilde\cF \left[\frac{4 (\hat h_\perp \cdot k_\perp) (k_\perp \cdot p_\perp)  + 2k_\perp^2 (\hat h_\perp \cdot p_\perp) - 8 (\hat h_\perp \cdot k_\perp)^2 (\hat h_\perp \cdot p_\perp)}{\Ma^2\Mb}\, \cH g D\right]
 \nn\\&
 = \tilde\cF \left[\frac{-4 k_{Tx}\, (\vec k_T\cdot \vec p_T) -2 k_T^2\, p_{Tx} + 8 k_{Tx}^2\, p_{Tx}}{\Ma^2\Mb}\, \cH g D\right]
\,.\end{align}
\end{subequations}
Each result is given in two forms. First, in a manifestly covariation form
with the unit vector $\hat h_\perp = q_\perp^\mu/q_T$, and secondly by making use
that in the factorization frame $\hat h_\perp = (0, -1, 0, 0)$ such that $\hat h \cdot k_\perp = k_{Tx}$.
At NLP, we need the additional combinations
\begin{subequations} \label{eq:cF_sol_NLP}
\begin{align}
 \cF[\cH\, g^{(1)} \, D^{(0')}] &
 = \tilde\cF \biggl[ \frac{-(\hat h \cdot k_\perp) p_\perp^2}{\Ma \Mb^2} \cH  g D \biggr]
 \nn\\&
 = \tilde\cF \biggl[ \frac{k_{Tx} p_T^2}{\Ma \Mb^2} \cH  g D \biggr]
\,,\\
 \cF[\cH\, g^{(0')} \, D^{(1)}] &
 = \tilde\cF \biggl[ \frac{-k_\perp^2 (\hat h \cdot p_\perp)}{\Ma^2 \Mb} \cH  g D \biggr]
 \nn\\&
 = \tilde\cF \biggl[ \frac{k_T^2 \, p_{Tx}}{\Ma^2 \Mb} \cH  g D \biggr]
\,,\\
 \cF[\cH\, g^{(1)} \, D^{(1')}] &
 = \cF\Biggl[  -\frac{2 (h_\perp \cdot k_\perp) (\hat h_\perp \cdot p_\perp) + (k_\perp \cdot p_\perp)}{\Ma \Mb} \frac{p_\perp^2}{\Mb^2} \cH g D \biggr]
 \nn\\&
 = \cF\Biggl[ \frac{2 k_{Tx} p_{Tx} - \kt \cdot \pt }{\Ma \Mb} \frac{p_T^2}{\Mb^2} \cH g D \biggr]
\,,\\
 \cF[\cH\, g^{(2)} \, D] &
 = \cF\biggl[ \frac{4 (\hat h_\perp \cdot k_\perp)^2 + 2 k_\perp^2}{\Ma^2} \cH g D \biggr]
 \nn\\&
 = \cF\biggl[ \frac{4 k_{Tx}^2 - 2 k_T^2}{\Ma^2} \cH g D \biggr]
 = \cF\biggl[ \frac{2 (k_{Tx}^2 - k_{Ty}^2)}{\Ma^2} \cH g D \biggr]
\,,\\
 \cF[\cH\, g^{(2)} \, D^{(0')}] &
 = \cF\biggl[ \frac{4 (\hat h_\perp \cdot k_\perp)^2 + 2 k_\perp^2}{\Ma^2} \frac{-p_\perp^2}{\Mb^2} \cH g D \biggr]
 \nn\\&
 = \cF\biggl[ \frac{4 k_{Tx}^2 - 2 k_T^2}{\Ma^2} \frac{p_T^2}{\Mb^2}  \cH g D \biggr]
 = \cF\biggl[ \frac{2 (k_{Tx}^2 - k_{Ty}^2)}{\Ma^2}  \frac{p_T^2}{\Mb^2} \cH g D \biggr]
\,,\\
 \cF[\cH\, g^{(2)} \, D^{(1')}] &
 = \tilde\cF \left[\frac{4 (\hat h_\perp \cdot k_\perp) (k_\perp \cdot p_\perp)  + 2k_\perp^2 (\hat h_\perp \cdot p_\perp) - 8 (\hat h_\perp \cdot k_\perp)^2 (\hat h_\perp \cdot p_\perp)}{\Ma^2\Mb} \frac{-p_\perp^2}{\Mb^2} \cH g D\right]
 \nn\\&
 = \tilde\cF \left[\frac{-4 k_{Tx}\, (\vec k_T\cdot \vec p_T) -2 k_T^2\, p_{Tx} + 8 k_{Tx}^2\, p_{Tx}}{\Ma^2\Mb} \frac{p_T^2}{\Mb^2}\, \cH g D\right]
\,.\end{align}
\end{subequations}
In addition, we also require expressions for the $\cF'$ transform
defined in \eq{cFp} as
\begin{align} \label{eq:cFp_rep}
  \cF'[\cH\, g^{(n)} \, D^{(m)}] &
  = 2 z \sum_f \cH_f(q^+ q^-) \int\frac{\df^2\bt}{(2\pi)^2} e^{-\img \qt \cdot \bt}
    (-\img\Ma b_T)^n (\img\Mb b_T)^m
    \nn\\&\hspace{3.5cm}\times
    \, g_f^{(n)}(x,b_T) \, D_f^{(m)}(z,b_T) + (f\to\bar f)
 \,.\end{align}
Specifically, we encounter the transformations
\begin{subequations} \label{eq:cFp_sol_NLP}
\begin{alignat}{2}
 \cF'[\cH g D] &
 = \cF[\cH g D]
 &&= \tilde\cF[\cH g D]
\,,\\
 \cF'[\cH g^{(0')} D] &
 = \cF[\cH g^{(0')} D]
 &&= \tilde\cF\biggl[ \frac{k_T^2}{\Ma^2} \cH g D \biggr]
\,,\\
 \cF'[\cH g D^{(0')}] &
 = \cF'[\cH g D^{(0')}]
 &&= \tilde\cF\biggl[ \frac{p_T^2}{\Ma^2} \cH g D \biggr]
\,,\\
 \cF'[\cH g^{(1)} D^{(1)}] &
 = \tilde\cF\biggl[ \frac{- k_\perp \cdot p_\perp}{\Ma \Mb} \cH g D\biggr]
 &&= \tilde\cF\biggl[ \frac{\kt \cdot \pt}{\Ma \Mb} \cH g D\biggr]
\,.\end{alignat}
\end{subequations}

\section{Transverse Gauge Links at LP and NLP}
\label{app:Twilsonlines}

In the main text, we have derived factorization formula at NLP for SIDIS where the various TMD PDFs, TMD FFs and soft functions were considered across any covariant gauge, which are gauges where the gluon fields vanish at infinity. This avoided the discussion of Wilson lines in the transverse direction at light-cone infinity that are needed for full gauge invariance of various definitions, and are of crucial importance in light-cone gauge~\cite{Ji:2002aa}.
In order to modify the definitions so that they are gauge invariant across all  gauges, it is natural to add gauge links connecting the points at lightcone infinity separated in the transverse direction. For the leading power results discussed in \sec{factorization_LP} this gives a staple shaped Wilson line path $\sqsupset$ with ends attached to the quark fields for the matrix elements $\hat B_{f/N}^{\beta\beta'}$, two links that extend the Wilson lines to infinity in the transverse direction for the two matrix elements in $\hat \cG_{h/f}^{\alpha\alpha'}$, in the form $\halfstaple$ and $\halfstapleu$ (which would combine to $\sqsupset$ if they were in the same matrix elements), and a closed six sided Wilson loop $\leftsoftstaple[1.5]$ for the soft function $S(b_T)$.  See \refcite{Collins:2011zzd} for a review.

We can justify the addition of the extra gauge links by starting from the factorized results in covariant gauge.
Since here the gauge fields vanish at infinity, 
all the TMD functions are identical to their original versions after adding the transverse connecting links (and no matter which specific path we choose).
Since they are now manifestly gauge invariant even at light-cone infinity, they are invariant even for gauges whose gauge field does not vanish at infinity, as long as we include a suitable regulator to deal with the issues rapidity/lightcone divergences consistently.
Although different choices for the path of the transverse Wilson lines should be equivalent, the most convenient one (and the one considered in literature) involves straight transverse Wilson lines. We can modify the LP definitions in \eq{def_B_hat} with these transverse Wilson lines inserted to give
\begin{align} \label{eq:def_B_hat_trans}
  \hat B_{f/N}^{\beta\beta'}(x, \bt) &
  = \theta(\w_a) \, \braket{N | \bar\chi_{n}^{\beta'}(b_\perp)\, T_n(b_\perp,0_\perp)\, \chi_{n,\w_a}^{\beta}(0) | N}
  \,,\nn\\
   \hat \cG_{h/f}^{\alpha'\alpha}(z, \bt) &
   = \frac{\theta(\w_b)}{2 z N_c} \SumInt_{X_\bn} \tr \braket{0 | T_\bn(\infty,b_\perp) \chi_{\bn}^{\alpha'}(b_\perp)\,  \,| h, X_\bn} \braket{h, X_\bn |\bar{\chi}_{\bn,-\w_b}^{\alpha}(0) T_\bn(0,\infty) | 0}
  \,,\nn\\
   S(b_T, b_s^+ b_s^-) &
   = \frac{1}{N_c} \tr \braket{0 | T_n^s(0,b_\perp)\,  \bigl( S_n^\dagger S_{\bn} \bigr)(b_s^+, b_s^-, b_\perp)\,T_\bn^s(b_\perp,0)\, \bigl(S_{\bn}^\dagger S_{n}\bigr)(0)| 0}
  \,,\end{align}
where $T_{n_i}(a_\perp,b_\perp)$ and $T_{n_i}^s(a_\perp,b_\perp)$ are collinear and soft transverse gauge links connecting two points at infinity, $\infty n_i^\mu+a_\perp^\mu$ and $\infty n_i^\mu+b_\perp^\mu$.

We may also want to consider such modification at the Lagrangian level, since for example otherwise the hard scattering operators are not gauge invariant unless we assume that the gauge fields vanish at infinity. This approach has been taken in Refs.~\cite{Idilbi:2010im,Garcia-Echevarria:2011ewi,GarciaEchevarria:2011rb}. There should be some gauge link connecting the $n_i$ (for $S_{n_i}$) and $\bn_i$ (for $W_{n_i}$) infinities ($i=1,2$), and we should make sure that the modified operator is invariant for both collinear and soft gauge transformations. 

A possible choice for this construction would be that, we assign a reference point at infinity, and make it invariant for all kinds of gauge transformations that we are considering ($n_i$ collinear and soft gauge transformations, which in SCET can be done independently).
It can be ensured that there is no transformation at this special point by using a global SU(3) transformation. We then connect this point to the endpoint of $W_{n_i}$ and $S_{n_i}$ with   Wilson lines $Z_{n_i}$ involving $n_i$-collinear gauge fields, and Wilson lines $U_{n_i}$ involving soft gauge fields,
\begin{align} \label{eq:LPtranslines}
  \bar\chi_{n_2} [S^\dagger_{n_2} S_{n_1}]  \chi_{n_1}\to \bar\chi_{n_2}Z_{n_2} [U^{\dagger}_{n_2} S^\dagger_{n_2} S_{n_1}U_{n_1} ] Z_{n_1}^\dagger \chi_{n_1}
\end{align}
Using the same logic as in the previous paragraph, the choices of path to the reference point does not matter and they should all be equivalent. 
We can then choose a path consistent with the choice of the transverse gauge link in the previous paragraph, so that for example  $Z_{n}(b_\perp)=T_n(b_\perp,\infty)$ and $Z_{n}^\dagger(0_\perp)=T_n(\infty,0_\perp)$, which give the proper Wilson lines for the fragmentation matrix element and together combine to $T_n(b_\perp,0_\perp)$ for the staple needed in the TMD PDF and the closed Wilson loop needed for the soft function,
\begin{align}
  Z_{n}(b_\perp) Z_{n}^\dagger(0_\perp) &= T_n(b_\perp,0_\perp)\,,
  & U_{n}(b_\perp) U_{n}^\dagger(0_\perp) &= T_n^s(b_\perp,0_\perp)\,.
\end{align}
The analysis in \refcite{GarciaEchevarria:2011rb} corresponds to the choice of $\infty n^\mu +\infty a_\perp^\mu$ as the reference point for these steps, and it was argued there that the choice of $a_\perp^\mu$ did not matter.

We now consider the generalization of this discussion of transverse Wilson lines to the operators that give non-zero contributions at NLP. 
Generalizing the LP analysis of transverse lines to the $\cB_{n_i\perp}$ operators given in \eq{Bf} is fairly straightforward. Each of the three $\omega_i$ momenta appearing in this operator can be written as Fourier transforms of fields at light-cone positions $b_i^+$ in position space. The quark and gluon fields that are next to each other and in the same $n_i$ collinear direction, get connected by a finite length Wilson line from (for eg.) $b_1^+$ to $b_3^+$. Thus the only transverse lines that are needed, occur in the parts of the operator that are next to the $S^\dagger_{\bn_1} S_{n_1}$, just like in the LP operator in \eq{LPtranslines}.  For this reason, the exact same type of modifications end up holding for the NLP quark-gluon-quark correlators, giving
\begin{align} \label{eq:def_Bb_hat_trans}
 \hat B_{\cB\,f/N}^{\rho \beta \beta'}(x, \xi, \bt) 
 &= \theta(\w_a)
   \MAe{N}{ \bar\chi_{\na}^{\beta'}(b_\perp) \,T_n(b_\perp,0_\perp)\, g{\cB}_{\na\perp,\,-\xi\w_a}^{\rho}(0)
            \,\chi_{\na,\,(1-\xi)\w_a}^{\beta}(0)}{N}
\,,\nn\\
 \hat \cG_{\cB \, h/f}^{\rho\,\alpha'\alpha}(z,\xi,\bt) 
& = \frac{1}{2z N_c} \theta(\w_b)
   \SumInt_{X_\bn} \tr
  \MAe{0}{ Z_\bn^\dagger(b_\perp)\chi_{\nb}^{\alpha'}(b_\perp)}{h,X_\bn}
   \nn\\&\qquad\times
   \MAe{h,X_\bn}{\bar\chi_{\nb,\,-(1-\xi)\w_b}^{\alpha}(0)\,g{\cB}_{\nb\perp,\,\xi\w_b}^{\rho}(0) Z_\bn(0_\perp)}{0}
\,.
\end{align}

For the case of $\cP_\perp$ operator contribution, we note  that $\cP_\perp$ must act on a gauge invariant object, so (suppressing Dirac matrices) we can modify the $\cP_\perp$ operators as
\begin{equation}
  \bar\chi_{n_2} [S^\dagger_{n_2} S_{n_1}]\,  \cP_{n_1\perp}\, \chi_{n_1}\to \bar\chi_{n_2}Z_{n_2} [U^{\dagger}_{n_2} S^\dagger_{n_2} S_{n_1}U_{n_1} ]\,  \cP_{n_1\perp}\, Z_{n_1}^\dagger \chi_{n_1}\,,
\end{equation}
with an analogous result for the term involving $\cP_{n_2\perp}^\dagger$.
Then, the modified $\hat B_{\cP\, f/N}$ becomes
\begin{align}  \label{eq:def_P_hat_trans}
  \hat B_{\cP f/N}^{\beta\beta'}(x,\bt)
  &\equiv\,\frac{\theta({\w_a})}{2Q} \biggl\{\MAe{N}{\bar\chi_{\na}^{\beta'}(b_\perp^\mu) Z_n(b_\perp)\,
  \left[\slashed\cP_\perp\slashed\bn\, Z_n^\dagger(0_\perp) \chi_{\na,\w_a}(0)\right]^{\beta}}{N}
  \nn\\&\qquad\qquad
  +\MAe{N}{\left[\bar\chi_{\na}(b_\perp^\mu)\,Z_n(b_\perp)\slashed\bn\,\slashed\cP_\perp^\dagger\right]^{\beta'} \,Z_n^\dagger(0_\perp)
  \chi_{\na,\w_a}^{\beta}(0)}{N}\biggr\}
  \nn \\ 
  &=\,\frac{-\img}{2Q}\frac{\partial}{\partial b_\perp^\rho}
  \left[ \gamma_\perp^\rho\, \slashed\bn\,,\,\hat B_{f/N}(x,\bt) \right]^{\beta\beta'}\,.
\end{align}
The second equality holds since $\cP_\perp$ acts like a derivative to the 2-dimensional transverse local points at $0_\perp$ or $b_\perp$. After rewriting $\cP_\perp$ as a derivative, it can be pulled out, and then $Z_n(b_\perp)$ and $Z_n(0_\perp)$ combine to give the $T_n(b_\perp,0_\perp)$ that appears in $\hat B_{f/N}$, so it is the $\hat B_{f/N}$ with the transverse link that appears on the last line of \eq{def_P_hat_trans}. Thus we find that relation between NLP and LP TMD PDFs in \eq{def_BP_f_b} is still valid in the presence of transverse Wilson lines. A similar story also applies to $\hat \cG_{\cP h/f}$ with transverse lines, and its relation to $\hat\cG_{h/f}$ with transverse lines, 
\begin{align} \label{eq:def_GP_f_trans}
 \hat\cG_{\cP\, h/f}^{\alpha'\alpha}(z,\bt) &
 \equiv \frac{\theta(\w_b)}{4z N_c Q} \SumInt_{X_\bn} \biggl\{
   \MAe{0}{\left[\slashed\cP_\perp\slashed n\, Z_\bn^\dagger(b_\perp) \chi_{\nb}(b_\perp)\right]^{\alpha'}}{h,X_\bn}
   \MAe{h,X_\bn}{\bar\chi_{\nb,-\w_b}^{\alpha}(0)Z_\bn(0_\perp)}{0}
 \nn\\&\hspace{2cm}
   + \MAe{0}{Z_\bn(b_\perp)\chi_{\nb}^{\alpha'}(b_\perp)}{h,X_\bn}
     \MAe{h,X_\bn}{\left[\bar\chi_{\nb,-\w_b}(0)\,Z_\bn^\dagger(0_\perp) \slashed n\,\slashed\cP_\perp^\dagger\right]^{\alpha}}{0}
 \biggr\}
 \nn \\&  
 = \frac{\img}{2Q}\frac{\partial}{\partial b_\perp^\rho}
   \left[ \gamma_\perp^\rho\, \slashed n\,,\,\hat\cG_{h/f}(x,\bt) \right]^{\alpha'\alpha}
\,.\end{align}
Hence we find that \eq{def_GP_f_b} also remains valid including transverse Wilson lines.  

In conclusion,  all the results presented in the main text remain valid with the inclusion of transverse Wilson lines as presented by the explicit formulas given here.

\section{Relation to Correlators in the Literature}
\label{app:notation_relation}

Our results heavily make use of the decompositions of quark-quark and quark-gluon-quark correlators
in the literature, which is required to make a connection to our results for the structure functions.
In our work, we use different conventions for the lightcone notation, and different definitions
of our correlators. In this appendix, we show how to relate our correlators to those in the literature,
and use this to obtain decompositions of our correlators since the general Lorentz structure of these correlators is identical to the ones that have been considered in the past. All our reference results for these decompositions are taken
from \refcite{Bacchetta:2006tn}, up to using certain combinations of the functions therein,
which we will point out explicitly.

\subsection{Lightcone Conventions}
\label{app:lightcone_relation}

Our conventions are based on two reference vectors $n^\mu$ and $\bn^\mu$
with $n^2 = \bn^2 = 0$ and $n \cdot \bn = 2$, and we define the lightcone coordinates
as $p^-  = \bn \cdot p$ and $p^+ = n \cdot p$, see \sec{factorization_frame}.
\Refcite{Bacchetta:2006tn} uses reference vectors $n_\pm^\mu$ with $n_+^2 = n_-^2 = 0$
and $n_+ \cdot n_- = 1$, and defines lightcone components as $\tilde p^\pm = n_\mp \cdot p$.
To distinguish the two conventions, in this section we will always denote their lightcone
coordinates using a tilde. The two conventions are related by
\begin{align} \label{eq:their_lc}
 n_+^\mu = \frac{n^\mu}{\sqrt 2} \,,\quad n_-^\mu = \frac{\bn^\mu}{\sqrt 2}
\,,\quad
 \tilde p^\pm = \frac{p^\mp}{\sqrt 2}
\,.\end{align}
The first two relations are fixed by having the incoming momentum $P^\mu$ be aligned
along the $n$ and $n_+$ direction, respectively.

\subsection{Quark-Quark Correlator}
\label{app:qq_relation}

\paragraph{The quark-quark correlator in  \refcite{Bacchetta:2006tn}}
is defined in their Eq.~(3.10) as
\begin{align} \label{eq:def_Phi_ij_1}
 \Phi_{ij}(x, \pt) &
 = \int\frac{\df \tilde\xi^- \df^2\vec \xi_T}{(2\pi)^3} e^{\img x \tilde P^+ \tilde \xi^-} e^{- \img \pt \cdot \vec\xi_T}
   \braket{N | \bar\psi_j(0) \, \cU^{n_-}_{(0,\infty)} \,\cU^{n_-}_{(\infty,\xi)} \, \psi_i(\xi) | N}
 \nn\\&
 = \int\frac{\df \xi^+ \df^2\vec \xi_T}{\sqrt2 (2\pi)^3} e^{\frac{\img}{2} x P^- \xi^+} e^{- \img \pt \cdot \vec\xi_T}
   \braket{N | \bar\psi_j(0) \cU^{n_-}_{(0,\infty)} \,\cU^{n_-}_{(\infty,\xi)} \psi_i(\xi) | N}
\,,\end{align}
where $ij$ are spinor indices, and the quark flavor is kept implicit.
The Wilson lines $\cU$ will be converted to our $W_n$ Wilson lines below.
In the first line, we use their lightcone convention, while in the second line we convert to our notation.
Next, we use that the correlator only depends on the relative position $0 - \xi = -\xi$
to shift the positions of the fields. Defining $b \equiv -\xi$, this yields%
\footnote{In recent work by Collins~\cite{Collins:2011zzd},
the fields are evaluated as $\bar\psi(w/2) W(w/2, -w/2) \psi(-w/2)$,
which by the same argument is identical to our convention.}
\begin{align} \label{eq:def_Phi_ij_3}
 \Phi_{ij}(x, \pt) &
 = \int\frac{\df b^+ \df^2\bt}{\sqrt2 (2\pi)^3} e^{-\frac{\img}{2} x P^- b^+} e^{\img \pt \cdot \bt}
   \braket{N | \bar\psi_j(b) \cU^{n_-}_{(b,\infty)} \,\cU^{n_-}_{(\infty,0)} \psi_i(0) | N}
\,.\end{align}
This ensures that the Fourier phase for the transverse integration has the same sign as in our convention.
This yields the position-space correlator
\begin{align} \label{eq:def_Phi_ij_4}
 \Phi_{ij}(x, \bt) &
 = \int\df^2\pt \, e^{-\img \bt \cdot \pt} \Phi_{ij}(x, \pt)
 \nn\\&
 = \frac{1}{\sqrt2} \int\frac{\df b^+}{2\pi} e^{-\frac{\img}{2} x P^- b^+}
   \braket{N | \bar\psi_j(b) \cU^{n_-}_{(b,\infty)} \,\cU^{n_-}_{(\infty,0)} \psi_i(0) | N}
\,.\end{align}
It remains to rewrite their Wilson lines $\cU^{n_-}$ in terms of our $W_n$.
The relation is
\begin{alignat}{2} \label{eq:rel_Wn}
 \cU^{n_-}_{(\infty,0)} &
 = P \exp\left[-\img g \int_{\infty}^0 \df s \, \bn \cdot A(s \bn )\right]
 &&= W_n^\dagger(0)
\,,\nn\\
 \cU^{n_-}_{(b,\infty)} &
 = P \exp\left[-\img g \int_{0}^\infty \df s \, \bn \cdot A(s \bn + b)\right]
 &&= W_n(b)
\,,\end{alignat}
where we note that the $P$s written here would be $\bar P$s in our notation, and we neglected transverse Wilson lines for brevity.
The first equality is the definition of \refcite{Bacchetta:2006tn},
converted to our conventions, and in the second step we used \eq{Wn_FT}.
Inserting \eq{rel_Wn} into \eq{def_Phi_ij_4} yields
\begin{align} \label{eq:def_Phi_ij_5}
 \Phi_{ij}(x, \bt) &
 = \frac{1}{\sqrt2} \int\frac{\df b^+}{2\pi} e^{-\frac{\img}{2} x P^- b^+}
   \braket{N | \bar\psi_j(b) W_n(b) W_n^\dagger(0) \psi_i(0) | N}
\,,\end{align}
where we remind the reader that $b^\mu = (b^+, 0, b_\perp)$.

\paragraph{Our beam function} is defined in \eq{def_B_hat} as
\begin{align}
 \hat B^{ij}(x = \w_a/P_N^-,\bt)&
 = \braket{N | \bar\chi_{n}^{j}(b_\perp) \chi_{n,\w_a}^{i}(0) | N}
\,,\end{align}
where here we label the spinor indices as $ij$ instead of $\beta\beta'$.
For consistency with \eq{def_Phi_ij_1}, we also and suppress the flavor and hadron labels,
as well as the explicit $\theta(\w_a)$ factor present in \eq{def_B_hat}.
Inserting the definitions of the collinear quark fields from \eq{chiB},
\begin{align} \label{eq:app:B_1}
 \chi_{n,\w}(x) &
 = \Bigl[\delta(\w - \bnP_n)\, W_n^\dagger(x)\, \xi_n(x) \Bigr]
\,,\nn\\
 \bar \chi_{n,\w}(x) &
 = (\chi_{n,-\w})^\dagger \gamma^0
 = \Bigl[\bar\xi_n(x) W_n(x) \delta(\w + \bnP_n^\dagger)\Bigr]
\,,\end{align}
the beam correlator becomes
\begin{align} \label{eq:app:B_2}
 \hat B^{ij}(x,\bt) &
 = \Braket{N | \bar\xi_n^j(b_\perp) W_n(b_\perp) \bigl[\delta(\w_a - \bnP_n)\, W_n^\dagger(0)\, \xi_n^i(0) \bigr]| N}
\,.\end{align}
The combination in square brackets has a fixed longitudinal momentum $\w_a$,
which can equivalently be expressed through the Fourier-transformed field
\begin{align} \label{eq:field_msp_bsp}
 \bigl[\delta(\w_a - \bnP_n)\, W_n^\dagger(0)\, \xi_n^i(0) \bigr] &
 = \int\frac{\df b^+}{4\pi} e^{\frac{\img}{2} \w_a (-b^+)} W_n^\dagger(-b^+)\, \xi_n^i(-b^+)
\,.\end{align}
This puts the $\xi_n$ field in \eq{app:B_2} to position $-b^+$
which we again shift into the first $\bar\xi_n$ field,
\begin{align} \label{eq:app:B_3}
 \hat B^{ij}(x,\bt) &
 = \int\frac{\df b^+}{4\pi} e^{-\frac{\img}{2} \w_a b^+}
   \Braket{N | \bar\xi_n^j(b) W_n(b) W_n^\dagger(0)\, \xi_n^i(0) | N}
\,,\end{align}
where now $b^\mu = (b^+, 0, b_\perp)$.
The $n$-collinear fields $\xi_n$ only consistent of the ``good components'' of the quark field,
which are projected out as
\begin{align} \label{eq:good_components}
 \xi_n^i = \frac{(\slashed{n} \slashed{\bn})^{ii'}}{4} \psi_{i'}
\,,\qquad
 \bar\xi_n^j = \bar\psi_{j'} \frac{(\slashed{\bn} \slashed{n})^{j'j}}{4}
\,,\end{align}
where we make the spinor indices explicit. Inserting this into \eq{app:B_3} yields
\begin{align} \label{eq:app:B_4}
 \hat B^{ij}(x,\bt) &
 = \frac{(\slashed{n} \slashed{\bn})^{ii'}}{4}
   \frac{(\slashed{\bn} \slashed{n})^{jj'}}{4}
   \int\frac{\df b^+}{4\pi} e^{-\frac{\img}{2} \w_a b^+}
   \Braket{N | \bar\psi_{j'}(b) W_n(b) W_n^\dagger(0)\, \psi_{i'}(0) | N}
\,.\end{align}
Comparing \eqs{def_Phi_ij_4}{app:B_4}, we finally obtain the desired relation
\begin{align} \label{eq:relation_B_phi}
 \hat B^{ij}(x,\bt) &
 = \frac{(\slashed{n} \slashed{\bn})^{ii'}}{4}
   \, \frac{\Phi_{i'j'}(x, \bt)}{\sqrt2} \,
   \frac{(\slashed{\bn} \slashed{n})^{j'j}}{4}
 \nn\\&
 = \frac{1}{4 \sqrt2}
   \bigl[\slashed{n}_+ \slashed{n}_- \Phi(x, \bt) \slashed{n}_- \slashed{n}_+ \bigr]^{ij}
\,.\end{align}
In the second line, we have made use of \eq{their_lc}.
Since the projection operators and constant factors commute
with taking the Fourier transform, one obtains the same
result for the Fourier-transformed correlators.

\paragraph{Decomposition.}
\Eq{relation_B_phi} allows us to easily obtain the spin-dependendent
decomposition of $\hat B^{ij}$ from that of $\Phi^{ij}$ given in
Eq.~(3.15) of \refcite{Bacchetta:2006tn},
\begin{align}
 \Phi(x, \kt) &
 = \biggl( f_1
   - f_{1T}^\perp \frac{\eps_\perp^{\rho\sigma} k_{\perp\rho} S_{\perp\sigma}}{M}
   + g_{1s} \gamma_5
   + h_{1T} \gamma_5\slashed{S}_\perp
   + h_{1s}^\perp \gamma^5 \frac{\slashed{k}_\perp}{M}
   + \img h_1^\perp \frac{\slashed{k}_\perp}{M}  \biggr) \frac{\slashed{n}_+}{2}
 \nn\\&\quad
 + (\mathrm{higher~twist})
\,,\end{align}
where all functions on the right-hand side are evaluated as $f_i \equiv f_i(x, k_T)$.
Inserting this into \eq{relation_B_phi} only requires to use
\begin{align}
 \frac{1}{4 \sqrt2} (\slashed{n}_+ \slashed{n}_-) \frac{\slashed{n}_+}{2} (\slashed{n}_- \slashed{n}_+)
  = \frac{\slashed{n}_+}{2 \sqrt2} = \frac{\slashed{n}}{4}
\,.\end{align}
We arrive at the desired result
\begin{align} \label{eq:relation_B_phi_decomposed}
 \hat B(x,\kt) &
 = \biggl( f_1
   - f_{1T}^\perp \frac{\eps_\perp^{\rho\sigma} k_{\perp\rho} S_{\perp\sigma}}{M}
   + g_{1s} \gamma_5
   + h_{1T} \gamma_5\slashed{S}_\perp
   + h_{1s}^\perp \frac{\gamma^5 \slashed{k}_\perp}{M}
   + \img h_1^\perp \frac{\slashed{k}_\perp}{M} \biggr) \frac{\slashed{n}}{4}
\,.\end{align}
Note that all higher-twist terms vanish because $\slashed{n}_\pm^2 = 0$,
and only the leading twist contribution has the $\slashed{n}_+$ structure
preventing one from commuting a $\slashed{n}_-$ past $\Phi$ in \eq{relation_B_phi}.
Since $h_{1s}^\perp$ is multiplied by a term proportional to $\slashed{k}_\perp$
in \eq{relation_B_phi_decomposed}, it contributes a linear and quadratic Lorentz structure.
It is useful to split off a traceless piece, as the tracelessness is preservered
by the Fourier transform we ultimately need to take. This leads to
\begin{align} \label{eq:relation_B_phi_decomposed_2}
 B_{f/N}(x, \kt) &
 = \biggl[ f_{1}
      - f_{1 T}^{\perp} \frac{\eps_\perp^{\rho \sigma} k_{\perp \rho} S_{\perp \sigma}}{M_N}
      + g_{1L} \, S_L \gamma_5
      - g_{1T} \frac{k_\perp \cdot S_\perp}{M_N} \gamma_5
      + h_1 \, \gamma_5 \slashed{S}_\perp
   \nn\\&\qquad
      + h_{1L}^\perp \, S_L\frac{ \gamma_5\,\slashed k_{\perp}}{M_N}
      - h_{1T}^\perp \,\frac{k_\perp^2}{M_N^2}  \Bigl( \frac12 g_\perp^{\mu\nu} - \frac{k_\perp^\mu  k_\perp^\nu}{k_\perp^2}  \Bigr) S_{\perp\,\mu} \gamma_\nu \gamma_5
      +\img h_{1}^{\perp} \frac{\slashed k_{\perp}}{M_N} \biggr] \frac{\slashed{n}}{4}
\,.\end{align}
Here, we defined the correlator~\cite{Bacchetta:2006tn}
\begin{align} \label{eq:h_1}
 h_1(x, k_T) = h_{1 T}(x,k_T) - \frac{k_\perp^2}{2M_N^2} h_{1T}^\perp(x, k_T)
\,.\end{align}
We have also replaced the shorthand notations $g_{1s}$ and $h_{1s}$ by \cite{Bacchetta:2006tn}
\begin{align} \label{eq:g_1s}
 g_{1s}(x,\kt) &= S_L\,g_{1L}(x, k_T) - \frac{k_\perp \cdot S_\perp}{M_N}\,g_{1T}(x, k_T)
\,,\nn\\
 h_{1s}^\perp(x,\kt) &= S_L\,h_{1L}^\perp(x, k_T) - \frac{k_\perp \cdot S_\perp}{M_N}\,h_{1T}^\perp(x, k_T)
\,.\end{align}

\subsection{Quark-Gluon-Quark Correlator}
\label{app:qgq_relation}

Here, we repeat the steps shown in \app{qq_relation} for the quark-gluon-quark correlators.
Most intermediate steps have been discussed in full detail in there,
and here we will limit ourselves to the key steps.

\subsubsection{TMDPDF}

We begin with the quark-gluon-quark correlator for the incoming hadron,
i.e.~corresponding to the TMDPDF. In \refcite{Bacchetta:2006tn}, it is defined as
\begin{align} \label{eq:qgq_Bacchetta}
 (\tilde\Phi_A^\rho)^{\beta\beta'}(x, \bt) &
 = \int\frac{\df b^+}{\sqrt2 (2\pi)} e^{-\frac{\img}{2} x \Pa^- b^+}
   \braket{h(P) | \bar\psi^{\beta'}(b) g A_T^\rho(0) \psi^\beta(0) | h(P)}
\,,\end{align}
where as usual $b = (b^+, 0, b_\perp)$.
Here, we have already converted to our conventions following the same steps as in \app{qq_relation}.
Note that $A_T^\rho$ is the transverse part of the gluon field, and hence $\rho$ is a transverse index.
For simplicity, we use light-cone gauge with $A^+ = 0$ but neglect transverse Wilson lines.
Note that \refcite{Bacchetta:2006tn} also defines a correlator $(\Phi_A^\rho)^{\beta\beta'}$
similar to \eq{qgq_Bacchetta} but with a covariant derivative $\img D^\mu$ instead of the gluon field.
Since its large component can be reduced to the leading-power correlator,
it is subtracted out to yield the purely transverse correlator in \eq{qgq_Bacchetta}.
In our approach, this is built in by using a collinear gluon field from the beginning.

We contrast this with our correlator as defined in \eq{def_Bb_hat},
\begin{align} \label{eq:def_Bb_hat_rep}
 \hat B_{\cB}^{\rho \beta \beta'}(x, \xi, \bt) &
 = \MAe{N}{ \bar\chi_{\na}^{\beta'}(b_\perp) \, g {\cB}_{\na\perp,\,-\xi\w_a}^{\rho}(0)
            \,\chi_{\na,\,(1-\xi)\w_a}^{\beta}(0)}{N}
 \\\nn&
 = \int\frac{\df b_1^+}{4\pi} \int\frac{\df b_2^+}{4\pi}
   e^{\frac{\img}{2} [ \xi b_1^+ +  (1-\xi) b_2^+] x \Pa^- }
   \MAe{N}{ \bar\chi_{\na}^{\beta'}(b_\perp) \, g {\cB}_{\na\perp}^{\rho}(-b_1^+)
            \,\chi_{\na}^{\beta}(-b_2^+)}{N}
\,.\end{align}
Here, we suppressed the overall $\theta(\w_a)$ for simplicity,
and in the second step we replaced the fields with fixed momentum fraction
by their Fourier transform as in \eq{field_msp_bsp}.
Note that only the $\bar\chi_n$ field has a transverse position,
while the other fields only have a longitudinal displacement.
In particular, the fields have distinct lightcone positions,
reflecting that one is sensitive to their individual momenta.

To relate \eq{def_Bb_hat_rep} to \eq{qgq_Bacchetta},
we integrate over $\xi$ which fixes $b_1^+ = b_2^+ \equiv -b^+$.
Shifting this position into the antiquark field, we obtain
\begin{align} \label{eq:Bb_hat_integrated}
 \int\!\df\xi\, \hat B_{\cB}^{\rho \beta \beta'}(x, \xi, \bt)
 = \frac{1}{x \Pa^-} \int\frac{\df b^+}{4\pi} e^{-\frac{\img}{2} b_2^+ x \Pa^- }
   \MAe{N}{ \bar\chi_{\na}^{\beta'}(b) \,g  {\cB}_{\na\perp}^{\rho}(0) \chi_{\na}^{\beta}(0)}{N}
\,,\end{align}
where now $b^\mu = (b^+, 0, b_\perp)$.
By replacing the good components $\chi_n$ of the collinear quarks with the projection
onto the full collinear fields as in \eq{good_components},  we arrive at
\begin{align} \label{eq:Bb_hat_integrated_2}
 \int\!\df\xi\, \hat B_{\cB}^{\rho \beta \beta'}(x, \xi, \bt)
 = \frac{(\slashed{n} \slashed{\bn})^{\beta\alpha}}{4}
   \frac{(\tilde\Phi_A^\rho)^{\alpha\alpha'}(x, \bt)}{\sqrt2 x \Pa^-}
   \frac{(\slashed{\bn} \slashed{n})^{\alpha'\beta'}}{4}
\,.\end{align}
We remark that here, we have neglected some Wilson lines, which were dropped in writing \eq{qgq_Bacchetta}.
In \eq{Bb_hat_integrated}, collinear quark and gluon fields contain Wilson lines
to yield gauge-invariant building blocks. Likewise, for a general gauge \refcite{Bacchetta:2006tn}
provides a gauge-invariant definition of \eq{qgq_Bacchetta}
that dresses the fields in there with Wilson lines.
Since anyway our Wilson line structure will differ, due to the appearance of the extra momentum fraction $\xi$, we do not bother to demonstrate explicitly that upon integration over $\xi$ the Wilson line structures match between the approaches. 
Both procedures should anyway be unique, up to choosing the Wilson line directions as outgoing or incoming 
as appropriate for SIDIS or Drell-Yan kinematics.

Finally, we remark that the projectors in \eq{Bb_hat_integrated_2} have no practical
effect, as the correlators $\tilde\Phi_A$ correlators in \refcite{Bacchetta:2006tn}
are parameterized in momentum space without writing out subleading power terms, with the form
\begin{align}
 \tilde \Phi_A^\rho(x, \kt) &
 = \frac{x \Ma}{2} \bigl\{ \cdots \bigr\} \frac{\slashed{n}_+}{2}
\,,\end{align}
where the ellipses contain the individual TMDs and will be presented in \eq{BB_decomposition}.
Inserting this into \eq{Bb_hat_integrated_2} and using that $n_+ = n/\sqrt2$, we obtain
\begin{align} \label{eq:Bb_hat_integrated_3}
 \int\!\df\xi\, \hat B_{\cB}^{\rho \beta \beta'}(x, \xi, \bt) &
 = \frac{\Ma}{4 \Pa^-} \bigl\{ \cdots \bigr\} \frac{\slashed{n}}{2}
\,.\end{align}

\subsubsection{TMDFF}

Next, we consider the quark-gluon-quark correlator for the fragmenting hadron,
i.e.~corresponding to the TMDPDF. In \refcite{Bacchetta:2006tn}, it is defined as%
\footnote{Note that the color average is not made explicit in \refcite{Bacchetta:2006tn}.
Following e.g.~\refcite{Collins:2011zzd}, we have reinstated it here.}
\begin{align}
 (\tilde \Delta_A^\rho)^{\alpha\alpha'}(z, \bt) &
 = \frac{1}{2z N_c} \int\frac{\df b^-}{\sqrt2 (2\pi)} e^{\frac{\img}{2} b^- (P_h^+/z)}
   \SumInt_X \braket{0 | g A_T^\rho(b) \psi_\alpha(b) | h X} \braket{h X | \bar\psi_{\alpha'}(0) | 0}
\,,\end{align}
where $b = (0, b^-, b_\perp)$, and as before we use lightcone gauge for simplicity.
We will also require the following conjugate operator:
\begin{align} \label{eq:DeltaA}
 \bigl[\gamma_0 \tilde \Delta_A^{\dagger\rho}(z, \bt) \gamma_0 \bigr]^{\alpha\alpha'} &
 = \frac{1}{2z N_c} \int\frac{\df b^-}{\sqrt2(2\pi)^2} e^{-\frac{\img}{2} b^- (P_h^+/z)}
   \nn\\&\quad\times
   \SumInt_X \braket{0 | \psi_{\alpha'}(0) | h X} \braket{h X | g A_T^\rho(b) \bar \psi_\alpha(b)  | 0}
\,.\end{align}
Our corresponding correlator is defined in \eq{BBfbar_hat} as
\begin{align} \label{eq:BBfbar_hat_rep}
 \hat \cG_{\cB \, h/f}^{\rho\,\alpha\alpha'}(z,\xi,\bt) &
 = \frac{1}{2z N_c}
   \SumInt_{X_\bn} \tr \MAe{0}{\chi_{\nb}^{\alpha}(b_\perp)}{h,X_\bn}
   \MAe{h,X_\bn}{\bar\chi_{\nb,\,-(1-\xi)\w_b}^{\alpha'}(0)\,g {\cB}_{\nb\perp,\,\xi\w_b}^{\rho}(0)}{0}
 \nn\\&
 = \frac{1}{2z N_c}
   \int\frac{\df b_1^-}{4\pi} \int\frac{\df b_2^-}{4\pi} e^{\frac{\img}{2} [(1-\xi) b_1^- + \xi b_2^-] (P_h^+/z)}
   \nn\\&\qquad\times
   \SumInt_{X_\bn} \tr \MAe{0}{\chi_{\nb}^{\alpha}(b_\perp)}{h,X_\bn}
   \MAe{h,X_\bn}{\bar\chi_{\nb}^{\alpha'}(b_1^-)\,g {\cB}_{\nb\perp}^{\rho}(b_2^-)}{0}
\,.\end{align}
Here, we suppress all $\theta$ functions for brevity, and in the second step
replaced the fields of fixed label momentum by their Fourier representation as in \eq{field_msp_bsp}.

To relate \eq{BBfbar_hat_rep} to \eq{DeltaA},
we integrate over $\xi$ which fixes $b_1^- = b_2^- \equiv -b^-$,
and we shift the field positions by $b^-$. This yields
\begin{align} \label{eq:BBfbar_hat_integrated}
 \int\df\xi\,\hat \cG_{\cB \, h/f}^{\rho\,\alpha\alpha'}(z,\xi,\bt) &
 = \frac{1}{2z N_c}
   \frac{1}{P_h^+/z} \int\frac{\df b^-}{4\pi} e^{-\frac{\img}{2} b^- (P_h^+/z)}
   \nn\\&\qquad\times
   \SumInt_{X_\bn} \tr \MAe{0}{\chi_{\nb}^{\alpha}(b)}{h,X_\bn}
   \MAe{h,X_\bn}{\bar\chi_{\nb}^{\alpha'}(0)\,g {\cB}_{\nb\perp}^{\rho}(0)}{0}
\,,\end{align}
where now $b = (0, b^-, b_\perp)$.
Comparing this to \eq{DeltaA} by using the quark projection in \eq{good_components}, we find that
\begin{align} \label{eq:BBfbar_hat_integrated_2}
 \int\df\xi\,\hat \cG_{\cB \, h/f}^{\rho\,\alpha\alpha'}(z,\xi,\bt) &
 = \frac{(\slashed{\bn}\slashed{n})^{\alpha\beta}}{4}
   \frac{\bigl[\gamma_0 \tilde \Delta_A^{\dagger \rho}(z, \bt) \gamma_0 \bigr]^{\beta\beta'}}{\sqrt2 P_h^+/z}
   \frac{(\slashed{n}\slashed{\bn})^{\beta'\alpha'}}{4}
\,.\end{align}
This result is consistent with \eq{Bb_hat_integrated_2} for the $B_{\cB}$ operator.
As in that case, here we do not explicitly verify the Wilson structures
in \eq{BBfbar_hat_integrated} and the correlator in \refcite{Bacchetta:2006tn}
for a generic gauge.

Finally, we remark that the projectors in \eq{BBfbar_hat_integrated_2} have no practical
effect, as the correlators $\tilde \Delta_A$ correlators in \refcite{Bacchetta:2006tn}
are parameterized in momentum space in a form that does not contain power suppressed terms, 
\begin{align}
 \tilde \Delta_A^\rho(z, \kt) = \frac{\Mb}{2z} \bigl\{ \cdots \bigr\} \frac{\slashed{n}_-}{2}
\,,\end{align}
where the ellipses contain the individual TMDs and will be presented in \eq{deltaA}.
Inserting this into \eq{BBfbar_hat_integrated_2} and using that $n_- = \bn/\sqrt2$, we obtain
\begin{align} \label{eq:BBfbar_hat_integrated_3}
 \int\df\xi\,\hat \cG_{\cB \, h/f}^{\rho\,\alpha\alpha'}(z,\xi,\bt) &
 = \frac{\Mb}{4 \Pb^+} \bigl\{ \cdots \bigr\}^\dagger \frac{\slashed{\bn}}{2}
\,.\end{align}
Here, care has to be taken because $\bigl\{ \cdots \bigr\}$ contains imaginary parts
and needs to be conjugated, as indicated.


\FloatBarrier
\addcontentsline{toc}{section}{References}
\bibliographystyle{jhep}
\bibliography{refs}

\end{document}